\documentclass{aa}  
\usepackage{lscape}
\usepackage{graphicx}
\usepackage{txfonts}
\usepackage{lipsum}
\usepackage{subcaption}         
\usepackage{lscape}             
\usepackage{placeins}           

\usepackage{graphicx}
\usepackage{txfonts}
\defcitealias{Poggianti2017a}{P17}
\defcitealias{Poggianti2016}{P16}
\newcommand{\sinopsis}{{\sc sinopsis}\xspace}

\begin{document}


     \title{The MUSE view of ram pressure stripped galaxies in clusters: the GASP sample}

\titlerunning{Ram pressure-stripped galaxies: the GASP sample}
 
   \author{Bianca M. Poggianti
          \inst{1}
          \and
          Benedetta Vulcani
          \inst{1}
          \and  
          Neven Tomicic
          \inst{2}
          \and
          Alessia Moretti
          \inst{1}
          \and
          Marco Gullieuszik
          \inst{1} 
          \and
          Cecilia Bacchini
          \inst{3}
          \and
          Jacopo Fritz
          \inst{4}
          \and
          Koshy George
          \inst{5}
          \and
          Myriam Gitti
          \inst{6,7}
          \and
          Alessandro Ignesti
          \inst{1}
          \and
          Augusto Lassen
          \inst{1}
          \and
          Antonino Marasco
          \inst{1}
          \and
          Mario Radovich
          \inst{1}
          \and
          Paolo Serra
          \inst{8}
          \and
          Rory Smith
          \inst{9}
          \and
          Stephanie Tonnesen
          \inst{10}
          \and
          Anna Wolter
          \inst{11}
          }

   \institute{INAF- Osservatorio astronomico di Padova, Vicolo Osservatorio 5, I-35122 Padova, Italy\\
              \email{bianca.poggianti@inaf.it}
         \and
Department of Physics, Faculty of Science, University of Zagreb, Bijenicka 32, 10 000 Zagreb, Croatia
        \and
        DARK, Niels Bohr Institute, University of Copenhagen, Jagtvej 155, 2200 Copenhagen, Denmark
        \and
        Instituto de Radioastronomia y Astrofisica, UNAM, Campus Morelia, AP 3-72, CP 58089, Mexico
        \and
        Faculty of Physics, Ludwig-Maximilians-Universität, Scheinerstr. 1, Munich, 81679, Germany
        \and
        INAF, Istituto di Radioastronomia di Bologna, via Piero Gobetti 101, 40129 Bologna, Italy
        \and
         Dipartimento di Fisica e Astronomia, Università di Bologna, via Piero Gobetti 93/2, 40129 Bologna, Italy
        \and
        INAF - Osservatorio Astronomico di Cagliari, Via della Scienza 5, I-09047 Selargius (CA), Italy
        \and
        Departamento de Física, Universidad Técnica Federico Santa María, Vicuña Mackenna 3939, San Joaquín, Santiago de Chile
        \and
        Flatiron Institute, CCA, 162 5th Avenue, New York, NY 10010, USA
        \and
        INAF—Osservatorio Astronomico di Brera, via Brera, 28, 20121, Milano, Italy
            \\
             }
 \authorrunning{Poggianti et al. }

   \date{Received September 15, 1996; accepted March 16, 1997}

  \abstract
{We present the full sample of 76 galaxies in 39 galaxy cluster fields at z=0.04-0.07 observed with VLT/MUSE by the GASP survey. Most of them (64) were observed as possible ram pressure stripped galaxies (stripping candidates) based on optical B-band images, while the remaining 12 were a control sample of both star-forming and passive galaxies.
Based on spatially resolved ionized gas and stellar kinematics, we assess the physical origin of the gas asymmetries and find that 89\% of the stripping candidates are confirmed by the VLT/MUSE data. In addition, also 3 of the 4 star-forming galaxies in the control sample show signs of ram pressure. These control galaxies display a ring of unusual emission line ratios, which we see also in field galaxies, possibly originating from the interaction with a hotter surrounding medium.
The stripped galaxies are classified into various classes corresponding to different degrees of stripping, from weakest stripping to strong and extreme (jellyfish galaxies) stripping, as well as truncated gas disks with gas left only in the galaxy center. 
Our results show that selecting cluster stripping candidates based on optical imaging yields a sample that is indeed largely dominated by galaxies affected by ram pressure at different stages and stripping strength, though some contamination is present, mostly by tidal processes. Strong ram pressure cases are found in galaxies over the whole range of stellar masses studied ($10^9 - 10^{11.5} M_\odot$) both in low-mass and high-mass clusters (cluster velocity dispersions $\sigma = 500 - 1100 \, \rm km \, s^{-1}$). 
We examine the possible connection between the progressive stages of stripping, up to the phase of a truncated gas disk, and the subsequent complete stripping of gas.
We discuss the incompleteness intrinsic to this and other methods of selection to obtain a complete census of ram pressure stripping in clusters.}


   \maketitle
%

\section{Introduction} \label{sec:intro}

Studies of galaxies in clusters in the past almost five decades have accumulated overwhelming evidence that galaxies infalling into clusters are subject to the effects of the ram pressure exerted by the hot intracluster medium on the galaxy interstellar medium \citep{Gunn1972}. Both observations at different wavelengths and hydrodynamical simulations have established that gas is stripped by ram pressure when it exceeds the gravitational force of the galaxy, thereby depriving the galaxy of its fuel for star formation.

Whether ram pressure can be the driving force of differences between cluster and field galaxy population is a long-standing question, both concerning the morphologies of galaxies \citep{Dressler1980a} and their star formation histories \citep{Poggianti1999}. The answer to this question requires knowing how many cluster galaxies are subject to ram pressure during their lifetime (see e.g. \citealt{Vulcani2022}), at each cosmological epoch. It also requires an understanding of the consequences of ram pressure on those "transient" galaxy properties that strongly depend on environment, such as star formation and morphological appearance \citep{Bosch2013}, versus more fundamental galaxy properties, such as stellar mass, whose distribution at least above $\sim 10^{10} M_{\odot}$ seems invariant with halo mass \citep{Vulcani2013}.

Although a conclusive answer to this question has not been reached, several efforts have improved our understanding of this phenomenon. On the simulation side, wind-tunnel simulations of individual galaxies have clarified several aspects of the stripping process \citep{Schulz2001, TonnesenBryan2009, TonnesenBryan2012, TonnesenBryan2021, Roediger2006, Roediger2014} 
and recently addressed a number of questions posed by observations \citep{Akerman2023, Akerman2024, Zhu2024}. Cosmological (magneto)hydrodynamical simulations can now provide large samples of stripped galaxies as a function of halo mass and epoch \citep{Bahe2015, Pillepich2018, Goller2023, Kulier2023}, thus having enough statistics to investigate timescales and global populations \citep{Rohr2023, Zinger2024}.

{\sl Directly observing} large samples of ram-pressure stripped galaxies is clearly key. At low redshift, unilaterally stripped (thus extraplanar) material with morphologies suggestive of ram pressure has been directly observed in HI, radio continuum, $\rm H\alpha$, CO, X-rays, optical and ultraviolet imaging, and integral-field spectroscopy.  These works include detailed individual galaxy studies as well as searches for ram pressure candidates in entire clusters or several clusters (see also \S5). An incomplete list of recent surveys at  wavelengths other than optical include VESTIGE \citep[$\rm H\alpha$ in Virgo,][]{Boselli2018, Boselli2020, Boselli2021, Boselli2022, Boselli2023}, VERTICO \citep[CO in Virgo,][]{ Brown2021, Brown2023}, the MeerKAT Fornax survey \citep[HI,][]{Serra2023, Serra2024}, the LOFAR surveys of \cite{Roberts2021a,Roberts2021b, Roberts2022b, Roberts2024} and \cite{Ignesti2023a}, and the UV surveys of \cite{Smith2010} and \cite{George2024}. 

In this paper, we use a sample that has been originally selected from optical B-band imaging.
Optical imaging searches have been carried out in Coma \citep{Roberts2020}, in WINGS/OmegaWINGS imaging of 71 galaxy clusters at z=0.04-0.07 (\citealt{Poggianti2016}, hereafter \citetalias{Poggianti2016}, and \citealt{Vulcani2022}) and UNIONS imaging of some SDSS fields \citep{Roberts2022a}. A recent search for ram pressure stripped galaxies in Fornax, Antlia and Hydra in S-PLUS imaging \citep{Gondhalekar2024} adopts a semi-automated pipeline using self-supervised learning to find stripping candidates, exploring the feasibility for more automated searches of candidates in large imaging surveys.

At higher redshifts (z=0.2-0.9), HST imaging has provided increasingly larger samples of stripping candidates \citep{Cortese2007, Owers2012, McPartland2016, Ebeling2014, EbelingKalita2019, Roman2019, Roman2021, Durret2021, Durret2022} and the first integral-field spectroscopy studies of cluster galaxies at z=0.3-0.5 have confirmed the importance of ram pressure at intermediate redshifts \citep{KalitaEbeling2019,  Moretti2022, Werle2022, Lee2022a, Lee2022b, Bellhouse2022, Vulcani2024}, with the currently holding Integral-Field spectroscopy (IFS) record of two galaxies at z=0.7 \citep{Boselli2019}.

Optical/UV studies of the stellar light can provide candidates, but only adding observations probing the gas in one of its phases (ionized, neutral, or molecular) can confirm the ram pressure stripping (RPS) nature of these objects, by finding and characterizing extraplanar gas. Generally speaking, though, the mere presence of extraplanar gas (it being HI, $\rm H\alpha$ or other phases) is not sufficient per se to prove ram pressure, although the morphology of the gas with respect to the stars can usually come a long way in identifying the physical process at work and distinguish RPS and other processes, for example from tidal interactions or outflows from internal feedback.
The most solid smoking gun for ram pressure at work, however, is to contrast the gas kinematics with the stellar kinematics as can be done in integral field spectroscopy (IFS), or at least have the gas kinematics as in HI studies.

IFS observations of individual galaxies  \citep{Merluzzi2013, Merluzzi2016, Fumagalli2014, Fossati2016, Consolandi2017} have shown the power of spatially resolving all the main gaseous and stellar properties, providing a detailed picture of where in the galaxy and how stripping proceeds, and what its consequences on the star formation activity are both in the disk and in stripped tails.

GASP (GAs Stripping Phenomena in galaxies, \citealt{Poggianti2017a}, from now on \citetalias{Poggianti2017a}) is an ESO Large Program with the MUSE spectrograph to study the physical processes that can remove gas from galaxies and their consequences for the galaxy star formation activity and evolution. GASP has observed 114 galaxies at z=0.04-0.07, of which 76 are in galaxy cluster fields and the remaining 38 are in groups, filaments and isolated. Several GASP papers have presented a detailed analysis of individual galaxies, or of a specific aspect of a subset of the sample. The non-cluster subsample has been presented in \cite{Vulcani2021}. In this paper we present an overview of all 76 galaxies selected in 39 cluster fields, showing the MUSE data and the variety of stripping stages in our sample (\S3.1). The success rate in identifying ram pressure at work and the characteristics of the galaxies undergoing ram pressure are discussed in \S3.2, while \S3.3 is dedicated to those galaxies with a truncated ionized gas disk. We present the properties of the star-forming galaxies of the control sample in \S4, 
and discuss the pros and cons of selecting ram pressure candidates based on optical imaging and other wavelengths in \S5.

In the following we use a \cite{Chabrier2003} Initial Mass Function and a standard $\Lambda$ cold dark matter cosmology with $\Omega_M = 0.3$, $\Omega_{\Lambda} = 0.7$ and $H_0 = 70 \rm \, km \, s^{-1} \, Mpc^{-1}$.


\section{Galaxy sample and data}\label{sec:analysis}
The GASP sample used in this paper comprises 64 stripping candidates from the atlas of \citetalias{Poggianti2016} and 12 galaxies that were selected as a control sample. Here we consider only the galaxies that make up the GASP cluster sample (see \citetalias{Poggianti2017a}), while field galaxies have been characterized separately in \cite{Vulcani2017c, Vulcani2018a, Vulcani2018c, Vulcani2019a, Vulcani2021}.

The 64 galaxies were selected from the P16 sample that used images and spectra from the WINGS/OmegaWINGS survey \citep{Fasano2006, Cava2009, Varela2009, Moretti2014, Gullieuszik2015, Moretti2017} to identify galaxies with a disturbed morphology: a visual inspection of B-band images unveiled the presence of tentacles of material that appear to be stripped from the galaxy most likely due to gas-only removal mechanisms. In some cases tails were extremely long and spectacular, in many others the evidence of debris tails and stripped material was less pronounced.  The candidates were assigned by P16 to five classes according to the visual evidence for stripping signatures, from extreme cases (JClass=5) to progressively weaker cases, down to the weakest ones (JClass=1). The selection was based only on the images, therefore a subset of the candidates did not have a known spectroscopic redshift. Instead, galaxies with morphologies clearly disturbed due to mergers or tidal interactions were removed from the sample, still retaining and flagging the most doubtful cases where either gas stripping or tidal forces, or both, might be at work. In fact, the eventual presence of tidal forces does not exclude the possibility that also gas stripping mechanisms, such as ram pressure, are at work, as it is sometimes observed \citep[e.g.][]{Fritz2017, Serra2024,Watson2025}. More details on the sample can be found in \citetalias{Poggianti2016} and \citetalias{Poggianti2017a}.

Galaxies in the control sample belonging to clusters were selected from the same WINGS/OmegaWINGS sample. This sample of 12 objects was assembled with the goal of contrasting the properties of stripping candidates with those of galaxies that show no optical evidence of ongoing gas removal.  We selected 12 previously known spectroscopic cluster members, of which 4 were selected to be star-forming (having emission lines in their OmegaWINGS spectra) morphologically-spiral galaxies that did not show any clear sign of extraplanar debris in the B-band WINGS+OmegaWINGS optical images. The remaining 8 control galaxies were selected to be passive galaxies devoid of ongoing star formation, as testified by the lack of emission lines in their OmegaWINGS fibre spectra. Of these 8 (5 spirals and 3 S0s by choice), 6 were previously known post-starburst (k+a) galaxies \citep{Paccagnella2017} and two had neither emission nor strong Balmer lines in absorption (spectral k-type, \citealt{Dressler1999}).  GASP k+a galaxies have been the subject of a dedicated publication \citep{Vulcani2020a} and will not be discussed in detail hereafter.

For all 76 galaxies, T-type morphologies (Hubble types) were assessed from V-band images using {\sc morphot} \citep[see][for details]{Vulcani2011, Fasano2012, Vulcani2023}, an automatic tool purposely devised in the framework of the WINGS project. {\sc morphot} was designed with the aim to reproduce as closely as possible visual morphological classifications.

The 76 galaxies were observed with MUSE/VLT as part of the GASP program to probe the physics of gas and stars and establish the main physical mechanisms, if any, acting on them. Details on observations, data reduction, and data analysis can be found in \citetalias{Poggianti2017a}. Relevant for this paper are stellar masses,  H$\alpha$ fluxes and the gas and stellar kinematics. 
To obtain  emission-only datacubes on which we measured line fluxes, we subtracted the stellar-only component of each spectrum derived with our spectrophotometric code \sinopsis \citep{Fritz2014, Fritz2017}.  In addition, \sinopsis provided us with spatially resolved estimates of the following stellar population properties: stellar masses; luminosity-weighted age; average star formation rate and total mass formed in four age bins (= star formation histories, SFH): young (ongoing star formation) =  $t < 2 \times 10^7$ yr, recent $= 2 \times 10^7 < t< 5.7 \times 10^8$ yr, intermediate-age = $5.7 \times 10^8 <t< 5.7 \times 10^9$ yr, and old =$> 5.7 \times 10^9$ yr. The latter outputs of \sinopsis will be used only in Sec. \ref{sec:truncated}.

To derive emission line fluxes, velocities, and velocity dispersions with associated errors we used {\sc kubeviz} \citep{Fossati2016}. Before performing the fits, we averaged filter the data cube in the spatial direction with a 5$\times$5 kernel, corresponding to our worst seeing conditions of 1\arcsec  = 0.7-1.3 kpc at the redshifts of our galaxies. 
To extract the stellar kinematics from the spectrum, we used the Penalized Pixel-Fitting ({\sc ppxf}) code \citep{Cappellari2004}, fitting the observed spectra with the stellar population templates by \cite{Vazdekis2010}. 

In \cite{Gullieuszik2020} we determined the galaxy disk boundaries computed from the map of the stellar continuum in the H$\alpha$ region and from the isophote with a surface 
brightness 1$\sigma$ above the average sky background level. Because of the (stellar and gaseous) emission from the stripped gas tails, this isophote does not have an elliptical symmetry. To obtain a symmetric isophote, we fitted an ellipse to the undisturbed side of the isophote and replaced the isophote on the disturbed side with the ellipse. Everything inside of this isophote represents the galaxy disk, the rest constitutes the galaxy tail.  


\section{Results}

\subsection{Assessing the physical process at work and estimating the level of ram pressure}

We consider a galaxy to be undergoing RPS if there is extraplanar ionized gas (i.e. $\rm H\alpha$ emission) preferentially on one side of the disk while the disk stellar kinematics is undisturbed. A chaotic stellar kinematics is considered a signature of strong tidal effects or mergers \citep[e.g.,][]{Mihos1993, Struck1999}.

\begin{figure*}
\centerline{\includegraphics[scale=0.18]
{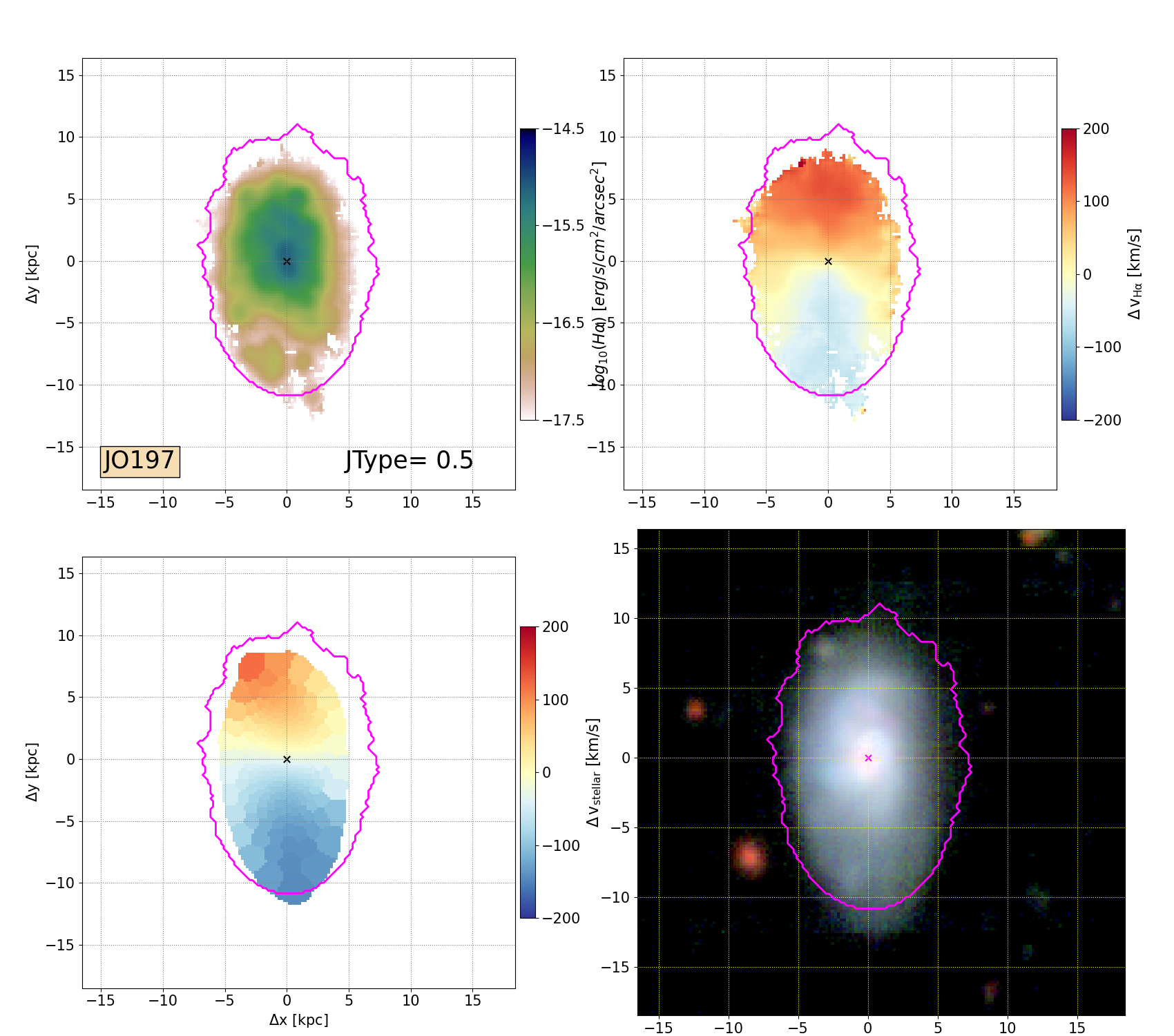}\includegraphics[scale=0.18]
{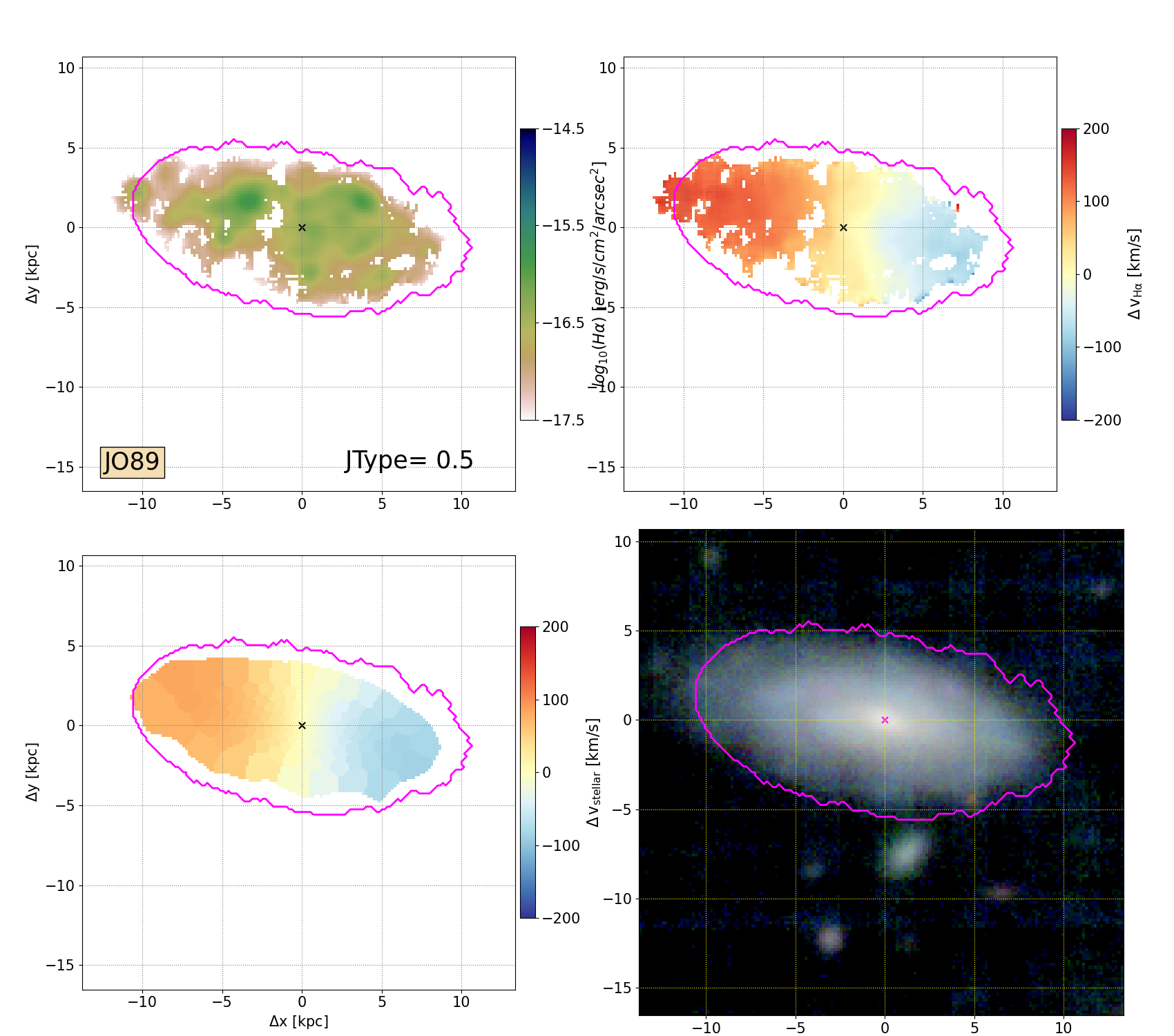}}
\centerline{\includegraphics[scale=0.25]
{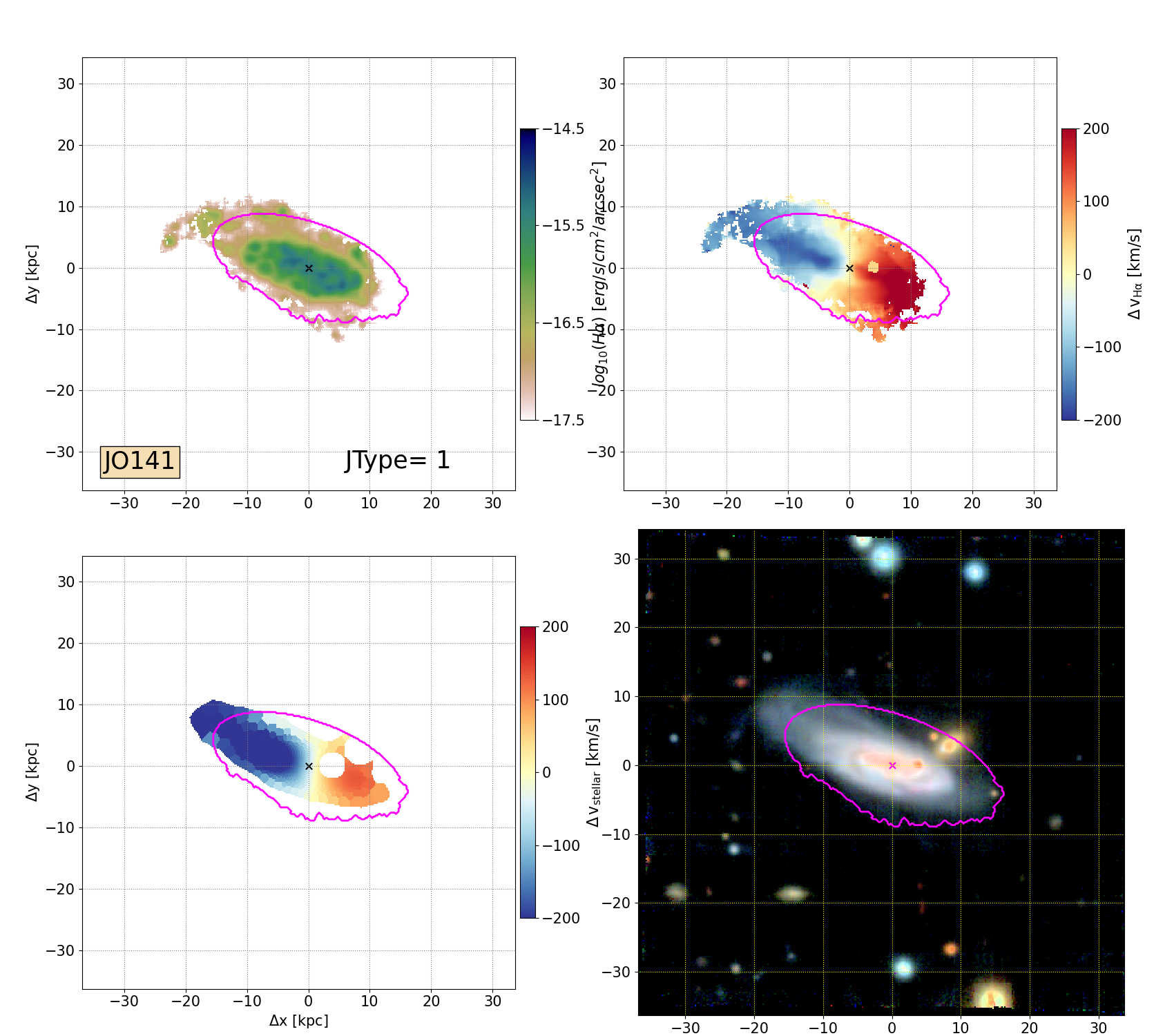}\includegraphics[scale=0.25]{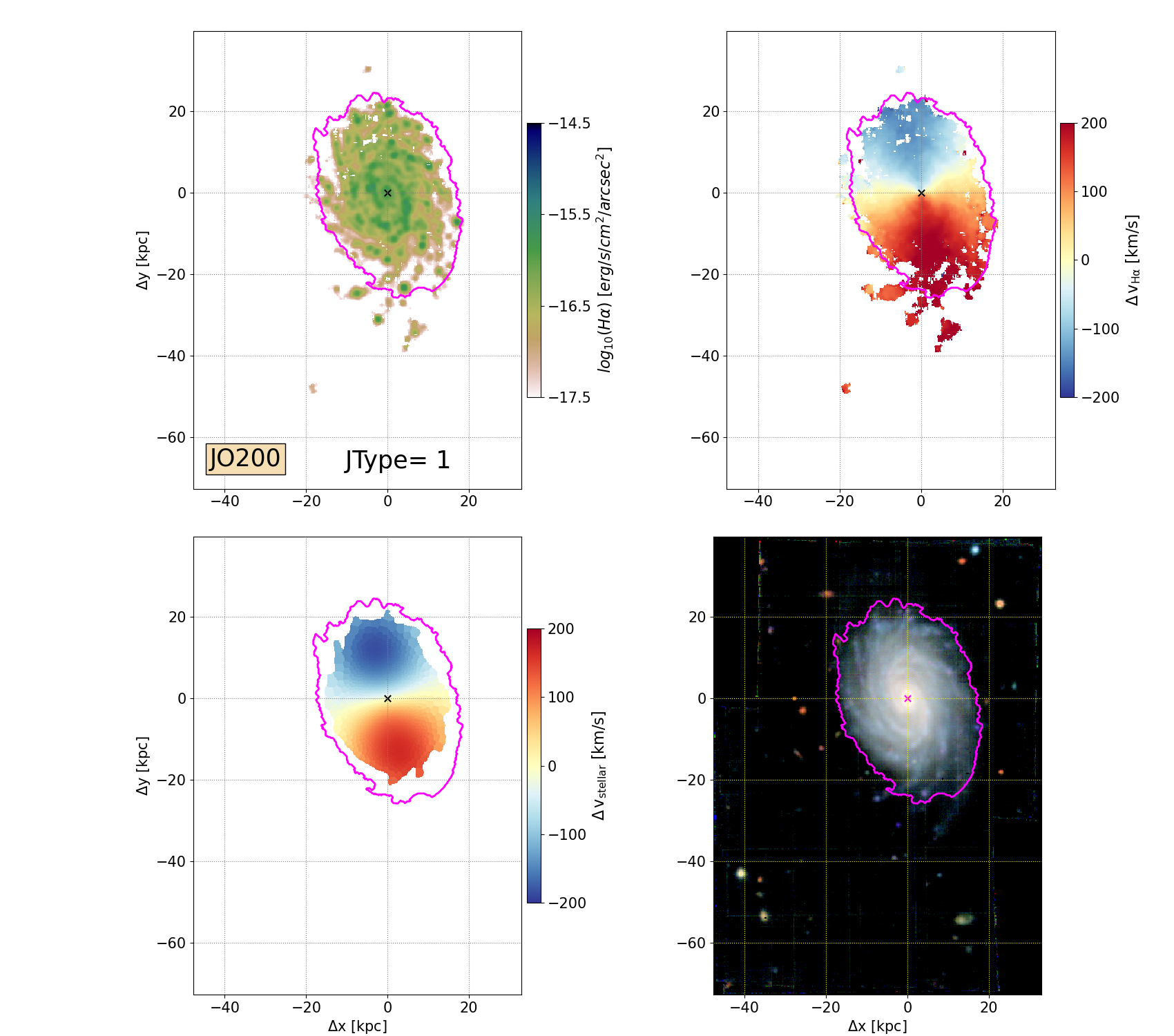}}
\caption{Illustrative examples of stripped galaxies of different JTypes: 
Top:  JType=0.5, mild stripping (JO197 and JO89);  Bottom JType=1, strong stripping (JO141 and JO200). For each galaxy, the H$\alpha$ flux (top left), the gas kinematics (top right), the stellar kinematics (bottom left) and the color composite image (bottom right) are shown. The pink line delimits the stellar disk as described in \S2.
\label{fig:Examples1}}
\end{figure*}

\begin{figure*}
\centerline{\includegraphics[scale=0.25]
{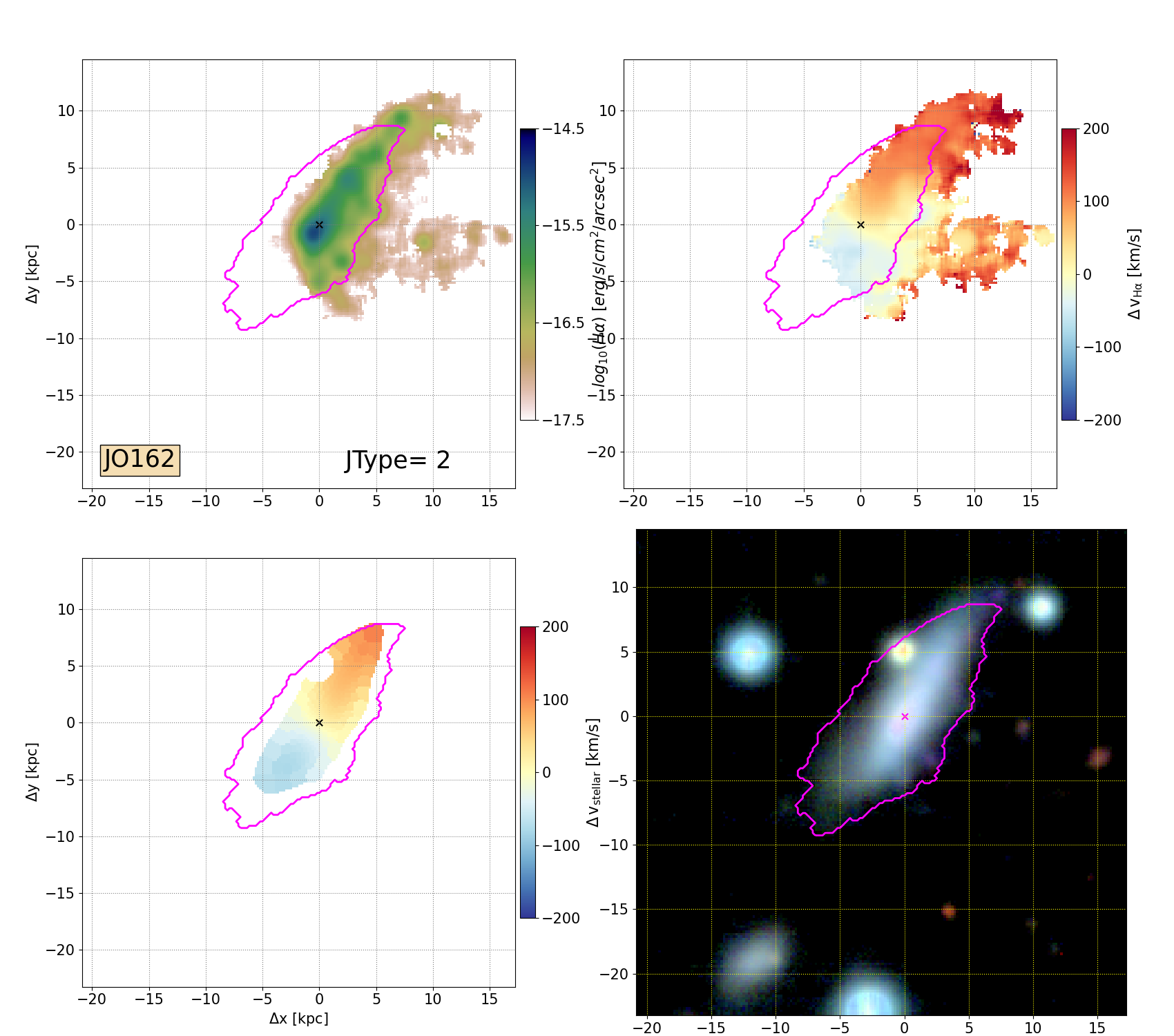}\includegraphics[scale=0.25]{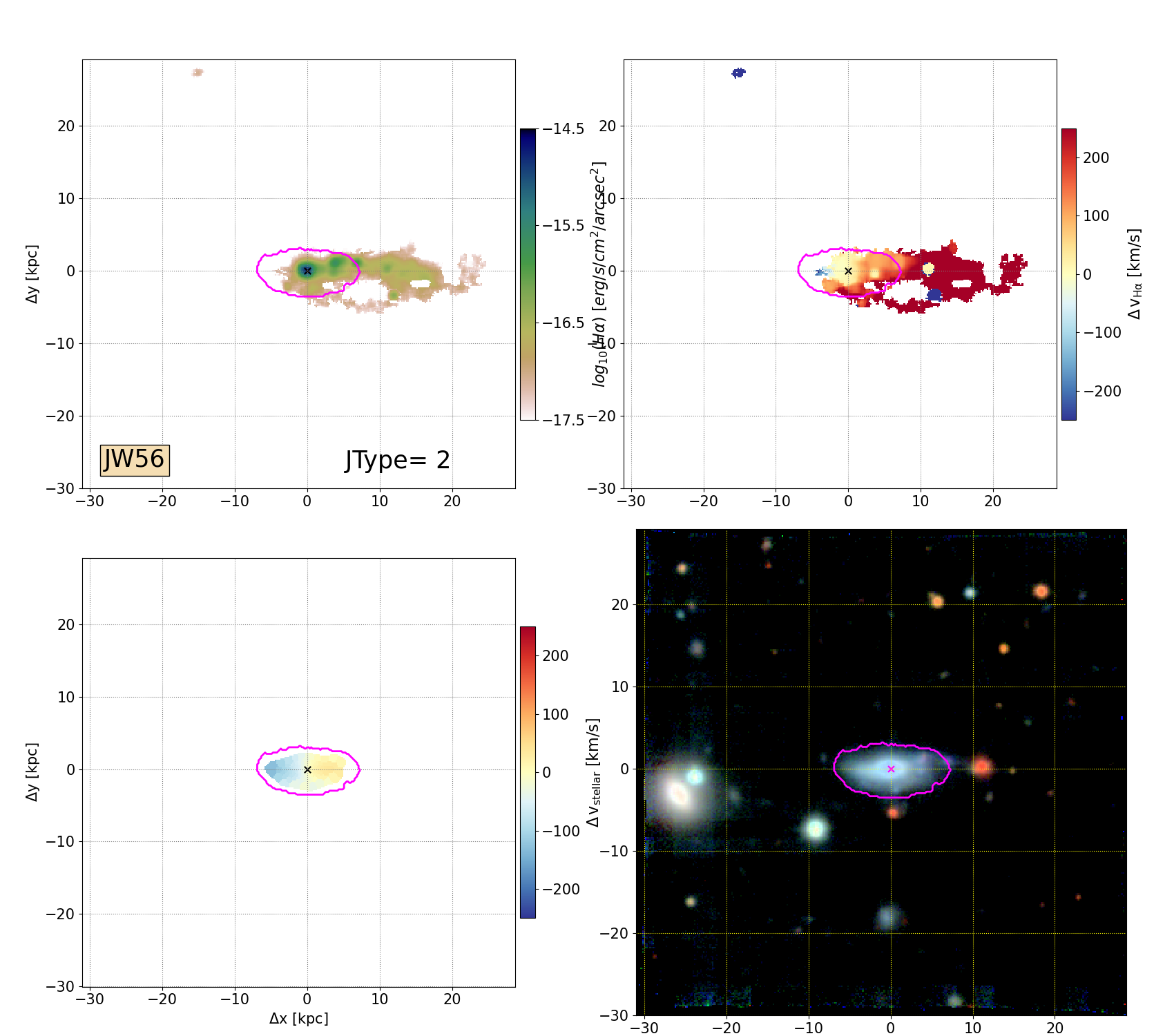}}
\centerline{\includegraphics[scale=0.25]
{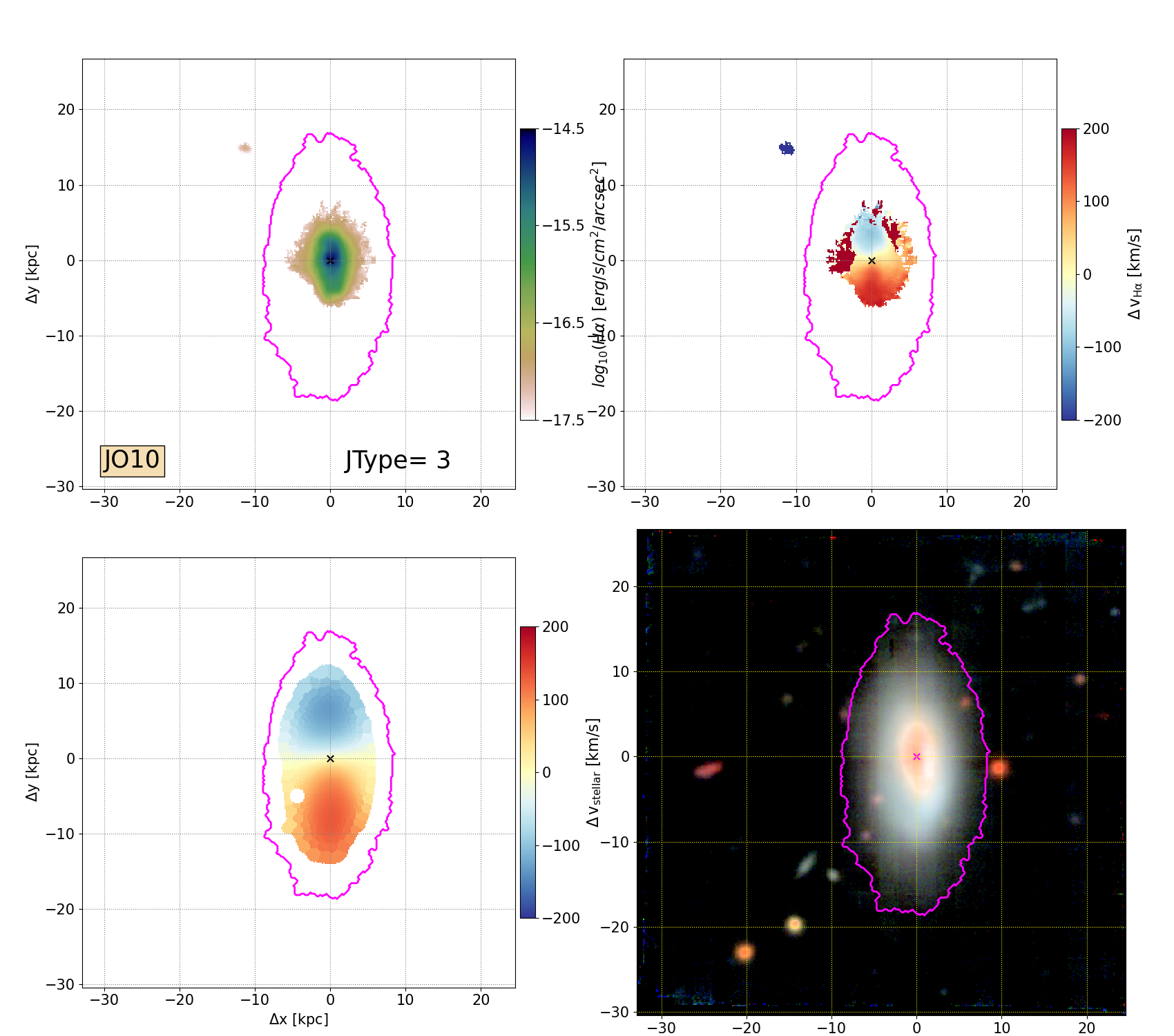}\includegraphics[scale=0.25]{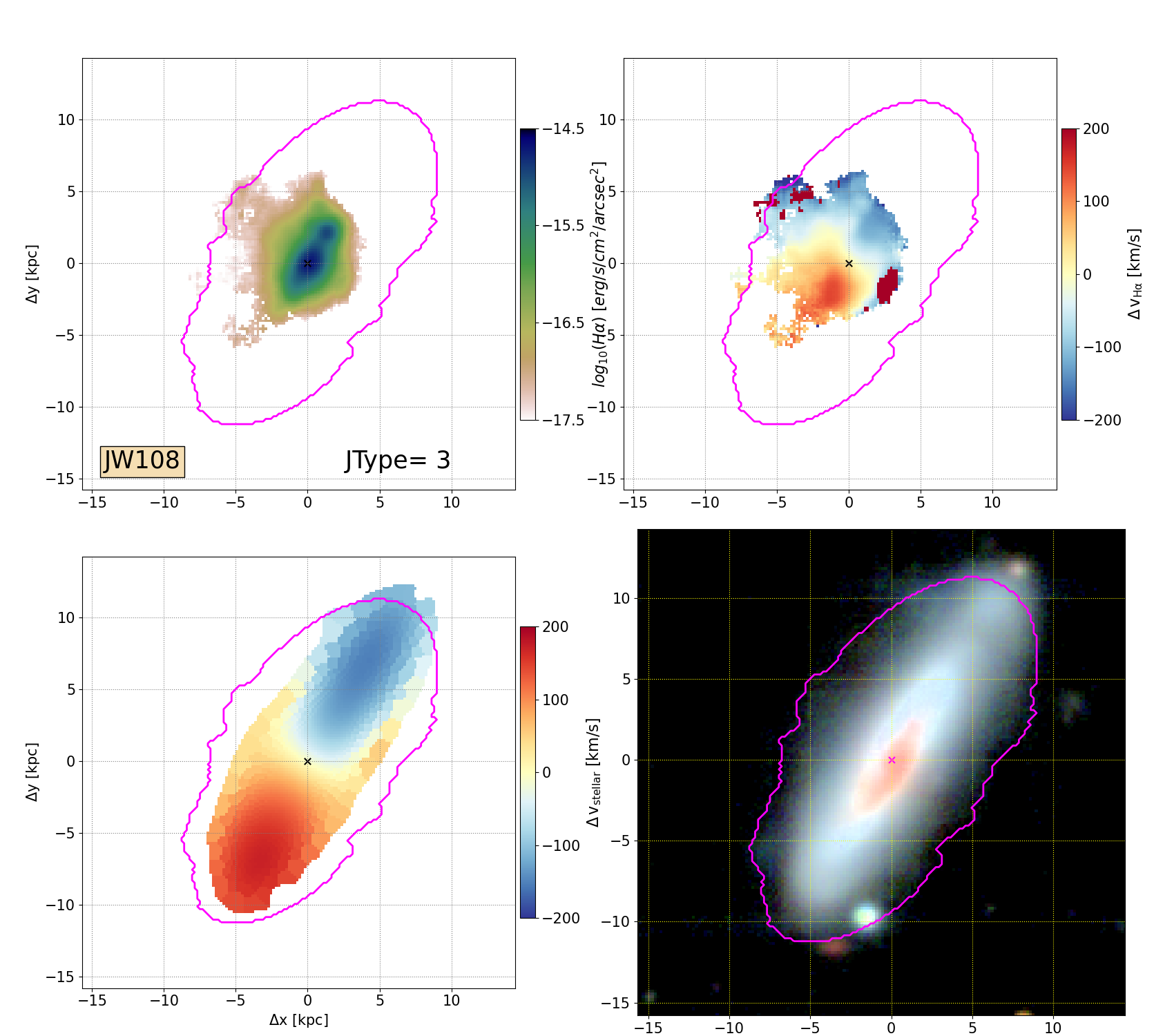}}
\caption{Illustrative examples of stripped galaxies of different JTypes: top JType=2, extreme stripping (JO162 and JW56), JType=3, truncated disks (JO10 and JW108). Panels and colors are as in Fig.\ref{fig:Examples1}. 
\label{fig:Examples2}}
\end{figure*}

\begin{table}
\caption{Stripping types scheme and number of galaxies in the stripping candidate sample}              
\label{tab:levels}      
\centering                                      
\begin{tabular}{c|c|r}          
\hline\hline                        
JType & Description & N of gal. \\    
\hline                                   
0    &  No signs of stripping & 2 \\
0.3  &  Weakest stripping & 4 \\
0.5  &  Mild stripping  &  12 \\
1    &  Strong stripping  & 20 \\
2    &  Extreme stripping  & 16 \\
3    &  Truncated disk  &  4 \\
4    &  Fully stripped &  1 \\
-9   &  Unknown &  3 \\
-99  &  Merger & 2 \\
\hline                                             
\end{tabular}
\tablefoot{
JType=2 "Extreme stripping" galaxies are also refereed to as jellyfish galaxies, and have a gas tail at least as long as the stellar disk diameter.}
\end{table}
 
All galaxies were inspected and classified according to the degree of ram pressure stripping (see Table~\ref{tab:levels}), based on the visible extension of the $\rm H\alpha$-emitting explanar (and unilateral) gas obtained from the MUSE datacubes.
The classification ranges from weakest and mild stripping (JTypes=0.3 and 0.5, respectively) with only weak signs of extraplanar gas, to strong stripping (JType=1) that show a significant gaseous tail, to extreme stripping (JType=2) that have a gas tail at least as long as the stellar disk diameter. The latter are the most striking cases which we refer to also as jellyfish galaxies. Finally, a JType=3 was assigned to galaxies that have a truncated $\rm H\alpha$ disk with gas present only in the central regions of the disk and little extraplanar gas. These are most probably in an advanced stage of stripping, and might have gone through one of the other classes during their evolution (see \S3.3). A JType=4 was assigned to galaxies that have no ionized gas left,  a JType=0 was given to those galaxies with undisturbed ionized gas distribution and thus no sign of stripping, while -9 represents unknown cases and -99 was assigned to mergers. 

We note that { both JType=0.3 and 0.5 galaxies are at a very early stage of stripping and are similar from the point of view of the strength of the tail. The only difference between them is that for the purposes of studying the galaxy disks, the JType=0.3 galaxies have often been used in previous GASP papers as control sample galaxies \citep{Vulcani2018a, Vulcani2019b, Franchetto2020, Bellhouse2021, Vulcani2020b, Tomicic2021a, Tomicic2021b, Franchetto2021a, Peluso2023} because the properties of the gas {\sl within their stellar disk} can be considered unaffected by stripping.} For the purposes of this paper, given the similarities of the extraplanar gas amount between JType=0.3 and JType=0.5, in the following figures we will consider these two classes together.

Figures \ref{fig:Examples1} and \ref{fig:Examples2} show two examples for each JType 0.5, 1, 2 and 3, presenting the MUSE $\rm H\alpha$ flux and velocity maps, the stellar velocity map, and an RGB image obtained combining the $g-r-i$ MUSE images. Figures for all galaxies are presented in the Appendix \ref{sec:Appendix}. Table \ref{tab:stripping} presents the JType classification of all stripping candidates together with other relevant quantities.

\begin{figure}
    \includegraphics[width=1\linewidth]{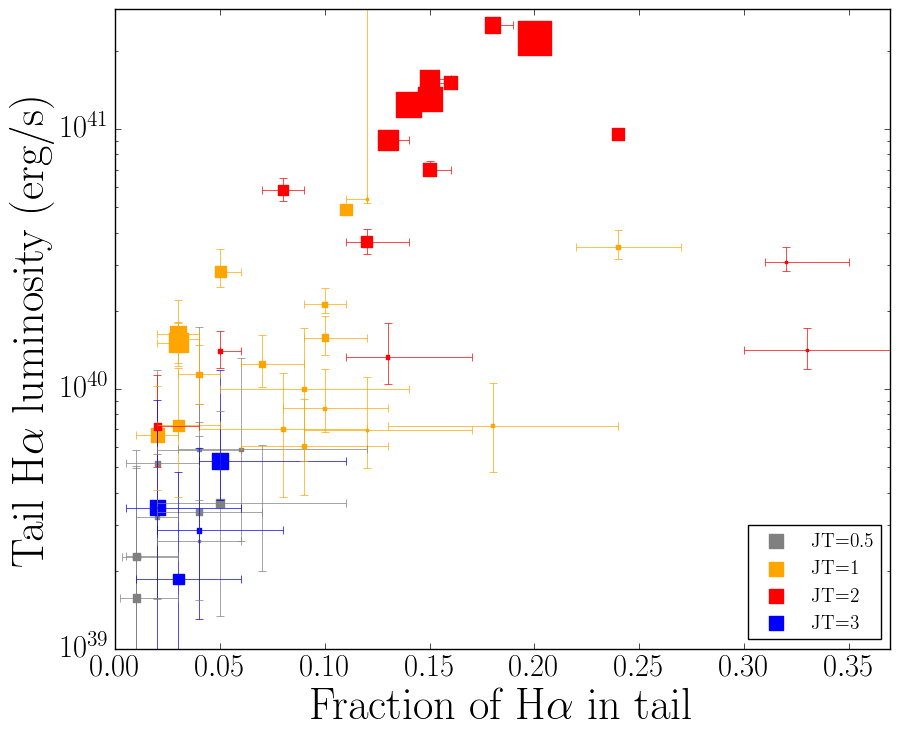}
\caption{Tail $\rm H\alpha$ luminosity versus fraction of $\rm H\alpha$ emission in the tail for different JTypes: 3=blue, 2=red, 1=orange, 0.5=grey. Values are computed for S/N $\rm H\alpha =4$ and errorbars are the range of values with cuts at S/N=3 and 5. The size of the points is proportional to the stellar mass.
\label{fig:Hafrac}}
\end{figure}

The classification described above is based on visual inspection of the $\rm H\alpha$ extension with respect to the stellar disk. For galaxies with $0.3\leq$JType$\leq3$, we also measured two quantities that should be linked with the degree of stripping: the fraction $f_{\rm H\alpha}^{out}$ defined as the percentage of $\rm H\alpha$ emission\footnote{The $\rm H\alpha$ emission is computed irrespective of the ionization mechanism of such emission, i.e. including emission powered by star formation, LINER and AGN according to BPT \citep{Baldwin1981} diagrams.} that is outside of the stellar disk,  and the total $\rm H\alpha$ luminosity $L_{\rm H\alpha}$ outside of the stellar disk. These quantities are shown in Fig.\ref{fig:Hafrac} for the different JTypes. A discussion of the stellar mass distribution of the different JTypes is deferred to \S3.3. At some level, $f_{\rm H\alpha}^{out}$ and $f_{\rm H\alpha}^{out}$ are linked, but they are not equivalent: for example, the two galaxies with the highest fraction of $\rm H\alpha$ in the tail (JO149 and JW56) are low-mass galaxies that have over 30\% of their total $\rm H\alpha$ emission in their tails, but have only a moderate tail $\rm H\alpha$ luminosity because their overall $\rm H\alpha$ luminosity is quite low.

As shown in Fig.\ref{fig:Hafrac}, above $f_{\rm H\alpha}^{out} \sim 0.1$ the great majority of stripped galaxies have a JType=2 and a $L_{\rm H\alpha} \geq 10^{40} \rm \, erg/s$. On the other extreme, all truncated disks have $f_{\rm H\alpha}^{out} < 0.05$ and $L_{\rm H\alpha} \leq 4 \times 10^{39} \rm \, erg/s$.  The JType=1 and 0.5 classes span a range of values that is intermediate between the two classes described above, with typical values of  $f_{\rm H\alpha}^{out} \geq 0.05$ and $L_{\rm H\alpha} \geq 5 \times 10^{39} \rm \, erg/s$ for JType=1 and $f_{\rm H\alpha}^{out}$ between 0.01 and  0.05 and $L_{\rm H\alpha}$ between $0.5$ and $5 \times 10^{39} \rm \, erg/s$ for JType=0.5. The visual discrimination between JTypes 1 and 2 is clearly driven more by the $\rm H\alpha$ tail length than by the amount of extraplanar ionized gas, although galaxies with the longest ionized gas tails (JType=2) tend to have the most luminous tails and the largest fractions of $\rm H\alpha$ emission residing in the tail.

\subsection{Comparison with stellar light classifications}

\begin{figure}
\includegraphics[scale=0.5]{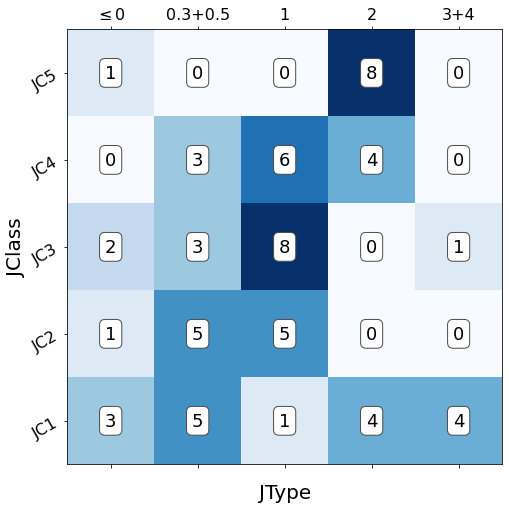}
\caption{Comparison between the original JClass by \citet{Poggianti2016} and the JType presented in Table \ref{tab:levels}, for the GASP galaxies constituting the stripping sample. Galaxies with JType$\leq$0 are considered as failures, galaxies with  JType=0.3 and 0.5 are the weakest and mild cases of stripping, galaxies with JType = 1 are cases of strong stripping, galaxies with JType =2 are cases of extreme stripping,  galaxies with JType=3 and 4 are galaxies at the final stages of stripping, i.e. truncated disks and fully stripped, respectively. The number of galaxies are given inside the boxes and the darkness of the color increases with the number of galaxies.
\label{fig:JClass-Jtype}}
\end{figure}

The JType classification shown above is independent of the "JClass" assigned in \citetalias{Poggianti2016}. While the latter was based on B-band stellar emission potentially coming from stars formed in the stripped gas, here we judge directly the ionized gas that is detected outside of the galaxy disk, which is a direct evidence for stripping and, in principle, is not related to the ionization mechanism of such gas, which can be star formation or other processes. It is therefore interesting to check a posteriori the correspondence between the MUSE (gas)-based GASP classification (JType) and the imaging-based \citetalias{Poggianti2016} classification (JClass), as shown in Fig.~\ref{fig:JClass-Jtype}.  In this plot, we consider together JType=3 and 4 galaxies, given their similar appearance in the B-band images. 

Overall, there is a quite good correspondence between JType and JClass for all types, from weakest to extreme stripping (0.3 to 2) and also for  truncated disks (JType=3). In fact, the latter are expected to have only weak signatures still visible in the images and are in fact dominated by JClass=1 objects. 
The JClass-JType correspondence is particularly striking for the JClass = 5: GASP has confirmed the most secure \citetalias{Poggianti2016} stripping candidates as jellyfish JType = 2 galaxies in 8 spectroscopically confirmed cluster members, the only mismatch being JO190, a foreground merger discussed in \S3.2.

We note two other interesting trends. First, there are a number of cases with very weak imaging signatures (JClass=1) that turn out to be spectacular cases (JType=2 and 1) when the MUSE observations of the gas are considered. 
Being able to detect directly the stripped gas is clearly a great advantage compared to trying to detect the stars that form within this gas. Second, we note that, vice-versa, a significant fraction of the weak MUSE cases (JTypes 0.3 and 0.5) were considered much more promising cases from the images (JClass 3 and 4). This is not due, as it  might be expected, to contamination in the images from superimposed background or foreground B-band sources.\footnote{The only case of such a superposition is a clear failure, JW105, described in the next section.} Inspecting again these cases, it turns out that the reason for the mismatch is that while the JClass was meant to give an assessment of how {\it secure} we were that the galaxy was subject to ram pressure, the JType focuses on the direct observation of extraplanar gas. The two aspects are correlated in the majority of cases, but not always because even the B-band morphology in the disk can be strong evidence for ram pressure at work without a long tail: for example, JO13 (a JClass=4 and a JType=0.5, see Appendix) has a clear half ring of very bright HII regions on one side of the disk, strongly suggesting the impact of ram pressure, though it has only little extraplanar gas.

\subsection{Success rate and properties of stripped galaxies}

\begin{figure*}
\centering
\includegraphics[scale=0.45]{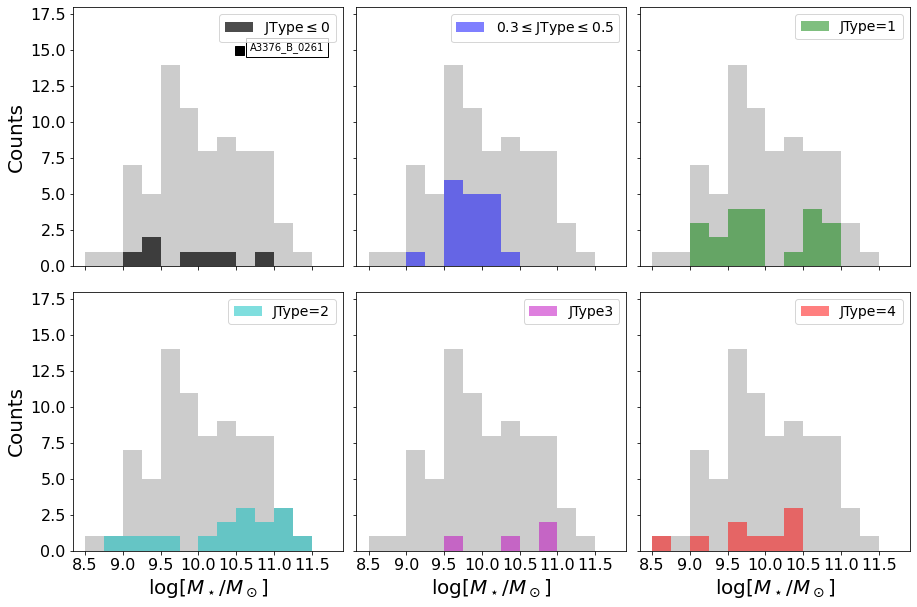}
\caption{Stellar mass distribution for galaxies of the different stripping types, as indicated in the legend. 
A3376\_B\_0261 is the only galaxy from the control sample with JType = 0 and it is reported as black square in the JType $\leq0$ panel. The grey distribution in the background represents the entire sample of 76 galaxies. 
\label{fig:masses}}
\end{figure*}

\begin{figure*}
\centering
\includegraphics[scale=0.45]{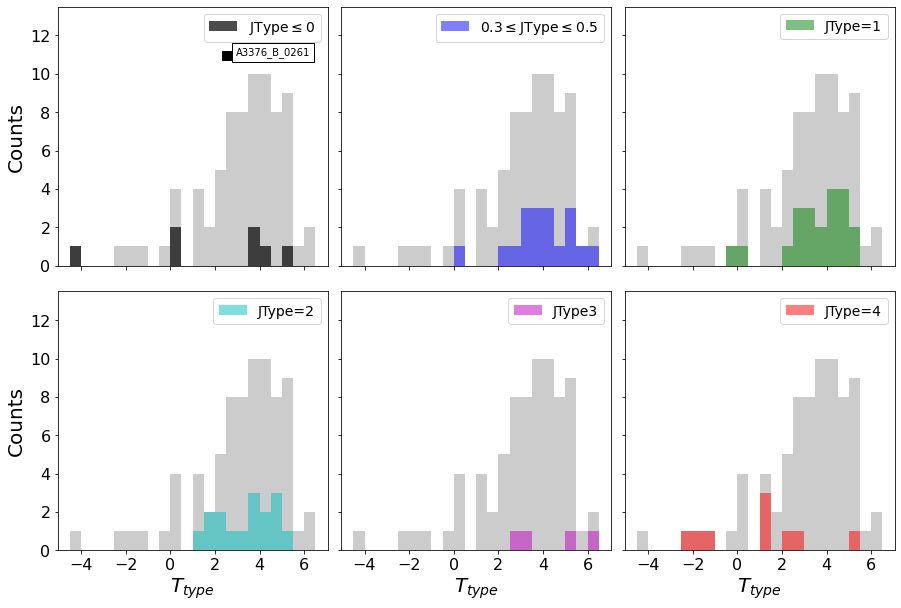}
\caption{Morphological type distribution for galaxies of the different stripping types, as indicated in the legend. Colors and symbols are as in Fig.\ref{fig:masses}. Ttypes and Hubble types are 6=Scd, 5=Sc, 4=Sbc, 3=Sb, 2=Sab, 1=Sa, 0=S0/a, -1=$\rm S0^{+}$, -2=S0, -4=E/S0, -5=E.
\label{fig:morph}}
\end{figure*}

Of the 64 stripping candidates, 56 have been confirmed to be subject to RPS which disturbs their gas in a detectable way according to the criteria described above. From now on, we name these "stripped galaxies". Three of them (JO24, JO73 and JO134\footnote{JO134 is subject to both ram pressure and a minor merger as discussed in detail in \cite{Vulcani2021}.}) turned out to be non-members of the main cluster targeted in the field, but are members of a background cluster or group that exerts ram pressure on them. In addition, JW36 is most likely the end product of stripping, being a blue comet-like galaxy devoid of ionized gas, with strong $\rm H\beta$ in absorption (plot not shown), with characteristics thus quite similar to IC3418 which is one of the prototypical jellyfish galaxies in the Coma cluster \citep{Hester2010}. Therefore, overall 89\% (57/64) of the stripping candidates have been confirmed by the GASP data.

Of the remaining 7 galaxies, one (JW105) is a chance superposition between a passive cluster member early-type galaxy and some background sources at $z\sim 0.38$ and $z\sim 0.15$; three galaxies show signs of a merger (the cluster member JO153 and the non-members JO20 and JO190, see \citealt{Vulcani2021}), two are likely interacting with a close neighbour (JO119 and JO157) and one (JW10) has an uncertain classification given the morphology of the $\rm H\alpha$-emitting extraplanar gas.

With a large sample of IFU-confirmed ram pressure stripped galaxies in a wide range of stripping stages in 39 different clusters it is possible to address questions regarding the intrinsic properties of the stripped galaxies and the occurrence of different levels of stripping in the various regions of clusters and in different types of clusters.

In what follows we also include three galaxies from the control sample, which will be discussed in detail in Sec.\ref{sec:control}. Briefly, only one control sample galaxy turned out to be truly undisturbed, while the other three star-forming galaxies of the control sample also show signs of stripping and have been classified in the same way as the stripping sample, following the classification scheme presented in Tab.\ref{tab:levels}. 

The stellar mass distribution of galaxies of different JTypes is shown in Fig.\ref{fig:masses}. 
Extreme and strong stripping cases (JTypes 2 and 1) are found in a very wide range of masses, from $6 \times 10^{8}$ to $3 \times 10^{11} M_{\odot}$. The most massive stripped galaxies ($\geq 10^{11} M_{\odot}$) are all spectacular jellyfish cases (JType=2) (see also \citealt{Luber2022}), but the majority of jellyfish galaxies in the GASP sample (9 out of 16) have masses $ \leq 4 \times 10^{10} M_{\odot}$. In Fig.\ref{fig:Hafrac} we also see that the JType=2 galaxies with the highest $\rm H\alpha$ tail luminosities ($\sim 10^{41} \rm \, erg \, s^{-1}$ and above) are massive/very massive.

Moreover, the masses of truncated disks range from $4.6 \times 10^{9}$ to $ 6.5 \times  10^{10} M_{\odot}$, showing that also this phase of stripping can be observed in galaxies over a wide mass range. 
Finally, it is noticeable that all the weakest and mild cases (JTypes 0.3 and 0.5) have masses below $ 2 \times  10^{10} M_{\odot}$. The median mass is 
$ \sim 8 \times  10^{9} M_{\odot}$ for JTypes = 0.3+0.5 and JType=1, while it is five times higher ($4 \times  10^{10} M_{\odot}$) for JType=2. Note that in this figure and the next there is an histogram also for JType=4 galaxies for completeness.

Given that the JType classification is influenced by visibility effects and observational biases, it is hard to draw conclusions regarding the {\sl occurrence} of different levels of ram pressure as a function of galaxy mass without risking to over-interpret the results. In fact, as discussed in detail in \cite{Gullieuszik2020}, the mass of the galaxy is only one of many factors that determine the level of stripping. However, the fact that strong stripping effects are observed over more than two orders of magnitude in mass is a solid result. 

Moving to galaxy Hubble types, the great majority of GASP stripped galaxies are spirals of types between Sab and Sc, as shown in Fig.\ref{fig:morph}, and no T-type segregation is observed going from weakest to extreme stripping. The fully stripped galaxy sample, instead, by selection included S0s and early spirals.

\begin{figure*}
\centerline{\includegraphics[scale=0.4]{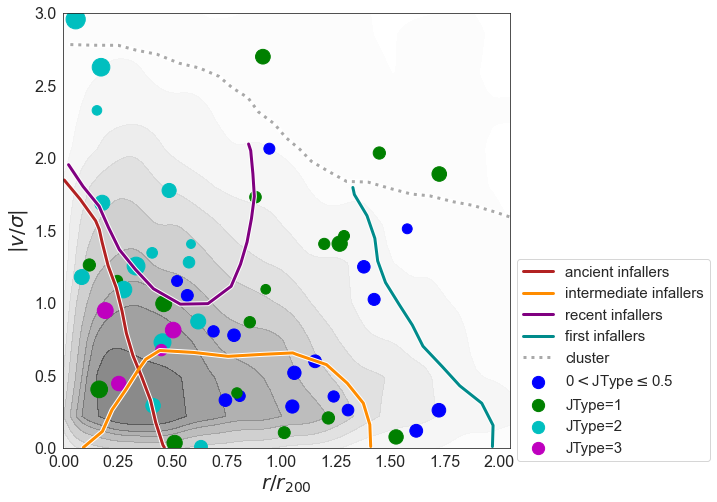}}
\caption{Stripping types distributed across the projected phase space diagram, as indicated in the legend. Overplotted are coloured lines that delimit the regions defined by Rhee et al. (2017): first (not fallen yet; turquoise), recent ($0 < t_{infall} < 3.63$ Gyr; purple), intermediate ($3.63 < t_{infall} < 6.45$ Gyr; yellow), and ancient ($6.45 < t_{infall} < 13.7$ Gyr; red) infallers. The dashed line indicates the limit of subhalos, to define galaxies bounded to the clusters. The size of the circles is proportional to the galaxy stellar mass, and masses are given in Table~2. The grey contours in the background represent the density of points from all spectroscopic cluster members in the WINGS/OmegaWINGS sample. The control sample galaxies are not shown here.
\label{fig:phasespace}}
\end{figure*}

Next, we present the location of the various JTypes in the projected position versus projected velocity phase-space diagram in Fig.\ref{fig:phasespace}. 
Distances of the galaxies are measured from the Brightest Cluster Galaxy, velocity dispersions are from \cite{Biviano2017} and \cite{Gullieuszik2020}. Several simulation studies have characterized the typical time since infall of galaxies located in different regions of the phase-space diagram. We use here the representation from \cite{Rhee2017}, who identified the regions where the majority of galaxies lie at a given epoch after they enter the cluster halo (see their Fig. 5). \cite{Rhee2017} separated galaxies into first (not fallen yet; turquoise), recent ($0 < t_{infall} < 3.63$ Gyr; purple), intermediate ($3.63 < t_{infall} < 6.45$ Gyr; yellow), and ancient ($6.45 < t_{infall} < 13.7$ Gyr; red) infallers. Obviously these numbers should be taken with caution, as each galaxy can be located beyond the corresponding region, and the observed radii and velocities, being projected quantities, are lower limits to the real values.
In \cite{Jaffe2018} we studied the phase-space diagram of a subset of the sample of this paper (those available at the time) and concluded that jellyfish galaxies are moving at very high speeds close to the cluster center and are a recently infallen population moving mostly on radial orbits. In \cite{Gullieuszik2020} we studied the location in the phase-space diagram of a subset of the galaxies of this paper as a function of the star formation rate in the tails, finding a good correlation between star formation and the phase-space region.

In Fig.~\ref{fig:phasespace} we now show the full sample and confirm the results from both \cite{Jaffe2018} and \cite{Gullieuszik2020}. The most striking segregation in the plot is the location of the extreme stripping (cyan circles) galaxies (which are also those with the highest $\rm H\alpha$ tail luminosities, and thus SFR in the tails. cfr. Fig.~3), which are all located in the region at high velocities and low clustercentric radii where recently infallen galaxies are mostly located, above the purple line. The strong stripping galaxies (green circles) occupy on average larger radii and lower velocities than JType=2 but are characterized by a large range of both of these quantities. They are spread in most regions of the plot, except the recently infallen region, and some of them are approaching at high velocities close or beyond the cluster virial radius.
JType =0.3+0.5 galaxies (blue circles) are found at projected radii between 0.5 and 1.7 $r/r_{200}$, thus avoiding the cluster central region, on average at larger radii than the previous types and in most cases at velocities below 1 $v/\sigma$.
Finally, the Jtypes=3 (truncated disks) are consistent with having entered the cluster longer ago than the other JTypes, being located in or close to the ancient infallers region. 

\begin{figure}
\centerline{\includegraphics[scale=0.4]{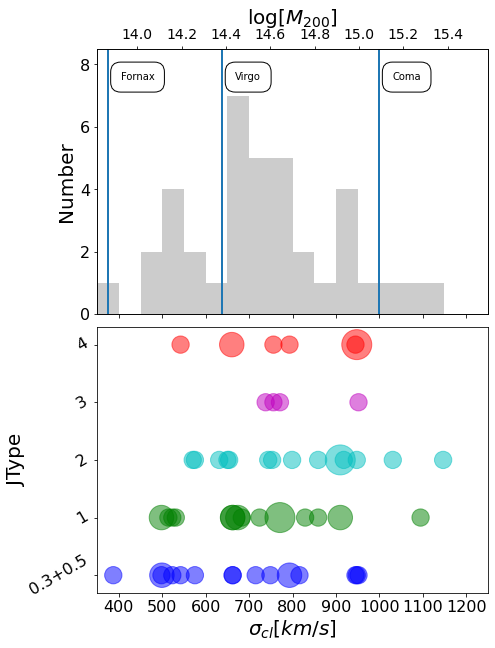}}
\caption{Top: Velocity dispersion distribution of the clusters hosting at least a ram pressure stripped galaxy (JType $\geq$0.3). Cluster masses, computed from the velocity dispersions assuming virialization, are shown in the top axis. Vertical lines show the velocity dispersion of largely studies clusters in the local Universe (from left to right: Fornax, Virgo, Coma). Bottom: Frequency of galaxies of different JType as a function of the velocity dispersion of the hosting cluster. Symbol size is proportional to the frequency. 
\label{fig:vdisp}}
\end{figure}

Overall, the location within the phase-space diagram appears to be more relevant than the cluster velocity dispersion, which can be considered as a proxy of the cluster mass. The distribution of velocity dispersions $\sigma$ of clusters with GASP stripped galaxies is shown in the top panel of Fig.~\ref{fig:vdisp}. Velocity dispersions range from 400 to above 1000 $\rm km \, s^{-1}$, thus encompassing clusters such as Fornax, Virgo and Coma, with most of the GASP clusters having intermediate $\sigma$ between 500 and 900 $\rm km \, s^{-1}$. Interestingly, weakest+mild (JType = 0.3+0.5), strong (JType = 1) and extreme (JType = 2) stripping cases are observed in the entire range of $\sigma$ from $\sim 550$ to $950 \, \rm km \, s^{-1}$. 
The GASP sample does not have mild cases in the most massive clusters ($\sigma > 1000 \, \rm km \, s^{-1}$) and extreme cases in very low mass clusters $\sigma < 550 \, \rm km \, s^{-1}$, but these two trends  could be due to low number statistics.
What the GASP data show clearly is that strong and extreme ram pressure can be effective also in clusters as small as Virgo or even smaller: being hosted in a very high mass clusters is not a necessary condition for jellyfishes and strong ram pressure.

\subsection{The truncated disks}\label{sec:truncated}

\begin{figure*}
\centerline{\includegraphics[scale=0.25]{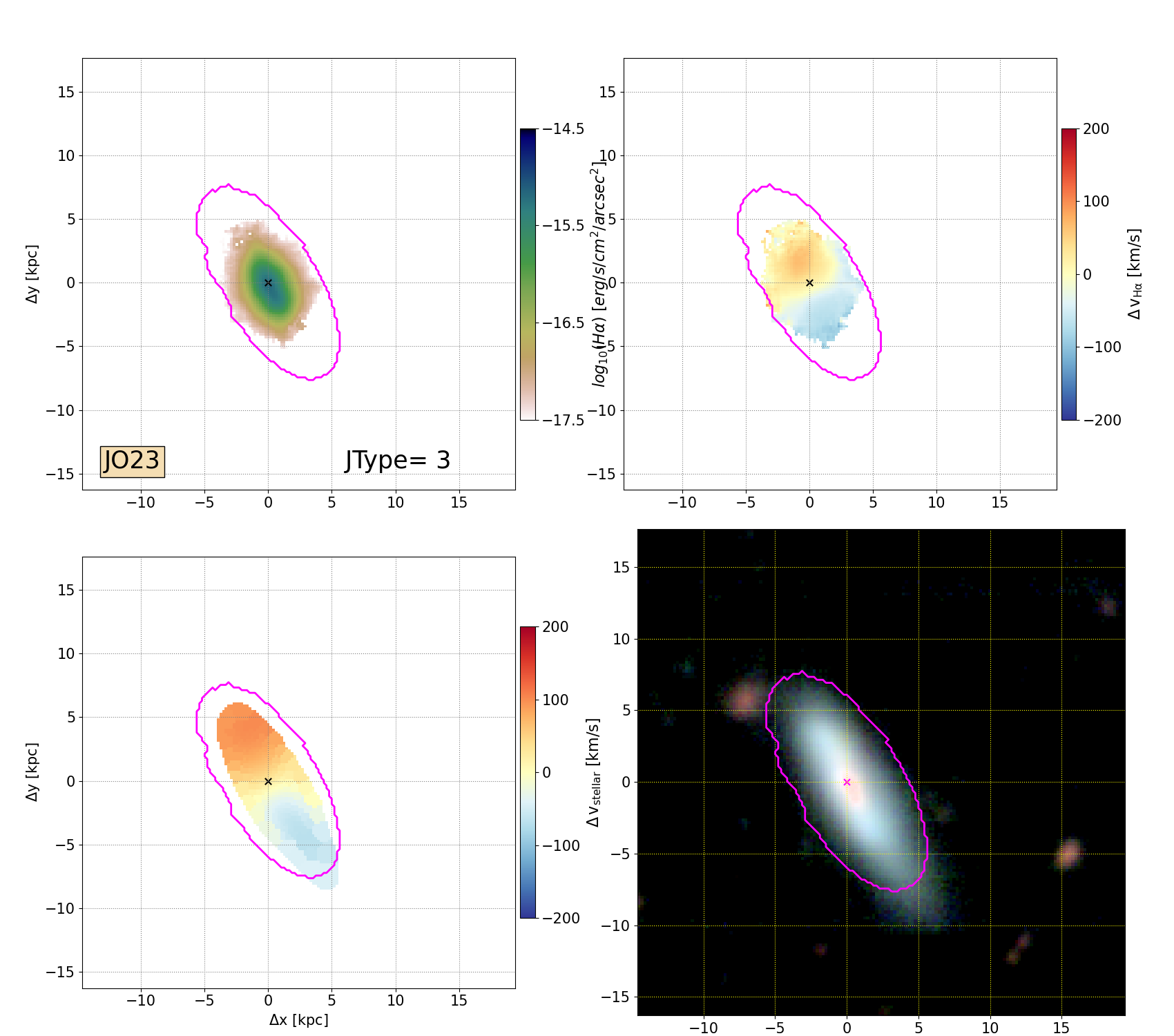}}
\centerline{\includegraphics[scale=0.4]{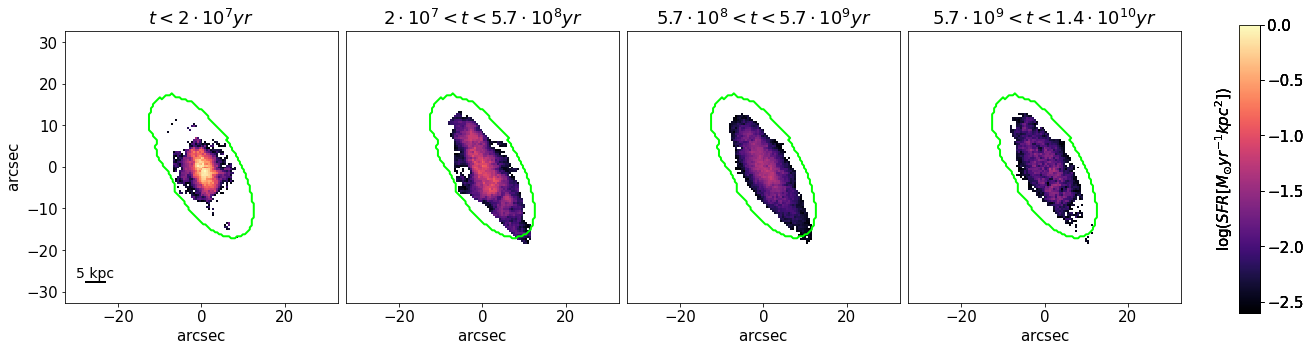}}
\caption{Example of a  JType=3 (truncated galaxy): JO23. Top: H$\alpha$ flux (top left), the gas kinematics (top right), the stellar kinematics (bottom left) and the color composite image (bottom right). Panels and colors are as in Fig.\ref{fig:Examples1}. Bottom: 
Stellar maps of different ages, illustrating the average star formation rate per kpc$^2$ during the last $2\times10^7$ yr (left), between $2\times10^7$ and $5.7\times10^8$yr (central left), between $5.7\times 10^8$ and $5.7 \times 10^9 $ yr (central right), and $>5.7 \times 10^9$ yr ago (right). The green ellipses show the contour that defines the stellar disk (see text for details).
\label{fig:SFH_truncated}}
\end{figure*}

Four of our stripped galaxies have a truncated $\rm H\alpha$ disk (JO10, JO23, JO36, JW108). All of them have ionized gas only left in the central region of the disk and less than 5\% of their $\rm H\alpha$ luminosity in small clouds or tendrils outside of the disk. For these reasons, they have been assigned a JType=3. One of them has been discussed in detail in \cite{Fritz2017} and the other three are shown in Fig.\ref{fig:Examples2} and Fig.\ref{fig:SFH_truncated}. 

As shown by numerous hydrodynamical simulations \cite[e.g.,][]{Roediger2005, Roediger2014, Kapferer2008, Kapferer2009, Tonnesen2011, TonnesenBryan2012, Zhu2024, Akerman2024} ram pressure strips the gas first from the outer regions of the disk where it is less gravitationally bound and then proceeds outside-in.
When a galaxy is hit by ram pressure face on, the stripping occurs in progressively inner cilindric annuli in an axisimmetric manner, but in the majority of cases there will be an angle between the wind direction and the disk axis. Depending on the viewing angle of the observer, in projection we will observe only one side of the galaxy devoid of gas as long as there is a significant tail. This is the case for example of some of the best studied galaxies in the GASP sample such as JO206 or JO204 or JW100 (see Appendix), which all have a "gas disk that is truncated" but still have a long tail.

What we define as "truncated disks" in this paper, instead, are galaxies that have very little extraplanar gas and are in an advanced stage of stripping occurring before the eventual stripping of all gas and after a phase in which the extraplanar gas tail was probably more prominent. 
This recent past can be recovered from the MUSE data, as shown in Fig.~\ref{fig:SFH_truncated} for JO23 as an example.
The RGB image of JO23  (middle right) clearly shows an elongation of the blue stellar light (stellar tail) towards the south-west, while the gas is left only in an approximately circular (when deprojected) region around the galaxy center. Analyzing the maps of star formation history (bottom panels), as derived by the code \sinopsis (see \S2), we can see where stars formed during four intervals in time. Currently (ongoing star formation rate, $2\times10^7$ yr), stars are forming only in the central region, where there is gas left.
Quite recently (between $2\times10^7$ and $5.7\times10^8$yr ago, and at a lower level also in the previous age bin) stars were forming both in a larger central area and along a tail to the south-west, where gas must still have been present. This also allows us to understand the projected direction of stripping. At older times ($> 5.7 \times 10^9$ yr ago) stars were forming in a disk regularly centered on the galaxy center, with no star-forming tails.
Unfortunately, given that the age resolution of the spectrophotometric reconstruction deteriorates at progressively older ages, we cannot age-date the beginning of the stripping more precisely than in these four age bins, but this analysis is already sufficient to see the progressive evolution of the star formation and, therefore, of the gas stripping.

Studying in a similar way from MUSE data the star formation history of galaxies in the subsequent stage of stripping, i.e. with no gas left, we have previously shown that the great majority of galaxies with post-starburst spectra in clusters are the end product of ram pressure and it has been possible to reconstruct the stripping(=quenching) histories
both in the GASP sample \citep{Vulcani2020a} and in a sample of galaxies in clusters at z=0.3-0.5 \citep{Werle2022}. 


\section{The control sample}\label{sec:control}

\begin{figure}
    \centering
    \includegraphics[width=1\linewidth]{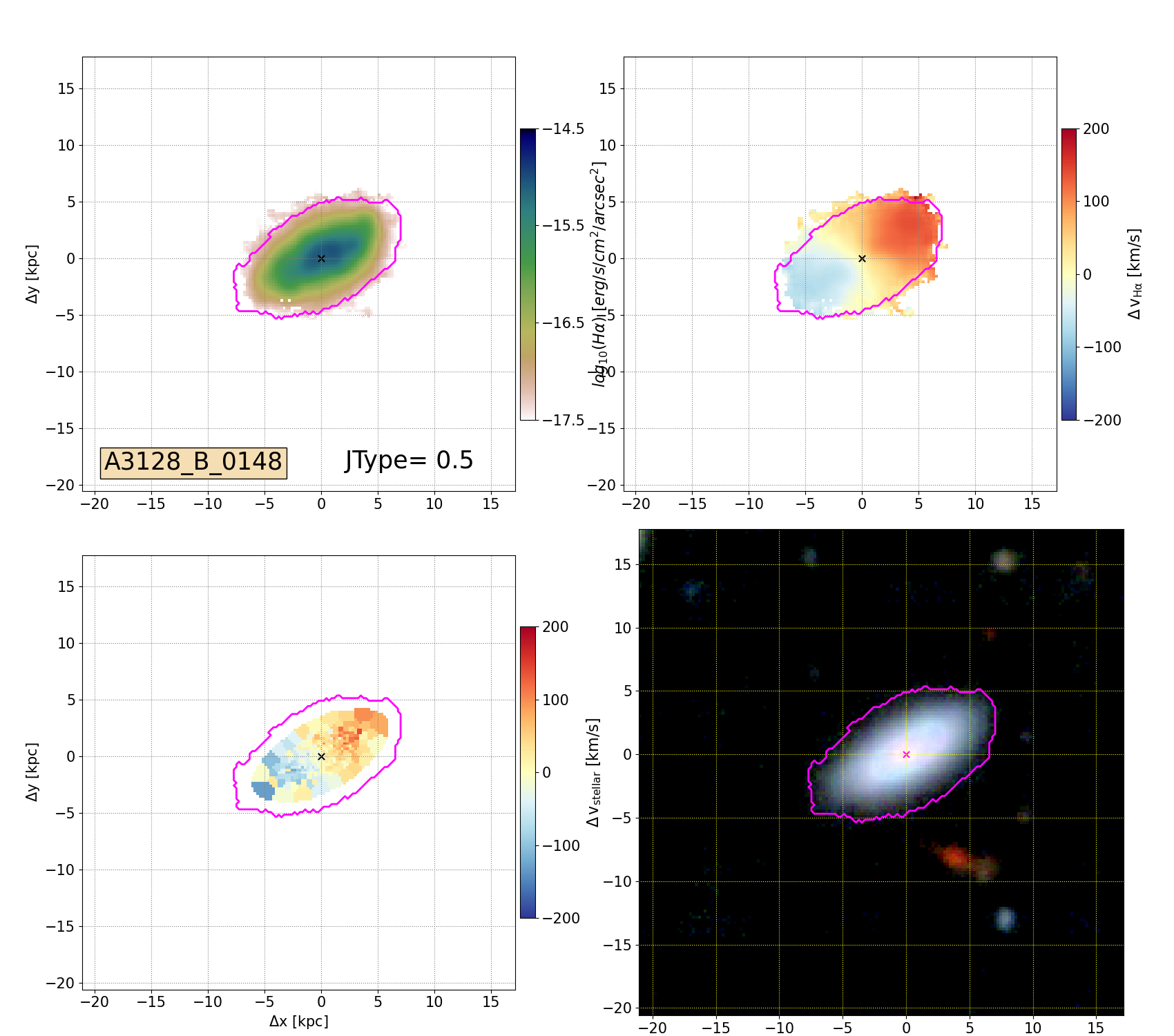}
    \caption{Illustrative example of the control sample galaxy A3128\_B\_0148. Panels and colors are as in Fig.\ref{fig:Examples1}.}
    \label{fig:control_sample}
\end{figure}

{ We remind the reader that, in addition to the stripping candidate galaxies discussed in the previous sections, GASP observed also 12 "control cluster galaxies" that did not show any sign of stripping from optical B-band imaging.} The main properties of control cluster galaxies are listed in Table \ref{tab:control}. 
As already mentioned, 8 of these galaxies were selected to be devoid of gas (therefore are JType=4) and a dedicated study has been presented in \cite{Vulcani2020a}. { Out of the remaining 4 star-forming control cluster galaxies that were not expected to be undergoing stripping based on B-band data}, 3 turn out to be stripped at some level { when observed by MUSE}, with some $\rm H\alpha$ emission outside and preferentially on one side of the stellar disk, as visible in Fig.\ref{fig:control_sample} and Appendix \ref{sec:Appendix}. Two of them are very weak cases of stripping and are classified as JType=0.5: A970\_B\_0338 has short $\rm H\alpha$ filaments extending to the West of the disk, and A3128\_B\_0148 has two small tendrils of gas coming out from the South-West side of the disk. The other galaxy, A3266\_B\_0257, is a JType=1 with detached gas clouds at the north-east side of the disk. The $\rm H\alpha$ fractions and luminosity outside of the disk are consistent with those of stripped galaxies of similar JType, that is $f_{\rm {H\alpha}^{out}} \sim 0.01-0.02$ in A970\_B\_0338 and A3128\_B\_0148 and 0.18 in A3266\_B\_0.257.
This confirms the well known fact that stripping signatures are much more visible when observing the ionized gas rather than the total optical light, and that cases such as these ones go miss when selecting from optical imaging.

\begin{figure*}
\centerline{\includegraphics[scale=0.25]{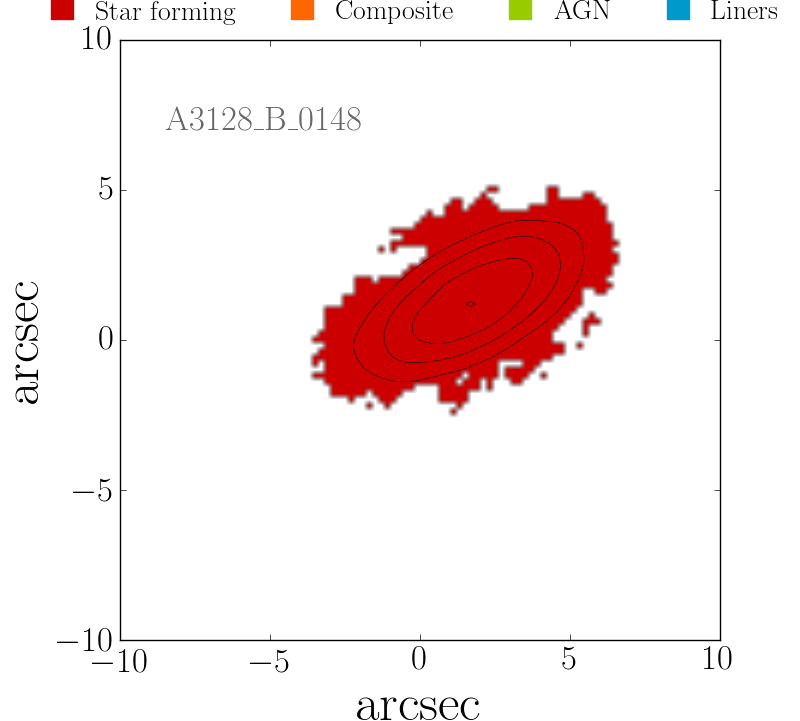}\includegraphics[scale=0.25]{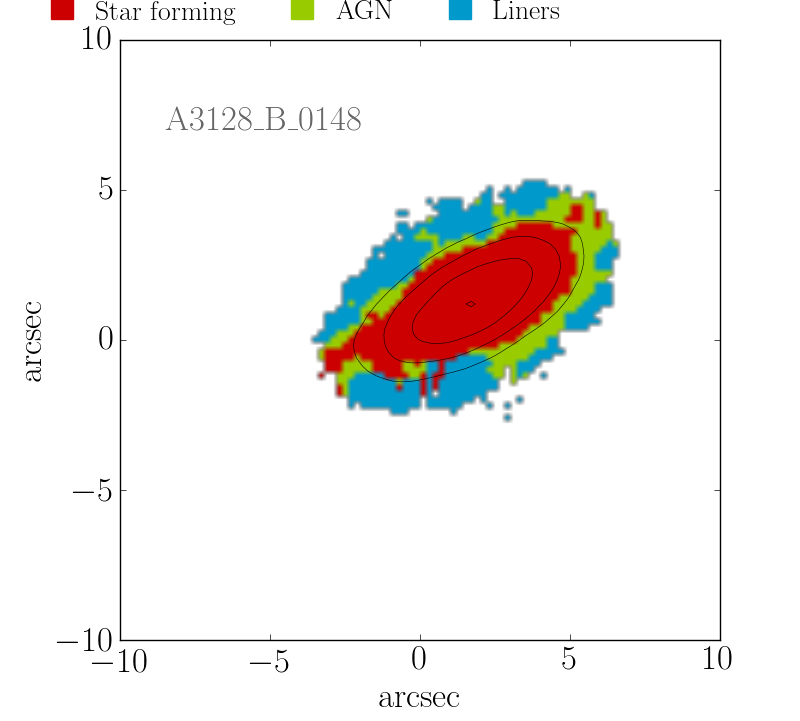}\includegraphics[scale=0.25]{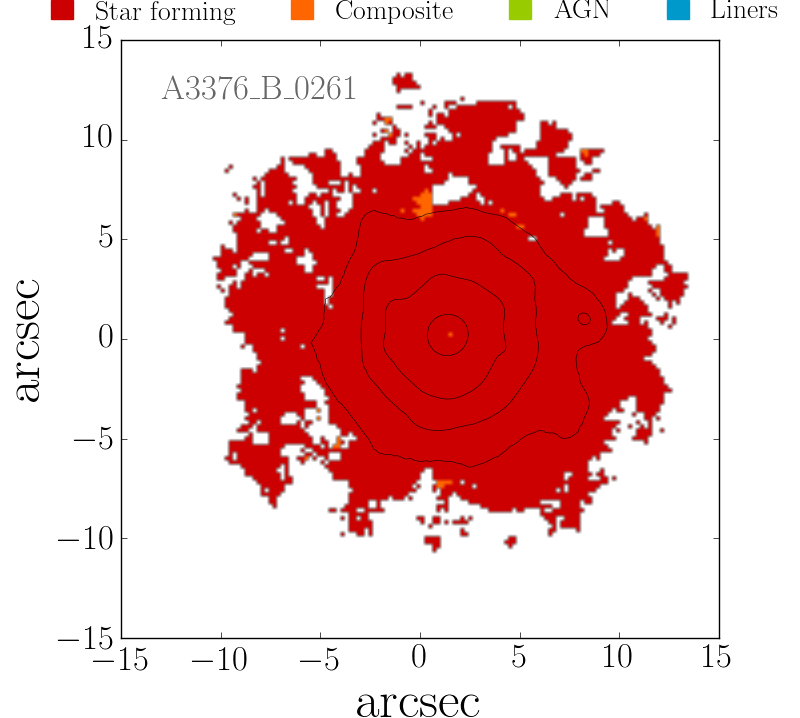}\includegraphics[scale=0.25]{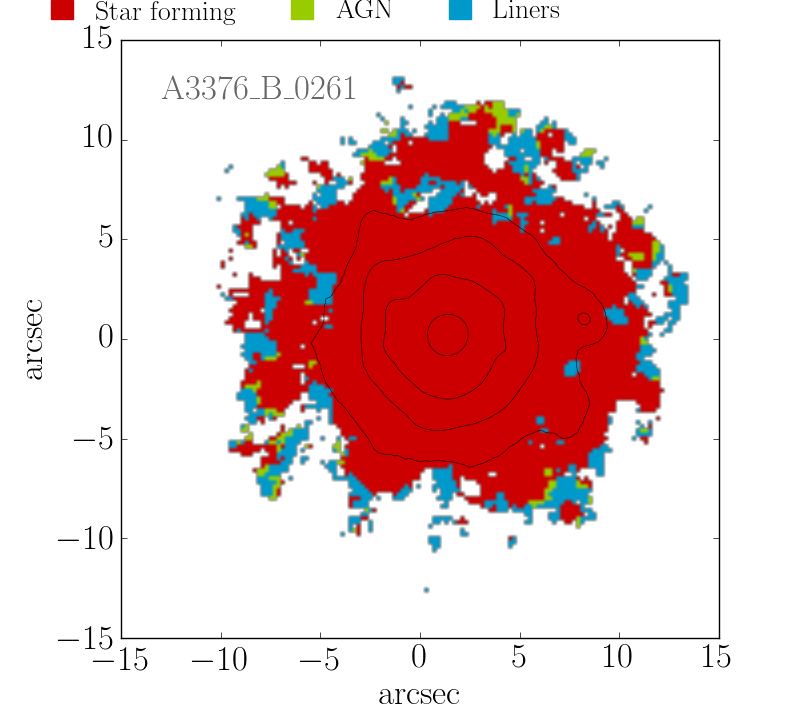}}
\centerline{\includegraphics[scale=0.25]{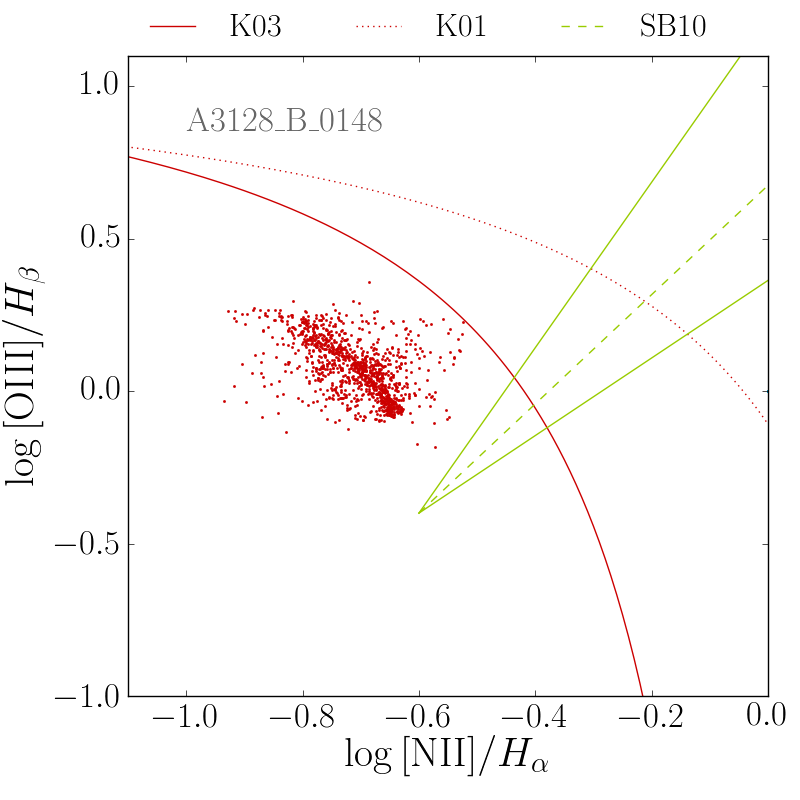}\includegraphics[scale=0.25]{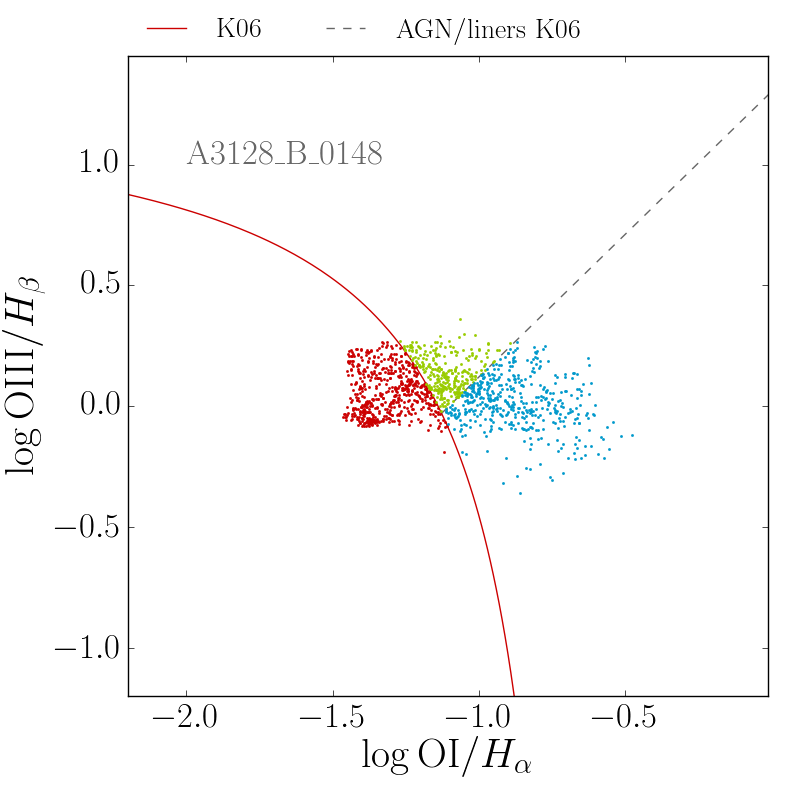}\includegraphics[scale=0.25]{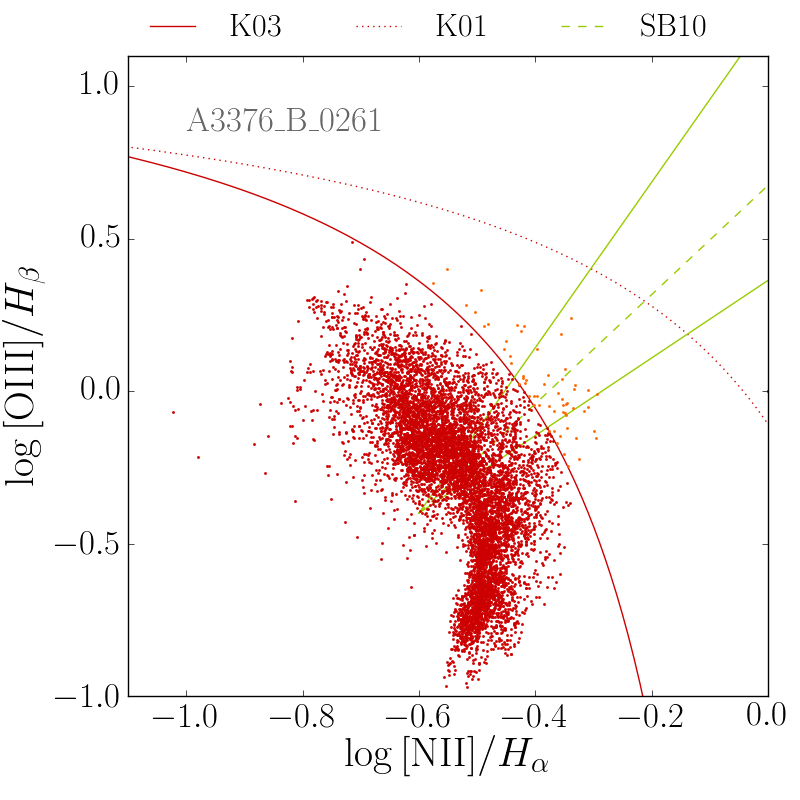}\includegraphics[scale=0.25]{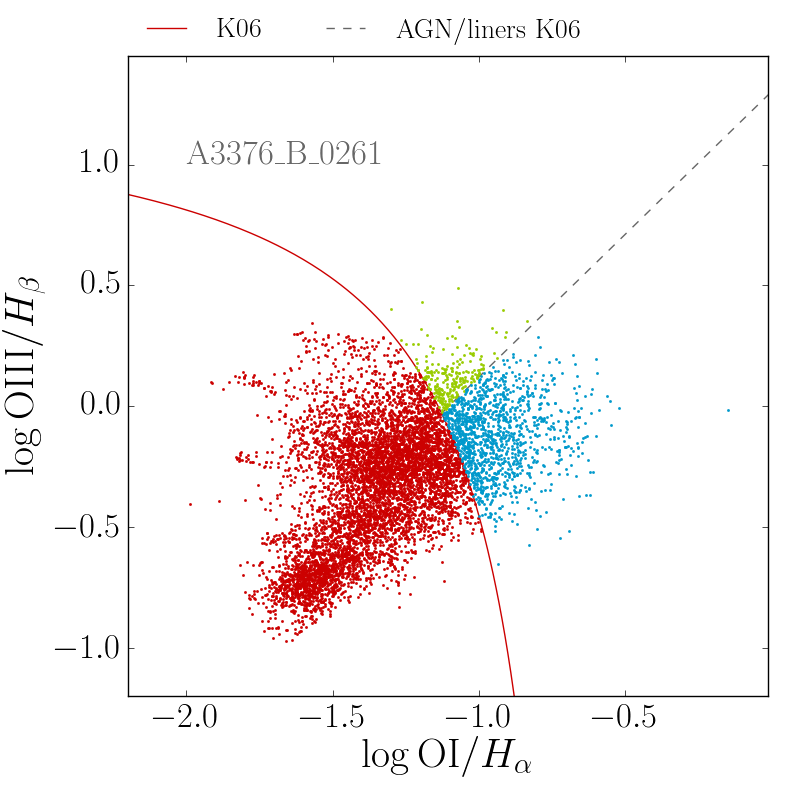}}
\centerline{\includegraphics[scale=0.25]{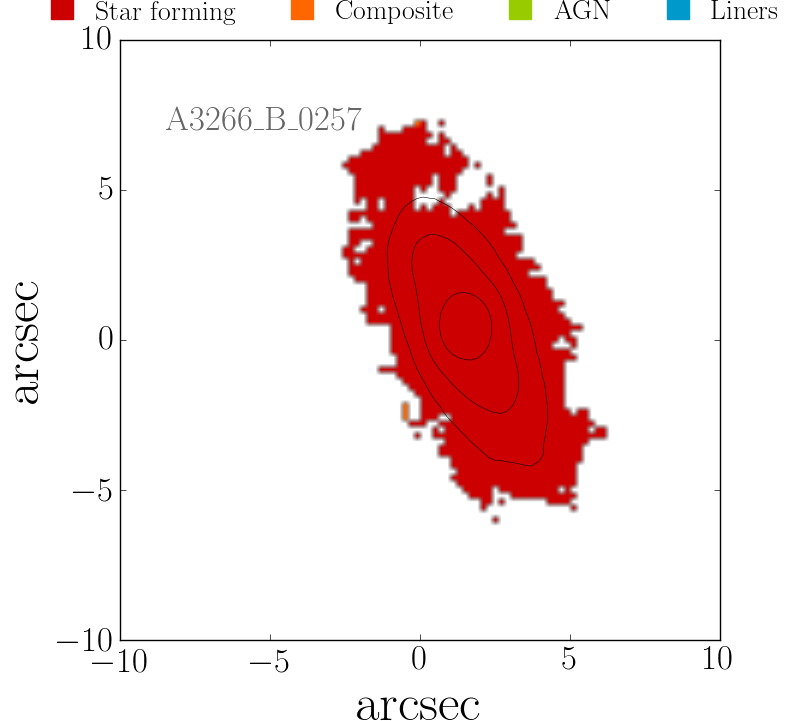}\includegraphics[scale=0.25]{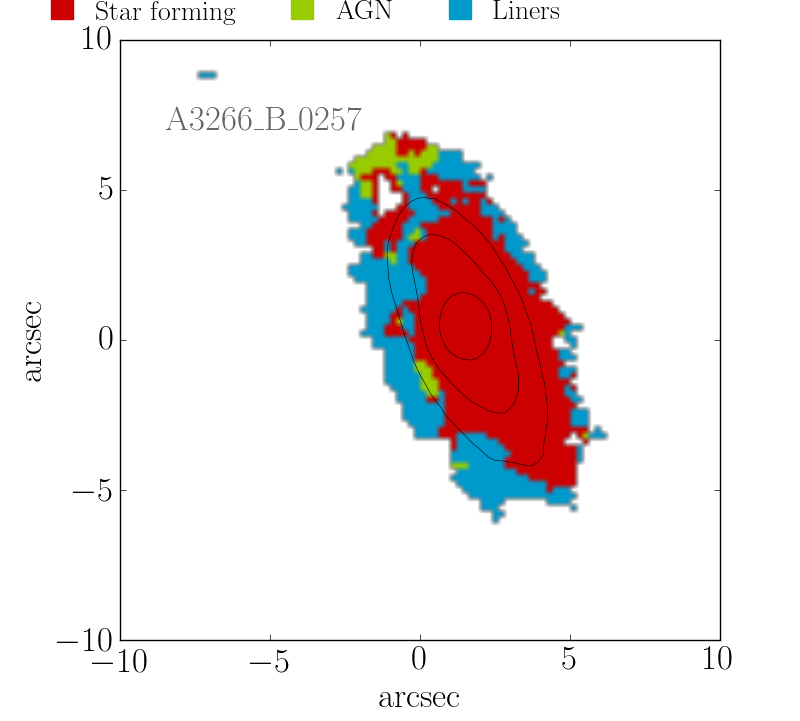}\includegraphics[scale=0.25]{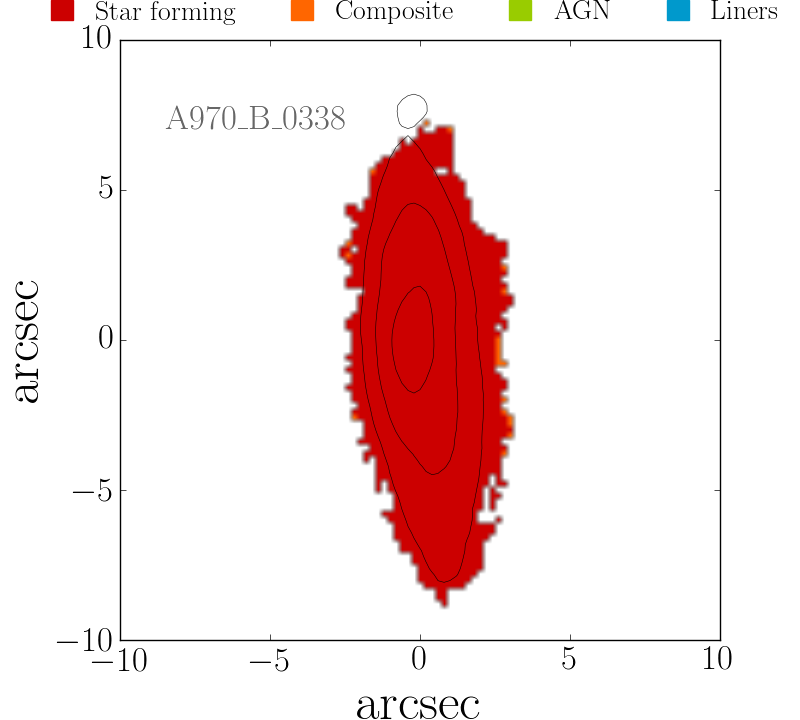}\includegraphics[scale=0.25]{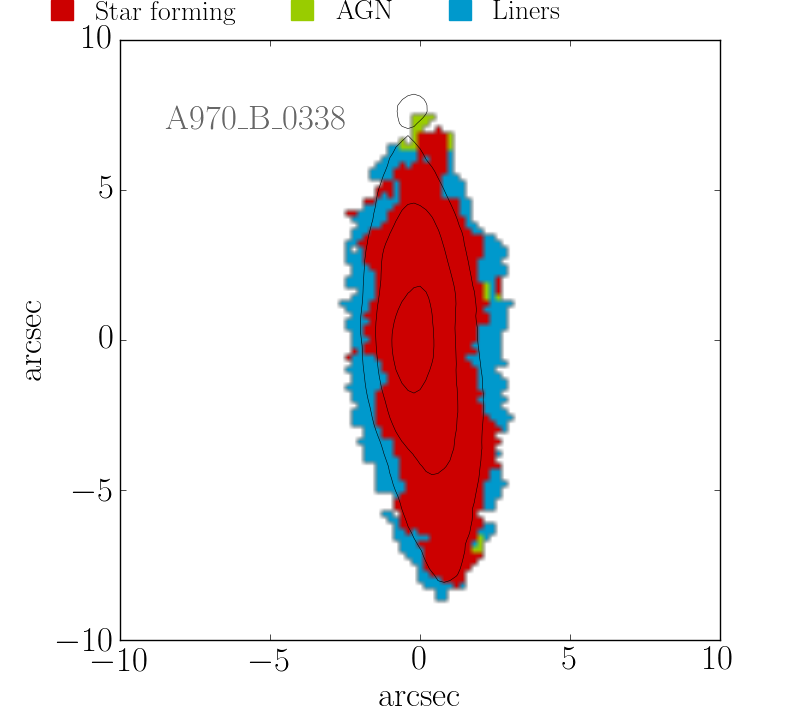}}
\centerline{\includegraphics[scale=0.25]{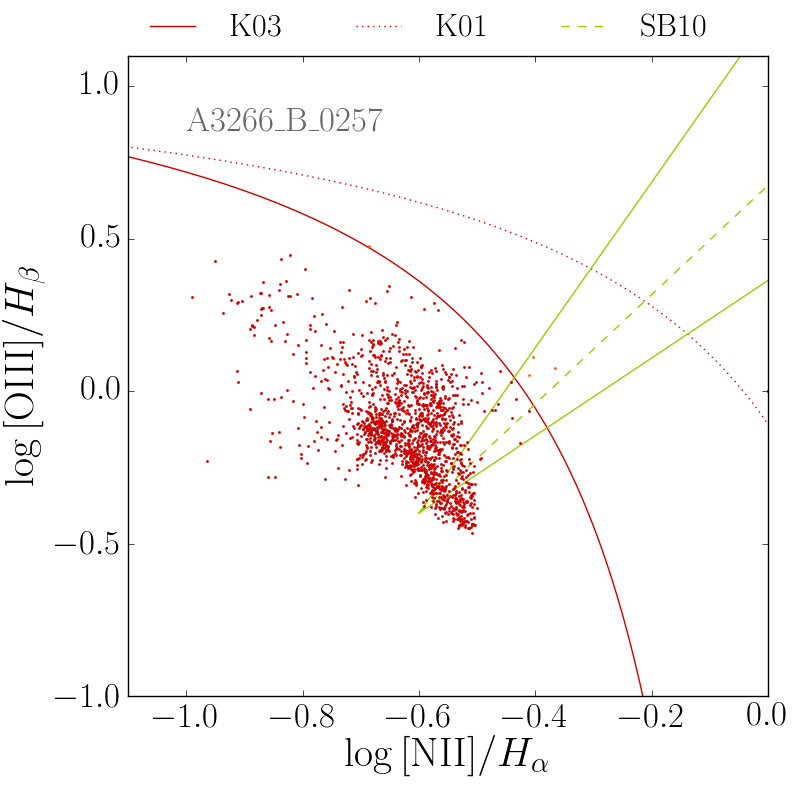}\includegraphics[scale=0.25]{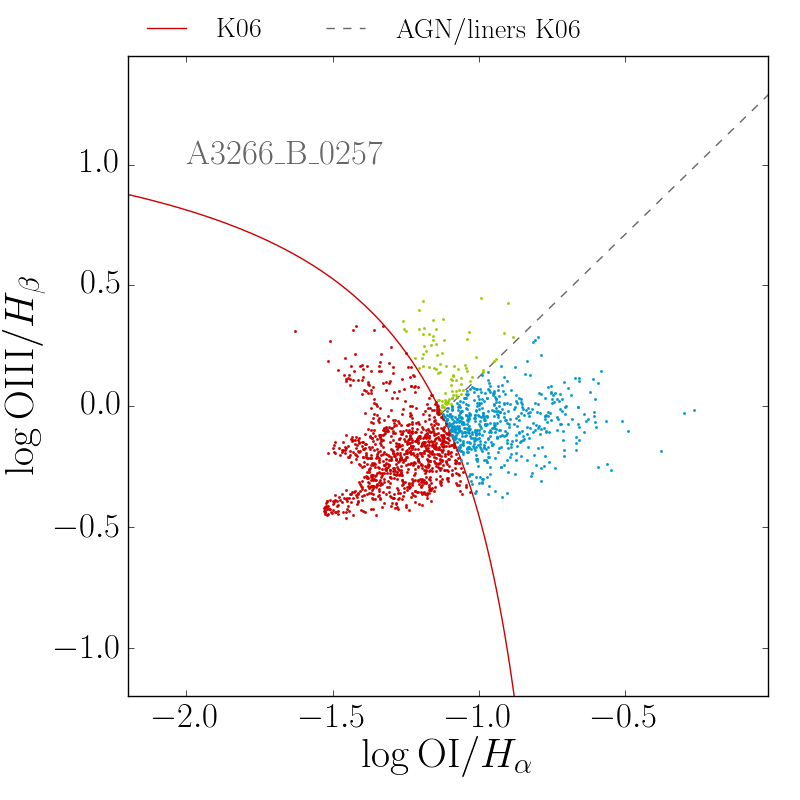}\includegraphics[scale=0.25]{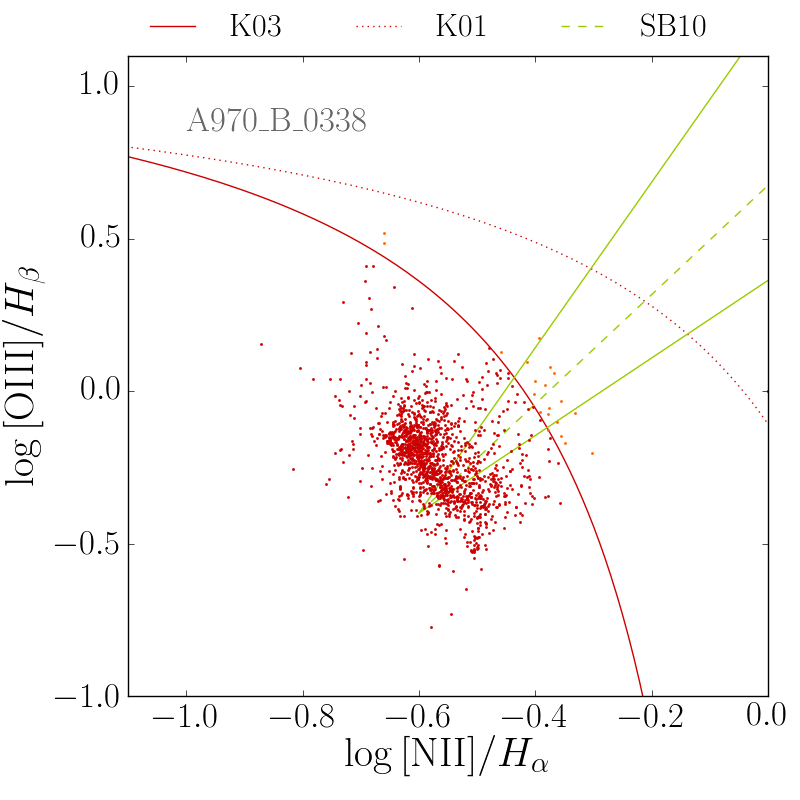}\includegraphics[scale=0.25]{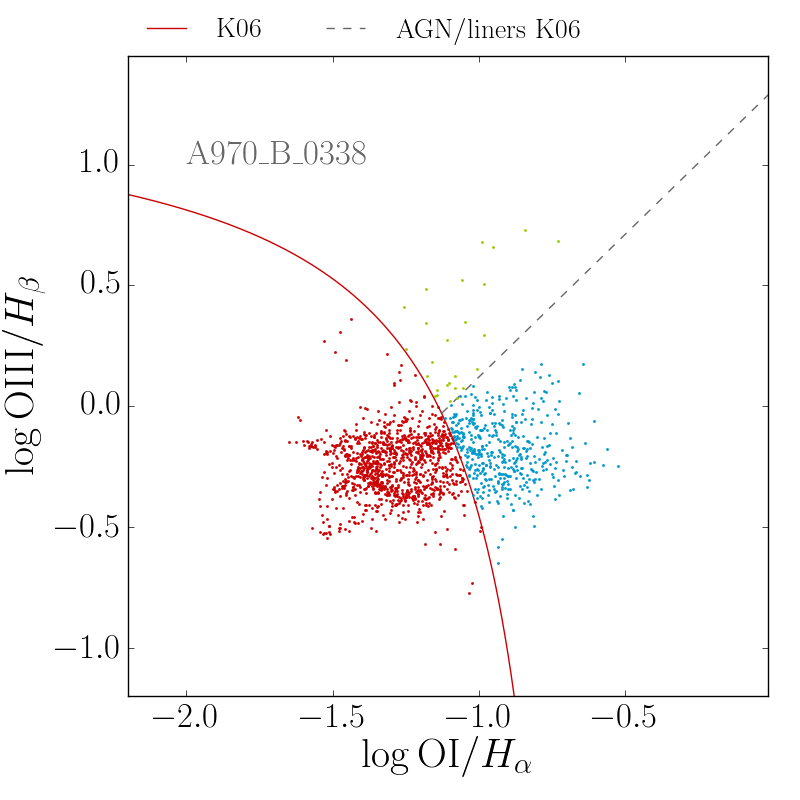}}
\caption{Diagnostic diagrams for [O III]5007/H$\beta$ vs. [N II]6583/H$\alpha$ and [OI]6300/H$\alpha$ for the control sample galaxies. For each galaxy, the top panel shows the  BPT line-ratio map, the bottom panel the corresponding spatially resolved BPT diagram. Lines are from \citet[][K03]{Kauffmann2003}, \citet[][K01]{Kewley2001}, and \citet[][SB10]{Sharp2010} to separate star-forming (red), composite (orange), AGN (green), and LINERS (cyan). Only spaxels with an S/N > 3 in all of the emission lines involved are shown. Adopting a S/N > 4 does not modify these plots.
\label{fig:BPT1}}
\end{figure*}

A notable property of these four star-forming galaxies is their combination of emission line ratios at the edges of their disk. Fig.\ref{fig:BPT1} shows the \cite{Baldwin1981} (BPT) diagrams and maps based on the [NII] and [OI] lines, which are commonly used to discriminate between regions powered by star formation, by a central AGN or by other mechanisms (LINER, shocks etc). While according to the [NII] diagram the ionizing source of the gas is star formation throughout the disk, there is a clear excess of [OI] emission at the disk edges, in a ring shape. Such an excess has been observed to be quite common in significant portions of the tails of stripped cluster galaxies \citep{Fossati2016,Poggianti2019,Tomicic2021b}, and it may arise  from a layer of mixed intracluster medium -interstellar medium gas at temperatures $\sim 1$keV, as reproduced by photoionization models published in \cite{Campitiello2021}.

{ The GASP sample includes also 14 control field galaxies (outside of clusters) lacking signs of stripping from B-band imaging. These are galaxies mostly in small groups and filaments which are not expected to be embedded in a very dense and hot intergalactic medium, therefore are not expected to be ram pressure stripped.
Interestingly}, the [OI] excess seen
in Fig.~\ref{fig:BPT1} is observed to occur in a very similar way also in our {\sl  field} control sample galaxies (plots not shown, sample presented in \citealt{Vulcani2018b}). This indicates that the unsual emission line ratio at the periphery of disks is {\sl not} a cluster-specific effect. 
We note that the regions with the [OI] excess are regions of diffuse ionized gas (DIG) (see the plots in \citealt{Tomicic2021a}).\footnote{We note that the vice-versa is not true, i.e. not all the DIG spaxels lie in the LINER region of the [OI] diagram.} 
{ We also stress that the [OI] excess observed in the tails of stripping galaxies and at the edges of control cluster and field galaxies of the GASP sample cannot originate from contamination by sky lines, nor it can be an artefact due to low S/N: retaining only spaxels with a much higher S/N cut (even above 4) the effect remains visible.}

The presence of diffuse ionized gas in the extraplanar regions up to several kpc from the disk, with enhanced low-ionization line ratios ([NII]/H$\alpha$, [SII]/H$\alpha$, [OI]/H$\alpha$) in disagreement with photoionization by OB stars, was already revealed a few decades ago by long-slit spectroscopy of a few edge-on galaxies \citep[see e.g.][]{Sokolowski1991}. Studies based on MaNGA \citep{Bundy2015} and SAMI \citep{Croom2012} data observed this excess in a few cases of edge-on galaxies \citep{Belfiore2016, Fogarty2012} but these previous integral field surveys did not have the necessary areal coverage to observe the whole LINER-like [OI] {\sl ring}.  
 
Different  ionization mechanisms  were proposed to reproduce such ratios: shocks originated e.g. in expanding structures such as superbubbles \citep{Rand1998,Collins2001}, or starburst-driven shocks \citep{Croom2012}; the so--called Turbulent Mixing Layers
\citep{Slavin1993}, where the line emitting regions is at the interface between cold ($T\le10^4$ K) and hot ($T > 10^5$ K) gas; hot evolved (pAGB) stars \citep{Belfiore2022}, or younger (age $<$ 25 Myr) stars in low-density gas \citep{McClymont2024},  producing a ionizing radiation harder than OB stars.
However, while in the cases cited above the excess occurs for all the BPT low-ionization line ratios,  in the GASP data we often observe an excess only in [OI]/H$\alpha$, whereas [NII]/H$\alpha$ and [SII]/H$\alpha$ are more in agreement with ionization by OB stars: this introduces a further challenge to the interpretation. Whether this can arise 
from the interaction with a surrounding medium hotter than the interstellar medium, which can be the circumgalactic and/or the intergalactic medium, with a mechanism similar to the one proposed for the stripped tails at the interface of the stripped gas and the intracluster medium, is a possibility that is worth exploring in the future. We defer the discussion of the [OI] excess issue to a separate paper (M. Radovich et al., in preparation), where a more detailed comparison of different photoionization models will be given.


\section{Discussion: identifying ram pressure using observations at different wavelengths}

In the previous sections, we have shown that an optically selected sample of stripping candidates obtained from blue-light imaging of galaxy clusters \citepalias{Poggianti2016} has a high success rate in identifying galaxies subject to RPS. As discussed in \cite{Vulcani2021}, this is not the case when selecting galaxies in the same way from general field samples: based on MUSE data as those used in this paper, among 27 GASP non-cluster galaxies in  \cite{Vulcani2021} we identified a plethora of physical mechanisms as the cause for the galaxy disturbance (galaxy-galaxy interactions, mergers, cosmic web enhancement, starvation, gas accretion and ram pressure in groups and filaments).
Together, these results show that  optical imaging selection is only effective when applied in cluster fields and requires deep and high-quality imaging.

Blue or UV imaging \citep{Smith2010, Vulcani2021, Durret2021, Durret2022, McPartland2016,  Ebeling2014, Cortese2007, Roman2019, Roberts2022a, Owers2012, George2024} is just one of the several methods used to identify RPS candidates. Observations of gas in different phases (molecular (typically CO, \citealt{Brown2021, Brown2023}), neutral (HI, \citealt{Serra2023, Serra2024, Ramatsoku2019, Ramatsoku2020, Deb2022, Deb2023}), ionized gas (through $\rm H\alpha$ imaging or integral field spectroscopy such as GASP, \citealt{Boselli2018, Moretti2022, Gondhalekar2024}), as well as
radio continuum and X-ray studies \citep{Sun2010, Roberts2021b, Roberts2022b, Sun2022, Ignesti2023b} all provide samples of ram pressure candidates, which are all affected by intrinsic incompleteness and observational biases. LOFAR and SKA-Low, in the coming years, will be efficient in finding radio tails on large areas of sky, also because the length of the radio tails increases at low frequencies \citep{Ignesti2022a, Roberts2024}. However, no observational method for identifying ram pressure ensures a complete, holistic view of the stripping phenomenon. An optical/ultraviolet imaging selection method assumes that at least some extraplanar star formation must be present in the stripped gas. This is not always the case, although, as we have shown in this paper, it turns out that with such an optical selection we end up observing a very diverse range of stripping strengths, and even cases with very little star formation in the tail can be identified \citep[see][]{Gullieuszik2020}. 

One might think that relying directly on a gas tracer (such, for example, $\rm H\alpha$ tracing the ionized gas, or an HI survey) could provide a "unbiased" complete sample. However, this is not the case, as demonstrated by the multi-wavelength observations of a sample like GASP. For example, there can be galaxies with prominent $\rm H\alpha$ tails and little neutral extraplanar gas \citep[see, e.g.,][]{Ramatsoku2020, Deb2022}.
In M. Ramatsoku et al. (in prep.) we start from a large sample of HI-selected galaxies that have neutral gas [HI] tails and we show that they do not always have significant star formation in the tail and thus do not display a visible extraplanar debris in optical+ultraviolet imaging. Radio continuum tails, so numerous in clusters and groups of galaxies \citep{Roberts2021a, Roberts2021b, Roberts2022b, Serra2024}, show the same behavior. 

It is thus important to realize that depending on the wavelength of selection, we obtain different samples, only partially overlapping. Much of the mismatch of the samples selected with different techniques can be attributed to different {\sl stages} of the stripping process. In GASP we have seen that the HI content of $\rm H\alpha$-selected jellyfish galaxies is generally already low \citep{Deb2022, Moretti2020b, Moretti2023}, whereas HI-rich galaxies with spectacular HI tails have not (yet?) developed a star-forming tail. While a detailed discussion of this is beyond the scope of this paper and is deferred to M. Ramatsoku et al. (in prep.), here it is important to consider whether the different JTypes we have discussed throughout this paper correspond to different stages of stripping.

Truncated disks are, for the reasons discussed in \S3.3, in an advanced stage that probably follows one of the other JTypes, and are the precursors of post-starburst cluster galaxies. For JTypes=0.3,0.5,1 and 2, it is harder to demonstrate an evolutionary sequence. Logically, such a sequence would go from an initial phase when the stripping has just begun (JType=0.3-0.5) to a subsequent stage with more prominent tails (JTypes=1 and 2). However, as we discussed when commenting Fig.~7, the location in the phase space indicates that the most spectacular tails in jellyfish galaxies form on first infall on radial orbits of gas-rich galaxies approaching with large velocities relative to the cluster (see also \citealt{Biviano2024}).\footnote{As discussed in \cite{Jaffe2018} and \cite{Luber2022}, more massive galaxies plunge deeper into the cluster before being stripped, and at that point they are in such a hostile environment that their tails become spectacular, although in the GASP sample there are also some lower mass galaxies with extreme tails that can be observed at quite low velocities radii, see Fig.~7.} As not all gas-rich galaxies infalling into clusters will have these orbital characteristics, in principle it might well be that some galaxies will never go through a JType=2 phase and will be stripped more gently. For example, most of JTypes=0.3-0.5 have relatively low velocities with respect to the cluster systemic velocity, although on their approach toward the cluster center they might gain a higher velocity moving upward and leftward in Fig.~7.

To conclude, most of the JTypes cannot be necessarily interpreted as different stages of an evolutionary sequence, and the reasons for the different tail morphologies and luminosities lies in the many factors determining the physics of the ram pressure phenomenon (the most important ones being galaxy mass, position in the phase-space diagram and cluster mass, see \citealt{Gullieuszik2020}) as well as the amount of gas in the galaxy when they fall into the cluster. 

\begin{table*}
{\tiny
\caption{Properties of galaxies in the stripping sample. Columns are: Galaxy ID, hosting cluster, coordinates, spectroscopic redshift, stellar mass, membership, JType, fraction and total luminosity of H$\alpha$ emission outside the galaxy disk, morphological type and references to the papers in which the galaxy is discussed. Hubble types are 6=Scd, 5=Sc, 4=Sbc, 3=Sb, 2=Sab, 1=Sa, 0=S0/a, -1=$\rm S0^{+}$, -2=S0, -4=E/S0, -5=E.}              
\label{tab:stripping}
\centering                                      
\addtolength{\tabcolsep}{-0.5em}
\begin{tabular}{cccccccccccl}          
\hline\hline                        
Name & Cluster & RA & DEC & z & $M_*$  &  Mem & JType & $f_{\rm H\alpha^{out}}$ & $\rm L_{H\alpha^{out}}$ & T-type  & References\\
&  & [J2000] & [J2000] &  & [$10^9 M_\odot$] &   &  &  & [$10^{39} erg/s/cm^2$] & & \\    
\hline                                   
  JO10 & A119 & 00:57:41.61 & -01:18:44.0 & 0.047055 & 57.2$\pm$8.5 &  1 & 3.0 & $0.02^{0.06}_{0.005}$ & $3.50^{9.1}_{0.7}$ & 3.0 & (14, 21, 30, 32, 35, 41, 59, 61, 65, 67, 69)\\
  JO102 & A3128 & 03:29:04.69 & -52:50:05.4 & 0.059411 & 10.2$\pm$1.9 & 1 & 0.5 & $0.01^{0.03}_{0.003}$ & $2.3^{5.8}_{0.5}$ & 2.7 & (14, 20, 21, 27, 30, 32, 35, 36, 41, 59, 61, 65, 69)\\
  JO112 & A3158 & 03:40:06.02 & -54:02:27.3 & 0.05829 & 4.1$\pm$0.7 &  1 & 0.5 & $0.02^{0.03}_{0.01}$ & $3.22^{5.63}_{1.56}$ & 3.7 & (14, 21, 30, 32, 35, 36, 41, 61, 65, 69)\\
  JO113 & A3158 & 03:41:49.17 & -53:24:13.7 & 0.055185 & 4.9$\pm$1.0 &  1 & 2.0 & $0.05^{0.06}_{0.05}$ & $14.0^{16.7}_{12.1}$ & 4.9 & (13, 14, 21, 27, 30, 32, 35, 36, 41, 59, 61, 65, 69)\\
  JO119 & A3395 & 06:29:59.10 & -54:47:38.7 & 0.049621 & 26.2$\pm$6.1 & 1 & 0.0 & --- & -- & 3.5 & (36, 65)\\
  JO123 & A3528b & 12:53:01.03 & -28:36:52.6 & 0.055028 & 7.5$\pm$1.6 &  1 & 0.5 & --- & -- & 5.2 &(14, 20, 30, 32, 35, 36, 41, 59, 61, 65,69)\\
  JO128 & A3530 & 12:54:56.84 & -29:50:11.2 & 0.049981 & 8.0$\pm$2.0 &  1 & 1.0 & $0.07^{0.09}_{0.06}$ & $12.5^{16.2}_{10.2}$ & 4.8 &(14, 20, 21, 27, 30, 32, 35, 36, 41, 59, 61, 65, 69)\\
  JO13 & A119 & 00:55:39.68 & -00:52:36.0 & 0.047851 & 6.6$\pm$1.1 &  1 & 0.5 & $0.02^{0.04}_{0.005}$ & $5.2^{11.9}_{1.6}$ & -- & (14, 21, 30, 32, 35, 36, 41, 61, 65, 69)\\
  JO134 & A3530 & 12:54:38.33 & -30:09:26.5 & 0.016585 & *1.1$\pm$0.3 & 0 & 1.0 & $0.12^{0.12}_{0.11}$ & $53.8^{520}_{52.0}$ & -- & (33, 65, 70)\\
  JO135 & A3532 & 12:57:04.30 & -30:22:30.3 & 0.054423 & 98.2$\pm$16.7 &  1 & 1.0 & $0.03^{0.03}_{0.02}$ & $15.1^{17.9}_{12.6}$ & 2.2 & (6, 13, 14, 19, 21, 27, 30, 32, 35, 36, 41, 59, 61, 69 \\
 & & & & & & & & & & &65)\\
  JO138 & A3532 & 12:56:58.51 & -30:06:06.3 & 0.057234 & 4.5$\pm$1.5 &  1 & 0.3 & $0.01^{0.09}_{0.00}$  & 4.72 & 4.2 & (14, 21, 27, 30, 32, 35, 36, 41, 59, 61, 65, 69)\\
  JO141 & A3532 & 12:58:38.38 & -30:47:32.2 & 0.058652 & 48.1$\pm$14.6 & 1 & 1.0 & $0.03^{0.04}_{0.03}$ & $15.6^{18.2}_{13.9}$ & 3.2 & (13, 14, 21, 27, 30, 32, 35, 36, 41, 59, 61, 65, 69)\\
  JO144 & A3556 & 13:24:32.43 & -31:06:59.0 & 0.051476 & 32.3$\pm$10.2 &  1 & 1.0 & $0.03^{0.05}_{0.02}$ & $7.25^{12.1}_{3.84}$ & 4.7 & (14, 21, 27, 30, 32, 35, 36, 41, 61, 65, 69)\\
  JO147 & A3558 & 13:26:49.73 & -31:23:45.5 & 0.050608 & 107.6$\pm$19.7 &  1 & 2.0 & $0.13^{0.14}_{0.14}$ & $91.1^{94.0}_{87.2}$ & 4.0 &(13, 14, 21, 27, 30, 32, 35, 36, 41, 59, 60, 61, 69 \\
 & & & & & & & & & & & 65, 67, 70, 71)\\
  JO149 & A3558 & 13:28:10.53 & -31:09:50.2 & 0.043824 & 0.6$\pm$0.2 &  1 & 2.0 & $0.32^{0.35}_{0.31}$ & $30.9^{35.3}_{28.6}$ & 4.9 & (14, 21, 27, 36, 41, 59, 61, 65, 70)\\
  JO153 & A3558 & 13:28:15.15 & -31:01:57.9 & 0.046865 & 2.6$\pm$0.6 &  1 & -99 & --- & 9.90 & 3.6 & (21, 36, 41, 59, 61, 65, 69)\\
  JO156 & A3558 & 13:28:34.46 & -31:01:26.8 & 0.05105 & 4.4$\pm$1.2 &  1 & 1.0 & $0.09^{0.13}_{0.06}$ & $6.03^{9.20}_{3.93}$ & 4.7 & (14, 21, 30, 32, 35, 41, 59, 61, 65, 69)\\
  JO157 & A3558 & 13:28:18.18 & -31:48:18.8 & 0.044641 & 13.1$\pm$2.9 &  1 & -9.0 & --- & -- & 4.2 & (36, 60, 65)\\
  JO159 & A3558 & 13:26:35.70 & -30:59:36.9 & 0.047963 & 6.6$\pm$1.6 &  1 & 1.0 & $0.10^{0.11}_{0.09}$ & $21.2^{24.5}_{19.7}$ & 2.8 & (14, 21, 27, 30, 32, 35, 36, 41, 59, 60, 61, 65, 69)\\
  JO160 & A3558 & 13:29:28.62 & -31:39:25.3 & 0.048275 & 11.5$\pm$2.9 &  1 & 2.0 & $0.02^{0.04}_{0.02}$ & $7.19^{11.3}_{5.0}$ & 2.5 & (13, 14, 21, 27, 30, 32, 35, 36, 41, 59, 61, 65, 69, 70)\\
  JO162 & A3560 & 13:31:29.92 & -33:03:19.6 & 0.045414 & 2.7$\pm$0.6 & 1 & 2.0 & $0.13^{0.17}_{0.11}$ & $13.3^{17.9}_{10.5}$ & 3.5 &(14, 21, 27, 30, 32, 35, 36, 41, 59, 61, 65, 67, 69)\\
  JO17 & A147 & 01:08:35.33 & +01:56:37.0 & 0.045082 & 14.4$\pm$2.8 & 1 & 0.5 & $0.01^{0.03}_{0.002}$ & $1.6^{5.1}_{0.3}$ & 3.4 & (14, 20, 21, 27, 30, 32, 35, 36, 41, 59, 61, 65, 69)\\
  JO171 & A3667 & 20:10:14.70 & -56:38:30.6 & 0.052121 & 40.8$\pm$5.7 &  1 & 2.0 & $0.24^{0.24}_{0.24}$ & $95.9^{98.0}_{93.6}$ & 3.9 & (5, 13, 14, 21, 27, 30, 32, 35, 36, 41, 59, 61,   \\
 & & & & & & & & & & & 65, 67, 69, 70)\\
  JO175 & A3716 & 20:51:17.60 & -52:49:21.8 & 0.04675 & 31.6$\pm$5.1 &  1 & 2.0 & $0.12^{0.14}_{0.11}$ & $36.9^{41.4}_{33.0}$ & 1.5 & (6, 13, 14, 19, 21, 27, 29, 30, 32, 35, 36, 41,  \\
 & & & & & & & & & & & 50, 51, 56, 59, 61, 63, 65, 67, 68, 69) \\
  JO179 & A3809 & 21:47:07.07 & -43:42:18.2 & 0.06182 & 3.3$\pm$0.9 &  1 & 0.5 & $0.04^{0.06}_{0.02}$ & $5.8^{8.7}_{3.7}$ & 4.0 & (14, 21, 30, 32, 35, 36, 41, 59, 61, 65, 69)\\
  JO180 & A3809 & 21:45:15.00 & -44:00:31.2 & 0.064665 & 9.6$\pm$2.1 & 1 & 0.5 & $0.04^{0.07}_{0.02}$ & $3.4^{6.6}_{1.5}$ & 5.0 & (14, 20, 21, 27, 30, 32, 35, 36, 41, 59, 61, 65, 69)\\
  JO181 & A3880 & 22:28:03.80 & -30:18:03.8 & 0.059928 & 1.2$\pm$0.3 & 1 & 1.0 & $0.12^{0.17}_{0.09}$ & $6.97^{11.1}_{4.97}$ & 4.7 & (14, 21, 27, 30, 32, 35, 36, 41, 59, 61, 65, 69)\\
  JO190 & A3880 & 22:26:53.62 & -30:53:11.1 & 0.013218 & *2.2$\pm$0.6 &  0 & -9.0 & --- & -- & -- & (33, 65, 70)\\
  JO194 & A4059 & 23:57:00.68 & -34:40:50.1 & 0.041951 & 150.0$\pm$31.1 &  1 & 2.0 & $0.15^{0.15}_{0.15}$ & $131.0^{133}_{129}$ & 4.5 & (6, 13, 14, 19, 21, 27, 29, 30, 32, 35, 36, 41,  \\
 & & & & & & & & & & & 44, 47, 57, 59, 61, 65, 66, 69)\\
  JO197 & A754 & 09:06:32.58 & -09:31:27.3 & 0.056163 & 10.9$\pm$2.6 & 1 & 0.5 & $0.01^{0.03}_{0.05}$ & $2.3^{5.0}_{0.8}$ & 3.2 & (14, 21, 27, 30, 32, 35, 36, 41, 59, 61, 65, 69)\\
  JO20 & A147 & 01:08:55.06 & +02:14:20.8 & 0.147063 & *71.5$\pm$12.6 &  0 & -99 & --- & -- & -- & (33, 65)\\
  JO200 & A85 & 00:42:05.03 & -09:32:03.8 & 0.052675 & 66.0$\pm$10.5 & 1 & 1.0 & $0.03^{0.04}_{0.02}$ & $16.3^{22.0}_{12.3}$ & 3.8 & (14, 21, 27, 29, 30, 32, 35, 36, 41, 59, 60, 61,  \\
 & & & & & & & & & & & 65, 69, 70)\\
  JO201 & A85 & 00:41:30.29 & -09:15:45.9 & 0.044631 & 62.1$\pm$8.0 & 1 & 2.0 & $0.18^{0.19}_{0.18}$ & $252^{262}_{242}$ & 1.8 & (2, 6, 10, 11, 13, 14, 15, 18, 19, 21, 26, 27, 29,  \\
 & & & & & & & & & & & 30, 31, 32, 35, 36, 41, 50, 51, 53, 56, 57, 59, 61,  \\
 & & & & & & & & & & & 63, 65, 67, 68, 69)\\
  JO204 & A957 & 10:13:46.83 & -00:54:51.0 & 0.042372 & 40.5$\pm$5.9 &  1 & 2.0 & $0.15^{0.16}_{0.15}$ & $69.9^{75.5}_{66.8}$ & 2.0 & (4, 6, 10, 13, 14, 19, 21, 25, 27, 30, 31, 32, 35,  \\
 & & & & & & & & & & & 36, 41, 50, 51, 53, 56, 59, 61, 63, 65, 67, 68, 69)\\
  JO205 & IIZW108 & 21:13:46.12 & +02:14:20.4 & 0.044743 & 3.3$\pm$0.8 & 1 & 0.3 & $0.01^{0.09}_{0.00}$ & 12.8 & 3.1 & (14, 20, 21, 27, 30, 32, 35, 36,  59, 61, 65, 69)\\
  JO206 & IIZW108 & 21:13:47.41 & +02:28:34.4 & 0.051089 & 91.2$\pm$9.3 & 1 & 2.0 & $0.15^{0.16}_{0.15}$ & $156^{161}_{160}$ & 4.1 & (1, 6, 10, 13, 14, 17, 19, 21, 26, 27, 28, 29, 30,   \\
 & & & & & & & & & & & 31, 32, 34, 35,  36, 40, 41, 43, 50, 51, 53, 56, 59, 68  \\
 & & & & & & & & & & & 61, 63, 65, 67, 69)\\
  JO23 & A151 & 01:08:08.10 & -15:30:41.8 & 0.055097 & 4.6$\pm$1.2 & 1 & 3.0 & $0.04^{0.08}_{0.02}$ & $2.86^{5.94}_{1.31}$ & 2.5 & (14, 21, 30, 32, 35, 41, 59, 61, 65, 69, 70)\\
  JO24 & A151 & 01:08:08.00 & -15:10:54.8 & 0.099373 & *16.8$\pm$2.1 &  0 & 1.0 & --- & 21.7 & -0.4 & (59, 65)\\
  JO27 & A151 & 01:10:48.56 & -15:04:41.6 & 0.049314 & 3.2$\pm$0.8 & 1 & 1.0 & $0.24^{0.27}_{0.22}$ & $35.3^{40.8}_{31.8}$ & 3.1 & (14, 21, 30, 32, 35, 36, 41, 59, 61, 65, 69)\\
  JO28 & A151 & 01:10:09.31 & -15:34:24.5 & 0.054292 & 2.3$\pm$0.4 & 1 & 1.0 & $0.18^{0.24}_{0.13}$ & $7.22^{10.6}_{4.78}$ & 4.4 & (14, 21, 27, 30, 32, 35, 36, 41, 59, 61, 65, 69)\\
  JO36 & A160 & 01:12:59.42 & +15:35:29.4 & 0.040743 & 65.0$\pm$11.6 & 1 & 3.0 & $0.05^{0.11}_{0.04}$ & $5.30^{11.8}_{3.74}$ & 5.0 & (3, 14, 21, 30, 32, 35, 41, 58, 59, 61, 65, 69, 70)\\
  JO41 & A1631a & 12:53:54.79 & -15:47:20.1 & 0.047729 & 16.0$\pm$2.5 &  1 & 0.5 & $0.05^{0.11}_{0.02}$ & $3.65^{8.3}_{1.3}$ & 3.7 & (14, 20, 21, 27, 30, 32, 35, 36, 41, 59, 61, 65, 69)\\
  JO45 & A168 & 01:13:16.58 & +00:12:05.8 & 0.042559 & 1.5$\pm$0.3 &  1 & 0.5 & $0.04^{0.02}_{0.06}$ & $7.5^{17.3}_{2.6}$ & 4.8 & (14, 20, 21, 27, 30, 32, 35, 36, 41, 59, 61, 65, 69)\\
  JO47 & A168 & 01:15:57.67 & +00:41:36.0 & 0.04274 & 4.0$\pm$0.6 &  1 & 1.0 & $0.08^{0.12}_{0.04}$ & $7.01^{11.5}_{3.83}$ & 3.9 & (14, 21, 27, 29, 30, 32, 35, 36, 41, 59, 61, 65, 69)\\
  JO49 & A168 & 01:14:43.85 & +00:17:10.1 & 0.045125 & 47.4$\pm$6.5 &  1 & 1.0 & $0.02^{0.03}_{0.01}$ & $6.66^{10.3}_{4.1}$ & 4.4 &(13, 14, 21, 27, 29, 30, 32, 35, 36, 41, 43, 59,  \\
 & & & & & & & & & & & 60, 61, 65, 69)\\
  JO5 & A1069 & 10:41:20.38 & -08:53:45.6 & 0.064783 & 18.7$\pm$4.4 &  1 & 0.3 & $0.02^{0.02}_{0.04}$ & -- & 3.7 & (14, 20, 21, 27, 30, 32, 35, 36, 59, 61, 65, 69)\\
  JO60 & A1991 & 14:53:51.57 & +18:39:06.4 & 0.062187 & 25.1$\pm$6.5 &  1 & 2.0 & $0.08^{0.09}_{0.07}$ & $58.4^{65.0}_{53.0}$ & 5.0 & (13, 14, 21, 27, 30, 32, 35, 36, 41, 43, 47, 57,  \\
 & & & & & & & & & & &  59, 60, 61, 65, 66, 69)\\
  JO68 & A2399 & 21:56:22.00 & -07:54:29.0 & 0.056117 & 9.9$\pm$2.1 & 1 & 0.3 & $0.01^{0.04}_{0.003}$ & $2.2^{6.3}_{0.4}$ & 5.0 & (14, 20, 21, 27, 30, 32, 35, 36, 59, 61, 65, 69)\\
  JO69 & A2399 & 21:57:19.20 & -07:46:43.8 & 0.054984 & 8.0$\pm$2.1 & 1 & 1.0 & $0.04^{0.05}_{0.03}$ & $11.4^{14.8}_{8.8}$ & 5.2 & (14, 21, 29, 30, 32, 35, 36, 41, 59, 61, 65, 69, 70)\\
  JO70 & A2399 & 21:56:04.07 & -07:19:38.0 & 0.057759 & 29.0$\pm$6.1 & 1 & 1.0 & $0.05^{0.06}_{0.05}$ & $28.3^{34.6}_{24.8}$ & 4.0 & (14, 21, 27, 29, 30, 32, 35, 36, 41, 59, 61, 65, 69, 70)\\
  JO73 & A2415 & 22:04:25.99 & -05:14:47.0 & 0.071251 & *10.6$\pm$2.6 &  0 & 1.0 & $0.10^{0.12}_{0.09}$ & $15.8^{19.1}_{13.5}$ & 4.1 & (14, 20, 27, 30, 32, 35, 36,  61, 65, 69)\\
  JO85 & A2589 & 23:24:31.36 & +16:52:05.3 & 0.035453 & 46.3$\pm$9.4 &  1 & 2.0 & $0.16^{0.16}_{0.15}$ & $151^{153}_{146}$ & 2.0 & (14, 21, 27, 29, 30, 32, 35, 36, 41, 43, 59, 61,   \\
 & & & & & & & & & & & 65, 69, 70)\\
\hline 
\end{tabular}
}
\end{table*}

\addtocounter{table}{-1} 

\begin{table*}
{\tiny
\caption{Continued. }              
\label{tab:stripping2}
\centering                                      
\addtolength{\tabcolsep}{-0.5em}
\begin{tabular}{cccccccccccl}          
\hline\hline                        
Name & Cluster & RA & DEC & z & $M_*$  &  Mem & JType & $f_{\rm H\alpha^{out}}$ & $\rm L_{H\alpha^{out}}$ & T-type  & References\\
&  & [J2000] & [J2000] &  & [$10^9 M_\odot$] &   &  &  & [$10^{39} erg/s/cm^2$] & & \\    
\hline         
  JO89 & A2593 & 23:26:00.60 & +14:18:26.3 & 0.04235 & 5.3$\pm$1.0 & 1 & 0.5 & $0.07^{0.07}_{0.02}$ & $2.0^{6.1}_{0.5}$ & 5.5 & (14, 20, 21, 27, 30, 32, 35, 36, 41, 59, 61, 65, 69)\\
  JO93 & A2593 & 23:23:11.74 & +14:54:05.0 & 0.036984 & 35.0$\pm$5.8 &  1 & 1.0 & $0.11^{0.11}_{0.11}$ & $49.1^{51.0}_{47.0}$ & 2.8 & (14, 21, 27, 29, 32, 35, 36, 41, 59, 61, 65, 69, 70)\\
  JO95 & A2657 & 23:44:26.66 & +09:06:55.8 & 0.043322 & 2.3$\pm$0.4 &  1 & 1.0 & $0.10^{0.13}_{0.08}$ & $8.5^{12.0}_{6.8}$ & 5.0 & (13, 14, 21, 27, 36, 41, 59, 61, 65)\\
  JW10 & A500 & 04:39:18.19 & -21:57:49.6 & 0.07179 & 9.8$\pm$2.4 & 1 & -9.0 & $0.12^{0.14}_{0.11}$ & $23.9^{26.8}_{21.5}$ & 5.2 & (14, 21, 30, 32, 35, 36, 65)\\
  JW100 & A2626 & 23:36:25.06 & +21:09:02.5 & 0.061891 & 293.2$\pm$70.2 & 1 & 2.0 & $0.20^{0.20}_{0.20}$ & $225^{227}_{223}$ & 1.4 & (6, 13, 14, 19, 21, 22, 23, 27, 30, 31, 32, 35, 36,   \\
 & & & & & & & & & & &37, 38, 39, 40, 41, 43, 50, 51, 53, 56, 57, 59, 61,   \\
 & & & & & & & & & & &63, 65, 66, 67, 68, 69, 72)\\
  JW105 & A2717 & 00:03:00.49 & -36:06:39.7 & 0.051505 & *1.5$\pm$0.2 & 1 & 0 & 0.00 & -- & -4.3  & (59, 65)\\
  JW108 & A3376 & 06:00:47.96 & -39:55:07.4 & 0.047688 & 30.2$\pm$7.7 & 1 & 3.0 & $0.03^{0.06}_{0.01}$ & $1.86^{4.81}_{0.41}$ & 6.1 & (14, 21, 30, 32, 35, 41, 59, 60, 61, 65, 66, 69)\\
  JW115 & A3497 & 12:00:47.95 & -31:13:41.6 & 0.072485 & 5.3$\pm$1.4 &  & 1.0 & $0.09^{0.14}_{0.05}$ & $10.0^{17.2}_{5.3}$ & 2.8 & (14, 21, 27, 30, 32, 35, 36, 41, 59, 61, 65, 69)\\
  JW29 & A1644 & 12:57:49.48 & -17:39:57.1 & 0.043119 & 3.2$\pm$0.6 & 1 & 0.5 & $0.06^{0.12}_{0.03}$ & $5.86^{13.2}_{2.6}$ & 6.1 & (14, 21, 30, 32, 35, 36, 41, 59, 61, 65, 69)\\
  JW36 & A1644 & 12:56:44.22 & -17:39:48.6 & 0.050062 & 0.5$\pm$0.1 & 1 & 4.0 & 0.00  & -- & 5.2 & (59, 65)\\
  JW39 & A1668 & 13:04:07.71 & +19:12:38.5 & 0.066319 & 161.4$\pm$28.5 &  1 & 2.0 & $0.14^{0.14}_{0.14}$  & $125^{129}_{120}$ & 3.5 & (13, 14, 21, 27, 29, 30, 32, 35, 36, 40, 41, 43, 47,  \\
 & & & & & & & & & & & 50, 51, 56, 57, 59, 61, 63, 65, 66, 67, 68, 69)\\
  JW56 & A1736 & 13:27:03.03 & -27:12:58.2 & 0.038664 & 1.1$\pm$0.3 & 1 & 2.0 & $0.33^{0.38}_{0.30}$ & $14.1^{17.2}_{12.0}$ & 3.0 & (14, 21, 27, 29, 30, 32, 35, 36, 41, 59, 61, 65, 67, 69, 70)\\
\hline 
\end{tabular}
\tablebib{{\scriptsize
(1)~\citet{Poggianti2017a}, (2)~\citet{Bellhouse2017}, (3)~\citet{Fritz2017}, (4)~\citet{Gullieuszik2017}, (5)~\citet{Moretti2018a}, (6)~\citet{Poggianti2017b}, 
(9)~\citet{Jaffe2018}, (10)~\citet{Moretti2018b}, (11)~\citet{George2018}, 
(13)~\citet{Poggianti2019a}, (14)~\citet{Vulcani2018b}, (15)~\citet{Bellhouse2019}, 
(17)~\citet{Ramatsoku2019}, (18)~\citet{George2019}, (19)~\citet{Radovich2019}, (20)~\citet{Vulcani2019b}, (21)~\citet{Gullieuszik2020}, (22)~\citet{Moretti2020a}, (23)~\citet{Poggianti2019b}, 
(25)~\citet{Deb2020}, (26)~\citet{Ramatsoku2020}, (27)~\citet{Franchetto2020}, (28)~\citet{Mueller2021a}, (29)~\citet{Bellhouse2021}, (30)~\citet{Vulcani2020b}, (31)~\citet{Moretti2020b}, (32)~\citet{Tomicic2021a}, (33)~\citet{Vulcani2021}, (34)~\citet{Campitiello2021}, (35)~\citet{Tomicic2021b}, (36)~\citet{Franchetto2021a}, (37)~\citet{Luber2022}, (38)~\citet{Ignesti2022a}, (39)~\citet{Deb2022}, (40)~\citet{Franchetto2021b}, (41)~\citet{Peluso2022}, (43)~\citet{Ignesti2022b}, (44)~\citet{Bartolini2022}, (45)~\citet{Sanchez2023}, 
(48)~\citet{George2023}, (50)~\citet{Giunchi2023a}, (51)~\citet{Gullieuszik2023}, (53)~\citet{Bacchini2023}, (56)~\citet{Giunchi2023b}, (57)~\citet{Tomicic2024}, (58)~\citet{Ignesti2023b}, (59)~\citet{Marasco2023}, (60)~\citet{Moretti2023}, (61)~\citet{Peluso2023}, (63)~\citet{Werle2024}, (65)~\citet{Salinas2024} (66)~\citet{George2024}, (67)~\citet{Ignesti2024}, (68)~\citet{Giunchi2025}, (69)~\citet{Peluso2025}, (70)~\citet{George2025},
(71)~\citet{Merluzzi2013},
(72)~\citet{Sun2022}.}}
\tablefoot{Stellar mass is computed within the galaxy disk defined in Sec\ref{sec:analysis}, except for galaxies with an asterisks, for which the galaxy disk was not measured due to the irregularities of the galaxy and the disk is defined as 1$\sigma$ above the background level on a 2D image of the near-H$\alpha$ continuum.}
}
\end{table*}

\begin{table*}
{\tiny
\caption{Properties of galaxies in the control sample. Columns are: Galaxy ID, hosting cluster, coordinates, spectroscopic redshift, stellar mass, membership, JType, fraction and total luminosity of H$\alpha$ emission outside the galaxy disk, morphological type, star forming classification and references to the GASP papers in which the galaxy is discussed. }              
\label{tab:control}
\centering                                      
\addtolength{\tabcolsep}{-0.5em}
\begin{tabular}{ccccccccccccl}          
\hline\hline                        
Name & Cluster & RA & DEC & z & $M_*$  &  Mem & JType & $f_{\rm H\alpha^{out}}$ & $\rm L_{H\alpha^{out}}$ & T-type (vis) & Comments & References \\
&  & [J2000] & [J2000] &  & [$10^9 M_\odot$] &   &  &  & [$10^{39 erg/s/cm^2}$] &  &\\    
\hline                                   
  A3128\_B\_0148 & A3128 & 03:27:31.09 & -52:59:07.7 & 0.057459 & 7.1$\pm$1.6 & 1 & 0.5 & $0.02^{0.04}_{0.01}$ & $3.2^{6.7}_{1.4}$ & 2 & SF & (14, 20, 27, 30, 32, 35, 36, 59,   \\
 & & & & & & & & & & & &61)\\
  A3266\_B\_0257 & A3266 & 04:27:52.58 & -60:54:11.6 & 0.058359 & 8.3$\pm$1.8 &  1 & 1.0 & $0.18^{0.20}_{0.17}$ & $23.5^{26.2}_{21.7}$ & 3.3 & SF & (14, 20, 27, 30, 32, 35, 36, 59,   \\
 & & & & & & & & & & & & 61)\\
  A3376\_B\_0261 & A3376 & 06:00:13.68 & -39:34:49.2 & 0.050596 & 34.0$\pm$6.5 &  1 & 0.0 & --- & --- & 2.5 & SF & (14, 20, 27, 30, 32, 35, 36, 59,  \\
 & & & & & & & & & & & & 60, 61)\\
  A970\_B\_0338 & A970 & 10:19:01.65 & -10:10:36.9 & 0.059026 & 11.6$\pm$2.1 &  1 & 0.5 & $0.01^{0.0}_{0.03}$ & $2.2^{5.4}_{0.5}$ & 4.3 & SF & (14, 20, 27, 30, 32, 35, 36, 59,   \\
 & & & & & & & & & & & &61)\\
  A1069\_B\_0103 & A1069 & 10:39:36.49 & -08:56:34.8 & 0.063479 & 21.5$\pm$3.6 &  1 & 4.0 & -- & -- & -1.2 & k & (24)\\
  A3128\_B\_0248 & A3128 & 03:29:23.42 & -52:26:02.9 & 0.053022 & 19.7$\pm$2.4  &  1 & 4.0 & -- & -- & -2.3 & k+a & (24)\\
  A3158\_11\_91 & A3158 & 03:41:16.75 & -53:24:00.6 & 0.061346 & 12.9$\pm$2.0 &  1 & 4.0 & -- & -- & 2.3 & a+k& (24)\\
  A3158\_B\_0223 & A3158 & 03:41:59.80 & -53:28:04.2 & 0.056209 & 20.4$\pm$2.4  & 1 & 4.0 & -- & -- & 1.1 & k & (24)\\
  A3158\_B\_0234 & A3158 & 03:42:24.68 & -53:29:26.0 & 0.066023 & 7.2$\pm$1.2  & 1 & 4.0 & -- & -- & 1.0 & a+k & (24)\\
  A3376\_B\_0214 & A3376 & 06:00:43.18 & -39:56:41.6 & 0.047285 & 5.4$\pm$1.1 &  1 & 4.0 & -- & -- & 1.2 & k+a & (24)\\
  A500\_22\_184 & A500 & 04:38:46.41 & -22:13:22.4 & 0.072477 & 3.6$\pm$0.5 & 1 & 4.0 & -- & -- & -1.9 & a+k & (24)\\
  A500\_F\_0152 & A500 & 04:38:21.25 & -22:13:02.2 & 0.069506 & 1.5$\pm$0.4  & 1 & 4.0 & -- & -- & 2.7 & a+k& (24)\\
\hline                                             
\end{tabular}
\tablebib{{\scriptsize
(14)~\citet{Vulcani2018b},  (20)~\citet{Vulcani2019b}, (24)~\citet{Vulcani2020a}, (27)~\citet{Franchetto2020}, (30)~\citet{Vulcani2020b},  (32)~\citet{Tomicic2021a},  (35)~\citet{Tomicic2021b}, (36)~\citet{Franchetto2021a},  (59)~\citet{Marasco2023}, (60)~\citet{Moretti2023}, (61)~\citet{Peluso2023}.}}
}
\end{table*}


\section{Summary}
We have presented the complete sample of cluster galaxies from the GASP survey, including 64 RPS candidates and 12 control (undisturbed) galaxies in 39 galaxy cluster fields at z=0.04-0.07.

Using MUSE integral-field spectroscopy of these galaxies, we have assessed whether the gas disturbed morphology originates from RPS based on ionized gas and stellar velocity maps. The visual classification of gas disturbance (=stripping strenght = JType), from very weak/weak to strong to extreme cases (jellyfishes) of ram pressure is well correlated with the fraction and the amount of $\rm H\alpha$ luminosity that is in a stripped tail, outside of the stellar disk.
GASP targets were selected from an optically-selected sample of stripping candidates based on B-band images. Comparing the imaging-based and the MUSE-based assessment of stripping strength, we find a very good correlation, and some interesting cases where the MUSE data show how superior is to observe directly the ionized gas to investigate not as much the existence, but the extent of the tail.

Of the 64 RPS imaging candidates, the MUSE data confirm 56+1\footnote{56 still have gas, and one has no gas left but a visible tail of recently formed stars.} of them. Thus, 89\% of the candidates have been confirmed, showing that a blue-light imaging selection in clusters, and only in clusters, is likely to yield confirmed cases in the vast majority of cases. Failures are mostly interacting/mergers and only one is a chance superposition of sources at different redshifts. 
We have discussed the properties of the whole sample of ram pressure stripped galaxies, including stellar mass and morphological distributions, and we present the location of the various JTypes in the phase-space diagram.
The velocity dispersions of clusters that host ram pressure stripped galaxies in the GASP sample range from less than 500 to over 1000 $\rm km \, s^{-1}$. We show that all stripping levels of strength, from weak to extreme, are present in clusters over the whole range of velocity dispersion, thus cluster masses.

Analyzing more in detail one of the JType classes, the one of truncated $\rm H\alpha$ disks, we conclude that they are in an advanced stage of stripping and show that their past stripping history can be studied from star formation history reconstruction.

Analyzing the star-forming galaxies of the control sample, we find that 3 out of 4 already show weak but detectable signs of stripping that were not identified in the imaging. A notable property of these galaxies, as well as all GASP control sample galaxies even in the field, is the fact that they display a ring of peculiar emission line ratios surrounding the disk: from BPT diagnostic diagrams, such rings appear as simply star-forming when using the [NII] line but have an excess of LINER-like [OI] emission. We mention possible causes of this phenomenon, which was not identified before as a ring and is common in the outskirts of all spirals, both in clusters and in the field.

Finally, we have discussed the biases and incompleteness of this and other methods to select ram pressure stripped galaxies, and we have reasoned that different JTypes do not always and necessarily correspond to an evolutionary sequence of different stages of RPS.

A constantly updated list of all the papers from the GASP survey can be found at
https://web.oapd.inaf.it/gasp/.

\begin{acknowledgements}
Based on observations collected at the European Organization for Astronomical Research in the Southern Hemisphere under ESO programme 196.B-0578. This project has received funding from the European Research Council (ERC) under the European Union's Horizon 2020 research and innovation programme (grant agreement No. 833824). We thank the INAF GO grant 2023 ``Identifying ram pressure induced unwinding arms in cluster spirals'' (PI Vulcani). RS acknowledges financial support from FONDECYT Regular projects 1230441 and 1241426, and also gratefully acknowledges financial support from ANID -- MILENIO -- NCN2024\_112. JF acknowledges financial support from the UNAM- DGAPA-PAPIIT IN110723 grant, México.
\end{acknowledgements}

\bibliography{gasp_all.bib}{}

\begin{thebibliography}{167}
\expandafter\ifx\csname natexlab\endcsname\relax\def\natexlab#1{#1}\fi

\bibitem[{{Akerman} {et~al.}(2023){Akerman}, {Tonnesen}, {Poggianti}, {Smith},
  \& {Marasco}}]{Akerman2023}
{Akerman}, N., {Tonnesen}, S., {Poggianti}, B.~M., {Smith}, R., \& {Marasco},
  A. 2023, \apj, 948, 18

\bibitem[{{Akerman} {et~al.}(2024){Akerman}, {Tonnesen}, {Poggianti}, {Smith},
  {Marasco}, {Kulier}, {M{\"u}ller}, \& {Vulcani}}]{Akerman2024}
{Akerman}, N., {Tonnesen}, S., {Poggianti}, B.~M., {et~al.} 2024, \mnras, 527,
  9505

\bibitem[{{Bacchini} {et~al.}(2023){Bacchini}, {Mingozzi}, {Poggianti},
  {Moretti}, {Gullieuszik}, {Marasco}, {Cervantes Sodi},
  {S{\'a}nchez-Garc{\'\i}a}, {Vulcani}, {Werle}, {Paladino}, \&
  {Radovich}}]{Bacchini2023}
{Bacchini}, C., {Mingozzi}, M., {Poggianti}, B.~M., {et~al.} 2023, \apj, 950,
  24

\bibitem[{{Bah{\'e}} \& {McCarthy}(2015)}]{Bahe2015}
{Bah{\'e}}, Y.~M. \& {McCarthy}, I.~G. 2015, \mnras, 447, 969

\bibitem[{{Baldwin} {et~al.}(1981){Baldwin}, {Phillips}, \&
  {Terlevich}}]{Baldwin1981}
{Baldwin}, J.~A., {Phillips}, M.~M., \& {Terlevich}, R. 1981, \pasp, 93, 5

\bibitem[{{Bartolini} {et~al.}(2022){Bartolini}, {Ignesti}, {Gitti},
  {Brighenti}, {Wolter}, {Moretti}, {Vulcani}, {Poggianti}, {Gullieuszik},
  {Fritz}, \& {Tomi{\v{c}}i{\'c}}}]{Bartolini2022}
{Bartolini}, C., {Ignesti}, A., {Gitti}, M., {et~al.} 2022, \apj, 936, 74

\bibitem[{{Belfiore} {et~al.}(2016){Belfiore}, {Maiolino}, {Maraston},
  {Emsellem}, {Bershady}, {Masters}, {Yan}, {Bizyaev}, {Boquien}, {Brownstein},
  {Bundy}, {Drory}, {Heckman}, {Law}, {Roman-Lopes}, {Pan}, {Stanghellini},
  {Thomas}, {Weijmans}, \& {Westfall}}]{Belfiore2016}
{Belfiore}, F., {Maiolino}, R., {Maraston}, C., {et~al.} 2016, \mnras, 461,
  3111

\bibitem[{{Belfiore} {et~al.}(2022){Belfiore}, {Santoro}, {Groves},
  {Schinnerer}, {Kreckel}, {Glover}, {Klessen}, {Emsellem}, {Blanc}, {Congiu},
  {Barnes}, {Boquien}, {Chevance}, {Dale}, {Kruijssen}, {Leroy}, {Pan},
  {Pessa}, {Schruba}, \& {Williams}}]{Belfiore2022}
{Belfiore}, F., {Santoro}, F., {Groves}, B., {et~al.} 2022, \aap, 659, A26

\bibitem[{{Bellhouse} {et~al.}(2017){Bellhouse}, {Jaff{\'e}}, {Hau}, {McGee},
  {Poggianti}, {Moretti}, {Gullieuszik}, {Bettoni}, {Fasano}, {D'Onofrio},
  {Fritz}, {Omizzolo}, {Sheen}, \& {Vulcani}}]{Bellhouse2017}
{Bellhouse}, C., {Jaff{\'e}}, Y.~L., {Hau}, G.~K.~T., {et~al.} 2017, \apj, 844,
  49

\bibitem[{{Bellhouse} {et~al.}(2019){Bellhouse}, {Jaff{\'e}}, {McGee},
  {Poggianti}, {Smith}, {Tonnesen}, {Fritz}, {Hau}, {Gullieuszik}, {Vulcani},
  {Fasano}, {Moretti}, {George}, {Bettoni}, {D'Onofrio}, {Omizzolo}, \&
  {Sheen}}]{Bellhouse2019}
{Bellhouse}, C., {Jaff{\'e}}, Y.~L., {McGee}, S.~L., {et~al.} 2019, \mnras,
  485, 1157

\bibitem[{{Bellhouse} {et~al.}(2021){Bellhouse}, {McGee}, {Smith}, {Poggianti},
  {Jaff{\'e}}, {Kraljic}, {Franchetto}, {Fritz}, {Vulcani}, {Tonnesen},
  {Roediger}, {Moretti}, {Gullieuszik}, \& {Shin}}]{Bellhouse2021}
{Bellhouse}, C., {McGee}, S.~L., {Smith}, R., {et~al.} 2021, \mnras, 500, 1285

\bibitem[{{Bellhouse} {et~al.}(2022){Bellhouse}, {Poggianti}, {Moretti},
  {Vulcani}, {Werle}, {Gullieuszik}, {Radovich}, {Jaff{\'e}}, {Fritz},
  {Ignesti}, {Bacchini}, {Tomi{\v{c}}i{\'c}}, {Richard}, \&
  {Soucail}}]{Bellhouse2022}
{Bellhouse}, C., {Poggianti}, B., {Moretti}, A., {et~al.} 2022, \apj, 937, 18

\bibitem[{{Biviano} {et~al.}(2017){Biviano}, {Moretti}, {Paccagnella},
  {Poggianti}, {Bettoni}, {Gullieuszik}, {Vulcani}, {Fasano}, {D'Onofrio},
  {Fritz}, \& {Cava}}]{Biviano2017}
{Biviano}, A., {Moretti}, A., {Paccagnella}, A., {et~al.} 2017, \aap, 607, A81

\bibitem[{{Biviano} {et~al.}(2024){Biviano}, {Poggianti}, {Jaff{\'e}},
  {Louren{\c{c}}o}, {Pizzuti}, {Moretti}, \& {Vulcani}}]{Biviano2024}
{Biviano}, A., {Poggianti}, B.~M., {Jaff{\'e}}, Y., {et~al.} 2024, \apj, 965,
  117

\bibitem[{{B{\"o}sch} {et~al.}(2013){B{\"o}sch}, {B{\"o}hm}, {Wolf},
  {Arag{\'o}n-Salamanca}, {Barden}, {Gray}, {Ziegler}, {Schindler}, \&
  {Balogh}}]{Bosch2013}
{B{\"o}sch}, B., {B{\"o}hm}, A., {Wolf}, C., {et~al.} 2013, \aap, 549, A142

\bibitem[{{Boselli} {et~al.}(2019){Boselli}, {Epinat}, {Contini},
  {Abril-Melgarejo}, {Boogaard}, {Pointecouteau}, {Ventou}, {Brinchmann},
  {Carton}, {Finley}, {Michel-Dansac}, {Soucail}, \&
  {Weilbacher}}]{Boselli2019}
{Boselli}, A., {Epinat}, B., {Contini}, T., {et~al.} 2019, \aap, 631, A114

\bibitem[{{Boselli} {et~al.}(2023){Boselli}, {Fossati}, {C{\^o}t{\'e}},
  {Cuillandre}, {Ferrarese}, {Gwyn}, {Amram}, {Ayromlou}, {Balogh}, {Bellusci},
  {Boquien}, {Gavazzi}, {Hensler}, {Longobardi}, {Nelson}, {Pillepich},
  {Roediger}, {Sanchez-Janssen}, {Sun}, \& {Trinchieri}}]{Boselli2023}
{Boselli}, A., {Fossati}, M., {C{\^o}t{\'e}}, P., {et~al.} 2023, \aap, 675,
  A123

\bibitem[{{Boselli} {et~al.}(2018){Boselli}, {Fossati}, {Ferrarese},
  {Boissier}, {Consolandi}, {Longobardi}, {Amram}, {Balogh}, {Barmby},
  {Boquien}, {Boulanger}, {Braine}, {Buat}, {Burgarella}, {Combes}, {Contini},
  {Cortese}, {C{\^o}t{\'e}}, {C{\^o}t{\'e}}, {Cuillandre}, {Drissen}, {Epinat},
  {Fumagalli}, {Gallagher}, {Gavazzi}, {Gomez-Lopez}, {Gwyn}, {Harris},
  {Hensler}, {Koribalski}, {Marcelin}, {McConnachie}, {Miville-Deschenes},
  {Navarro}, {Patton}, {Peng}, {Plana}, {Prantzos}, {Robert}, {Roediger},
  {Roehlly}, {Russeil}, {Salome}, {Sanchez-Janssen}, {Serra}, {Spekkens},
  {Sun}, {Taylor}, {Tonnesen}, {Vollmer}, {Willis}, {Wozniak}, {Burdullis},
  {Devost}, {Mahoney}, {Manset}, {Petric}, {Prunet}, \&
  {Withington}}]{Boselli2018}
{Boselli}, A., {Fossati}, M., {Ferrarese}, L., {et~al.} 2018, \aap, 614, A56

\bibitem[{{Boselli} {et~al.}(2020){Boselli}, {Fossati}, {Longobardi},
  {Boissier}, {Boquien}, {Braine}, {C{\^o}t{\'e}}, {Cuillandre}, {Epinat},
  {Ferrarese}, {Gavazzi}, {Gwyn}, {Hensler}, {Plana}, {Roehlly}, {Schimd},
  {Sun}, \& {Trinchieri}}]{Boselli2020}
{Boselli}, A., {Fossati}, M., {Longobardi}, A., {et~al.} 2020, \aap, 634, L1

\bibitem[{{Boselli} {et~al.}(2022){Boselli}, {Fossati}, {Longobardi},
  {Kianfar}, {Dametto}, {Amram}, {Anderson}, {Andreani}, {Boissier}, {Boquien},
  {Buat}, {Consolandi}, {Cortese}, {C{\^o}t{\'e}}, {Cuillandre}, {Ferrarese},
  {Galbany}, {Gavazzi}, {Gwyn}, {Hensler}, {Hutchings}, {Peng}, {Postma},
  {Roediger}, {Roehlly}, {Serra}, \& {Trinchieri}}]{Boselli2022}
{Boselli}, A., {Fossati}, M., {Longobardi}, A., {et~al.} 2022, \aap, 659, A46

\bibitem[{{Boselli} {et~al.}(2021){Boselli}, {Lupi}, {Epinat}, {Amram},
  {Fossati}, {Anderson}, {Boissier}, {Boquien}, {Consolandi}, {C{\^o}t{\'e}},
  {Cuillandre}, {Ferrarese}, {Galbany}, {Gavazzi}, {G{\'o}mez-L{\'o}pez},
  {Gwyn}, {Hensler}, {Hutchings}, {Kuncarayakti}, {Longobardi}, {Peng},
  {Plana}, {Postma}, {Roediger}, {Roehlly}, {Schimd}, {Trinchieri}, \&
  {Vollmer}}]{Boselli2021}
{Boselli}, A., {Lupi}, A., {Epinat}, B., {et~al.} 2021, \aap, 646, A139

\bibitem[{{Brown} {et~al.}(2023){Brown}, {Roberts}, {Thorp}, {Ellison},
  {Zabel}, {Wilson}, {Bah{\'e}}, {Bisaria}, {Bolatto}, {Boselli}, {Chung},
  {Cortese}, {Catinella}, {Davis}, {Jim{\'e}nez-Donaire}, {Lagos}, {Lee},
  {Parker}, {Smith}, {Spekkens}, {Stevens}, {Villanueva}, \&
  {Watts}}]{Brown2023}
{Brown}, T., {Roberts}, I.~D., {Thorp}, M., {et~al.} 2023, \apj, 956, 37

\bibitem[{{Brown} {et~al.}(2021){Brown}, {Wilson}, {Zabel}, {Davis}, {Boselli},
  {Chung}, {Ellison}, {Lagos}, {Stevens}, {Cortese}, {Bah{\'e}}, {Bisaria},
  {Bolatto}, {Cashmore}, {Catinella}, {Chown}, {Diemer}, {Elahi}, {Hani},
  {Jim{\'e}nez-Donaire}, {Lee}, {Leidig}, {Mok}, {Olsen}, {Parker}, {Roberts},
  {Smith}, {Spekkens}, {Thorp}, {Tonnesen}, {Vienneau}, {Villanueva}, {Vogel},
  {Wadsley}, {Welker}, \& {Yoon}}]{Brown2021}
{Brown}, T., {Wilson}, C.~D., {Zabel}, N., {et~al.} 2021, \apjs, 257, 21

\bibitem[{{Bundy} {et~al.}(2015){Bundy}, {Bershady}, {Law}, {Yan}, {Drory},
  {MacDonald}, {Wake}, {Cherinka}, {S{\'a}nchez-Gallego}, {Weijmans}, {Thomas},
  {Tremonti}, {Masters}, {Coccato}, {Diamond-Stanic}, {Arag{\'o}n-Salamanca},
  {Avila-Reese}, {Badenes}, {Falc{\'o}n-Barroso}, {Belfiore}, {Bizyaev},
  {Blanc}, {Bland-Hawthorn}, {Blanton}, {Brownstein}, {Byler}, {Cappellari},
  {Conroy}, {Dutton}, {Emsellem}, {Etherington}, {Frinchaboy}, {Fu}, {Gunn},
  {Harding}, {Johnston}, {Kauffmann}, {Kinemuchi}, {Klaene}, {Knapen},
  {Leauthaud}, {Li}, {Lin}, {Maiolino}, {Malanushenko}, {Malanushenko}, {Mao},
  {Maraston}, {McDermid}, {Merrifield}, {Nichol}, {Oravetz}, {Pan}, {Parejko},
  {Sanchez}, {Schlegel}, {Simmons}, {Steele}, {Steinmetz}, {Thanjavur},
  {Thompson}, {Tinker}, {van den Bosch}, {Westfall}, {Wilkinson}, {Wright},
  {Xiao}, \& {Zhang}}]{Bundy2015}
{Bundy}, K., {Bershady}, M.~A., {Law}, D.~R., {et~al.} 2015, \apj, 798, 7

\bibitem[{{Campitiello} {et~al.}(2021){Campitiello}, {Ignesti}, {Gitti},
  {Brighenti}, {Radovich}, {Wolter}, {Tomi{\v{c}}i{\'c}}, {Bellhouse},
  {Poggianti}, {Moretti}, {Vulcani}, {Jaff{\'e}}, {Paladino}, {M{\"u}ller},
  {Fritz}, {Louren{\c{c}}o}, \& {Gullieuszik}}]{Campitiello2021}
{Campitiello}, M.~G., {Ignesti}, A., {Gitti}, M., {et~al.} 2021, \apj, 911, 144

\bibitem[{{Cappellari} \& {Emsellem}(2004)}]{Cappellari2004}
{Cappellari}, M. \& {Emsellem}, E. 2004, \pasp, 116, 138

\bibitem[{{Cava} {et~al.}(2009){Cava}, {Bettoni}, {Poggianti}, {Couch},
  {Moles}, {Varela}, {Biviano}, {D'Onofrio}, {Dressler}, {Fasano}, {Fritz},
  {Kj{\ae}rgaard}, {Ramella}, \& {Valentinuzzi}}]{Cava2009}
{Cava}, A., {Bettoni}, D., {Poggianti}, B.~M., {et~al.} 2009, \aap, 495, 707

\bibitem[{{Chabrier}(2003)}]{Chabrier2003}
{Chabrier}, G. 2003, \pasp, 115, 763

\bibitem[{{Collins} \& {Rand}(2001)}]{Collins2001}
{Collins}, J.~A. \& {Rand}, R.~J. 2001, \apj, 551, 57

\bibitem[{{Consolandi} {et~al.}(2017){Consolandi}, {Gavazzi}, {Fossati},
  {Fumagalli}, {Boselli}, {Yagi}, \& {Yoshida}}]{Consolandi2017}
{Consolandi}, G., {Gavazzi}, G., {Fossati}, M., {et~al.} 2017, \aap, 606, A83

\bibitem[{{Cortese} {et~al.}(2007){Cortese}, {Marcillac}, {Richard},
  {Bravo-Alfaro}, {Kneib}, {Rieke}, {Covone}, {Egami}, {Rigby}, {Czoske}, \&
  {Davies}}]{Cortese2007}
{Cortese}, L., {Marcillac}, D., {Richard}, J., {et~al.} 2007, \mnras, 376, 157

\bibitem[{{Croom} {et~al.}(2012){Croom}, {Lawrence}, {Bland-Hawthorn},
  {Bryant}, {Fogarty}, {Richards}, {Goodwin}, {Farrell}, {Miziarski}, {Heald},
  {Jones}, {Lee}, {Colless}, {Brough}, {Hopkins}, {Bauer}, {Birchall}, {Ellis},
  {Horton}, {Leon-Saval}, {Lewis}, {L{\'o}pez-S{\'a}nchez}, {Min}, {Trinh}, \&
  {Trowland}}]{Croom2012}
{Croom}, S.~M., {Lawrence}, J.~S., {Bland-Hawthorn}, J., {et~al.} 2012, \mnras,
  421, 872

\bibitem[{{Deb} {et~al.}(2020){Deb}, {Verheijen}, {Gullieuszik}, {Poggianti},
  {van Gorkom}, {Ramatsoku}, {Serra}, {Moretti}, {Vulcani}, {Bettoni},
  {Jaff{\'e}}, {Tonnesen}, \& {Fritz}}]{Deb2020}
{Deb}, T., {Verheijen}, M. A.~W., {Gullieuszik}, M., {et~al.} 2020, \mnras,
  494, 5029

\bibitem[{{Deb} {et~al.}(2022){Deb}, {Verheijen}, {Poggianti}, {Moretti}, {van
  der Hulst}, {Vulcani}, {Ramatsoku}, {Serra}, {Healy}, {Gullieuszik},
  {Bacchini}, {Ignesti}, {M{\"u}ller}, {Zabel}, {Luber}, {Jaff{\"e}}, \&
  {Gitti}}]{Deb2022}
{Deb}, T., {Verheijen}, M. A.~W., {Poggianti}, B.~M., {et~al.} 2022, \mnras,
  516, 2683

\bibitem[{{Deb} {et~al.}(2023){Deb}, {Verheijen}, \& {van der Hulst}}]{Deb2023}
{Deb}, T., {Verheijen}, M.~A.~W., \& {van der Hulst}, J.~M. 2023, \aap, 676,
  A118

\bibitem[{{Dressler}(1980)}]{Dressler1980a}
{Dressler}, A. 1980, \apjs, 42, 565

\bibitem[{{Dressler} {et~al.}(1999){Dressler}, {Smail}, {Poggianti}, {Butcher},
  {Couch}, {Ellis}, \& {Oemler}}]{Dressler1999}
{Dressler}, A., {Smail}, I., {Poggianti}, B.~M., {et~al.} 1999, \apjs, 122, 51

\bibitem[{{Durret} {et~al.}(2021){Durret}, {Chiche}, {Lobo}, \&
  {Jauzac}}]{Durret2021}
{Durret}, F., {Chiche}, S., {Lobo}, C., \& {Jauzac}, M. 2021, \aap, 648, A63

\bibitem[{{Durret} {et~al.}(2022){Durret}, {Degott}, {Lobo}, {Ebeling},
  {Jauzac}, \& {Tam}}]{Durret2022}
{Durret}, F., {Degott}, L., {Lobo}, C., {et~al.} 2022, \aap, 662, A84

\bibitem[{{Ebeling} \& {Kalita}(2019)}]{EbelingKalita2019}
{Ebeling}, H. \& {Kalita}, B.~S. 2019, \apj, 882, 127

\bibitem[{{Ebeling} {et~al.}(2014){Ebeling}, {Ma}, \& {Barrett}}]{Ebeling2014}
{Ebeling}, H., {Ma}, C.-J., \& {Barrett}, E. 2014, \apjs, 211, 21

\bibitem[{{Fasano} {et~al.}(2006){Fasano}, {Marmo}, {Varela}, {D'Onofrio},
  {Poggianti}, {Moles}, {Pignatelli}, {Bettoni}, {Kj{\ae}rgaard}, {Rizzi},
  {Couch}, \& {Dressler}}]{Fasano2006}
{Fasano}, G., {Marmo}, C., {Varela}, J., {et~al.} 2006, \aap, 445, 805

\bibitem[{{Fasano} {et~al.}(2012){Fasano}, {Vanzella}, {Dressler}, {Poggianti},
  {Moles}, {Bettoni}, {Valentinuzzi}, {Moretti}, {D'Onofrio}, {Varela},
  {Couch}, {Kj{\ae}rgaard}, {Fritz}, {Omizzolo}, \& {Cava}}]{Fasano2012}
{Fasano}, G., {Vanzella}, E., {Dressler}, A., {et~al.} 2012, \mnras, 420, 926

\bibitem[{{Fogarty} {et~al.}(2012){Fogarty}, {Bland-Hawthorn}, {Croom},
  {Green}, {Bryant}, {Lawrence}, {Richards}, {Allen}, {Bauer}, {Birchall},
  {Brough}, {Colless}, {Ellis}, {Farrell}, {Goodwin}, {Heald}, {Hopkins},
  {Horton}, {Jones}, {Lee}, {Lewis}, {L{\'o}pez-S{\'a}nchez}, {Miziarski},
  {Trowland}, {Leon-Saval}, {Min}, {Trinh}, {Cecil}, {Veilleux}, \&
  {Kreimeyer}}]{Fogarty2012}
{Fogarty}, L. M.~R., {Bland-Hawthorn}, J., {Croom}, S.~M., {et~al.} 2012, \apj,
  761, 169

\bibitem[{{Fossati} {et~al.}(2016){Fossati}, {Fumagalli}, {Boselli}, {Gavazzi},
  {Sun}, \& {Wilman}}]{Fossati2016}
{Fossati}, M., {Fumagalli}, M., {Boselli}, A., {et~al.} 2016, \mnras, 455, 2028

\bibitem[{{Franchetto} {et~al.}(2021{\natexlab{a}}){Franchetto}, {Mingozzi},
  {Poggianti}, {Vulcani}, {Bacchini}, {Gullieuszik}, {Moretti},
  {Tomi{\v{c}}i{\'c}}, \& {Fritz}}]{Franchetto2021b}
{Franchetto}, A., {Mingozzi}, M., {Poggianti}, B.~M., {et~al.}
  2021{\natexlab{a}}, \apj, 923, 28

\bibitem[{{Franchetto} {et~al.}(2021{\natexlab{b}}){Franchetto}, {Tonnesen},
  {Poggianti}, {Vulcani}, {Gullieuszik}, {Moretti}, {Smith}, {Ignesti},
  {Bacchini}, {McGee}, {Tomi{\v{c}}i{\'c}}, {Mingozzi}, {Wolter}, \&
  {M{\"u}ller}}]{Franchetto2021a}
{Franchetto}, A., {Tonnesen}, S., {Poggianti}, B.~M., {et~al.}
  2021{\natexlab{b}}, \apjl, 922, L6

\bibitem[{{Franchetto} {et~al.}(2020){Franchetto}, {Vulcani}, {Poggianti},
  {Gullieuszik}, {Mingozzi}, {Moretti}, {Tomi{\v{c}}i{\'c}}, {Fritz},
  {Bettoni}, \& {Jaff{\'e}}}]{Franchetto2020}
{Franchetto}, A., {Vulcani}, B., {Poggianti}, B.~M., {et~al.} 2020, \apj, 895,
  106

\bibitem[{{Fritz} {et~al.}(2017){Fritz}, {Moretti}, {Gullieuszik}, {Poggianti},
  {Bruzual}, {Vulcani}, {Nicastro}, {Jaff{\'e}}, {Cervantes Sodi}, {Bettoni},
  {Biviano}, {Fasano}, {Charlot}, {Bellhouse}, \& {Hau}}]{Fritz2017}
{Fritz}, J., {Moretti}, A., {Gullieuszik}, M., {et~al.} 2017, \apj, 848, 132

\bibitem[{{Fritz} {et~al.}(2014){Fritz}, {Poggianti}, {Cava}, {Moretti},
  {Varela}, {Bettoni}, {Couch}, {D'Onofrio D'Onofrio}, {Dressler}, {Fasano},
  {Kj{\ae}rgaard}, {Marziani}, {Moles}, \& {Omizzolo}}]{Fritz2014}
{Fritz}, J., {Poggianti}, B.~M., {Cava}, A., {et~al.} 2014, \aap, 566, A32

\bibitem[{{Fumagalli} {et~al.}(2014){Fumagalli}, {Fossati}, {Hau}, {Gavazzi},
  {Bower}, {Sun}, \& {Boselli}}]{Fumagalli2014}
{Fumagalli}, M., {Fossati}, M., {Hau}, G.~K.~T., {et~al.} 2014, \mnras, 445,
  4335

\bibitem[{{George} {et~al.}(2019){George}, {Poggianti}, {Bellhouse},
  {Radovich}, {Fritz}, {Paladino}, {Bettoni}, {Jaff{\'e}}, {Moretti},
  {Gullieuszik}, {Vulcani}, {Fasano}, {Stalin}, {Subramaniam}, \&
  {Tandon}}]{George2019}
{George}, K., {Poggianti}, B.~M., {Bellhouse}, C., {et~al.} 2019, \mnras, 487,
  3102

\bibitem[{{George} {et~al.}(2018){George}, {Poggianti}, {Gullieuszik},
  {Fasano}, {Bellhouse}, {Postma}, {Moretti}, {Jaff{\'e}}, {Vulcani},
  {Bettoni}, {Fritz}, {C{\^o}t{\'e}}, {Ghosh}, {Hutchings}, {Mohan},
  {Sreekumar}, {Stalin}, {Subramaniam}, \& {Tandon}}]{George2018}
{George}, K., {Poggianti}, B.~M., {Gullieuszik}, M., {et~al.} 2018, \mnras,
  479, 4126

\bibitem[{{George} {et~al.}(2024){George}, {Poggianti}, {Omizzolo}, {Vulcani},
  {C{\^o}t{\'e}}, {Postma}, {Smith}, {Jaffe}, {Gullieuszik}, {Moretti},
  {Subramaniam}, {Sreekumar}, {Ghosh}, {Tandon}, \& {Hutchings}}]{George2024}
{George}, K., {Poggianti}, B.~M., {Omizzolo}, A., {et~al.} 2024, arXiv
  e-prints, arXiv:2409.10586

\bibitem[{{George} {et~al.}(2023){George}, {Poggianti}, {Tomi{\v{c}}i{\'c}},
  {Postma}, {C{\^o}t{\'e}}, {Fritz}, {Ghosh}, {Gullieuszik}, {Hutchings},
  {Moretti}, {Omizzolo}, {Radovich}, {Sreekumar}, {Subramaniam}, {Tandon}, \&
  {Vulcani}}]{George2023}
{George}, K., {Poggianti}, B.~M., {Tomi{\v{c}}i{\'c}}, N., {et~al.} 2023,
  \mnras, 519, 2426

\bibitem[{{George} {et~al.}(2025){George}, {Poggianti}, {Vulcani},
  {Gullieuszik}, {Postma}, {Fritz}, {C{\^o}t{\'e}}, {Jaffe}, {Moretti},
  {Ignesti}, {Peluso}, {Tomi{\'c}i{\'c}}, {Subramaniam}, {Ghosh}, \&
  {Tandon}}]{George2025}
{George}, K., {Poggianti}, B.~M., {Vulcani}, B., {et~al.} 2025, arXiv e-prints,
  arXiv:2505.15066

\bibitem[{{Giunchi} {et~al.}(2023{\natexlab{a}}){Giunchi}, {Gullieuszik},
  {Poggianti}, {Moretti}, {Werle}, {Scarlata}, {Zanella}, {Vulcani}, \&
  {Calzetti}}]{Giunchi2023a}
{Giunchi}, E., {Gullieuszik}, M., {Poggianti}, B.~M., {et~al.}
  2023{\natexlab{a}}, \apj, 949, 72

\bibitem[{{Giunchi} {et~al.}(2023{\natexlab{b}}){Giunchi}, {Poggianti},
  {Gullieuszik}, {Moretti}, {Werle}, {Zanella}, {Vulcani}, {Tonnesen},
  {Calzetti}, {Bellhouse}, {Scarlata}, \& {Bacchini}}]{Giunchi2023b}
{Giunchi}, E., {Poggianti}, B.~M., {Gullieuszik}, M., {et~al.}
  2023{\natexlab{b}}, \apj, 958, 73

\bibitem[{{Giunchi} {et~al.}(2025){Giunchi}, {Scarlata}, {Werle}, {Poggianti},
  {Moretti}, {Gullieuszik}, {Vulcani}, {Ignesti}, {Marasco}, {Zanella}, \&
  {Wolter}}]{Giunchi2025}
{Giunchi}, E., {Scarlata}, C., {Werle}, A., {et~al.} 2025, \aap, 696, A228

\bibitem[{{G{\"o}ller} {et~al.}(2023){G{\"o}ller}, {Joshi}, {Rohr}, {Zinger},
  \& {Pillepich}}]{Goller2023}
{G{\"o}ller}, J., {Joshi}, G.~D., {Rohr}, E., {Zinger}, E., \& {Pillepich}, A.
  2023, \mnras, 525, 3551

\bibitem[{{Gondhalekar} {et~al.}(2024){Gondhalekar}, {Chies-Santos}, {de
  Souza}, {Queiroz}, {Lopes}, {Ferrari}, {Azevedo}, {Monteiro-Pereira},
  {Overzier}, {Smith Castelli}, {Jaff{\'e}}, {Haack}, {Rahna}, {Shen}, {Mu},
  {Lima-Dias}, {Barbosa}, {Oliveira Schwarz}, {Riffel}, {Jimenez-Teja},
  {Grossi}, {Mendes de Oliveira}, {Schoenell}, {Ribeiro}, \&
  {Kanaan}}]{Gondhalekar2024}
{Gondhalekar}, Y., {Chies-Santos}, A.~L., {de Souza}, R.~S., {et~al.} 2024,
  \mnras, 532, 270

\bibitem[{{Gullieuszik} {et~al.}(2023){Gullieuszik}, {Giunchi}, {Poggianti},
  {Moretti}, {Scarlata}, {Calzetti}, {Werle}, {Zanella}, {Radovich},
  {Bellhouse}, {Bettoni}, {Franchetto}, {Fritz}, {Jaff{\'e}}, {McGee},
  {Mingozzi}, {Omizzolo}, {Tonnesen}, {Verheijen}, \&
  {Vulcani}}]{Gullieuszik2023}
{Gullieuszik}, M., {Giunchi}, E., {Poggianti}, B.~M., {et~al.} 2023, \apj, 945,
  54

\bibitem[{{Gullieuszik} {et~al.}(2015){Gullieuszik}, {Poggianti}, {Fasano},
  {Zaggia}, {Paccagnella}, {Moretti}, {Bettoni}, {D'Onofrio}, {Couch},
  {Vulcani}, {Fritz}, {Omizzolo}, {Baruffolo}, {Schipani}, {Capaccioli}, \&
  {Varela}}]{Gullieuszik2015}
{Gullieuszik}, M., {Poggianti}, B., {Fasano}, G., {et~al.} 2015, \aap, 581, A41

\bibitem[{{Gullieuszik} {et~al.}(2020){Gullieuszik}, {Poggianti}, {McGee},
  {Moretti}, {Vulcani}, {Tonnesen}, {Roediger}, {Jaff{\'e}}, {Fritz},
  {Franchetto}, {Omizzolo}, {Bettoni}, {Radovich}, \&
  {Wolter}}]{Gullieuszik2020}
{Gullieuszik}, M., {Poggianti}, B.~M., {McGee}, S.~L., {et~al.} 2020, \apj,
  899, 13

\bibitem[{{Gullieuszik} {et~al.}(2017){Gullieuszik}, {Poggianti}, {Moretti},
  {Fritz}, {Jaff{\'e}}, {Hau}, {Bischko}, {Bellhouse}, {Bettoni}, {Fasano},
  {Vulcani}, {D'Onofrio}, \& {Biviano}}]{Gullieuszik2017}
{Gullieuszik}, M., {Poggianti}, B.~M., {Moretti}, A., {et~al.} 2017, \apj, 846,
  27

\bibitem[{{Gunn} \& {Gott}(1972)}]{Gunn1972}
{Gunn}, J.~E. \& {Gott}, III, J.~R. 1972, \apj, 176, 1

\bibitem[{{Hester} {et~al.}(2010){Hester}, {Seibert}, {Neill}, {Wyder}, {Gil de
  Paz}, {Madore}, {Martin}, {Schiminovich}, \& {Rich}}]{Hester2010}
{Hester}, J.~A., {Seibert}, M., {Neill}, J.~D., {et~al.} 2010, \apjl, 716, L14

\bibitem[{{Ignesti} {et~al.}(2023{\natexlab{a}}){Ignesti}, {Brienza},
  {Vulcani}, {Poggianti}, {Marasco}, {Smith}, {Hardcastle}, {Botteon},
  {Roberts}, {Fritz}, {Paladino}, {Gitti}, {Wolter}, {Tomi{\v{c}}i{\'c}},
  {McGee}, {Moretti}, {Gullieuszik}, \& {Drabent}}]{Ignesti2023b}
{Ignesti}, A., {Brienza}, M., {Vulcani}, B., {et~al.} 2023{\natexlab{a}}, \apj,
  956, 122

\bibitem[{{Ignesti} {et~al.}(2024){Ignesti}, {Brunetti}, {Gullieuszik},
  {Akerman}, {Marasco}, {Poggianti}, {Li}, {Vulcani}, {Gitti}, {Moretti},
  {Giunchi}, {Tomi{\v{c}}i{\'c}}, {Bacchini}, {Paladino}, {Radovich}, \&
  {Wolter}}]{Ignesti2024}
{Ignesti}, A., {Brunetti}, G., {Gullieuszik}, M., {et~al.} 2024, arXiv
  e-prints, arXiv:2411.07034

\bibitem[{{Ignesti} {et~al.}(2023{\natexlab{b}}){Ignesti}, {Vulcani},
  {Botteon}, {Poggianti}, {Giunchi}, {Smith}, {Brunetti}, {Roberts}, {van
  Weeren}, \& {Rajpurohit}}]{Ignesti2023a}
{Ignesti}, A., {Vulcani}, B., {Botteon}, A., {et~al.} 2023{\natexlab{b}}, \aap,
  675, A118

\bibitem[{{Ignesti} {et~al.}(2022{\natexlab{a}}){Ignesti}, {Vulcani},
  {Poggianti}, {Moretti}, {Shimwell}, {Botteon}, {van Weeren}, {Roberts},
  {Fritz}, {Tomi{\v{c}}i{\'c}}, {Peluso}, {Paladino}, {Gitti}, {M{\"u}ller},
  {McGee}, \& {Gullieuszik}}]{Ignesti2022b}
{Ignesti}, A., {Vulcani}, B., {Poggianti}, B.~M., {et~al.} 2022{\natexlab{a}},
  \apj, 937, 58

\bibitem[{{Ignesti} {et~al.}(2022{\natexlab{b}}){Ignesti}, {Vulcani},
  {Poggianti}, {Paladino}, {Shimwell}, {Healy}, {Gitti}, {Bacchini}, {Moretti},
  {Radovich}, {van Weeren}, {Roberts}, {Botteon}, {M{\"u}ller}, {McGee},
  {Fritz}, {Tomi{\v{c}}i{\'c}}, {Werle}, {Mingozzi}, {Gullieuszik}, \&
  {Verheijen}}]{Ignesti2022a}
{Ignesti}, A., {Vulcani}, B., {Poggianti}, B.~M., {et~al.} 2022{\natexlab{b}},
  \apj, 924, 64

\bibitem[{{Jaff{\'e}} {et~al.}(2018){Jaff{\'e}}, {Poggianti}, {Moretti},
  {Gullieuszik}, {Smith}, {Vulcani}, {Fasano}, {Fritz}, {Tonnesen}, {Bettoni},
  {Hau}, {Biviano}, {Bellhouse}, \& {McGee}}]{Jaffe2018}
{Jaff{\'e}}, Y.~L., {Poggianti}, B.~M., {Moretti}, A., {et~al.} 2018, \mnras,
  476, 4753

\bibitem[{{Kalita} \& {Ebeling}(2019)}]{KalitaEbeling2019}
{Kalita}, B.~S. \& {Ebeling}, H. 2019, \apj, 887, 158

\bibitem[{{Kapferer} {et~al.}(2008){Kapferer}, {Kronberger}, {Ferrari},
  {Riser}, \& {Schindler}}]{Kapferer2008}
{Kapferer}, W., {Kronberger}, T., {Ferrari}, C., {Riser}, T., \& {Schindler},
  S. 2008, \mnras, 389, 1405

\bibitem[{{Kapferer} {et~al.}(2009){Kapferer}, {Sluka}, {Schindler}, {Ferrari},
  \& {Ziegler}}]{Kapferer2009}
{Kapferer}, W., {Sluka}, C., {Schindler}, S., {Ferrari}, C., \& {Ziegler}, B.
  2009, \aap, 499, 87

\bibitem[{{Kauffmann} {et~al.}(2003){Kauffmann}, {Heckman}, {Tremonti},
  {Brinchmann}, {Charlot}, {White}, {Ridgway}, {Brinkmann}, {Fukugita}, {Hall},
  {Ivezi{\'c}}, {Richards}, \& {Schneider}}]{Kauffmann2003}
{Kauffmann}, G., {Heckman}, T.~M., {Tremonti}, C., {et~al.} 2003, \mnras, 346,
  1055

\bibitem[{{Kewley} {et~al.}(2001){Kewley}, {Heisler}, {Dopita}, \&
  {Lumsden}}]{Kewley2001}
{Kewley}, L.~J., {Heisler}, C.~A., {Dopita}, M.~A., \& {Lumsden}, S. 2001,
  \apjs, 132, 37

\bibitem[{{Kulier} {et~al.}(2023){Kulier}, {Poggianti}, {Tonnesen}, {Smith},
  {Ignesti}, {Akerman}, {Marasco}, {Vulcani}, {Moretti}, \&
  {Wolter}}]{Kulier2023}
{Kulier}, A., {Poggianti}, B., {Tonnesen}, S., {et~al.} 2023, \apj, 954, 177

\bibitem[{{Lee} {et~al.}(2022{\natexlab{a}}){Lee}, {Lee}, {Mun}, {Cho}, \&
  {Kang}}]{Lee2022b}
{Lee}, J.~H., {Lee}, M.~G., {Mun}, J.~Y., {Cho}, B.~S., \& {Kang}, J.
  2022{\natexlab{a}}, \apj, 940, 24

\bibitem[{{Lee} {et~al.}(2022{\natexlab{b}}){Lee}, {Lee}, {Mun}, {Cho}, \&
  {Kang}}]{Lee2022a}
{Lee}, J.~H., {Lee}, M.~G., {Mun}, J.~Y., {Cho}, B.~S., \& {Kang}, J.
  2022{\natexlab{b}}, \apjl, 931, L22

\bibitem[{{Luber} {et~al.}(2022){Luber}, {M{\"u}ller}, {van Gorkom},
  {Poggianti}, {Vulcani}, {Franchetto}, {Bacchini}, {Bettoni}, {Deb}, {Fritz},
  {Gullieuszik}, {Ignesti}, {Jaffe}, {Moretti}, {Paladino}, {Ramatsoku},
  {Serra}, {Smith}, {Tomicic}, {Tonnesen}, {Verheijen}, \&
  {Wolter}}]{Luber2022}
{Luber}, N., {M{\"u}ller}, A., {van Gorkom}, J.~H., {et~al.} 2022, \apj, 927,
  39

\bibitem[{{Marasco} {et~al.}(2023){Marasco}, {Poggianti}, {Fritz}, {Werle},
  {Vulcani}, {Moretti}, {Gullieuszik}, \& {Kulier}}]{Marasco2023}
{Marasco}, A., {Poggianti}, B.~M., {Fritz}, J., {et~al.} 2023, \mnras, 525,
  5359

\bibitem[{{McClymont} {et~al.}(2024){McClymont}, {Tacchella}, {Smith},
  {Kannan}, {Maiolino}, {Belfiore}, {Hernquist}, {Li}, \&
  {Vogelsberger}}]{McClymont2024}
{McClymont}, W., {Tacchella}, S., {Smith}, A., {et~al.} 2024, \mnras, 532, 2016

\bibitem[{{McPartland} {et~al.}(2016){McPartland}, {Ebeling}, {Roediger}, \&
  {Blumenthal}}]{McPartland2016}
{McPartland}, C., {Ebeling}, H., {Roediger}, E., \& {Blumenthal}, K. 2016,
  \mnras, 455, 2994

\bibitem[{{Merluzzi} {et~al.}(2016){Merluzzi}, {Busarello}, {Dopita}, {Haines},
  {Steinhauser}, {Bourdin}, \& {Mazzotta}}]{Merluzzi2016}
{Merluzzi}, P., {Busarello}, G., {Dopita}, M.~A., {et~al.} 2016, \mnras, 460,
  3345

\bibitem[{{Merluzzi} {et~al.}(2013){Merluzzi}, {Busarello}, {Dopita}, {Haines},
  {Steinhauser}, {Mercurio}, {Rifatto}, {Smith}, \& {Schindler}}]{Merluzzi2013}
{Merluzzi}, P., {Busarello}, G., {Dopita}, M.~A., {et~al.} 2013, \mnras, 429,
  1747

\bibitem[{{Mihos} {et~al.}(1993){Mihos}, {Bothun}, \& {Richstone}}]{Mihos1993}
{Mihos}, J.~C., {Bothun}, G.~D., \& {Richstone}, D.~O. 1993, \apj, 418, 82

\bibitem[{{Moretti} {et~al.}(2017){Moretti}, {Gullieuszik}, {Poggianti},
  {Paccagnella}, {Couch}, {Vulcani}, {Bettoni}, {Fritz}, {Cava}, {Fasano},
  {D'Onofrio}, \& {Omizzolo}}]{Moretti2017}
{Moretti}, A., {Gullieuszik}, M., {Poggianti}, B., {et~al.} 2017, \aap, 599,
  A81

\bibitem[{{Moretti} {et~al.}(2018{\natexlab{a}}){Moretti}, {Paladino},
  {Poggianti}, {D'Onofrio}, {Bettoni}, {Gullieuszik}, {Jaff{\'e}}, {Vulcani},
  {Fasano}, {Fritz}, \& {Torstensson}}]{Moretti2018b}
{Moretti}, A., {Paladino}, R., {Poggianti}, B.~M., {et~al.} 2018{\natexlab{a}},
  \mnras, 480, 2508

\bibitem[{{Moretti} {et~al.}(2020{\natexlab{a}}){Moretti}, {Paladino},
  {Poggianti}, {Serra}, {Ramatsoku}, {Franchetto}, {Deb}, {Gullieuszik},
  {Tomi{\v{c}}i{\'c}}, {Mingozzi}, {Vulcani}, {Radovich}, {Bettoni}, \&
  {Fritz}}]{Moretti2020b}
{Moretti}, A., {Paladino}, R., {Poggianti}, B.~M., {et~al.} 2020{\natexlab{a}},
  \apjl, 897, L30

\bibitem[{{Moretti} {et~al.}(2020{\natexlab{b}}){Moretti}, {Paladino},
  {Poggianti}, {Serra}, {Roediger}, {Gullieuszik}, {Tomi{\v{c}}i{\'c}},
  {Radovich}, {Vulcani}, {Jaff{\'e}}, {Fritz}, {Bettoni}, {Ramatsoku}, \&
  {Wolter}}]{Moretti2020a}
{Moretti}, A., {Paladino}, R., {Poggianti}, B.~M., {et~al.} 2020{\natexlab{b}},
  \apj, 889, 9

\bibitem[{{Moretti} {et~al.}(2014){Moretti}, {Poggianti}, {Fasano}, {Bettoni},
  {D'Onofrio}, {Fritz}, {Cava}, {Varela}, {Vulcani}, {Gullieuszik}, {Couch},
  {Omizzolo}, {Valentinuzzi}, {Dressler}, {Moles}, {Kj{\ae}rgaard},
  {Smareglia}, \& {Molinaro}}]{Moretti2014}
{Moretti}, A., {Poggianti}, B.~M., {Fasano}, G., {et~al.} 2014, \aap, 564, A138

\bibitem[{{Moretti} {et~al.}(2018{\natexlab{b}}){Moretti}, {Poggianti},
  {Gullieuszik}, {Mapelli}, {Jaff{\'e}}, {Fritz}, {Biviano}, {Fasano},
  {Bettoni}, {Vulcani}, \& {D'Onofrio}}]{Moretti2018a}
{Moretti}, A., {Poggianti}, B.~M., {Gullieuszik}, M., {et~al.}
  2018{\natexlab{b}}, \mnras, 475, 4055

\bibitem[{{Moretti} {et~al.}(2022){Moretti}, {Radovich}, {Poggianti},
  {Vulcani}, {Gullieuszik}, {Werle}, {Bellhouse}, {Bacchini}, {Fritz},
  {Soucail}, {Richard}, {Franchetto}, {Tomi{\v{c}}i{\'c}}, \&
  {Omizzolo}}]{Moretti2022}
{Moretti}, A., {Radovich}, M., {Poggianti}, B.~M., {et~al.} 2022, \apj, 925, 4

\bibitem[{{Moretti} {et~al.}(2023){Moretti}, {Serra}, {Bacchini}, {Paladino},
  {Ramatsoku}, {Poggianti}, {Vulcani}, {Deb}, {Gullieuszik}, {Fritz}, \&
  {Wolter}}]{Moretti2023}
{Moretti}, A., {Serra}, P., {Bacchini}, C., {et~al.} 2023, \apj, 955, 153

\bibitem[{{M{\"u}ller} {et~al.}(2021){M{\"u}ller}, {Poggianti}, {Pfrommer},
  {Adebahr}, {Serra}, {Ignesti}, {Sparre}, {Gitti}, {Dettmar}, {Vulcani}, \&
  {Moretti}}]{Mueller2021a}
{M{\"u}ller}, A., {Poggianti}, B.~M., {Pfrommer}, C., {et~al.} 2021, Nature
  Astronomy, 5, 159

\bibitem[{{Owers} {et~al.}(2012){Owers}, {Couch}, {Nulsen}, \&
  {Randall}}]{Owers2012}
{Owers}, M.~S., {Couch}, W.~J., {Nulsen}, P.~E.~J., \& {Randall}, S.~W. 2012,
  \apjl, 750, L23

\bibitem[{{Paccagnella} {et~al.}(2017){Paccagnella}, {Vulcani}, {Poggianti},
  {Fritz}, {Fasano}, {Moretti}, {Jaff{\'e}}, {Biviano}, {Gullieuszik},
  {Bettoni}, {Cava}, {Couch}, \& {D'Onofrio}}]{Paccagnella2017}
{Paccagnella}, A., {Vulcani}, B., {Poggianti}, B.~M., {et~al.} 2017, \apj, 838,
  148

\bibitem[{{Peluso} {et~al.}(2023){Peluso}, {Radovich}, {Moretti}, {Mingozzi},
  {Vulcani}, {Poggianti}, {Marasco}, \& {Gullieuszik}}]{Peluso2023}
{Peluso}, G., {Radovich}, M., {Moretti}, A., {et~al.} 2023, \apj, 958, 147

\bibitem[{{Peluso} {et~al.}(2022){Peluso}, {Vulcani}, {Poggianti}, {Moretti},
  {Radovich}, {Smith}, {Jaff{\'e}}, {Crossett}, {Gullieuszik}, {Fritz}, \&
  {Ignesti}}]{Peluso2022}
{Peluso}, G., {Vulcani}, B., {Poggianti}, B.~M., {et~al.} 2022, \apj, 927, 130

\bibitem[{{Peluso} {et~al.}(2025){Peluso}, {Vulcani}, {Radovich}, {Moretti},
  {Poggianti}, {Watson}, {Acharyya}, {Lassen}, {Gullieuszik}, {Fritz},
  {Ignesti}, {Tomicic}, {Delvecchio}, \& {Khoram}}]{Peluso2025}
{Peluso}, G., {Vulcani}, B., {Radovich}, M., {et~al.} 2025, arXiv e-prints,
  arXiv:2504.18972

\bibitem[{{Pillepich} {et~al.}(2018){Pillepich}, {Springel}, {Nelson}, {Genel},
  {Naiman}, {Pakmor}, {Hernquist}, {Torrey}, {Vogelsberger}, {Weinberger}, \&
  {Marinacci}}]{Pillepich2018}
{Pillepich}, A., {Springel}, V., {Nelson}, D., {et~al.} 2018, \mnras, 473, 4077

\bibitem[{{Poggianti} {et~al.}(2016){Poggianti}, {Fasano}, {Omizzolo},
  {Gullieuszik}, {Bettoni}, {Moretti}, {Paccagnella}, {Jaff{\'e}}, {Vulcani},
  {Fritz}, {Couch}, \& {D'Onofrio}}]{Poggianti2016}
{Poggianti}, B.~M., {Fasano}, G., {Omizzolo}, A., {et~al.} 2016, \aj, 151, 78

\bibitem[{{Poggianti} {et~al.}(2019{\natexlab{a}}){Poggianti}, {Gullieuszik},
  {Tonnesen}, {Moretti}, {Vulcani}, {Radovich}, {Jaff{\'e}}, {Fritz},
  {Bettoni}, {Franchetto}, {Fasano}, {Bellhouse}, \&
  {Omizzolo}}]{Poggianti2019}
{Poggianti}, B.~M., {Gullieuszik}, M., {Tonnesen}, S., {et~al.}
  2019{\natexlab{a}}, \mnras, 482, 4466

\bibitem[{{Poggianti} {et~al.}(2019{\natexlab{b}}){Poggianti}, {Gullieuszik},
  {Tonnesen}, {Moretti}, {Vulcani}, {Radovich}, {Jaff{\'e}}, {Fritz},
  {Bettoni}, {Franchetto}, {Fasano}, {Bellhouse}, \&
  {Omizzolo}}]{Poggianti2019a}
{Poggianti}, B.~M., {Gullieuszik}, M., {Tonnesen}, S., {et~al.}
  2019{\natexlab{b}}, \mnras, 482, 4466

\bibitem[{{Poggianti} {et~al.}(2019{\natexlab{c}}){Poggianti}, {Ignesti},
  {Gitti}, {Wolter}, {Brighenti}, {Biviano}, {George}, {Vulcani},
  {Gullieuszik}, {Moretti}, {Paladino}, {Bettoni}, {Franchetto}, {Jaff{\'e}},
  {Radovich}, {Roediger}, {Tomi{\v{c}}i{\'c}}, {Tonnesen}, {Bellhouse},
  {Fritz}, \& {Omizzolo}}]{Poggianti2019b}
{Poggianti}, B.~M., {Ignesti}, A., {Gitti}, M., {et~al.} 2019{\natexlab{c}},
  \apj, 887, 155

\bibitem[{{Poggianti} {et~al.}(2017{\natexlab{a}}){Poggianti}, {Jaff{\'e}},
  {Moretti}, {Gullieuszik}, {Radovich}, {Tonnesen}, {Fritz}, {Bettoni},
  {Vulcani}, {Fasano}, {Bellhouse}, {Hau}, \& {Omizzolo}}]{Poggianti2017b}
{Poggianti}, B.~M., {Jaff{\'e}}, Y.~L., {Moretti}, A., {et~al.}
  2017{\natexlab{a}}, \nat, 548, 304

\bibitem[{{Poggianti} {et~al.}(2017{\natexlab{b}}){Poggianti}, {Moretti},
  {Gullieuszik}, {Fritz}, {Jaff{\'e}}, {Bettoni}, {Fasano}, {Bellhouse}, {Hau},
  {Vulcani}, {Biviano}, {Omizzolo}, {Paccagnella}, {D'Onofrio}, {Cava},
  {Sheen}, {Couch}, \& {Owers}}]{Poggianti2017a}
{Poggianti}, B.~M., {Moretti}, A., {Gullieuszik}, M., {et~al.}
  2017{\natexlab{b}}, \apj, 844, 48

\bibitem[{{Poggianti} {et~al.}(1999){Poggianti}, {Smail}, {Dressler}, {Couch},
  {Barger}, {Butcher}, {Ellis}, \& {Oemler}}]{Poggianti1999}
{Poggianti}, B.~M., {Smail}, I., {Dressler}, A., {et~al.} 1999, \apj, 518, 576

\bibitem[{{Radovich} {et~al.}(2019){Radovich}, {Poggianti}, {Jaff{\'e}},
  {Moretti}, {Bettoni}, {Gullieuszik}, {Vulcani}, \& {Fritz}}]{Radovich2019}
{Radovich}, M., {Poggianti}, B., {Jaff{\'e}}, Y.~L., {et~al.} 2019, \mnras,
  486, 486

\bibitem[{{Ramatsoku} {et~al.}(2020){Ramatsoku}, {Serra}, {Poggianti},
  {Moretti}, {Gullieuszik}, {Bettoni}, {Deb}, {Franchetto}, {van Gorkom},
  {Jaff{\'e}}, {Tonnesen}, {Verheijen}, {Vulcani}, {Andati}, {de Blok},
  {J{\'o}zsa}, {Kamphuis}, {Kleiner}, {Maccagni}, {Makhathini}, {Moln{\'a}r},
  {Ramaila}, {Smirnov}, \& {Thorat}}]{Ramatsoku2020}
{Ramatsoku}, M., {Serra}, P., {Poggianti}, B.~M., {et~al.} 2020, \aap, 640, A22

\bibitem[{{Ramatsoku} {et~al.}(2019){Ramatsoku}, {Serra}, {Poggianti},
  {Moretti}, {Gullieuszik}, {Bettoni}, {Deb}, {Fritz}, {van Gorkom},
  {Jaff{\'e}}, {Tonnesen}, {Verheijen}, {Vulcani}, {Hugo}, {J{\'o}zsa},
  {Maccagni}, {Makhathini}, {Ramaila}, {Smirnov}, \& {Thorat}}]{Ramatsoku2019}
{Ramatsoku}, M., {Serra}, P., {Poggianti}, B.~M., {et~al.} 2019, \mnras, 487,
  4580

\bibitem[{{Rand}(1998)}]{Rand1998}
{Rand}, R.~J. 1998, \apj, 501, 137

\bibitem[{{Rhee} {et~al.}(2017){Rhee}, {Smith}, {Choi}, {Yi}, {Jaff{\'e}},
  {Candlish}, \& {S{\'a}nchez-J{\'a}nssen}}]{Rhee2017}
{Rhee}, J., {Smith}, R., {Choi}, H., {et~al.} 2017, \apj, 843, 128

\bibitem[{{Roberts} \& {Parker}(2020)}]{Roberts2020}
{Roberts}, I.~D. \& {Parker}, L.~C. 2020, \mnras, 495, 554

\bibitem[{{Roberts} {et~al.}(2022{\natexlab{a}}){Roberts}, {Parker}, {Gwyn},
  {Hudson}, {Carlberg}, {McConnachie}, {Cuillandre}, {Chambers}, {Duc},
  {Furusawa}, {Gavazzi}, {Hill}, {Huber}, {Ibata}, {Kilbinger}, {Mei},
  {Mellier}, {Miyazaki}, {Oguri}, \& {Wainscoat}}]{Roberts2022a}
{Roberts}, I.~D., {Parker}, L.~C., {Gwyn}, S., {et~al.} 2022{\natexlab{a}},
  \mnras, 509, 1342

\bibitem[{{Roberts} {et~al.}(2024){Roberts}, {van Weeren}, {Lal}, {Sun},
  {Chen}, {Ignesti}, {Br{\"u}ggen}, {Lyskova}, {Venturi}, \&
  {Yagi}}]{Roberts2024}
{Roberts}, I.~D., {van Weeren}, R.~J., {Lal}, D.~V., {et~al.} 2024, \aap, 683,
  A11

\bibitem[{{Roberts} {et~al.}(2021{\natexlab{a}}){Roberts}, {van Weeren},
  {McGee}, {Botteon}, {Drabent}, {Ignesti}, {Rottgering}, {Shimwell}, \&
  {Tasse}}]{Roberts2021a}
{Roberts}, I.~D., {van Weeren}, R.~J., {McGee}, S.~L., {et~al.}
  2021{\natexlab{a}}, \aap, 650, A111

\bibitem[{{Roberts} {et~al.}(2021{\natexlab{b}}){Roberts}, {van Weeren},
  {McGee}, {Botteon}, {Ignesti}, \& {Rottgering}}]{Roberts2021b}
{Roberts}, I.~D., {van Weeren}, R.~J., {McGee}, S.~L., {et~al.}
  2021{\natexlab{b}}, \aap, 652, A153

\bibitem[{{Roberts} {et~al.}(2022{\natexlab{b}}){Roberts}, {van Weeren},
  {Timmerman}, {Botteon}, {Gendron-Marsolais}, {Ignesti}, \&
  {Rottgering}}]{Roberts2022b}
{Roberts}, I.~D., {van Weeren}, R.~J., {Timmerman}, R., {et~al.}
  2022{\natexlab{b}}, \aap, 658, A44

\bibitem[{{Roediger} {et~al.}(2006){Roediger}, {Br{\"u}ggen}, \&
  {Hoeft}}]{Roediger2006}
{Roediger}, E., {Br{\"u}ggen}, M., \& {Hoeft}, M. 2006, \mnras, 371, 609

\bibitem[{{Roediger} {et~al.}(2014){Roediger}, {Br{\"u}ggen}, {Owers},
  {Ebeling}, \& {Sun}}]{Roediger2014}
{Roediger}, E., {Br{\"u}ggen}, M., {Owers}, M.~S., {Ebeling}, H., \& {Sun}, M.
  2014, \mnras, 443, L114

\bibitem[{{Roediger} \& {Hensler}(2005)}]{Roediger2005}
{Roediger}, E. \& {Hensler}, G. 2005, \aap, 433, 875

\bibitem[{{Rohr} {et~al.}(2023){Rohr}, {Pillepich}, {Nelson}, {Zinger},
  {Joshi}, \& {Ayromlou}}]{Rohr2023}
{Rohr}, E., {Pillepich}, A., {Nelson}, D., {et~al.} 2023, \mnras, 524, 3502

\bibitem[{{Roman-Oliveira} {et~al.}(2021){Roman-Oliveira}, {Chies-Santos},
  {Ferrari}, {Lucatelli}, \& {Rodr{\'\i}guez Del Pino}}]{Roman2021}
{Roman-Oliveira}, F., {Chies-Santos}, A.~L., {Ferrari}, F., {Lucatelli}, G., \&
  {Rodr{\'\i}guez Del Pino}, B. 2021, \mnras, 500, 40

\bibitem[{{Roman-Oliveira} {et~al.}(2019){Roman-Oliveira}, {Chies-Santos},
  {Rodr{\'\i}guez del Pino}, {Arag{\'o}n-Salamanca}, {Gray}, \&
  {Bamford}}]{Roman2019}
{Roman-Oliveira}, F.~V., {Chies-Santos}, A.~L., {Rodr{\'\i}guez del Pino}, B.,
  {et~al.} 2019, \mnras, 484, 892

\bibitem[{{Salinas} {et~al.}(2024){Salinas}, {Jaff{\'e}}, {Smith}, {Shinn},
  {Crossett}, {Gullieuszik}, {Gonz{\'a}lez-Tor{\`a}}, {Piraino-Cerda},
  {Poggianti}, {Vulcani}, {Biviano}, {Louren{\c{c}}o}, {Bilton}, {Kelkar}, \&
  {Calder{\'o}n-Castillo}}]{Salinas2024}
{Salinas}, V., {Jaff{\'e}}, Y.~L., {Smith}, R., {et~al.} 2024, \mnras, 533, 341

\bibitem[{{S{\'a}nchez-Garc{\'\i}a} {et~al.}(2023){S{\'a}nchez-Garc{\'\i}a},
  {Cervantes Sodi}, {Fritz}, {Moretti}, {Poggianti}, {George}, {Gullieuszik},
  {Vulcani}, {Fasano}, \& {Tawfeek}}]{Sanchez2023}
{S{\'a}nchez-Garc{\'\i}a}, O., {Cervantes Sodi}, B., {Fritz}, J., {et~al.}
  2023, \apj, 945, 99

\bibitem[{{Schulz} \& {Struck}(2001)}]{Schulz2001}
{Schulz}, S. \& {Struck}, C. 2001, \mnras, 328, 185

\bibitem[{{Serra} {et~al.}(2023){Serra}, {Maccagni}, {Kleiner}, {Moln{\'a}r},
  {Ramatsoku}, {Loni}, {Loi}, {de Blok}, {Bryan}, {Dettmar}, {Frank}, {van
  Gorkom}, {Govoni}, {Iodice}, {J{\'o}zsa}, {Kamphuis}, {Kraan-Korteweg},
  {Loubser}, {Murgia}, {Oosterloo}, {Peletier}, {Pisano}, {Smith}, {Trager}, \&
  {Verheijen}}]{Serra2023}
{Serra}, P., {Maccagni}, F.~M., {Kleiner}, D., {et~al.} 2023, \aap, 673, A146

\bibitem[{{Serra} {et~al.}(2024){Serra}, {Oosterloo}, {Kamphuis}, {Jozsa}, {de
  Blok}, {Bryan}, {van Gorkom}, {Iodice}, {Kleiner}, {Loni}, {Loubser},
  {Maccagni}, {Molnar}, {Peletier}, {Pisano}, {Ramatsoku}, {Smith},
  {Verheijen}, \& {Zabel}}]{Serra2024}
{Serra}, P., {Oosterloo}, T.~A., {Kamphuis}, P., {et~al.} 2024, arXiv e-prints,
  arXiv:2407.09082

\bibitem[{{Sharp} \& {Bland-Hawthorn}(2010)}]{Sharp2010}
{Sharp}, R.~G. \& {Bland-Hawthorn}, J. 2010, \apj, 711, 818

\bibitem[{{Slavin} {et~al.}(1993){Slavin}, {Shull}, \& {Begelman}}]{Slavin1993}
{Slavin}, J.~D., {Shull}, J.~M., \& {Begelman}, M.~C. 1993, \apj, 407, 83

\bibitem[{{Smith} {et~al.}(2010){Smith}, {Lucey}, {Hammer}, {Hornschemeier},
  {Carter}, {Hudson}, {Marzke}, {Mouhcine}, {Eftekharzadeh}, {James},
  {Khosroshahi}, {Kourkchi}, \& {Karick}}]{Smith2010}
{Smith}, R.~J., {Lucey}, J.~R., {Hammer}, D., {et~al.} 2010, \mnras, 408, 1417

\bibitem[{{Sokolowski} {et~al.}(1991){Sokolowski}, {Bland-Hawthorn}, \&
  {Cecil}}]{Sokolowski1991}
{Sokolowski}, J., {Bland-Hawthorn}, J., \& {Cecil}, G. 1991, \apj, 375, 583

\bibitem[{{Struck}(1999)}]{Struck1999}
{Struck}, C. 1999, \physrep, 321, 1

\bibitem[{{Sun} {et~al.}(2010){Sun}, {Donahue}, {Roediger}, {Nulsen}, {Voit},
  {Sarazin}, {Forman}, \& {Jones}}]{Sun2010}
{Sun}, M., {Donahue}, M., {Roediger}, E., {et~al.} 2010, \apj, 708, 946

\bibitem[{{Sun} {et~al.}(2021){Sun}, {Ge}, {Luo}, {Yagi}, {J{\'a}chym},
  {Boselli}, {Fossati}, {Nulsen}, {Yoshida}, \& {Gavazzi}}]{Sun2022}
{Sun}, M., {Ge}, C., {Luo}, R., {et~al.} 2021, Nature Astronomy, 6, 270

\bibitem[{{Tomi{\v{c}}i{\'c}} {et~al.}(2021{\natexlab{a}}){Tomi{\v{c}}i{\'c}},
  {Vulcani}, {Poggianti}, {Mingozzi}, {Werle}, {Bettoni}, {Franchetto},
  {Gullieuszik}, {Moretti}, {Fritz}, \& {Bellhouse}}]{Tomicic2021a}
{Tomi{\v{c}}i{\'c}}, N., {Vulcani}, B., {Poggianti}, B.~M., {et~al.}
  2021{\natexlab{a}}, \apj, 907, 22

\bibitem[{{Tomi{\v{c}}i{\'c}} {et~al.}(2021{\natexlab{b}}){Tomi{\v{c}}i{\'c}},
  {Vulcani}, {Poggianti}, {Werle}, {M{\"u}ller}, {Mingozzi}, {Gullieuszik},
  {Wolter}, {Radovich}, {Moretti}, {Franchetto}, {Bellhouse}, \&
  {Fritz}}]{Tomicic2021b}
{Tomi{\v{c}}i{\'c}}, N., {Vulcani}, B., {Poggianti}, B.~M., {et~al.}
  2021{\natexlab{b}}, \apj, 922, 131

\bibitem[{{Tomi{\v{c}}i{\'c}} {et~al.}(2024){Tomi{\v{c}}i{\'c}}, {Werle},
  {Vulcani}, {Ignesti}, {Moretti}, {Wolter}, {George}, {Poggianti}, \&
  {Gullieuszik}}]{Tomicic2024}
{Tomi{\v{c}}i{\'c}}, N., {Werle}, A., {Vulcani}, B., {et~al.} 2024, \apj, 976,
  90

\bibitem[{{Tonnesen} \& {Bryan}(2009)}]{TonnesenBryan2009}
{Tonnesen}, S. \& {Bryan}, G.~L. 2009, \apj, 694, 789

\bibitem[{{Tonnesen} \& {Bryan}(2012)}]{TonnesenBryan2012}
{Tonnesen}, S. \& {Bryan}, G.~L. 2012, \mnras, 422, 1609

\bibitem[{{Tonnesen} \& {Bryan}(2021)}]{TonnesenBryan2021}
{Tonnesen}, S. \& {Bryan}, G.~L. 2021, \apj, 911, 68

\bibitem[{{Tonnesen} {et~al.}(2011){Tonnesen}, {Bryan}, \&
  {Chen}}]{Tonnesen2011}
{Tonnesen}, S., {Bryan}, G.~L., \& {Chen}, R. 2011, \apj, 731, 98

\bibitem[{{Varela} {et~al.}(2009){Varela}, {D'Onofrio}, {Marmo}, {Fasano},
  {Bettoni}, {Cava}, {Couch}, {Dressler}, {Kj{\ae}rgaard}, {Moles},
  {Pignatelli}, {Poggianti}, \& {Valentinuzzi}}]{Varela2009}
{Varela}, J., {D'Onofrio}, M., {Marmo}, C., {et~al.} 2009, \aap, 497, 667

\bibitem[{{Vazdekis} {et~al.}(2010){Vazdekis}, {S{\'a}nchez-Bl{\'a}zquez},
  {Falc{\'o}n-Barroso}, {Cenarro}, {Beasley}, {Cardiel}, {Gorgas}, \&
  {Peletier}}]{Vazdekis2010}
{Vazdekis}, A., {S{\'a}nchez-Bl{\'a}zquez}, P., {Falc{\'o}n-Barroso}, J.,
  {et~al.} 2010, \mnras, 404, 1639

\bibitem[{{Vulcani} {et~al.}(2020{\natexlab{a}}){Vulcani}, {Fritz},
  {Poggianti}, {Bettoni}, {Franchetto}, {Moretti}, {Gullieuszik}, {Jaff{\'e}},
  {Biviano}, {Radovich}, \& {Mingozzi}}]{Vulcani2020a}
{Vulcani}, B., {Fritz}, J., {Poggianti}, B.~M., {et~al.} 2020{\natexlab{a}},
  \apj, 892, 146

\bibitem[{{Vulcani} {et~al.}(2017){Vulcani}, {Moretti}, {Poggianti}, {Fasano},
  {Fritz}, {Gullieuszik}, {Duc}, {Jaff{\'e}}, \& {Bettoni}}]{Vulcani2017c}
{Vulcani}, B., {Moretti}, A., {Poggianti}, B.~M., {et~al.} 2017, \apj, 850, 163

\bibitem[{{Vulcani} {et~al.}(2024){Vulcani}, {Moretti}, {Poggianti},
  {Radovich}, {Werle}, {Gullieuszik}, {Fritz}, {Bacchini}, \&
  {Richard}}]{Vulcani2024}
{Vulcani}, B., {Moretti}, A., {Poggianti}, B.~M., {et~al.} 2024, \aap, 682,
  A117

\bibitem[{{Vulcani} {et~al.}(2011){Vulcani}, {Poggianti},
  {Arag{\'o}n-Salamanca}, {Fasano}, {Rudnick}, {Valentinuzzi}, {Dressler},
  {Bettoni}, {Cava}, {D'Onofrio}, {Fritz}, {Moretti}, {Omizzolo}, \&
  {Varela}}]{Vulcani2011}
{Vulcani}, B., {Poggianti}, B.~M., {Arag{\'o}n-Salamanca}, A., {et~al.} 2011,
  \mnras, 412, 246

\bibitem[{{Vulcani} {et~al.}(2023){Vulcani}, {Poggianti}, {Gullieuszik},
  {Moretti}, {Fritz}, {Bettoni}, {Facciolli}, {Fasano}, \&
  {Omizzolo}}]{Vulcani2023}
{Vulcani}, B., {Poggianti}, B.~M., {Gullieuszik}, M., {et~al.} 2023, \apj, 949,
  73

\bibitem[{{Vulcani} {et~al.}(2018{\natexlab{a}}){Vulcani}, {Poggianti},
  {Gullieuszik}, {Moretti}, {Tonnesen}, {Jaff{\'e}}, {Fritz}, {Fasano}, \&
  {Bettoni}}]{Vulcani2018b}
{Vulcani}, B., {Poggianti}, B.~M., {Gullieuszik}, M., {et~al.}
  2018{\natexlab{a}}, \apjl, 866, L25

\bibitem[{{Vulcani} {et~al.}(2018{\natexlab{b}}){Vulcani}, {Poggianti},
  {Jaff{\'e}}, {Moretti}, {Fritz}, {Gullieuszik}, {Bettoni}, {Fasano},
  {Tonnesen}, \& {McGee}}]{Vulcani2018c}
{Vulcani}, B., {Poggianti}, B.~M., {Jaff{\'e}}, Y.~L., {et~al.}
  2018{\natexlab{b}}, \mnras, 480, 3152

\bibitem[{{Vulcani} {et~al.}(2021){Vulcani}, {Poggianti}, {Moretti},
  {Franchetto}, {Bacchini}, {McGee}, {Jaff{\'e}}, {Mingozzi}, {Werle},
  {Tomi{\v{c}}i{\'c}}, {Fritz}, {Bettoni}, {Wolter}, \&
  {Gullieuszik}}]{Vulcani2021}
{Vulcani}, B., {Poggianti}, B.~M., {Moretti}, A., {et~al.} 2021, \apj, 914, 27

\bibitem[{{Vulcani} {et~al.}(2019{\natexlab{a}}){Vulcani}, {Poggianti},
  {Moretti}, {Franchetto}, {Gullieuszik}, {Fritz}, {Bettoni}, {Tonnesen},
  {Radovich}, {Jaff{\'e}}, {McGee}, {Bellhouse}, \& {Fasano}}]{Vulcani2019b}
{Vulcani}, B., {Poggianti}, B.~M., {Moretti}, A., {et~al.} 2019{\natexlab{a}},
  \mnras, 488, 1597

\bibitem[{{Vulcani} {et~al.}(2019{\natexlab{b}}){Vulcani}, {Poggianti},
  {Moretti}, {Gullieuszik}, {Fritz}, {Franchetto}, {Fasano}, {Bettoni}, \&
  {Jaff{\'e}}}]{Vulcani2019a}
{Vulcani}, B., {Poggianti}, B.~M., {Moretti}, A., {et~al.} 2019{\natexlab{b}},
  \mnras, 487, 2278

\bibitem[{{Vulcani} {et~al.}(2018{\natexlab{c}}){Vulcani}, {Poggianti},
  {Moretti}, {Mapelli}, {Fasano}, {Fritz}, {Jaff{\'e}}, {Bettoni},
  {Gullieuszik}, \& {Bellhouse}}]{Vulcani2018a}
{Vulcani}, B., {Poggianti}, B.~M., {Moretti}, A., {et~al.} 2018{\natexlab{c}},
  \apj, 852, 94

\bibitem[{{Vulcani} {et~al.}(2013){Vulcani}, {Poggianti}, {Oemler}, {Dressler},
  {Arag{\'o}n-Salamanca}, {De Lucia}, {Moretti}, {Gladders}, {Abramson}, \&
  {Halliday}}]{Vulcani2013}
{Vulcani}, B., {Poggianti}, B.~M., {Oemler}, A., {et~al.} 2013, \aap, 550, A58

\bibitem[{{Vulcani} {et~al.}(2022){Vulcani}, {Poggianti}, {Smith}, {Moretti},
  {Jaff{\'e}}, {Gullieuszik}, {Fritz}, \& {Bellhouse}}]{Vulcani2022}
{Vulcani}, B., {Poggianti}, B.~M., {Smith}, R., {et~al.} 2022, \apj, 927, 91

\bibitem[{{Vulcani} {et~al.}(2020{\natexlab{b}}){Vulcani}, {Poggianti},
  {Tonnesen}, {McGee}, {Moretti}, {Fritz}, {Gullieuszik}, {Jaff{\'e}},
  {Franchetto}, {Tomi{\v{c}}i{\'c}}, {Mingozzi}, {Bettoni}, \&
  {Wolter}}]{Vulcani2020b}
{Vulcani}, B., {Poggianti}, B.~M., {Tonnesen}, S., {et~al.} 2020{\natexlab{b}},
  \apj, 899, 98

\bibitem[{{Watson} {et~al.}(2024){Watson}, {Vulcani}, {Werle}, {Poggianti},
  {Gullieuszik}, {Trenti}, {Wang}, \& {Roy}}]{Watson2025}
{Watson}, P.~J., {Vulcani}, B., {Werle}, A., {et~al.} 2024, arXiv e-prints,
  arXiv:2409.15215

\bibitem[{{Werle} {et~al.}(2024){Werle}, {Giunchi}, {Poggianti}, {Gullieuszik},
  {Moretti}, {Zanella}, {Tonnesen}, {Fritz}, {Vulcani}, {Bacchini}, {Akerman},
  {Kulier}, {Tomicic}, {Smith}, \& {Wolter}}]{Werle2024}
{Werle}, A., {Giunchi}, E., {Poggianti}, B., {et~al.} 2024, \aap, 682, A162

\bibitem[{{Werle} {et~al.}(2022){Werle}, {Poggianti}, {Moretti}, {Bellhouse},
  {Vulcani}, {Gullieuszik}, {Radovich}, {Fritz}, {Ignesti}, {Richard},
  {Soucail}, {Bruzual}, {Charlot}, {Mingozzi}, {Bacchini}, {Tomicic}, {Smith},
  {Kulier}, {Peluso}, \& {Franchetto}}]{Werle2022}
{Werle}, A., {Poggianti}, B., {Moretti}, A., {et~al.} 2022, \apj, 930, 43

\bibitem[{{Zhu} {et~al.}(2024){Zhu}, {Tonnesen}, \& {Bryan}}]{Zhu2024}
{Zhu}, J., {Tonnesen}, S., \& {Bryan}, G.~L. 2024, \apj, 960, 54

\bibitem[{{Zinger} {et~al.}(2024){Zinger}, {Joshi}, {Pillepich}, {Rohr}, \&
  {Nelson}}]{Zinger2024}
{Zinger}, E., {Joshi}, G.~D., {Pillepich}, A., {Rohr}, E., \& {Nelson}, D.
  2024, \mnras, 527, 8257

\end{thebibliography}
\bibliographystyle{aa}

\begin{appendix}

\begin{figure*}
\section{All galaxies in the sample} \label{sec:Appendix}   
In this Appendix, we show the H$\alpha$ flux, the gas kinematics, the stellar kinematics and the color composite image for all galaxies in the sample. 
Figure \ref{fig:App1} shows the control sample, Figure \ref{fig:App2} shows the stripping sample, in alphabetical order.
\centerline{\includegraphics[scale=0.2]
{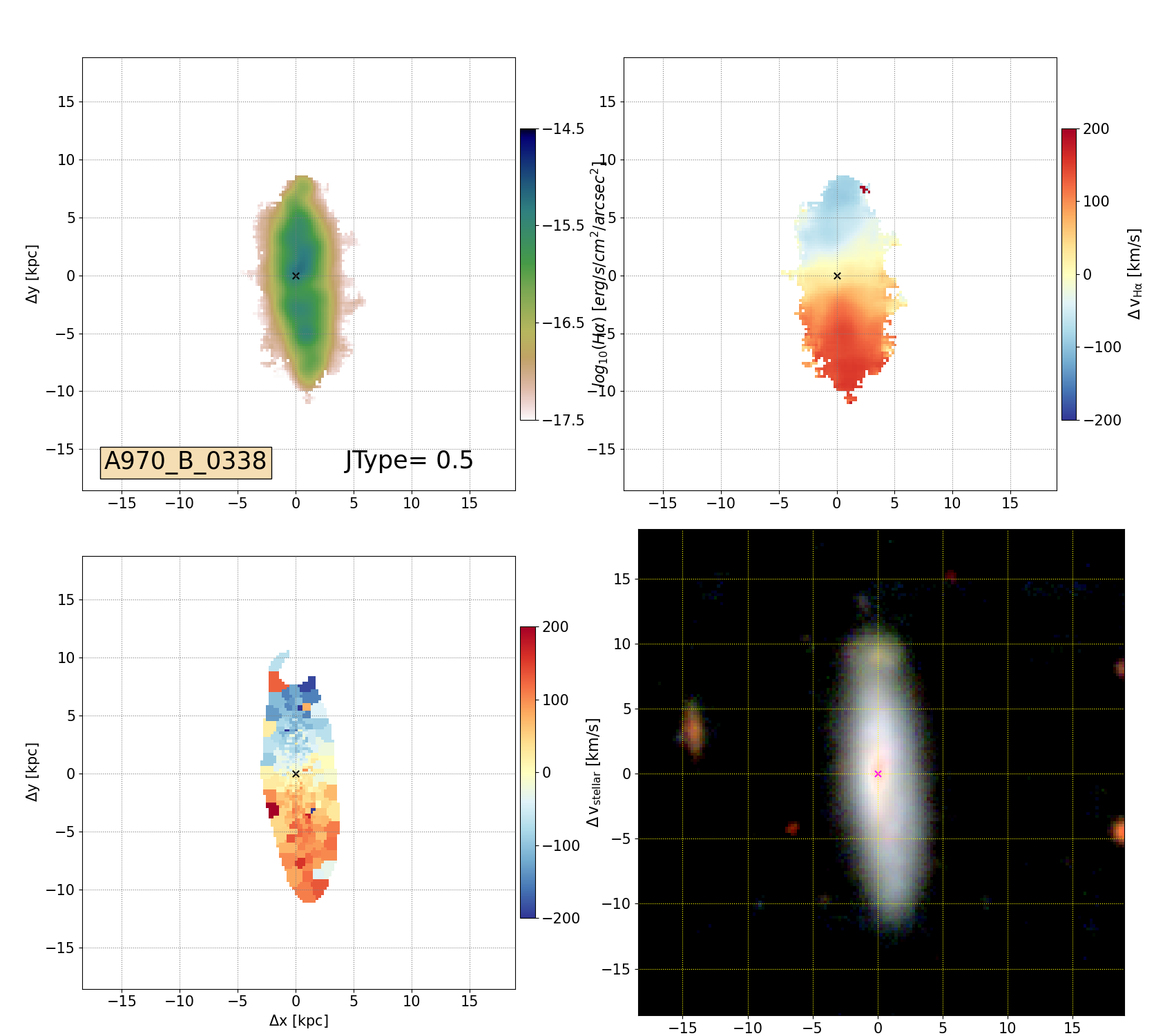}\includegraphics[scale=0.2]{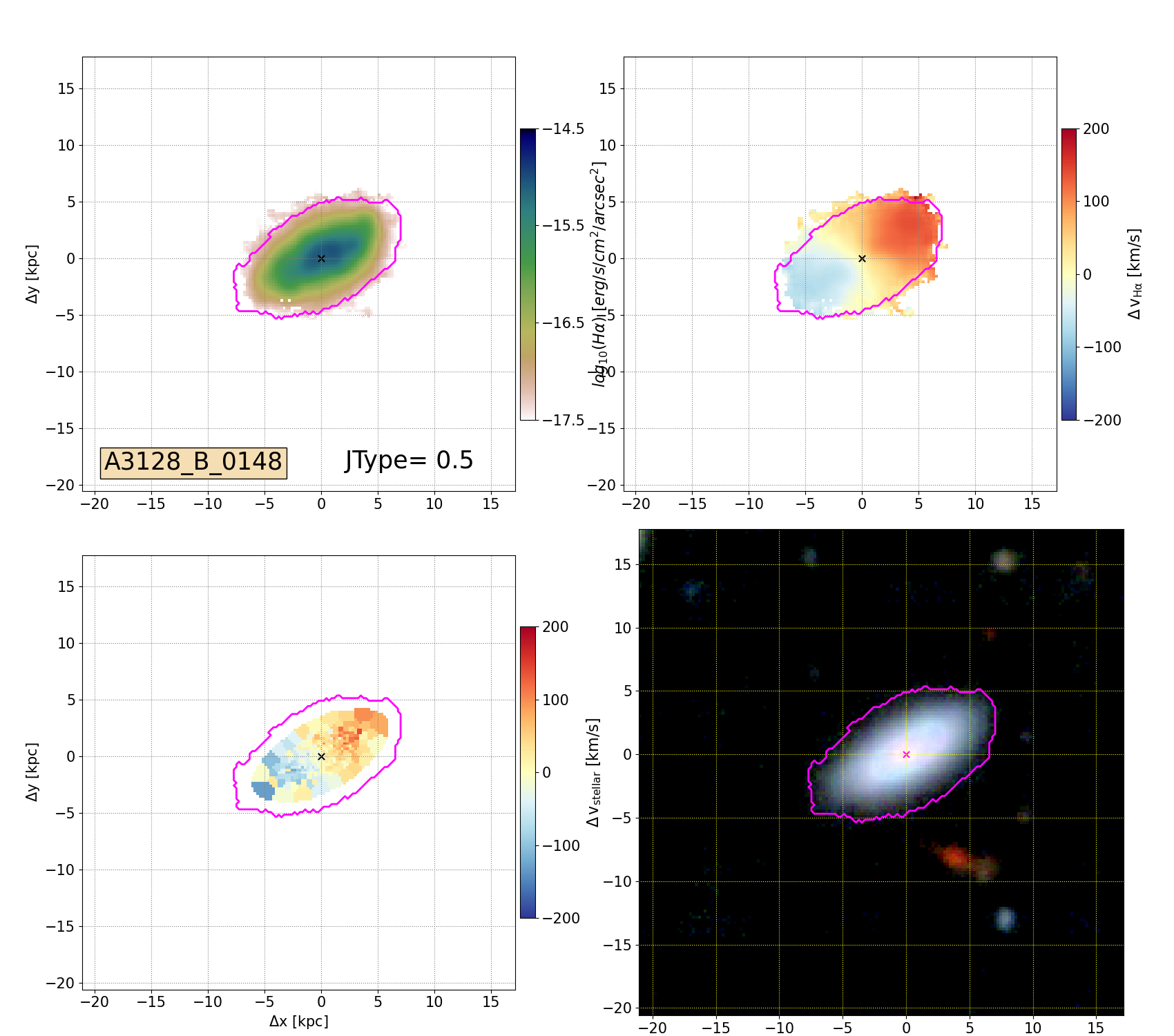}}
\centerline{\includegraphics[scale=0.2]
{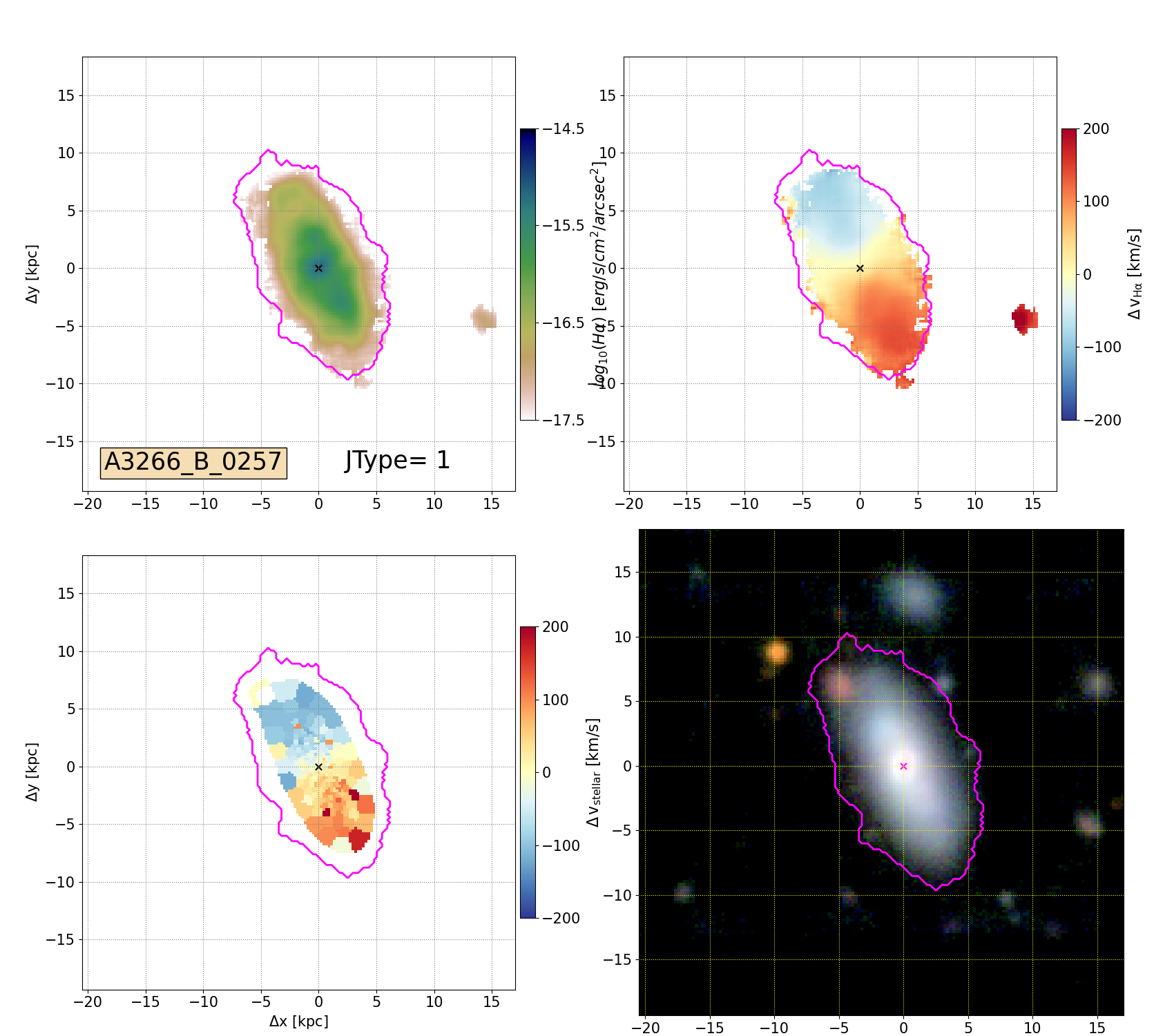}\includegraphics[scale=0.2]{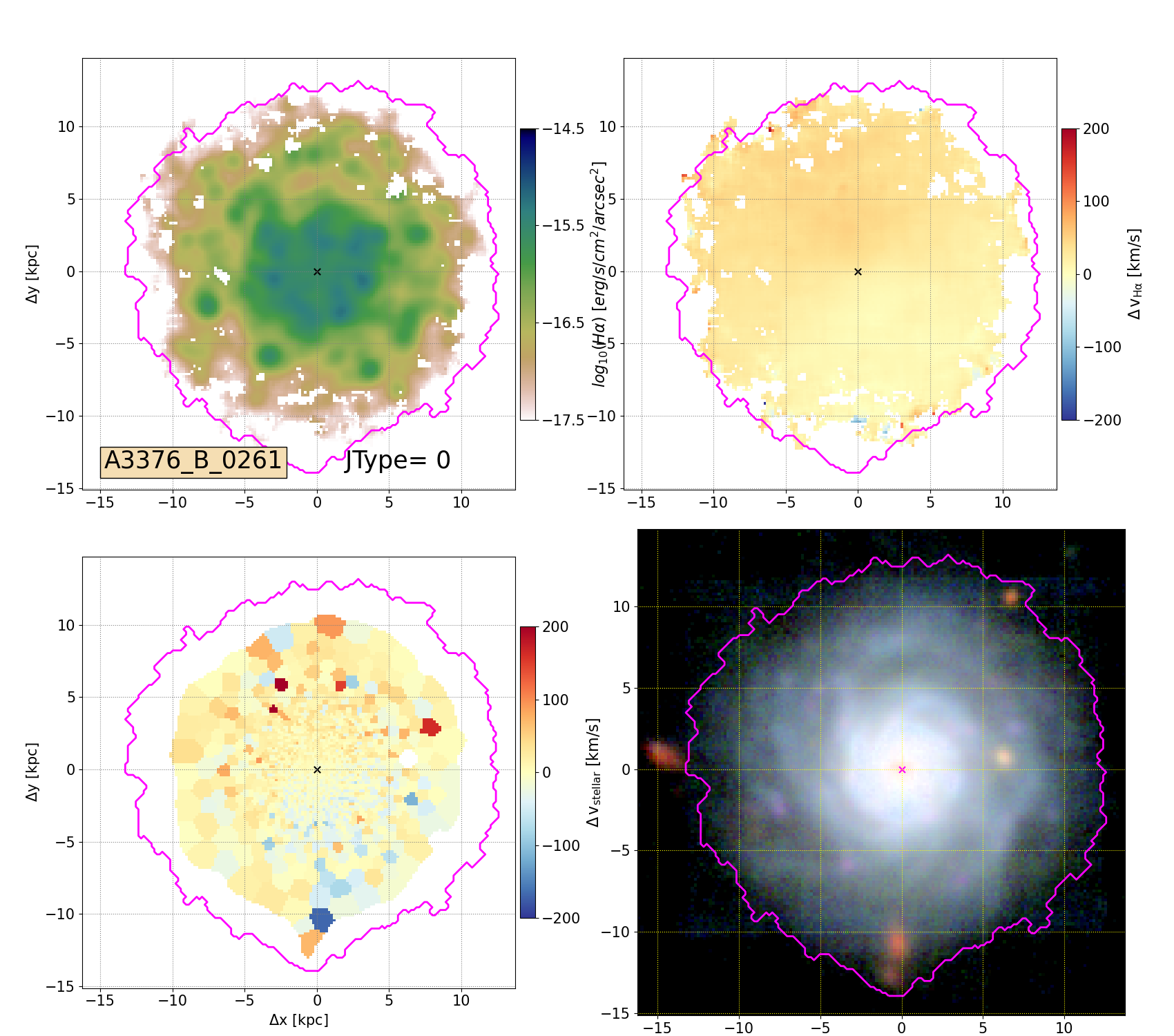}}
\caption{Control sample galaxies. Panels and colors are as in Fig.\ref{fig:Examples1}.
\label{fig:App1}}
\end{figure*}

\begin{figure*}
\centerline{\includegraphics[scale=0.2]
{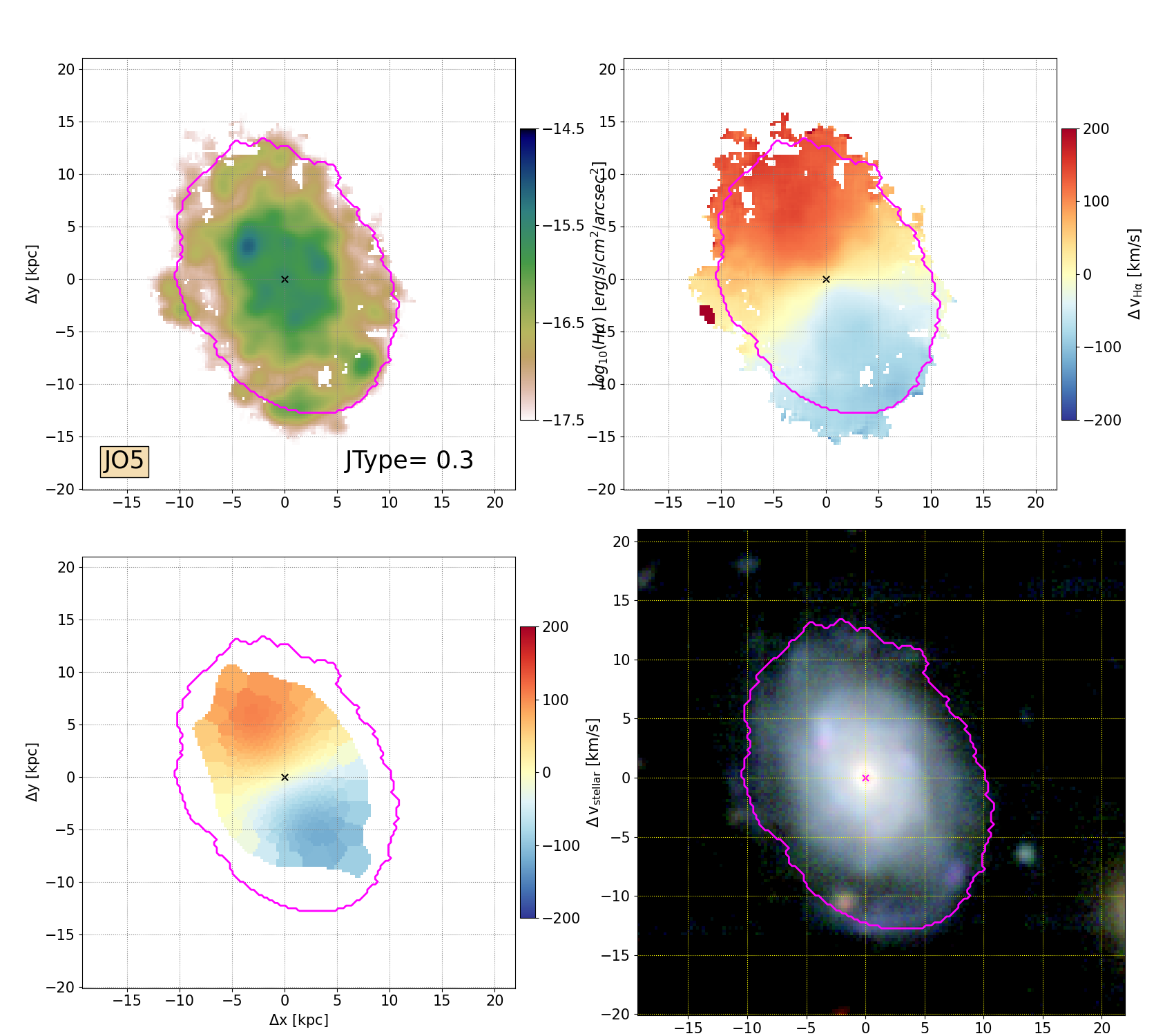}\includegraphics[scale=0.2]{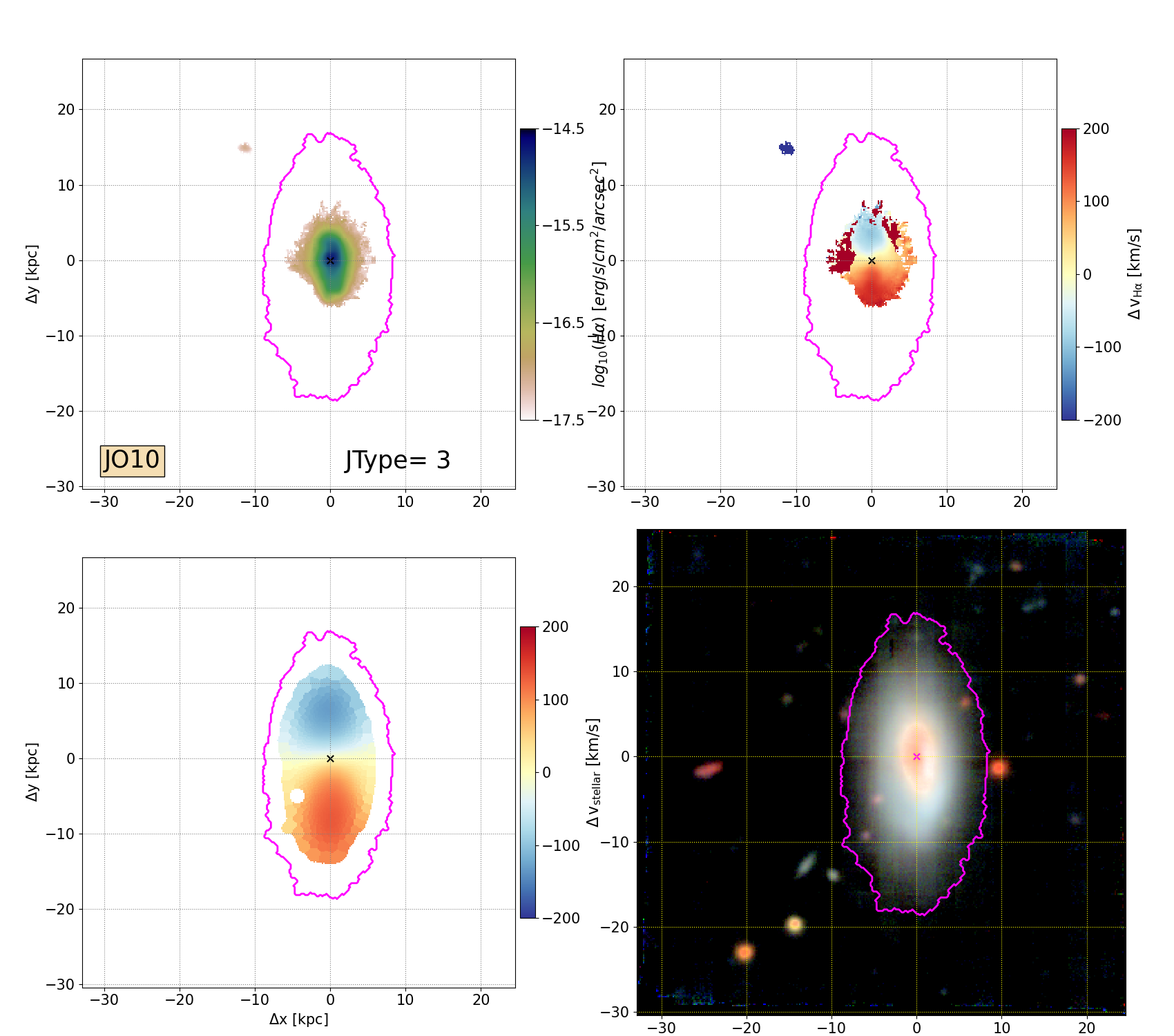}}
\centerline{\includegraphics[scale=0.2]
{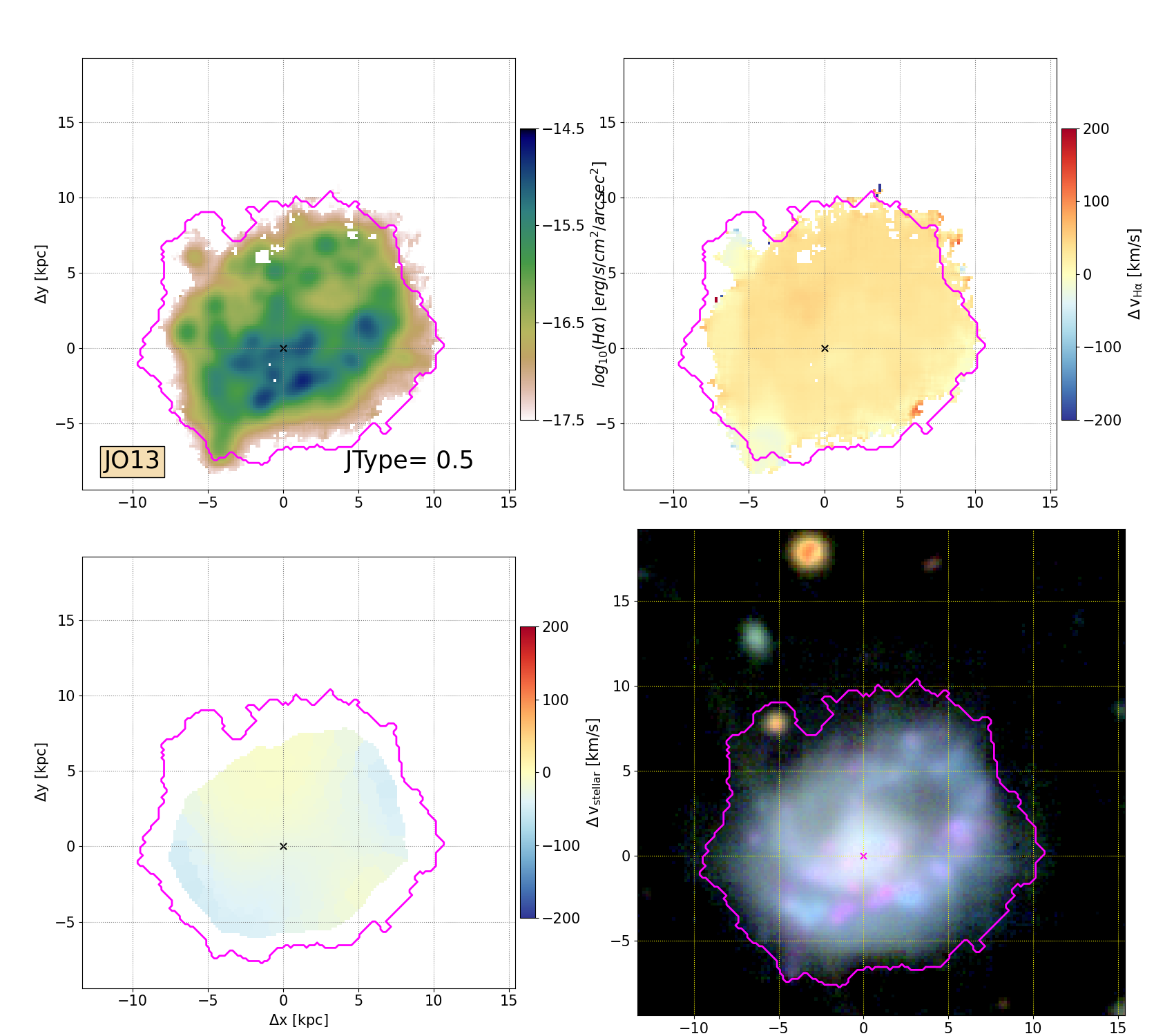}\includegraphics[scale=0.2]{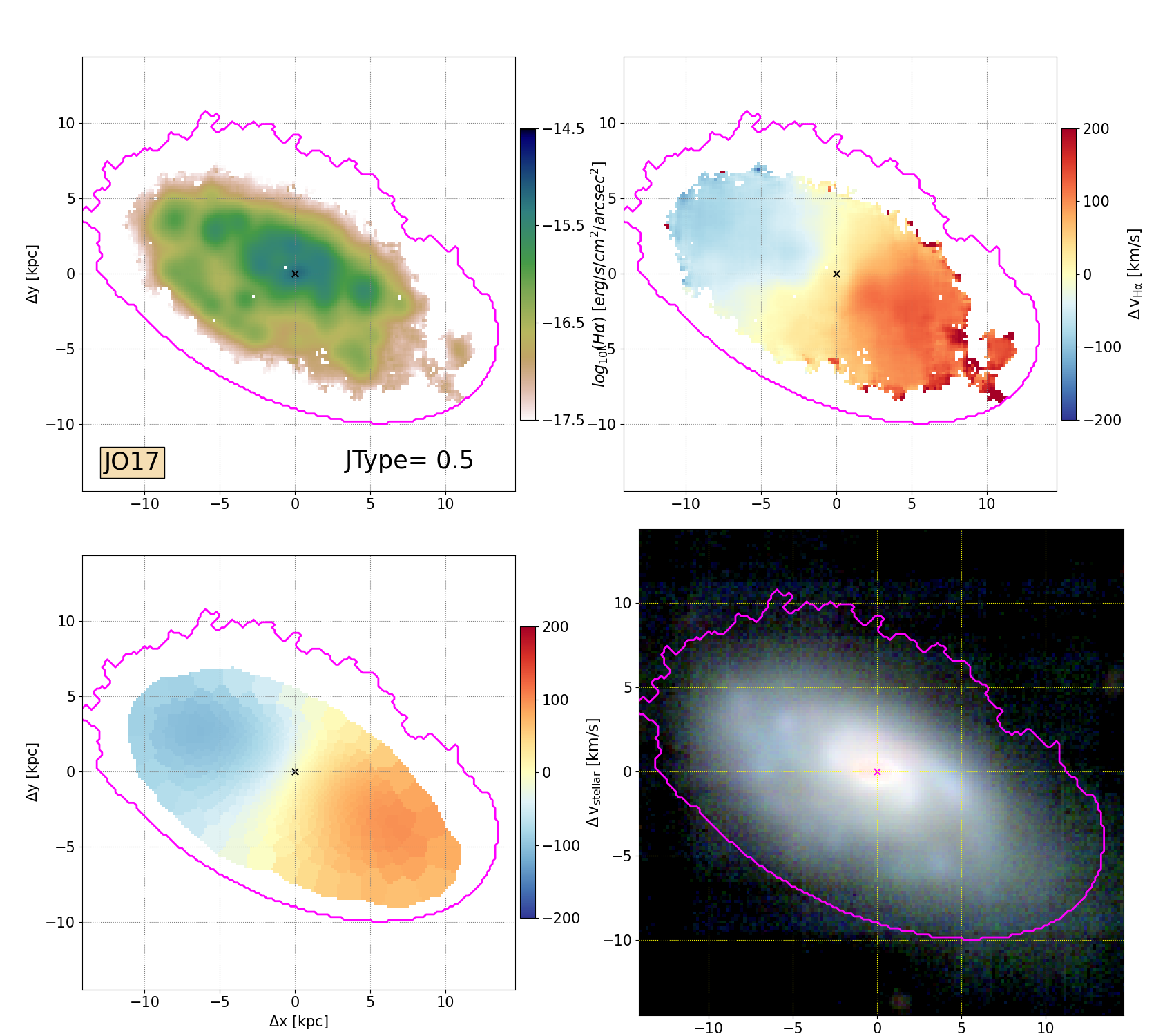}}
\centerline{\includegraphics[scale=0.2]
{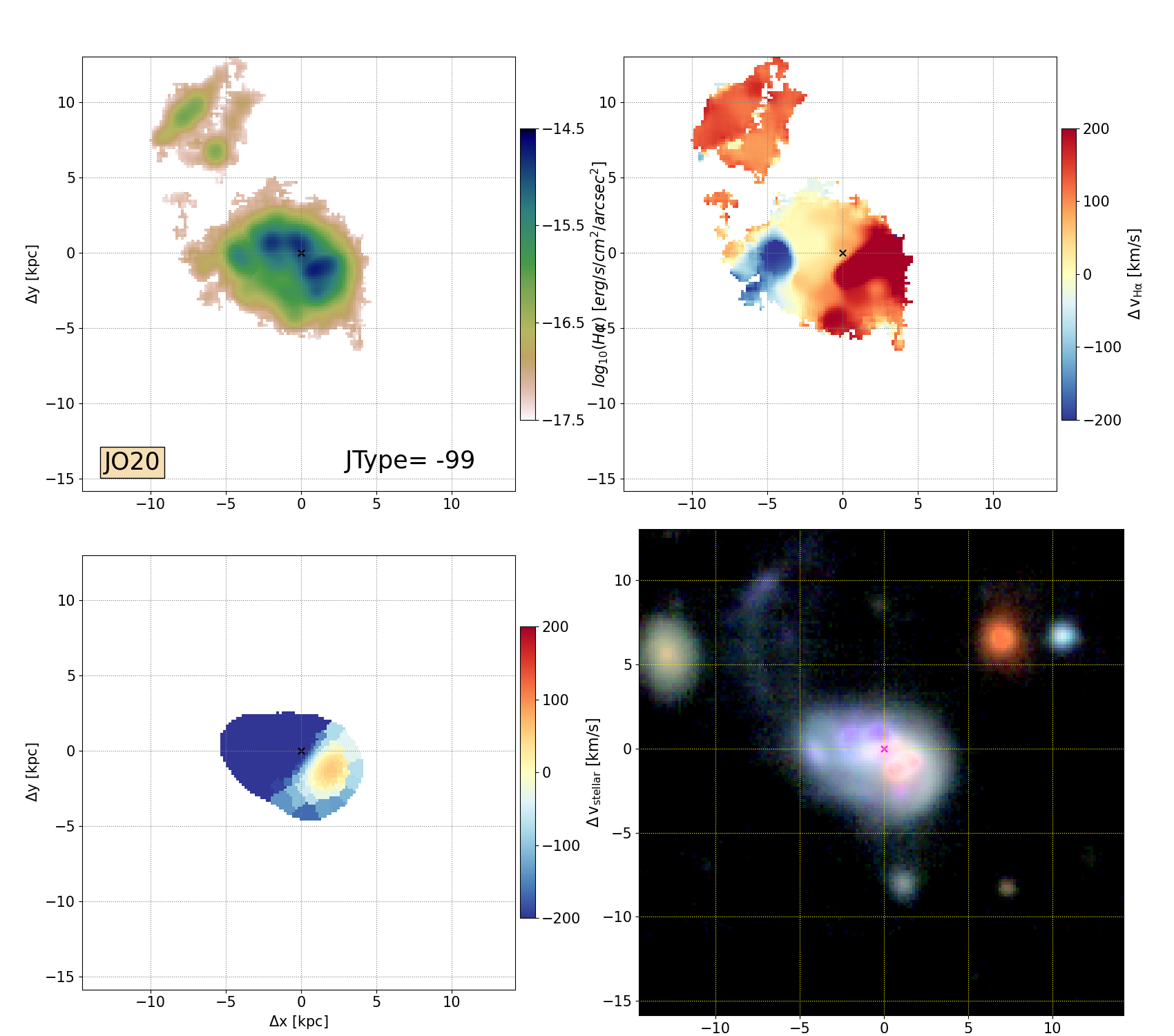}\includegraphics[scale=0.2]{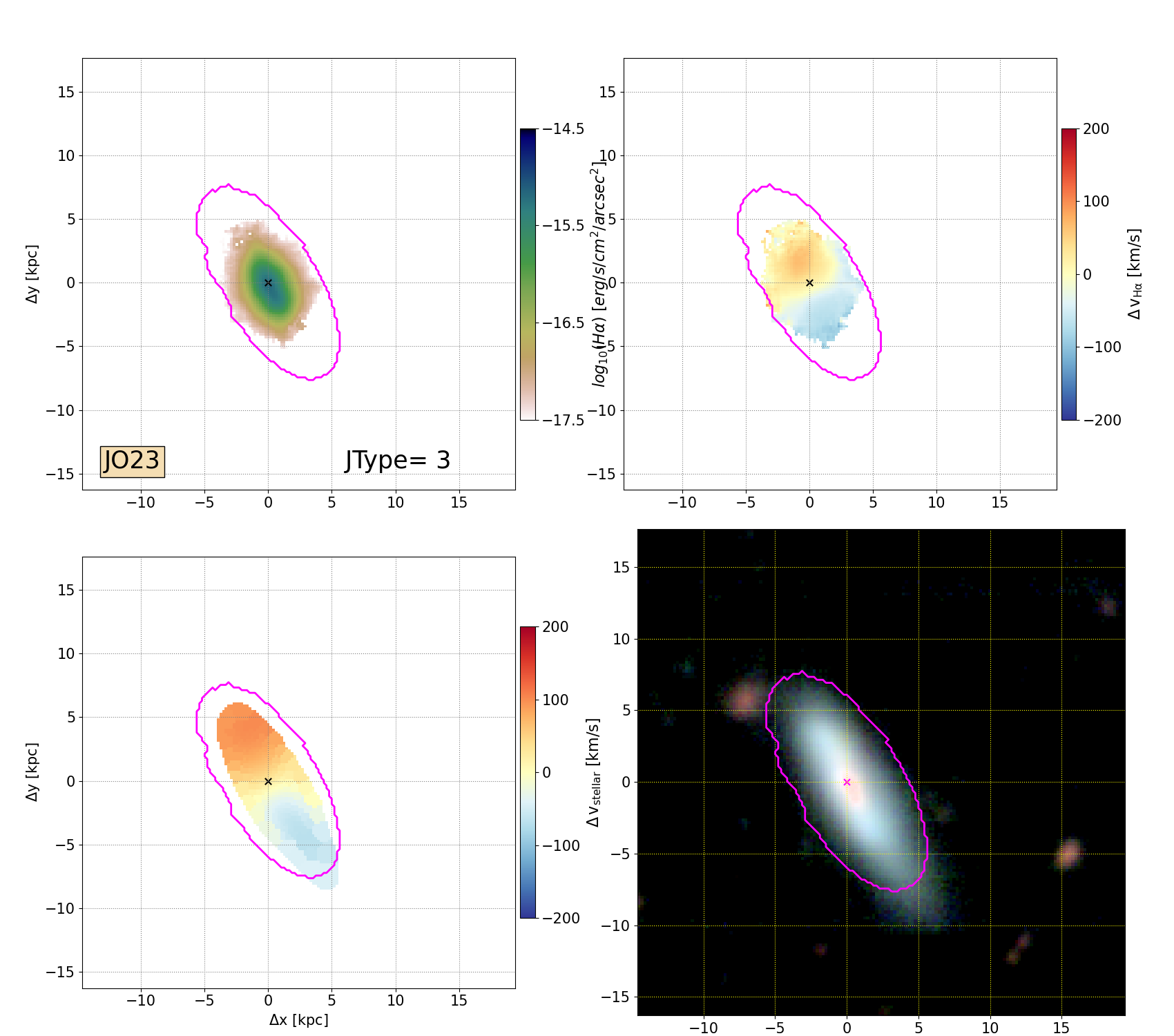}}
\caption{Stripping candidates. Panels and colors are as in Fig.\ref{fig:Examples1}.
\label{fig:App2}}
\end{figure*}

\addtocounter{figure}{-1} 

\begin{figure*}
\centerline{\includegraphics[scale=0.2]
{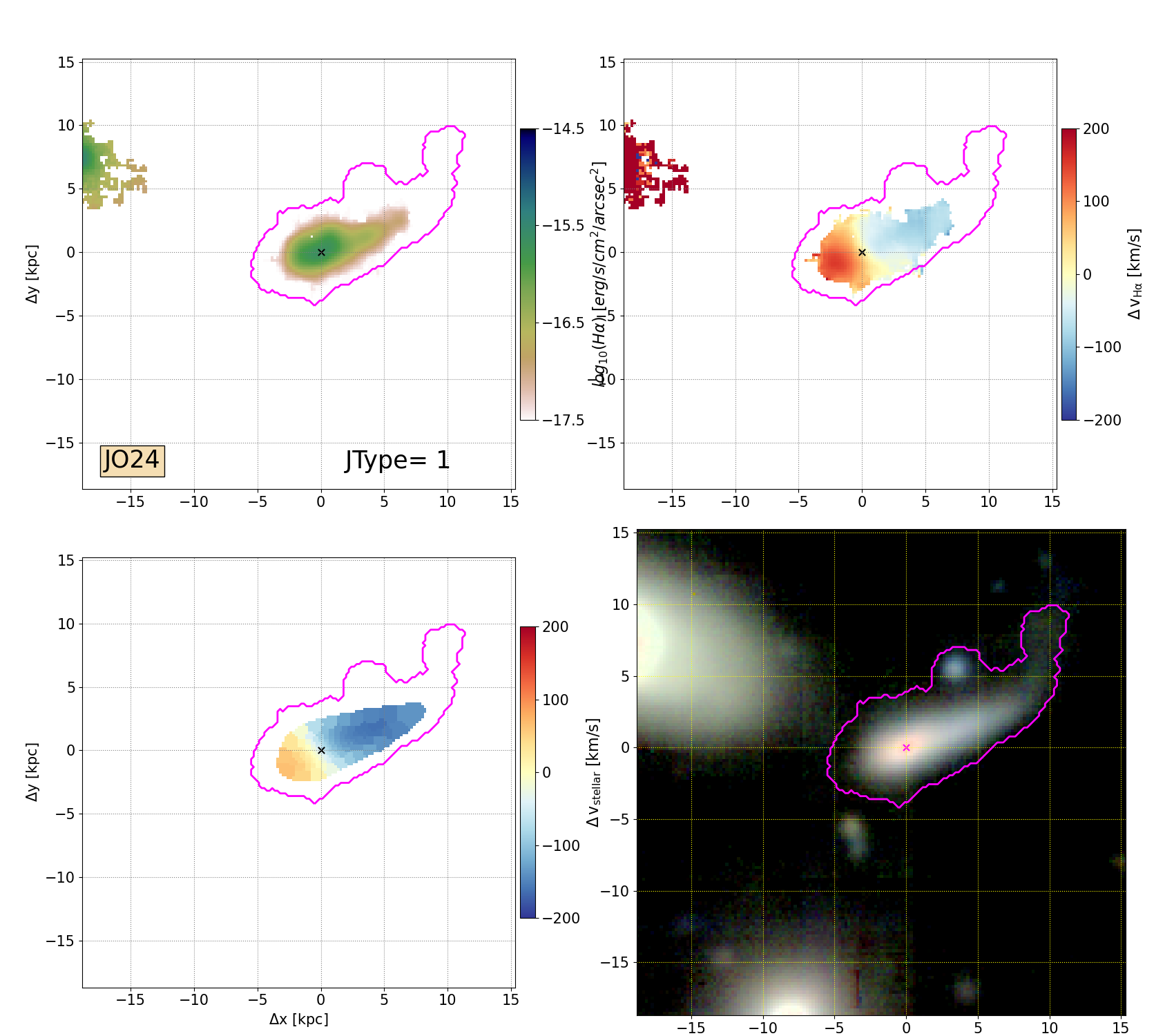}\includegraphics[scale=0.2]{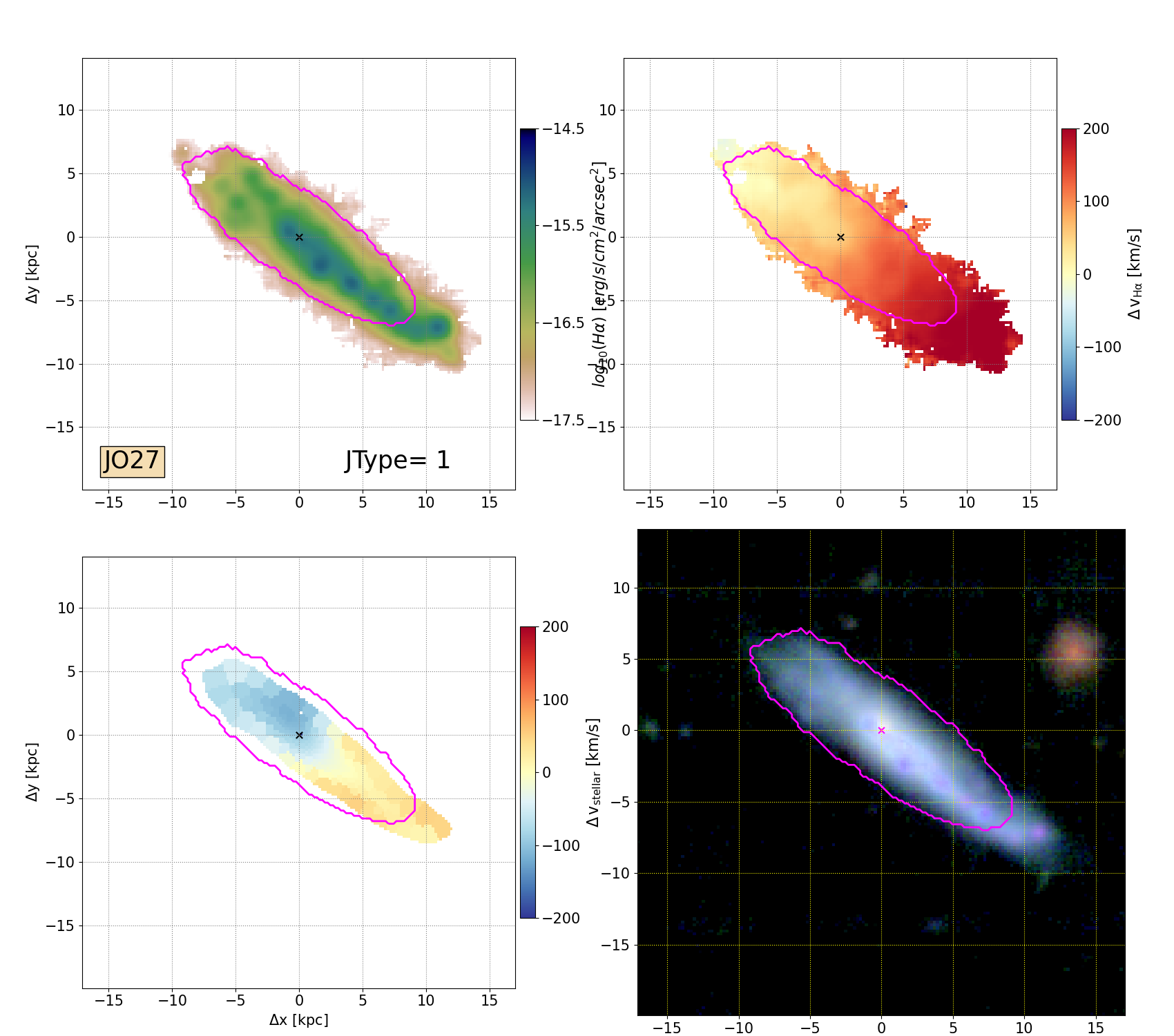}}
\centerline{\includegraphics[scale=0.2]
{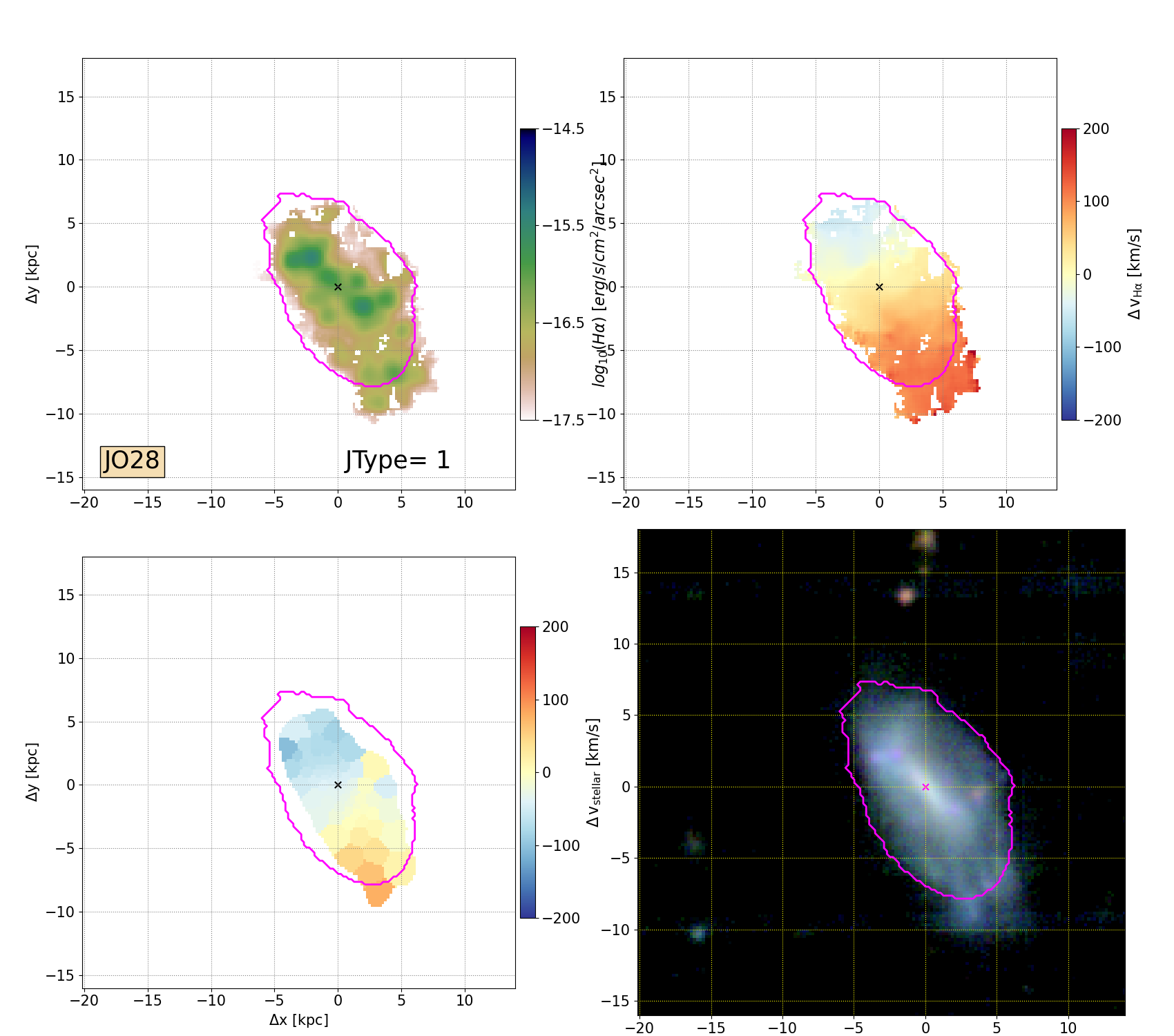}\includegraphics[scale=0.2]{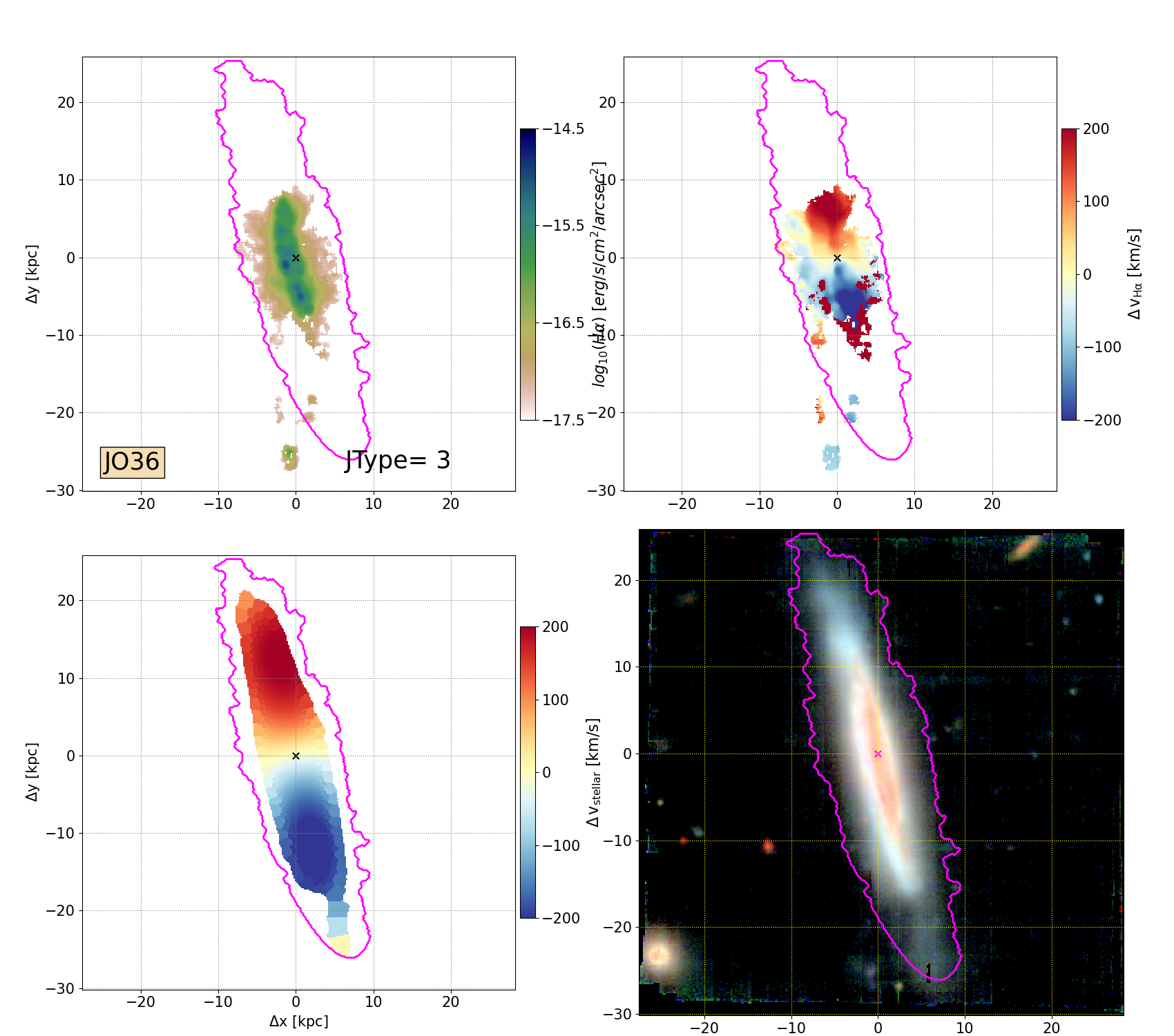}}
\centerline{\includegraphics[scale=0.2]
{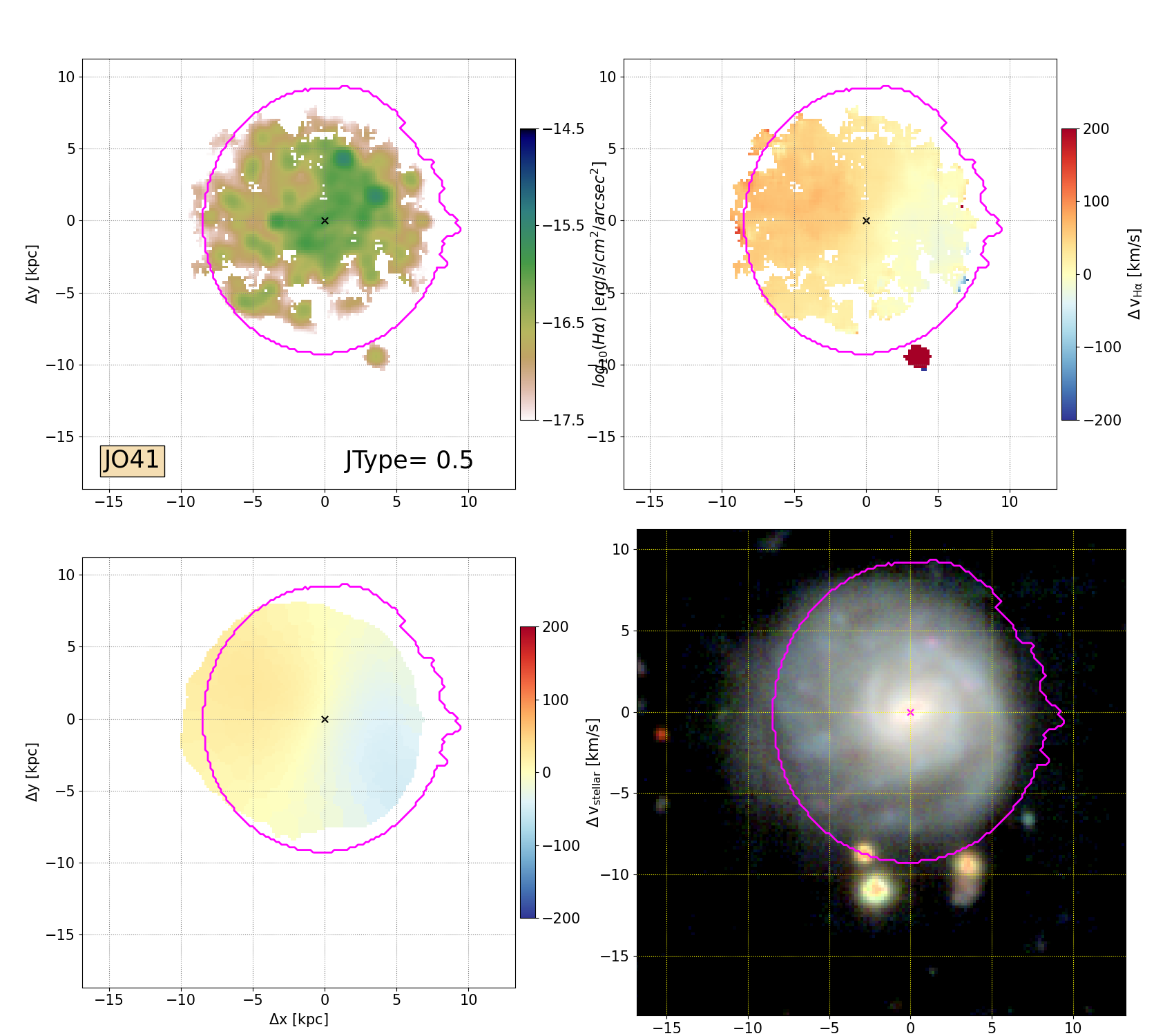}\includegraphics[scale=0.2]{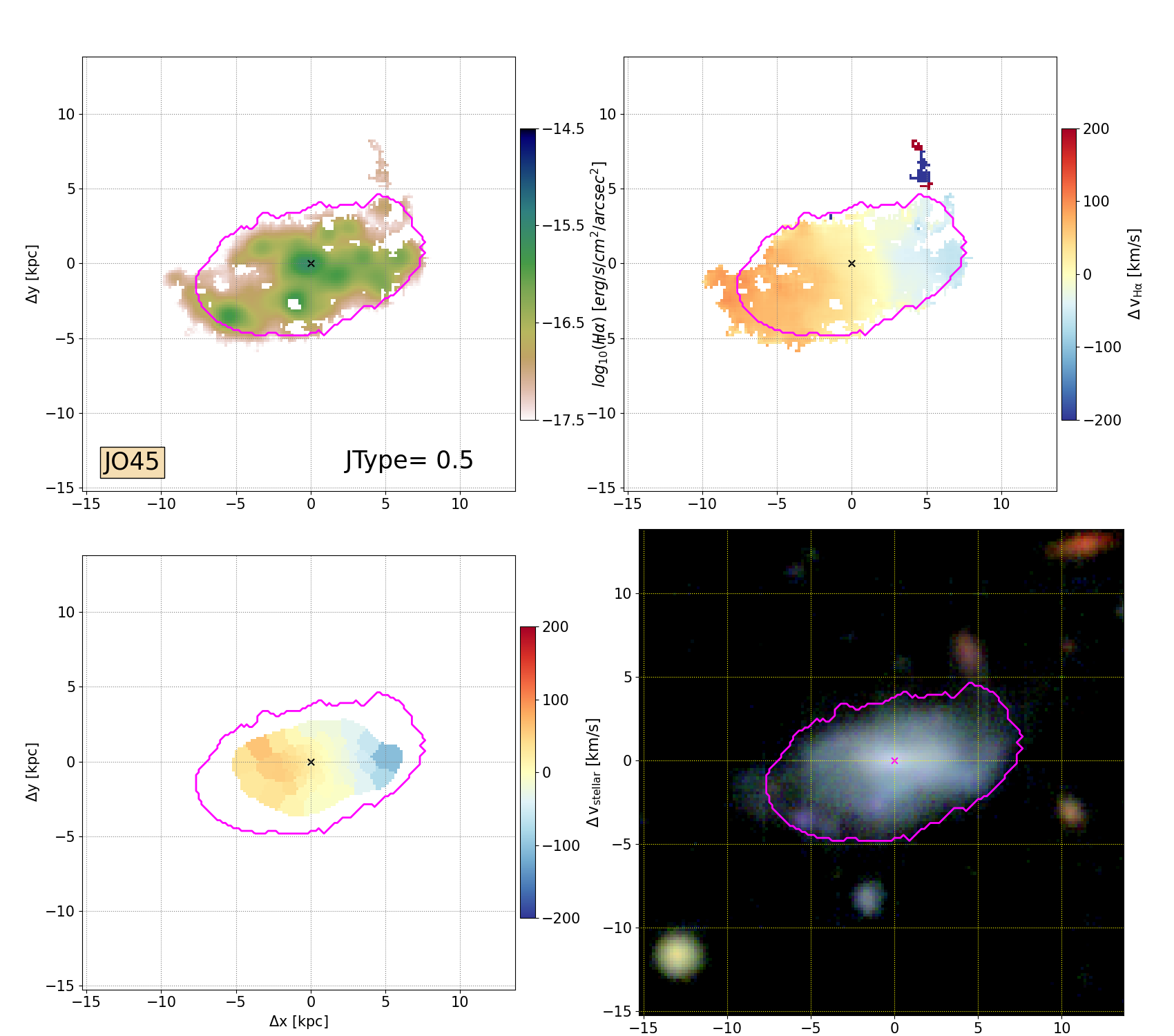}}
\caption{Continued. Stripping candidates.}
\end{figure*}

\addtocounter{figure}{-1} 

\begin{figure*}
\centerline{\includegraphics[scale=0.2]
{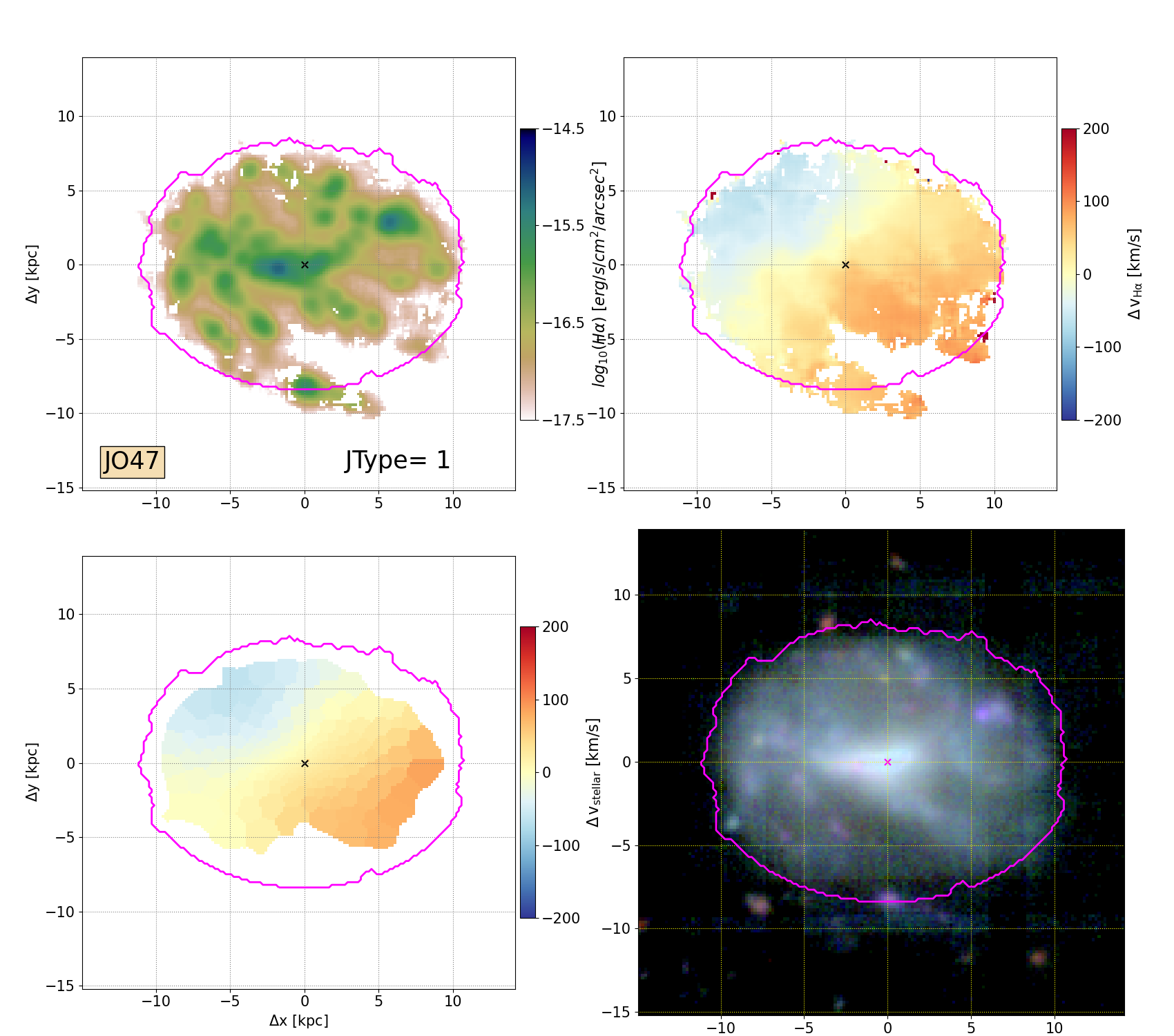}\includegraphics[scale=0.2]{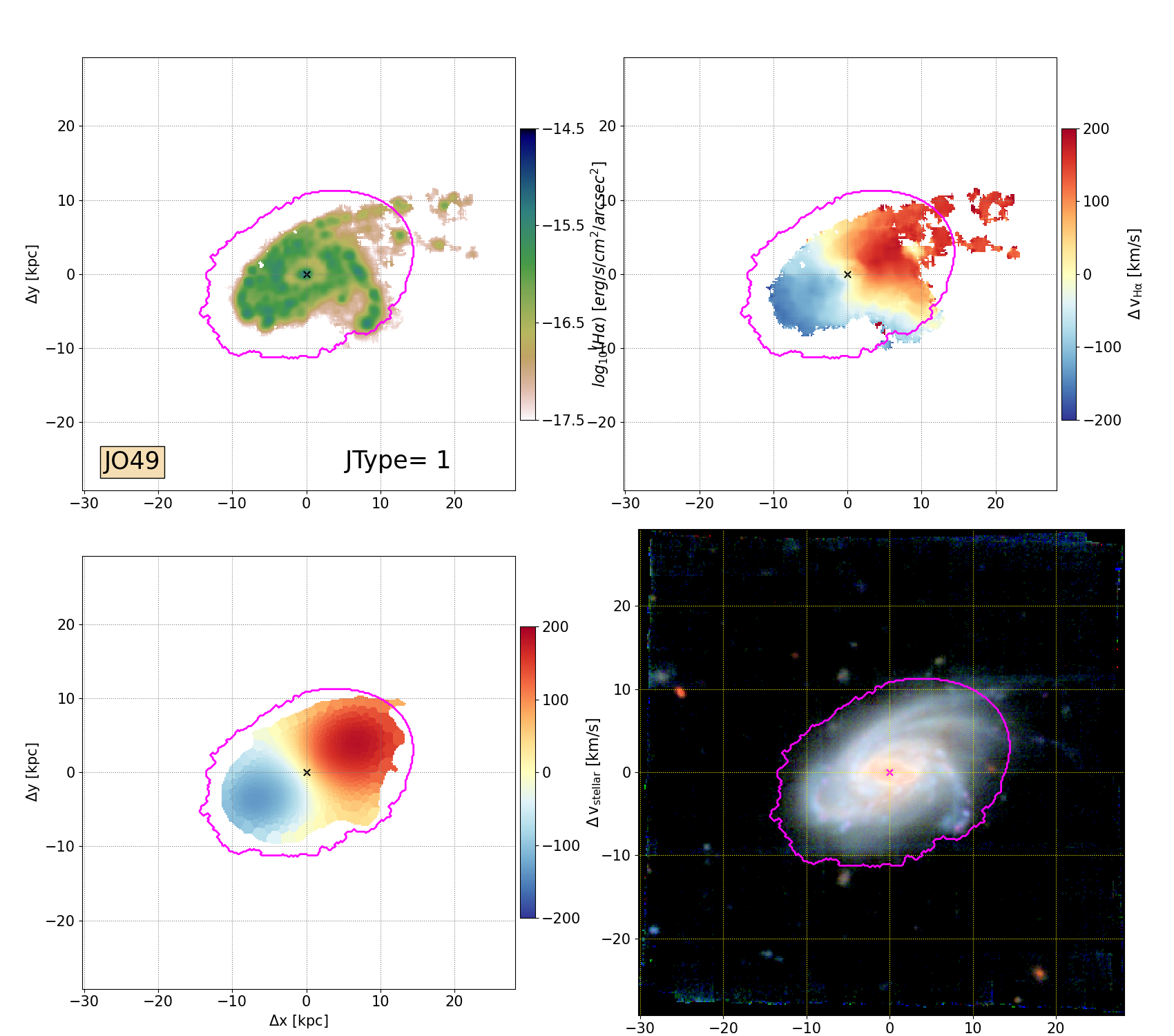}}
\centerline{\includegraphics[scale=0.2]
{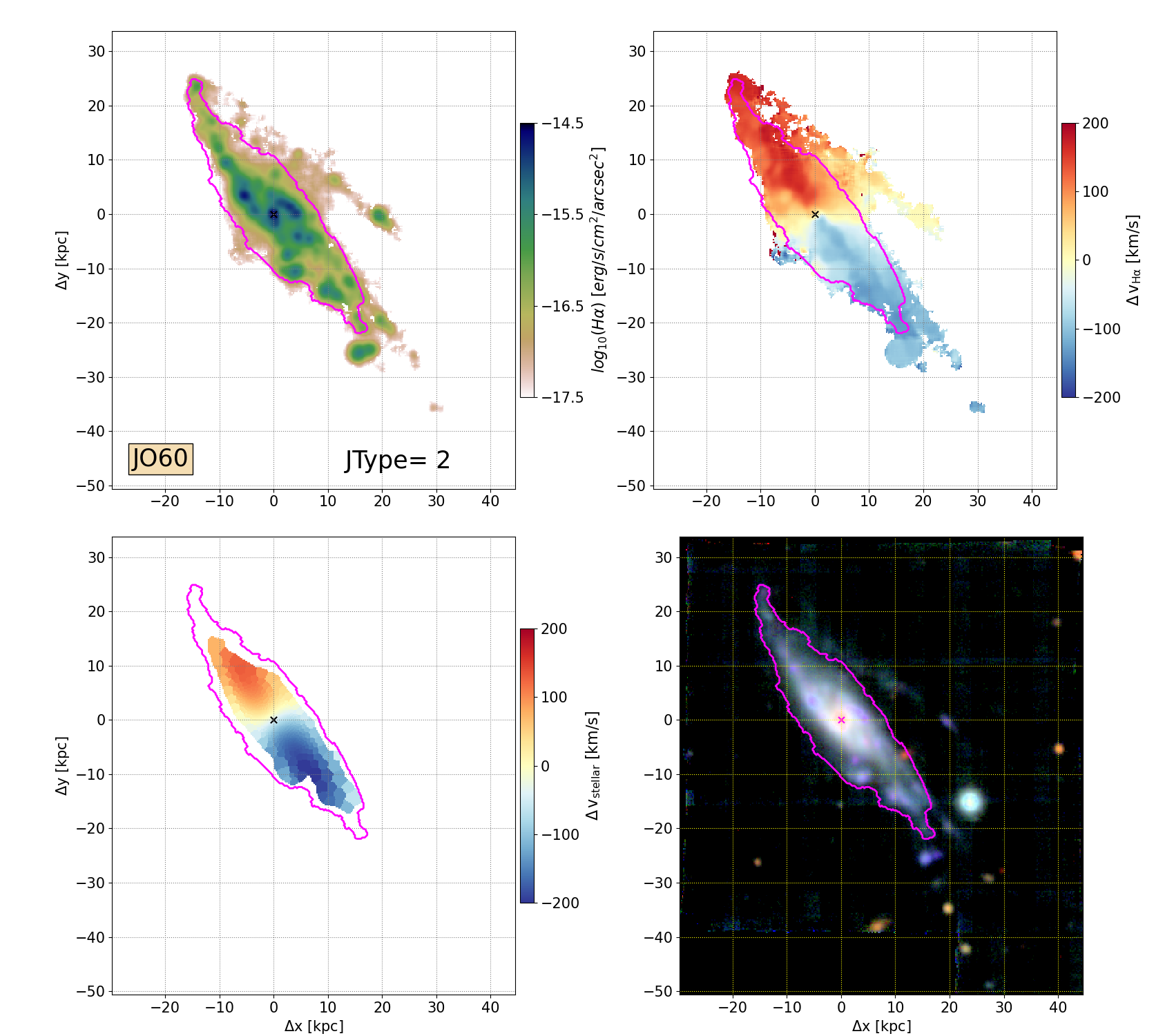}\includegraphics[scale=0.2]{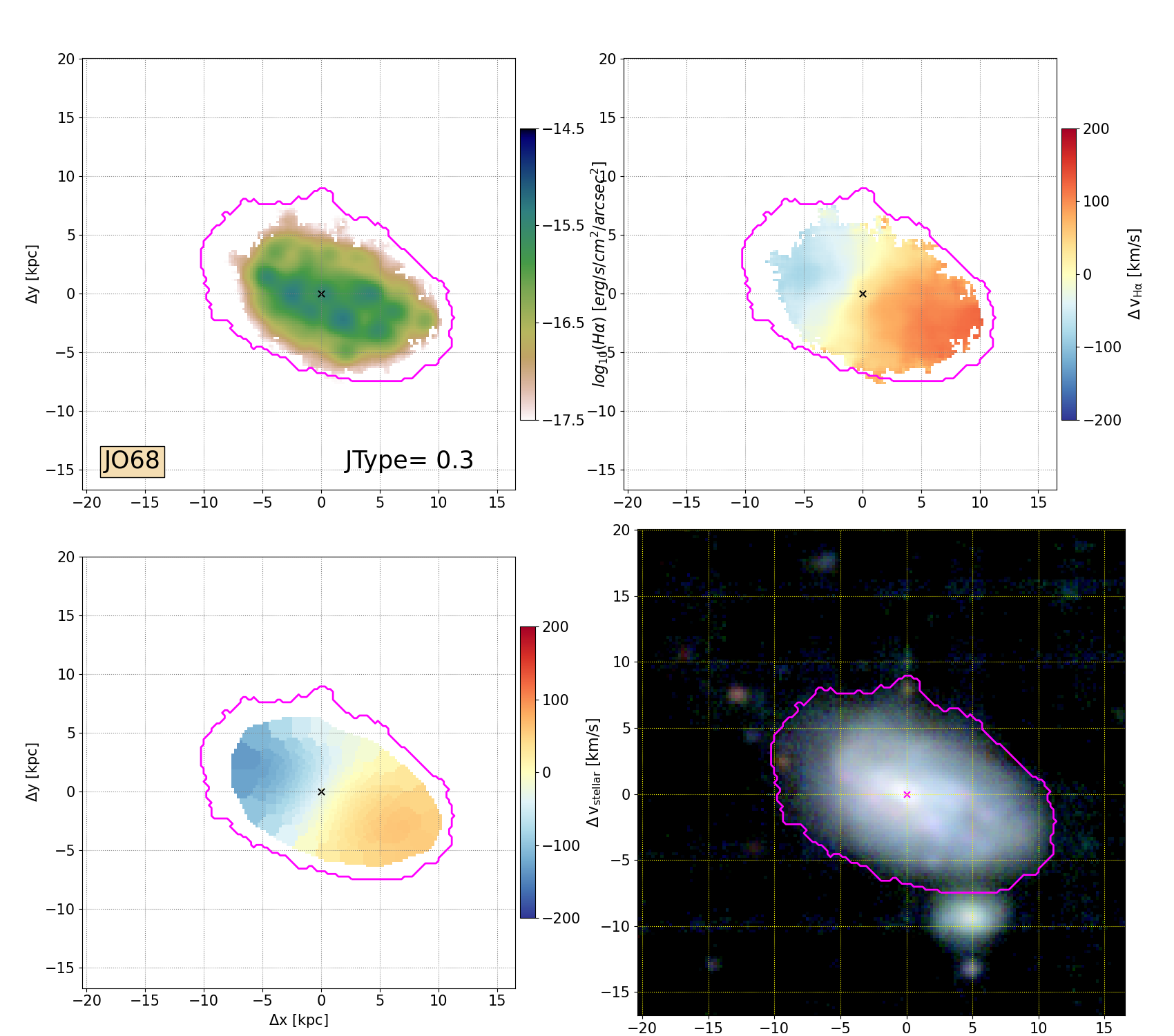}}
\centerline{\includegraphics[scale=0.2]
{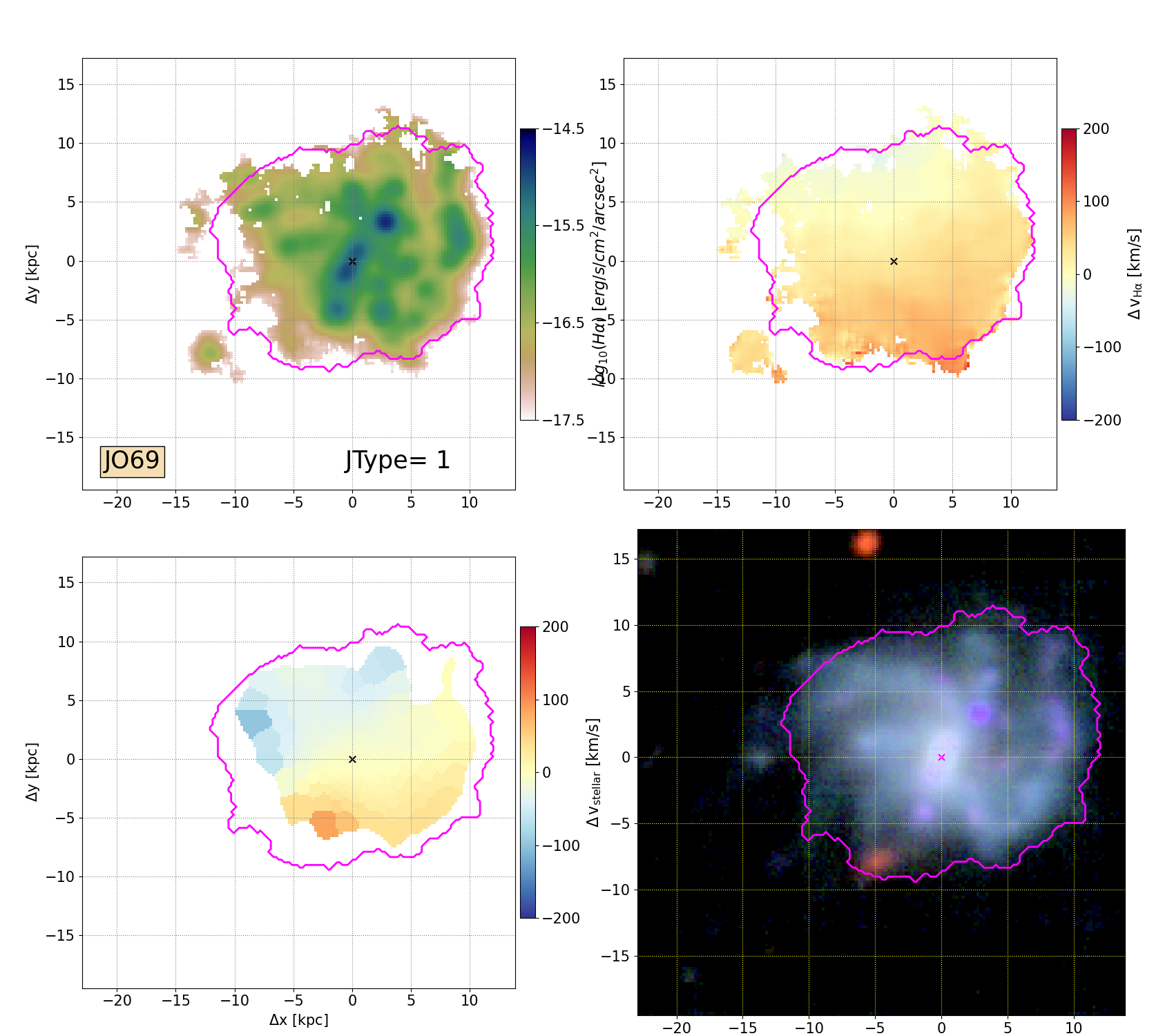}\includegraphics[scale=0.2]{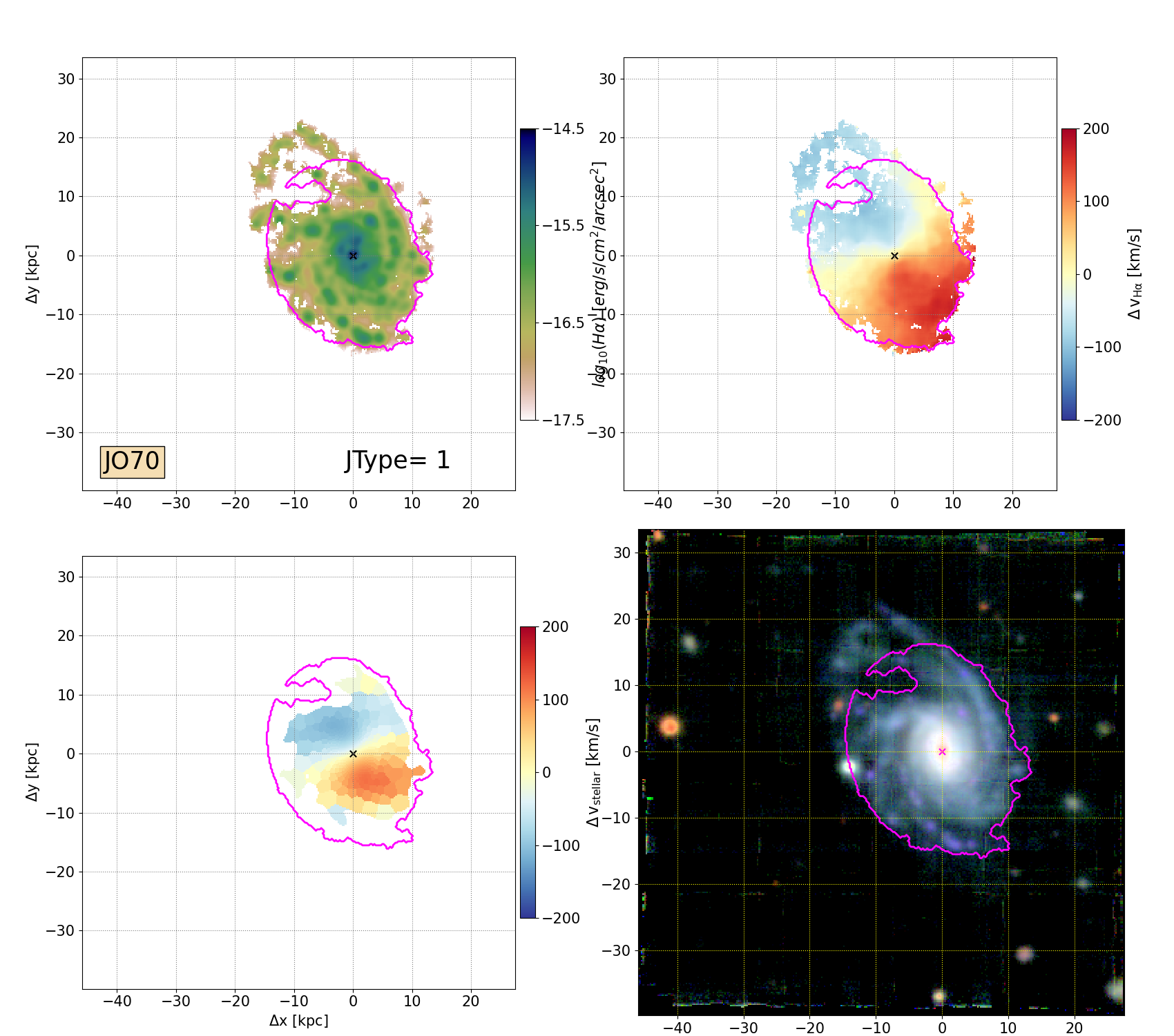}}
\caption{Continued. Stripping candidates.}
\end{figure*}

\addtocounter{figure}{-1}

\begin{figure*}
\centerline{\includegraphics[scale=0.2]
{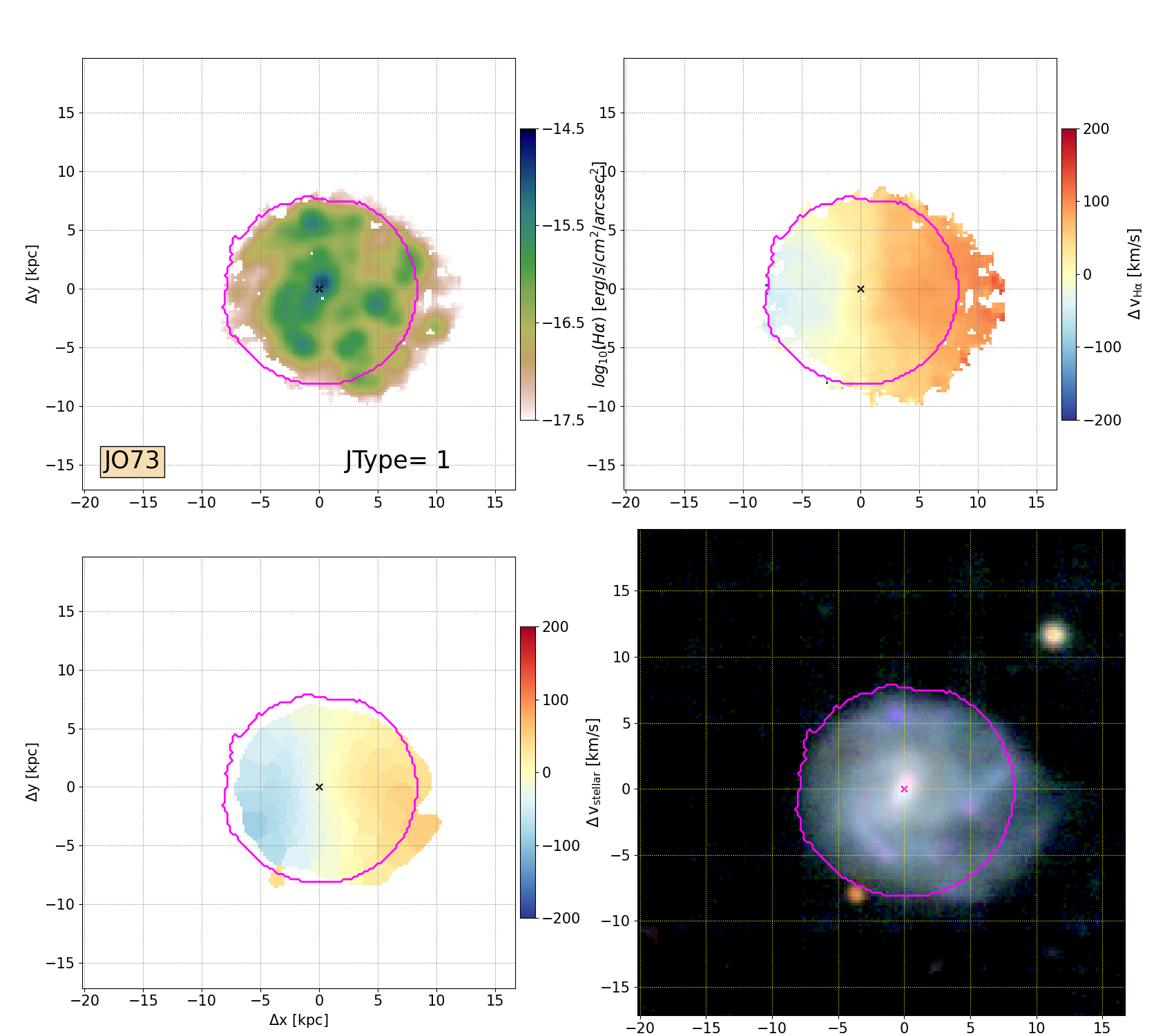}\includegraphics[scale=0.2]{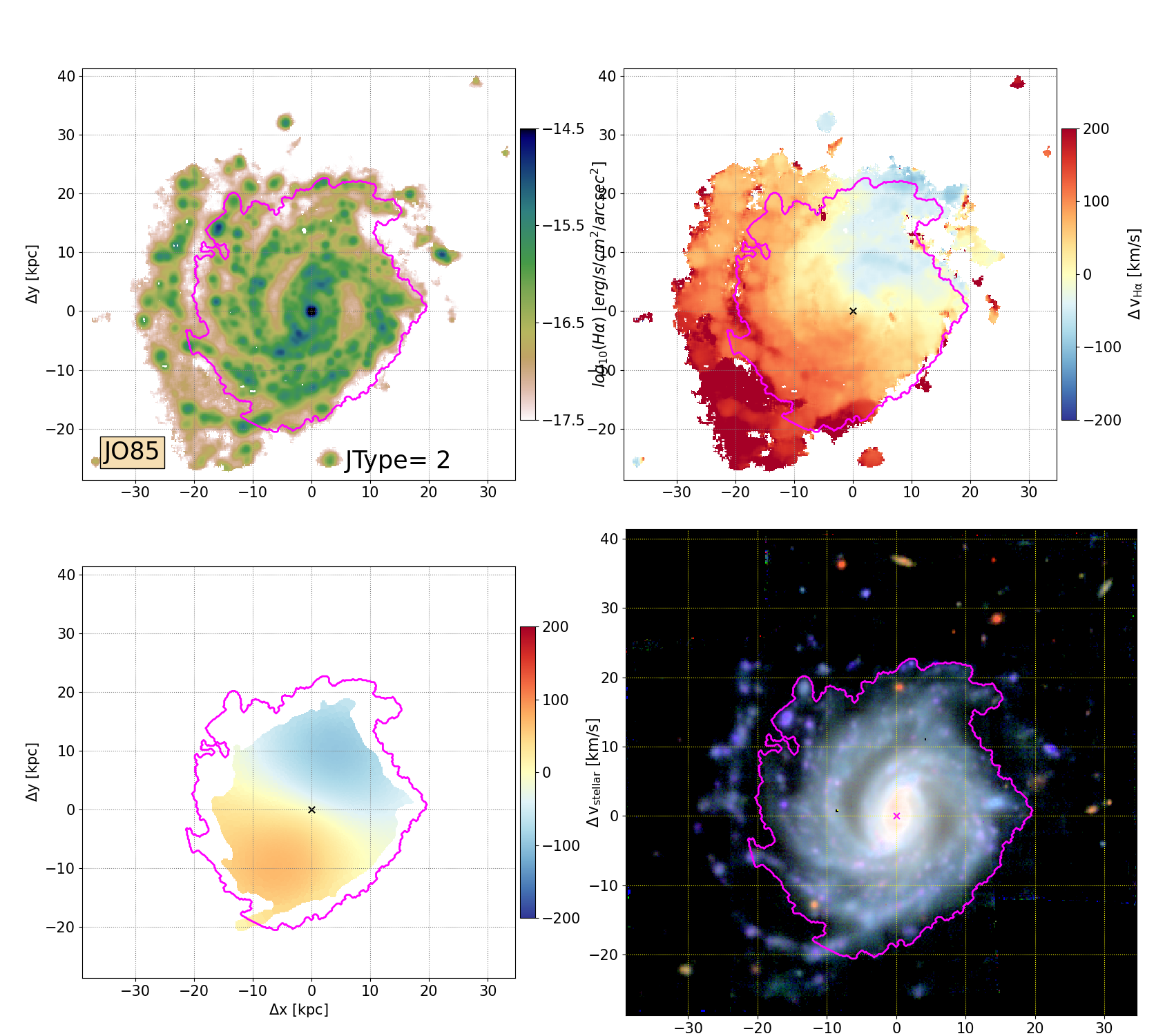}}
\centerline{\includegraphics[scale=0.2]
{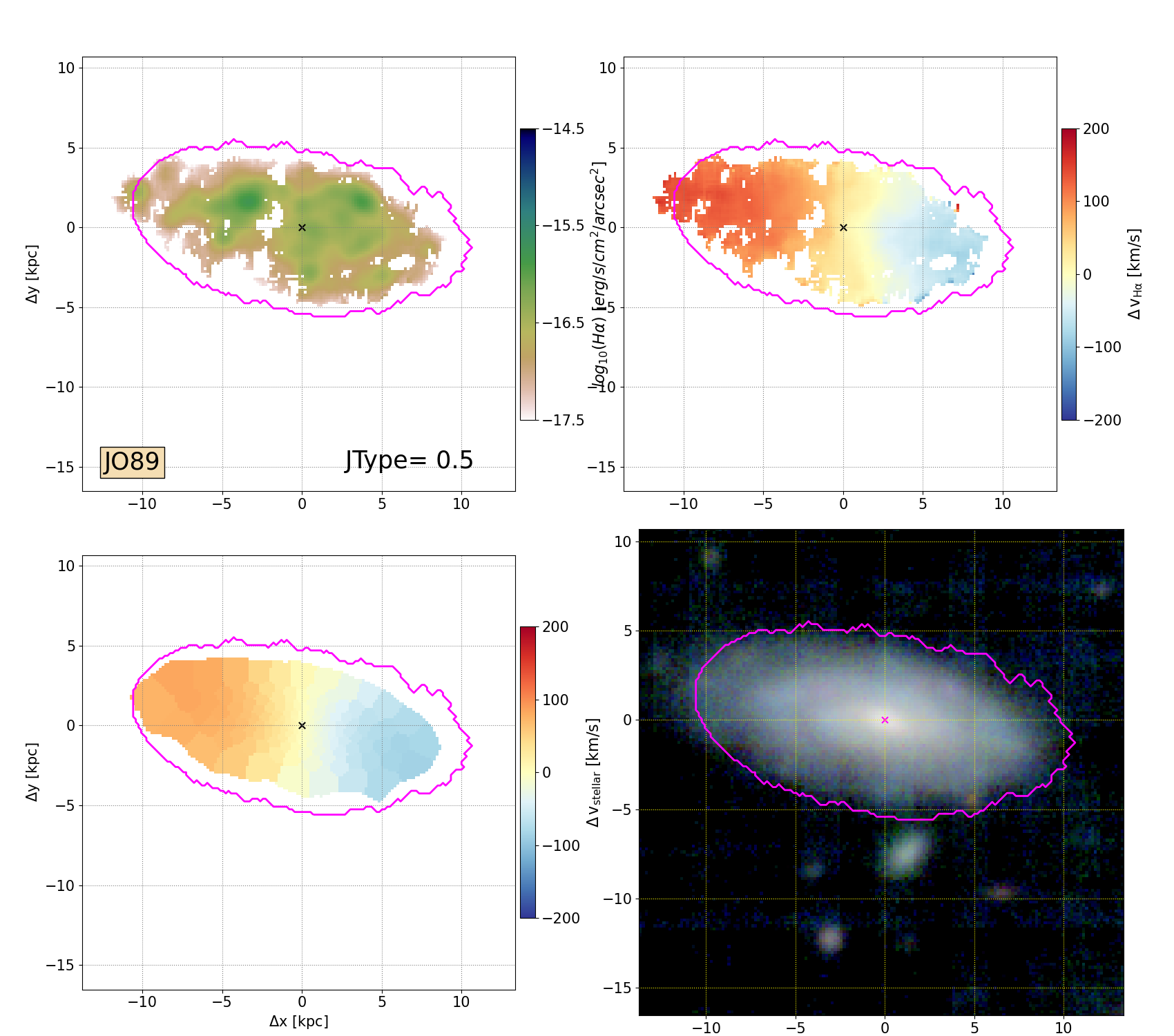}\includegraphics[scale=0.2]{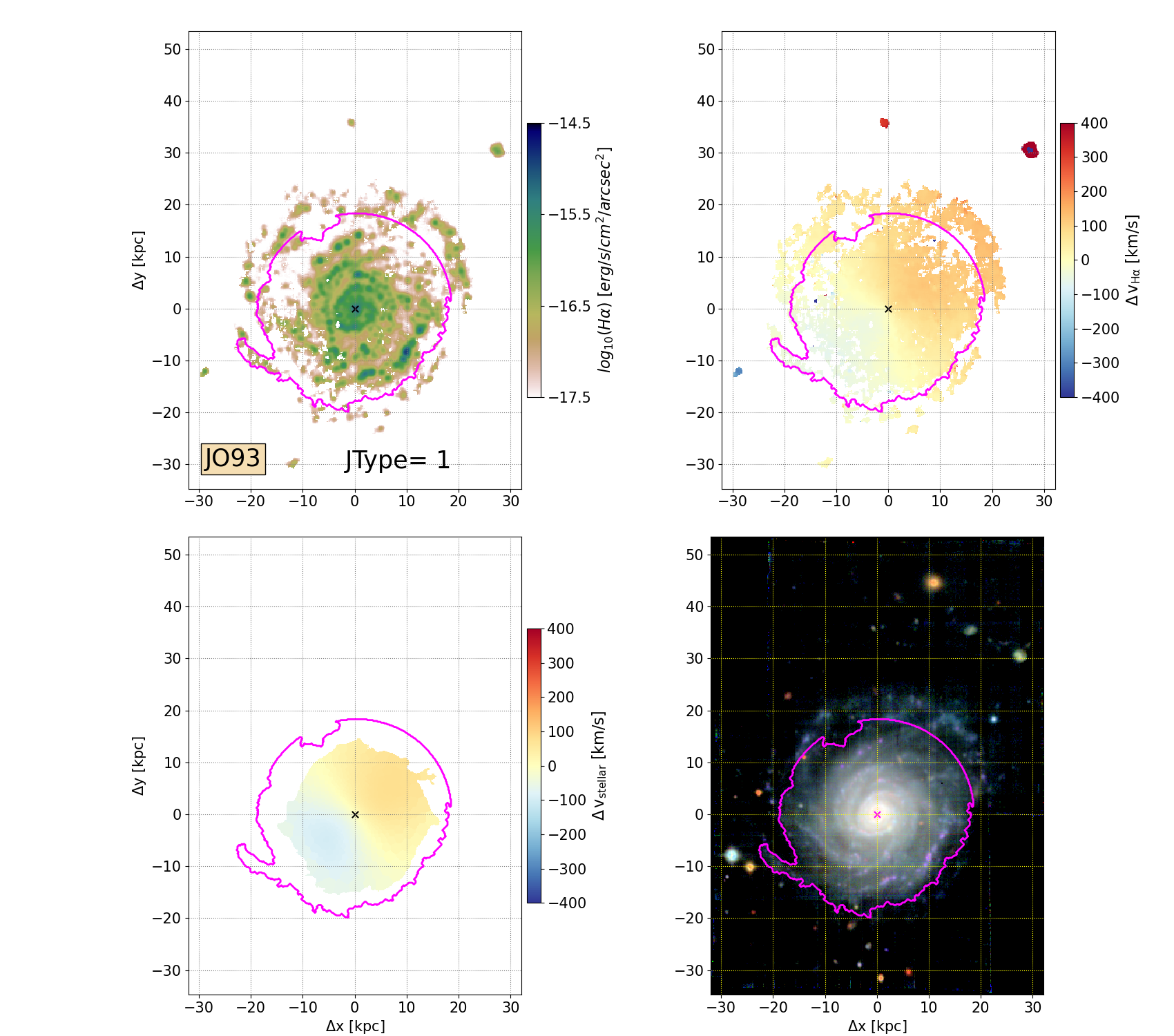}}
\centerline{\includegraphics[scale=0.2]
{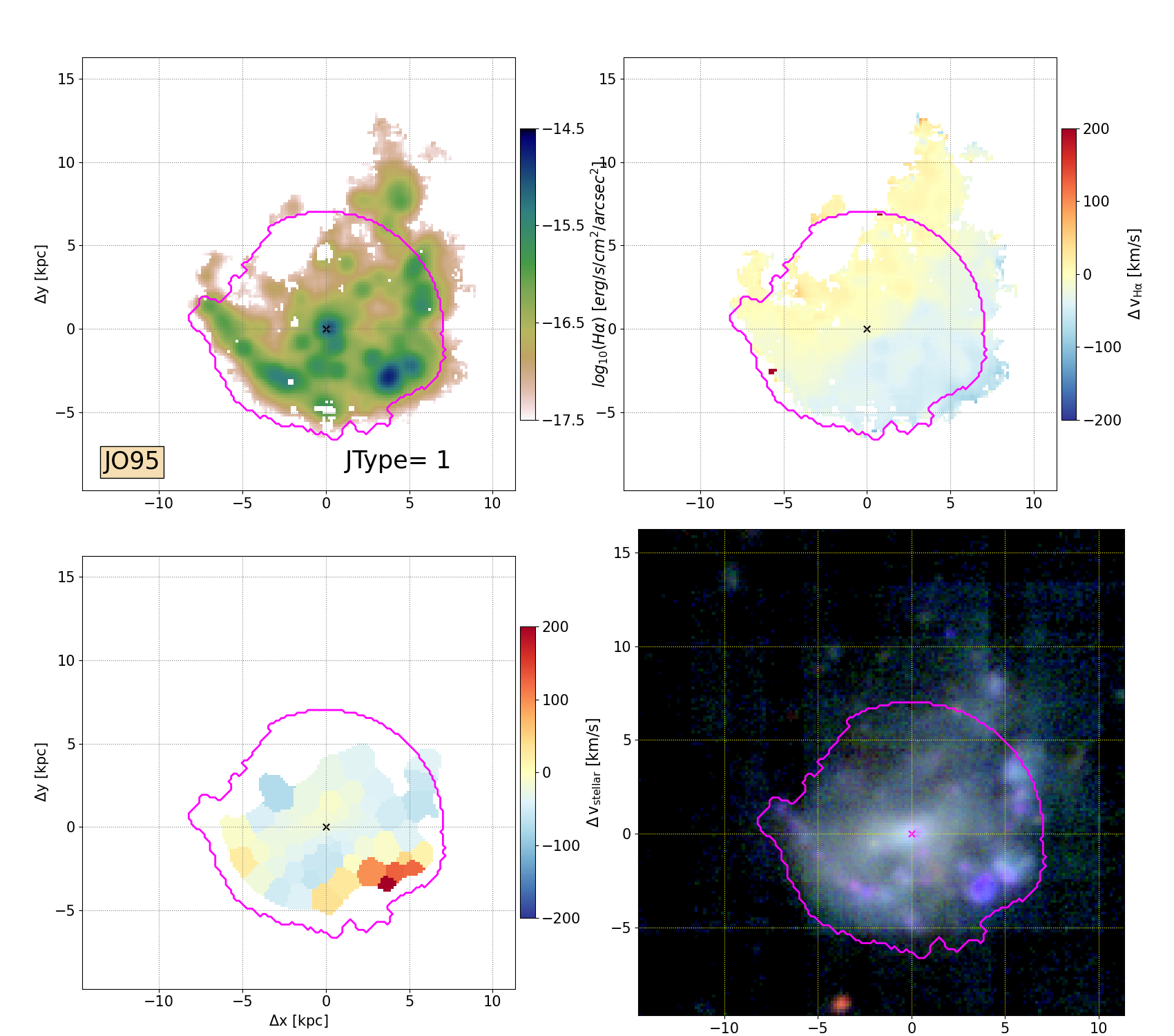}\includegraphics[scale=0.2]{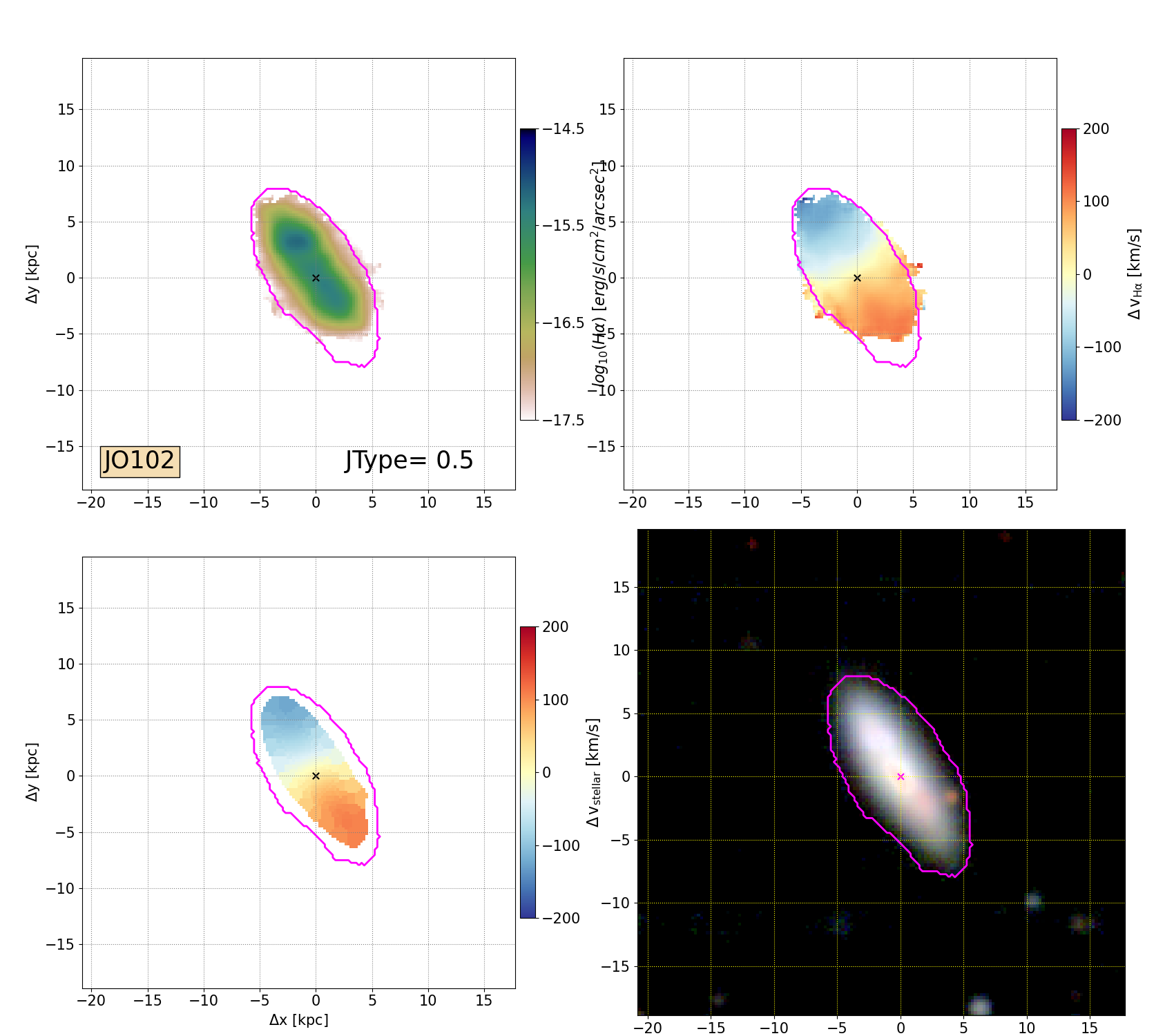}}
\caption{Continued. Stripping candidates.}
\end{figure*}

\addtocounter{figure}{-1}

\begin{figure*}
\centerline{\includegraphics[scale=0.2]
{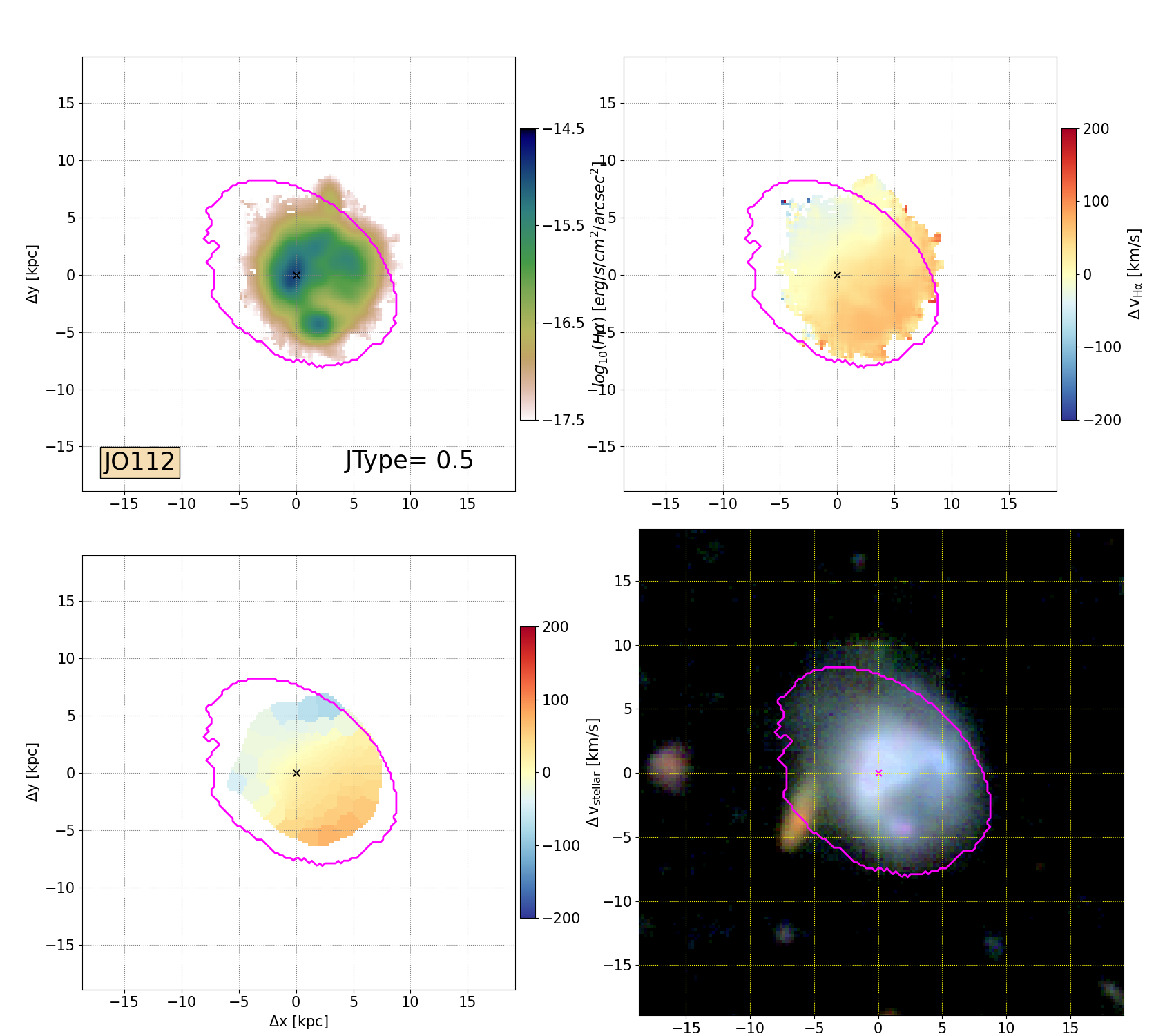}\includegraphics[scale=0.2]{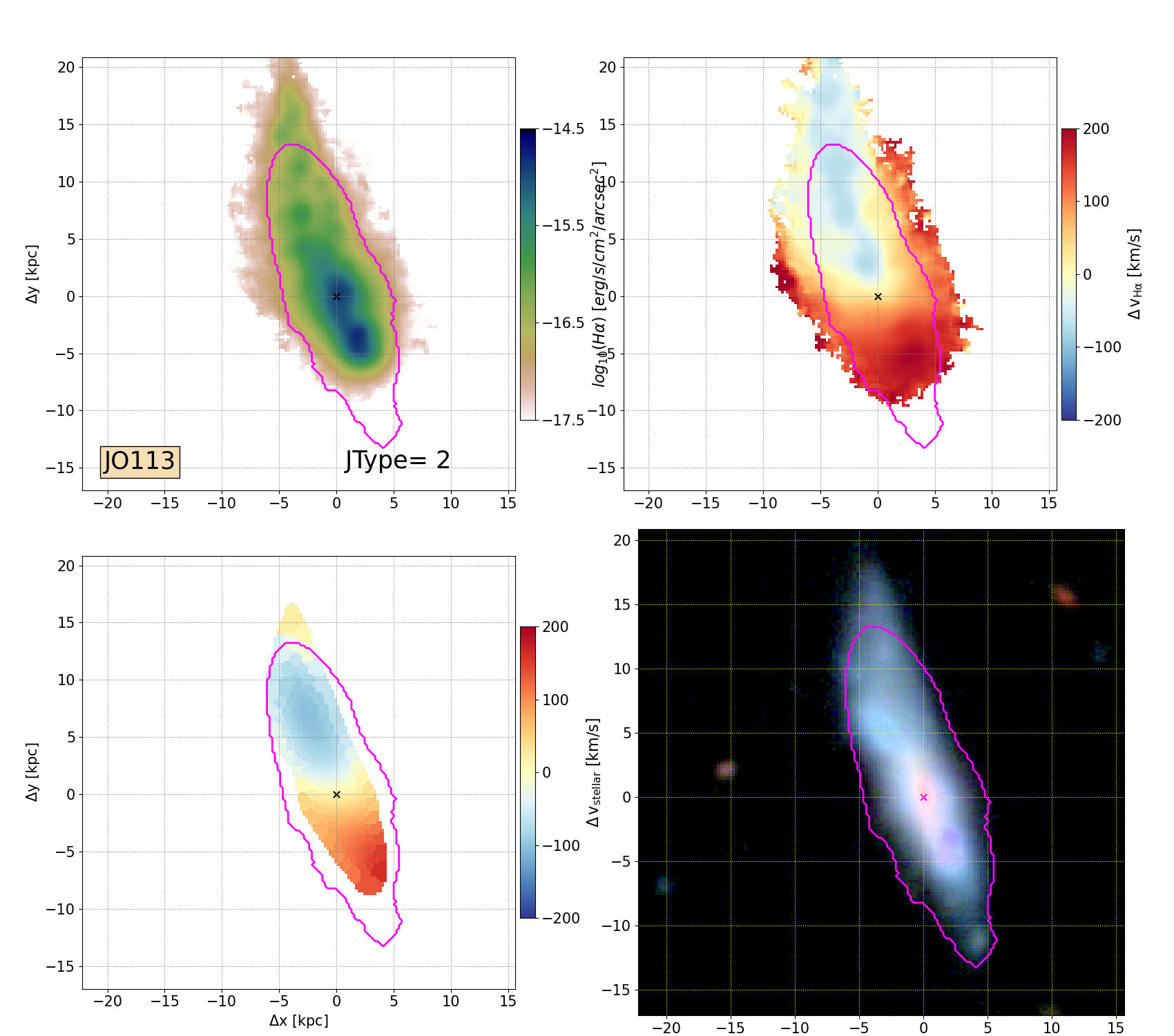}}
\centerline{\includegraphics[scale=0.2]
{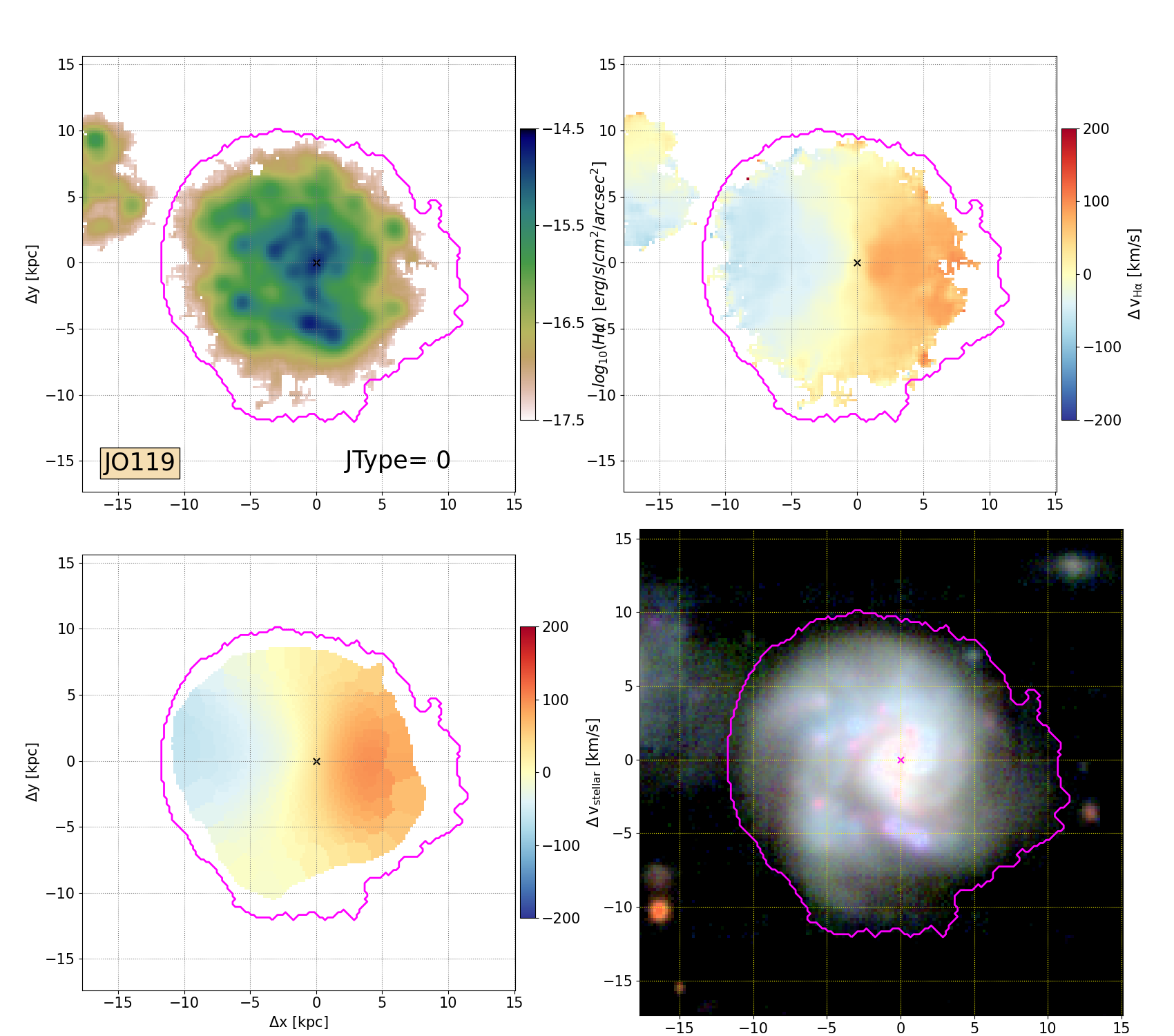}\includegraphics[scale=0.2]{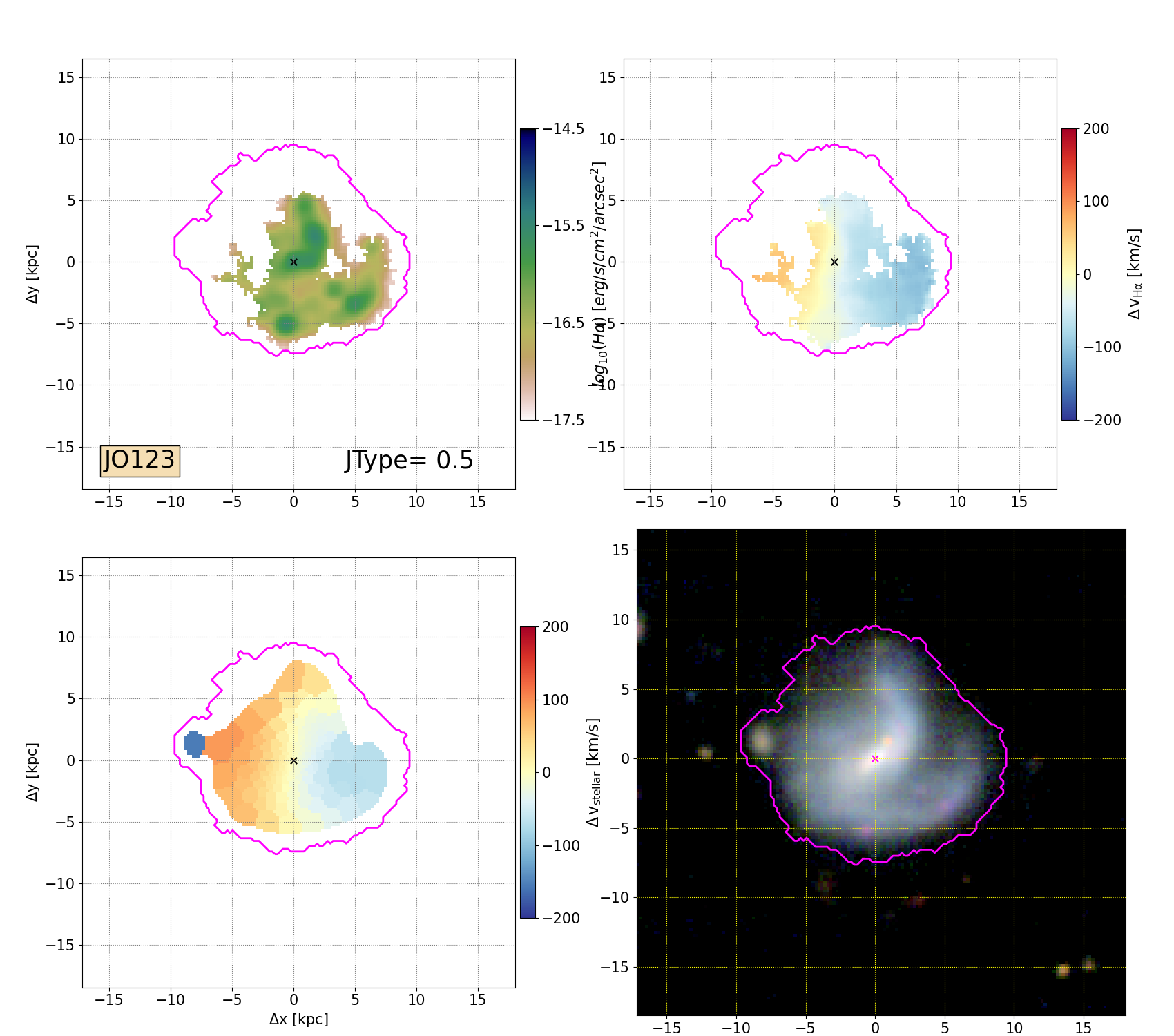}}
\centerline{\includegraphics[scale=0.2]
{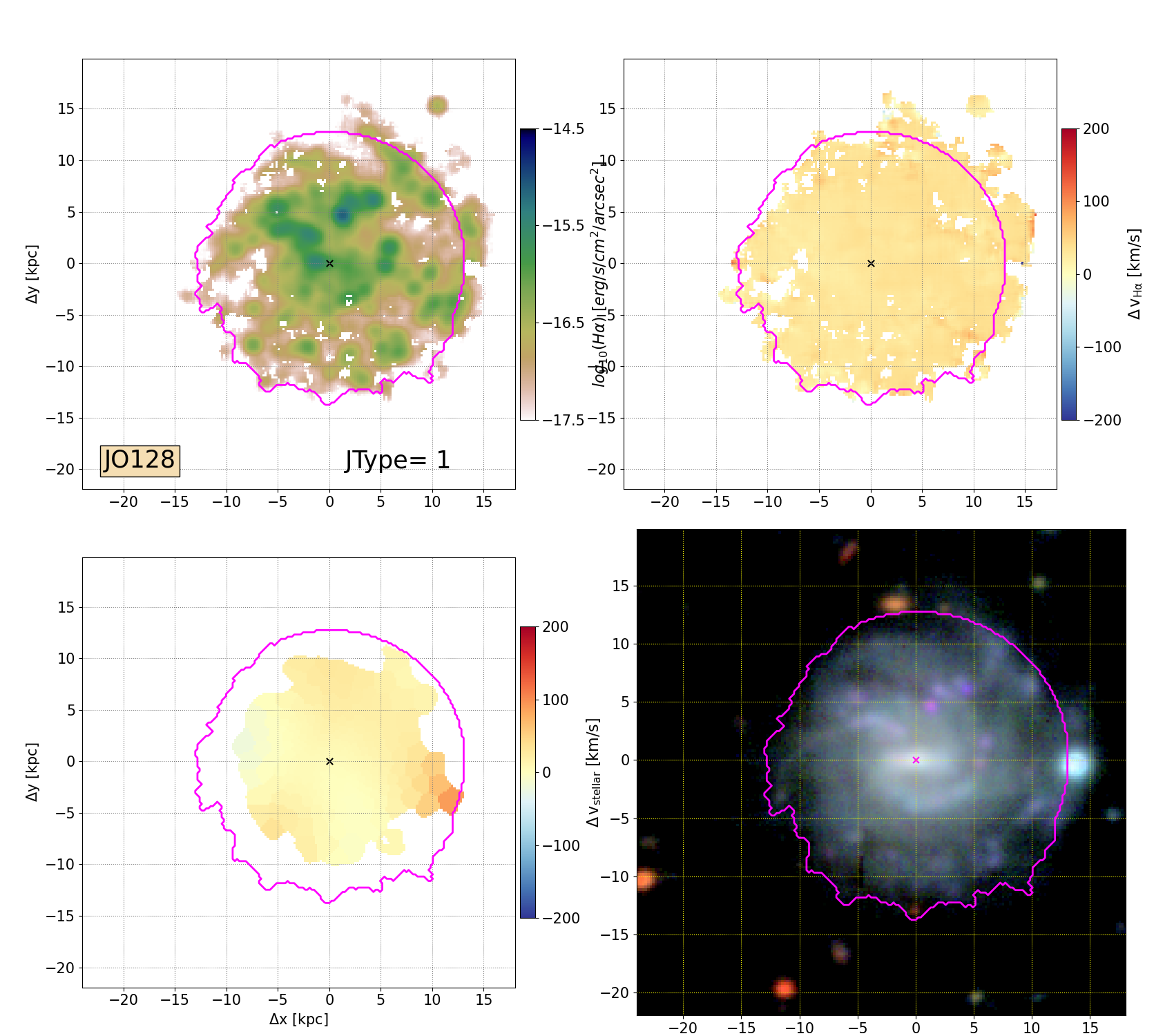}\includegraphics[scale=0.14]{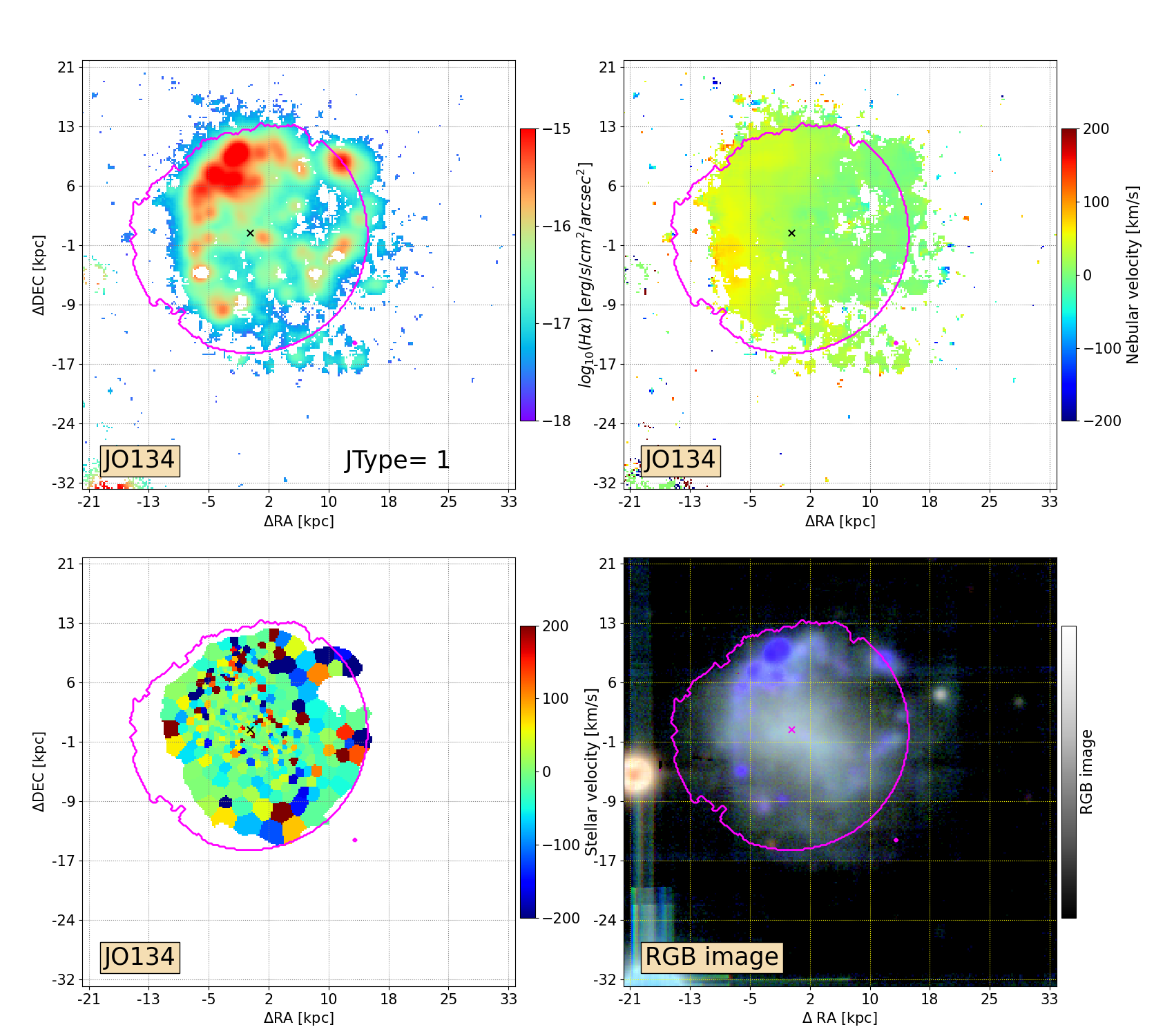}}
\caption{Continued. Stripping candidates.}
\end{figure*}

\addtocounter{figure}{-1}

\begin{figure*}
\centerline{\includegraphics[scale=0.2]
{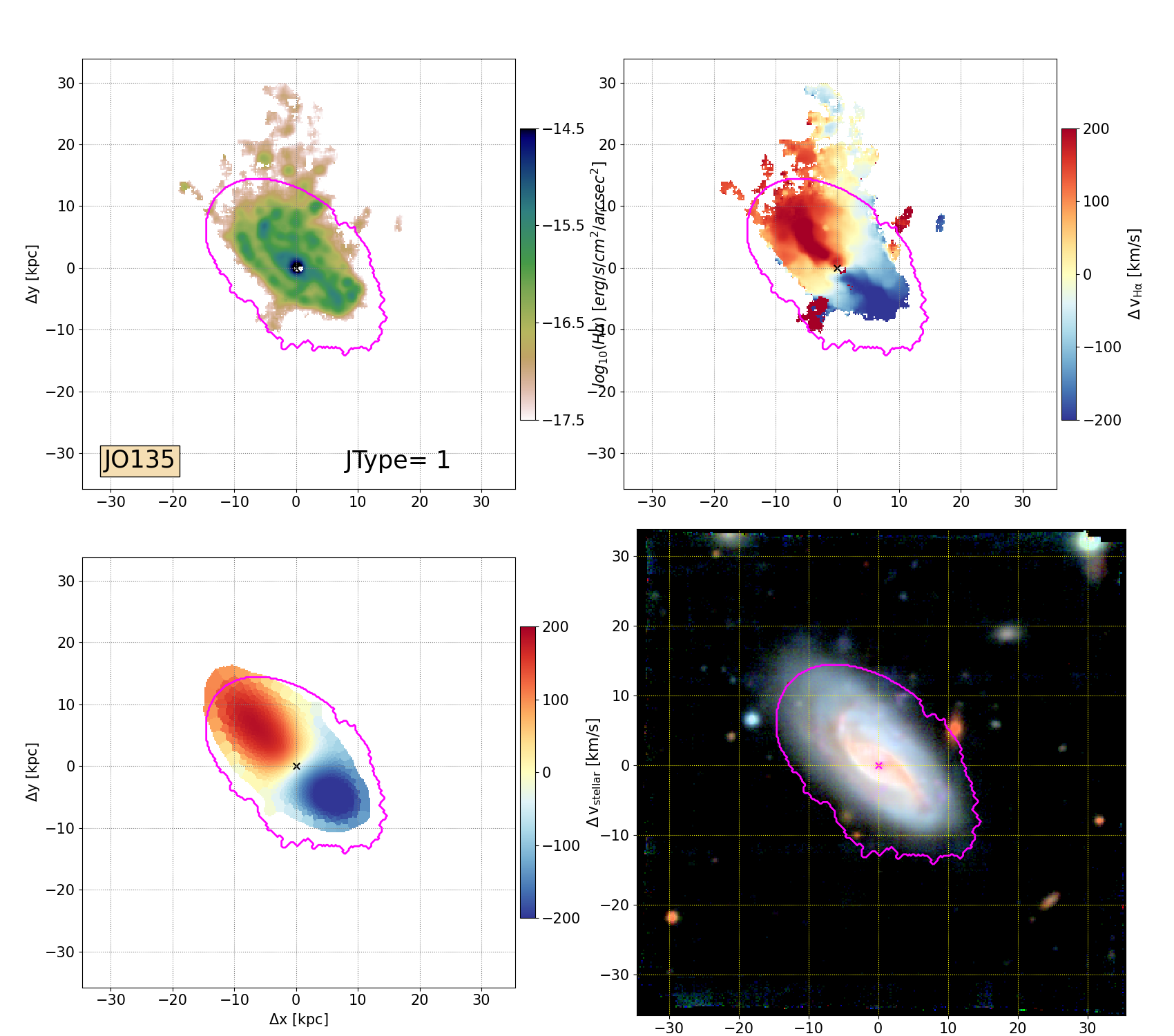}\includegraphics[scale=0.2]{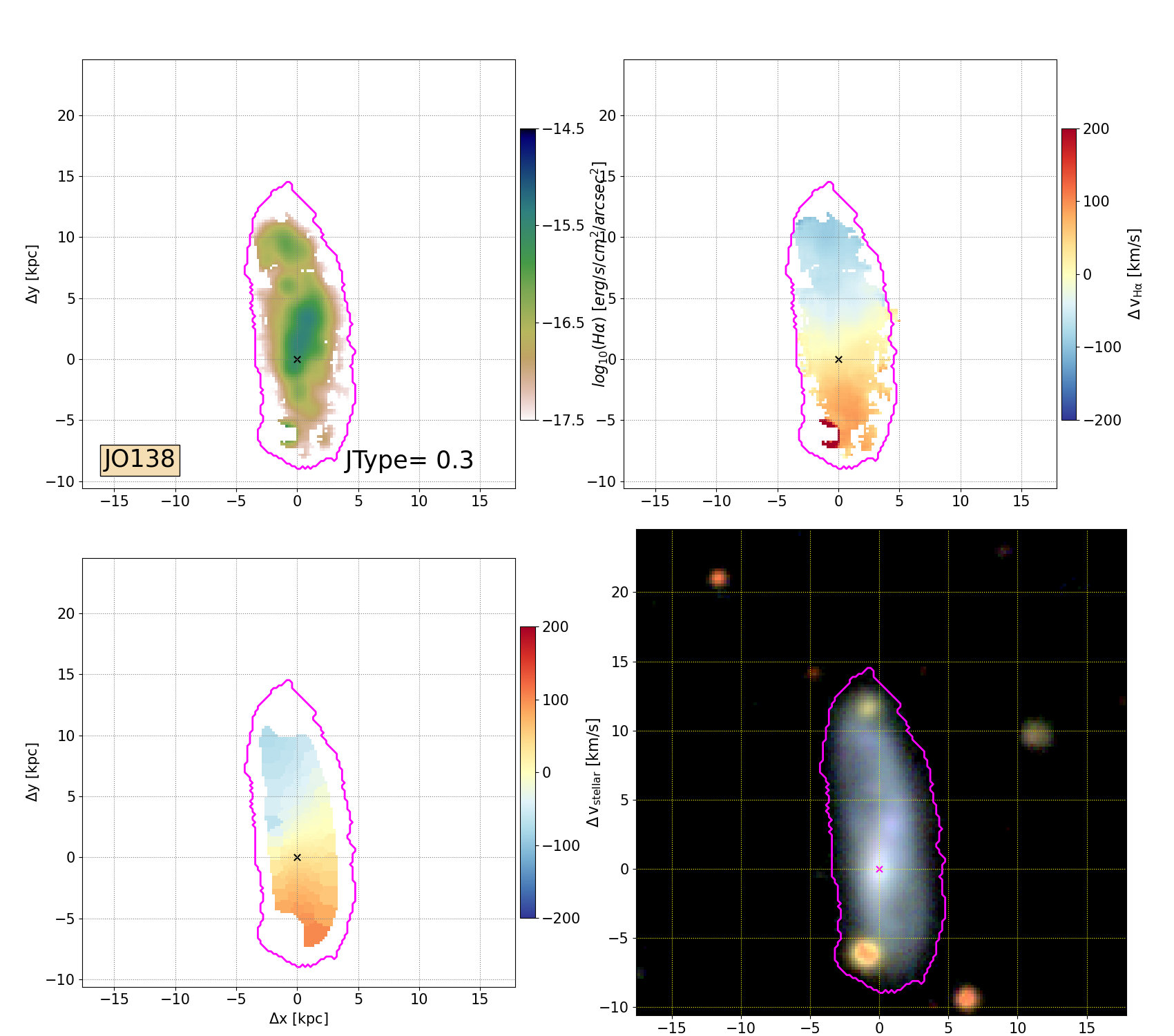}}
\centerline{\includegraphics[scale=0.2]
{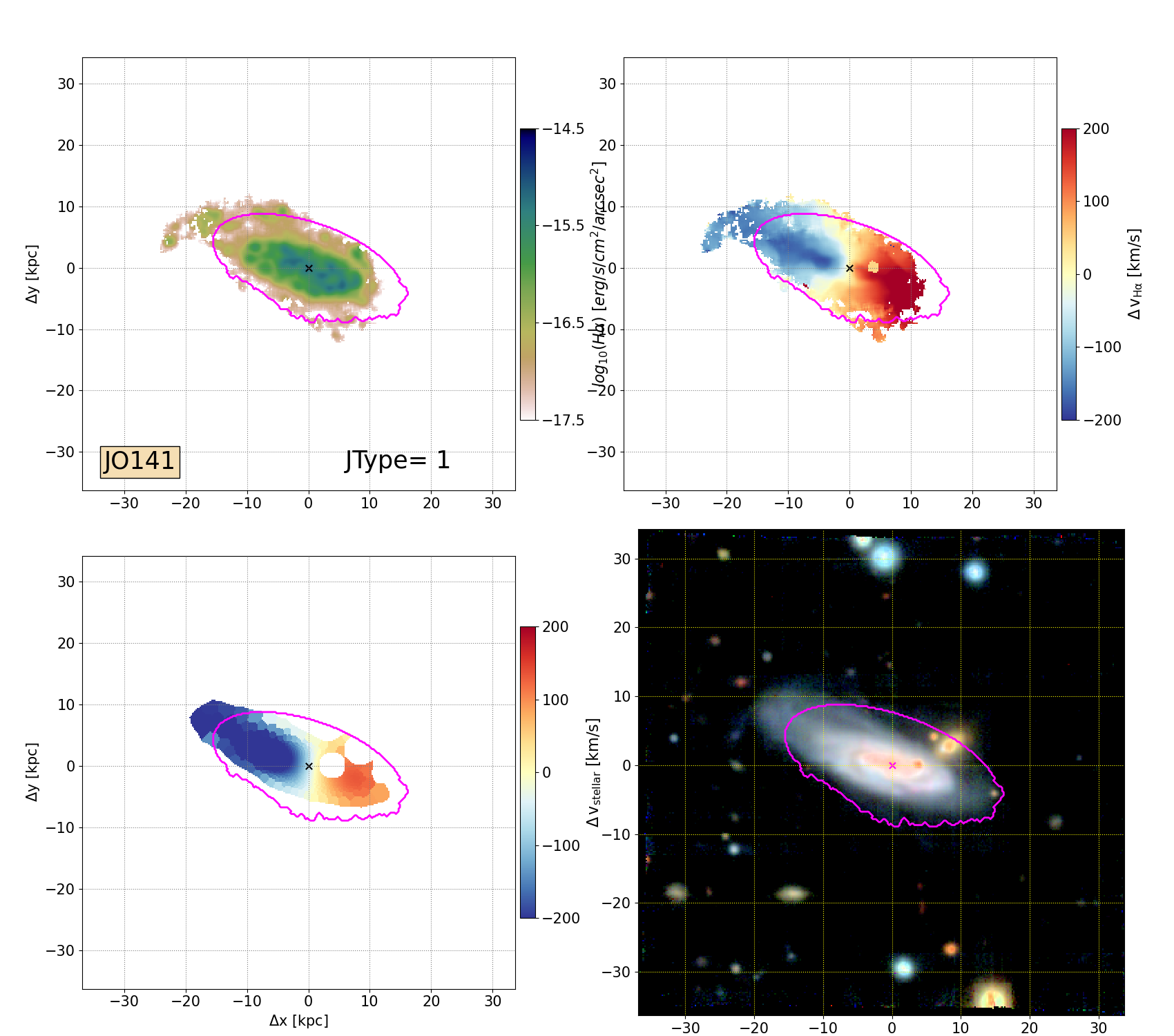}\includegraphics[scale=0.2]{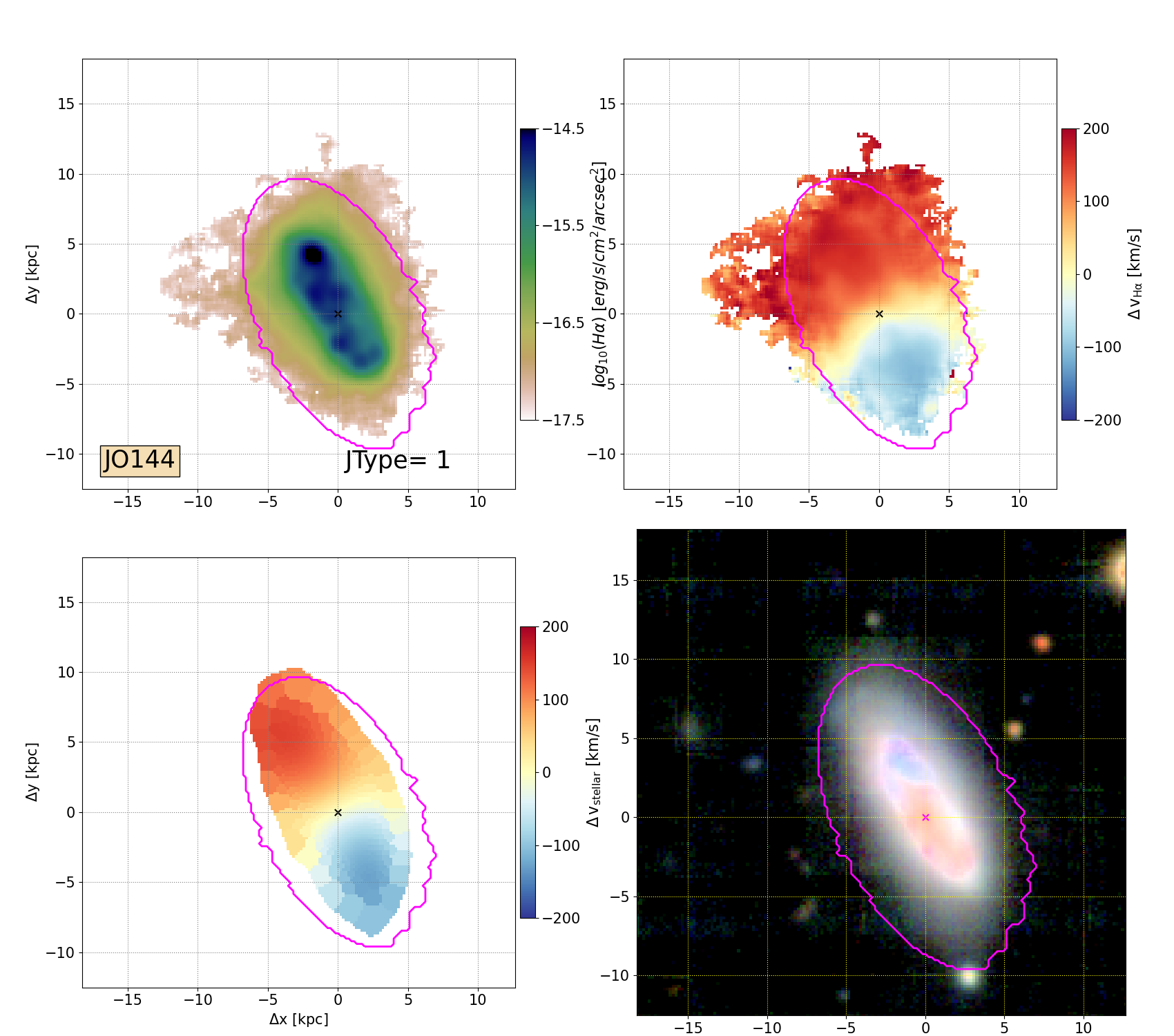}}
\centerline{\includegraphics[scale=0.2]
{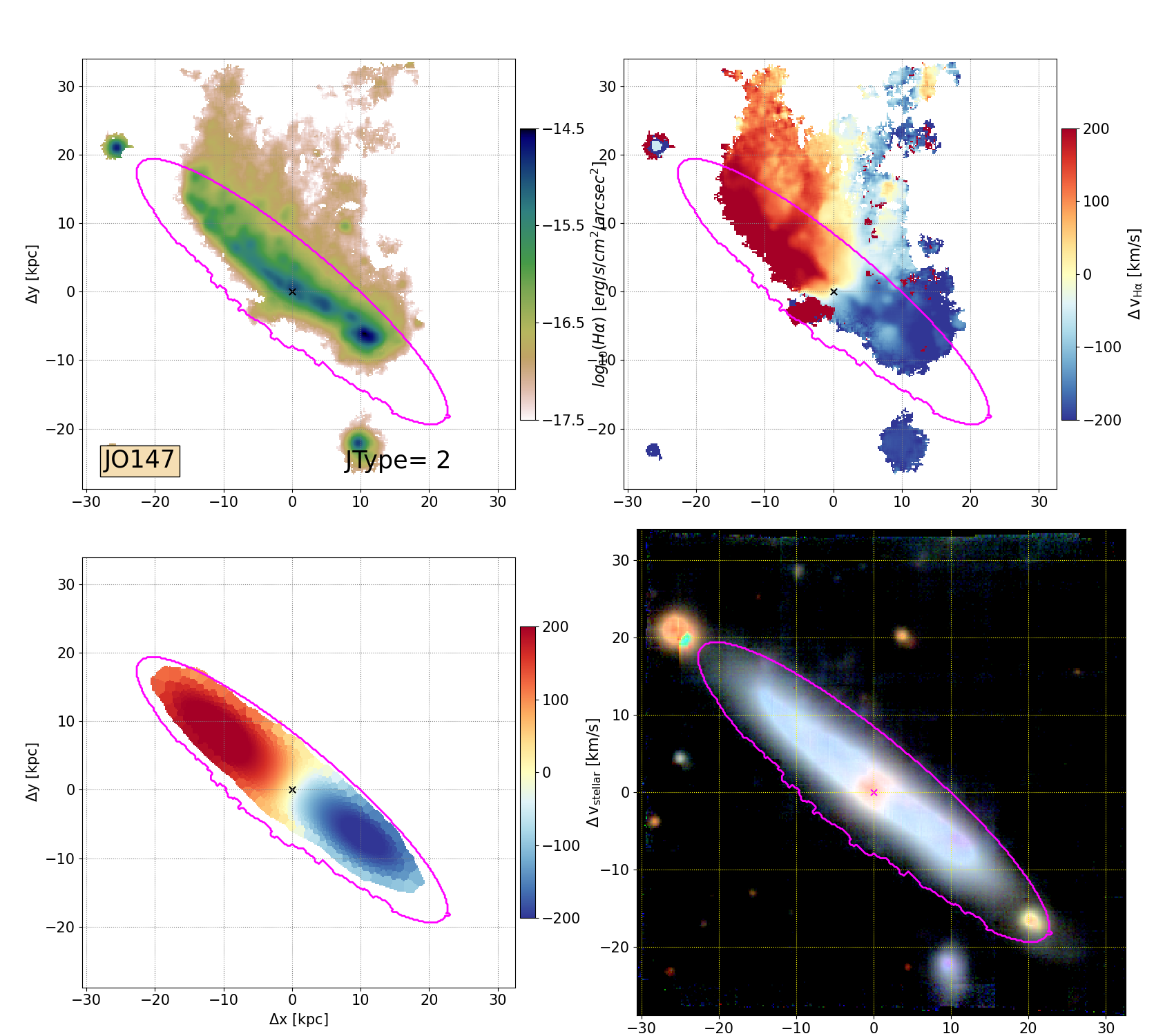}\includegraphics[scale=0.2]{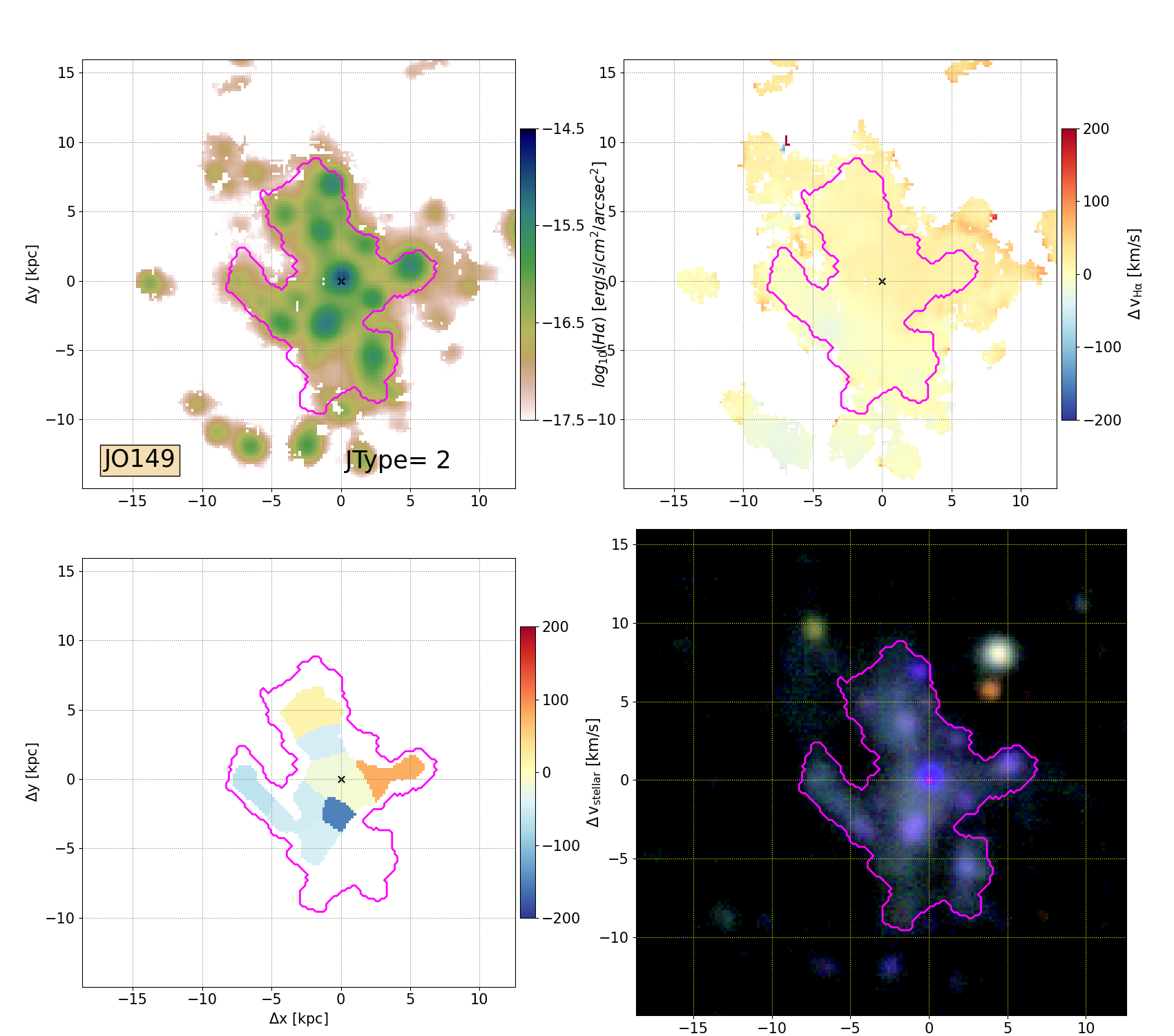}}
\caption{Continued. Stripping candidates. }
\end{figure*}

\addtocounter{figure}{-1}

\begin{figure*}
\centerline{\includegraphics[scale=0.2]
{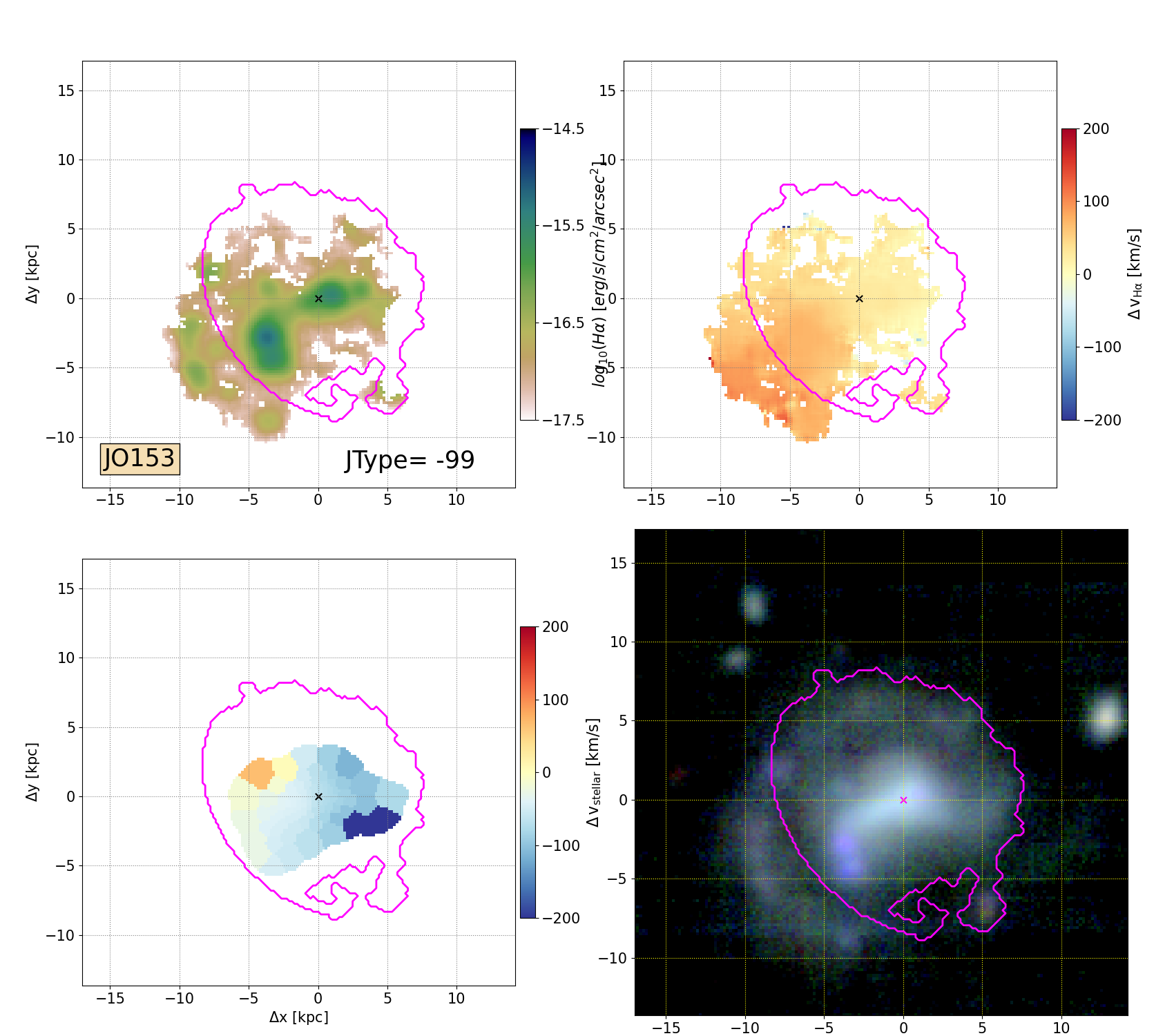}\includegraphics[scale=0.2]{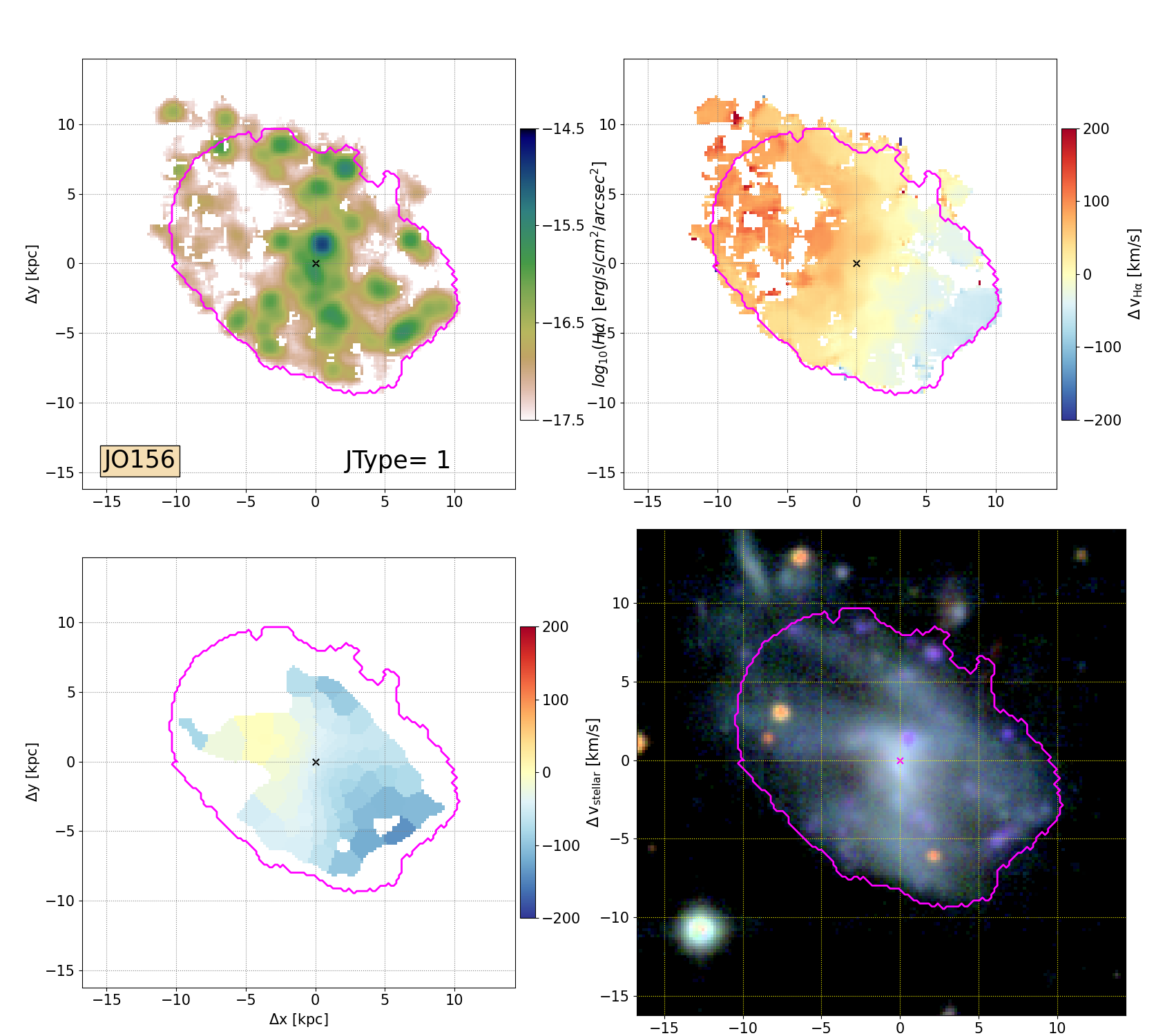}}
\centerline{\includegraphics[scale=0.2]
{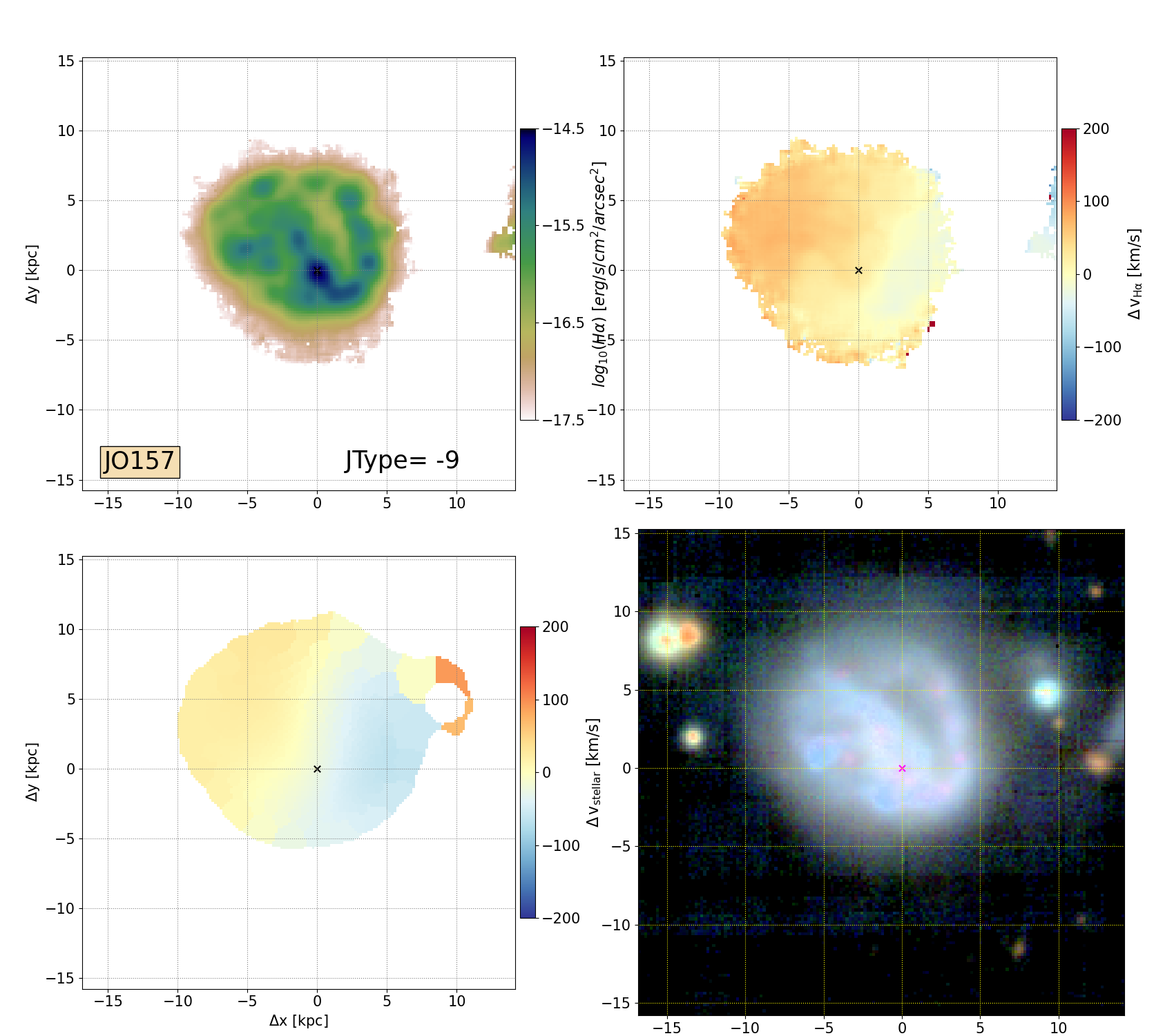}\includegraphics[scale=0.2]{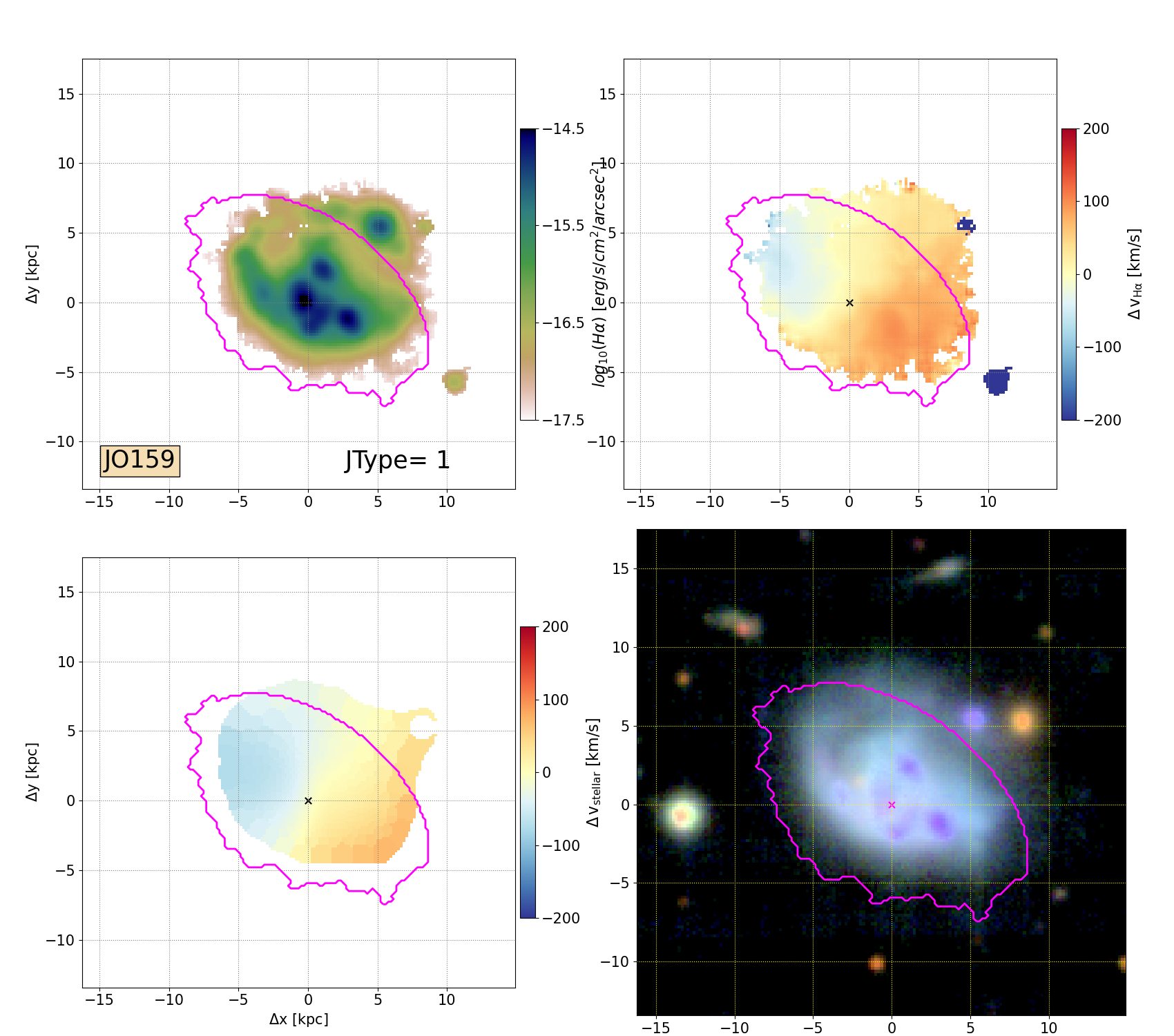}}
\centerline{\includegraphics[scale=0.2]
{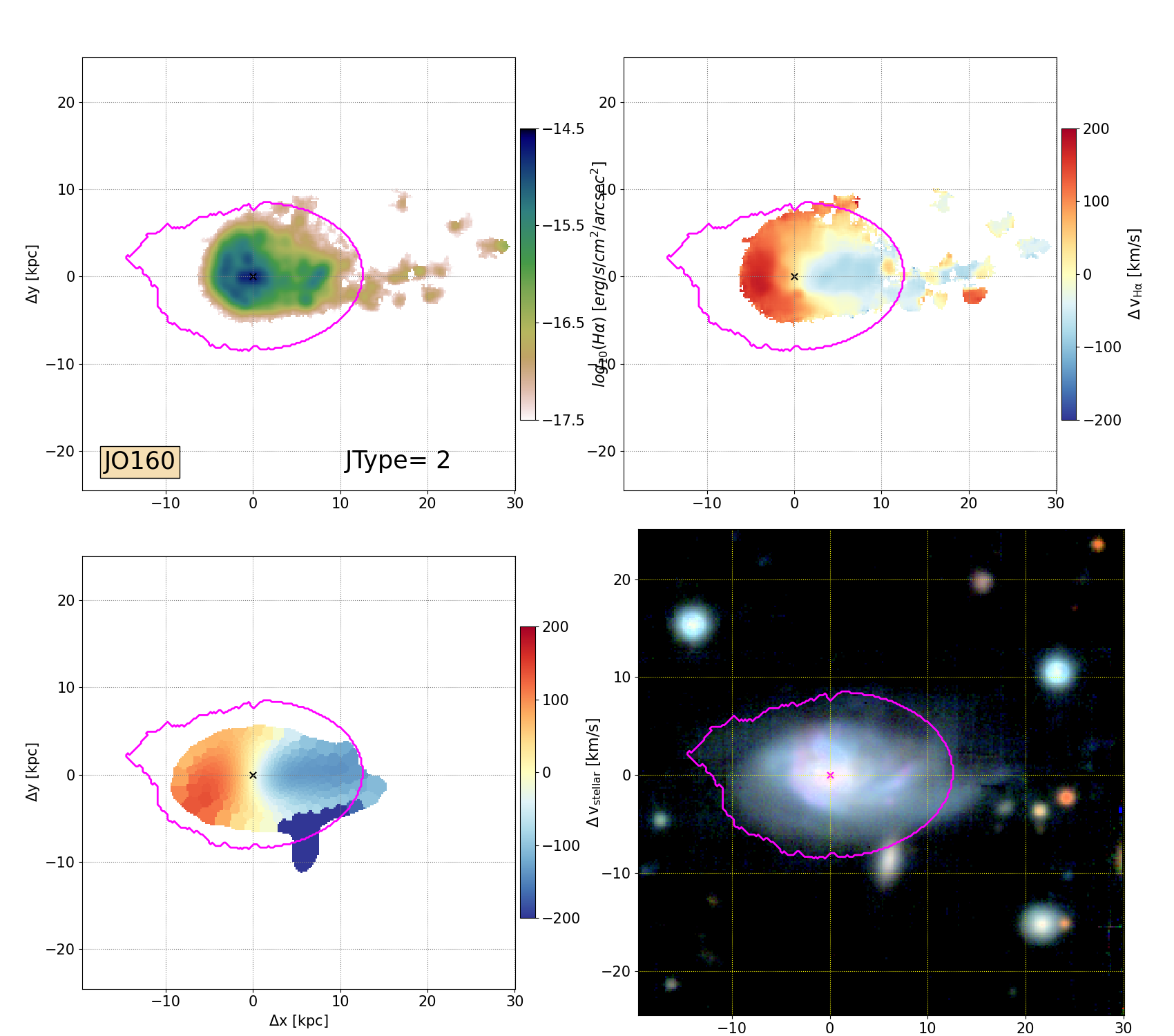}\includegraphics[scale=0.2]{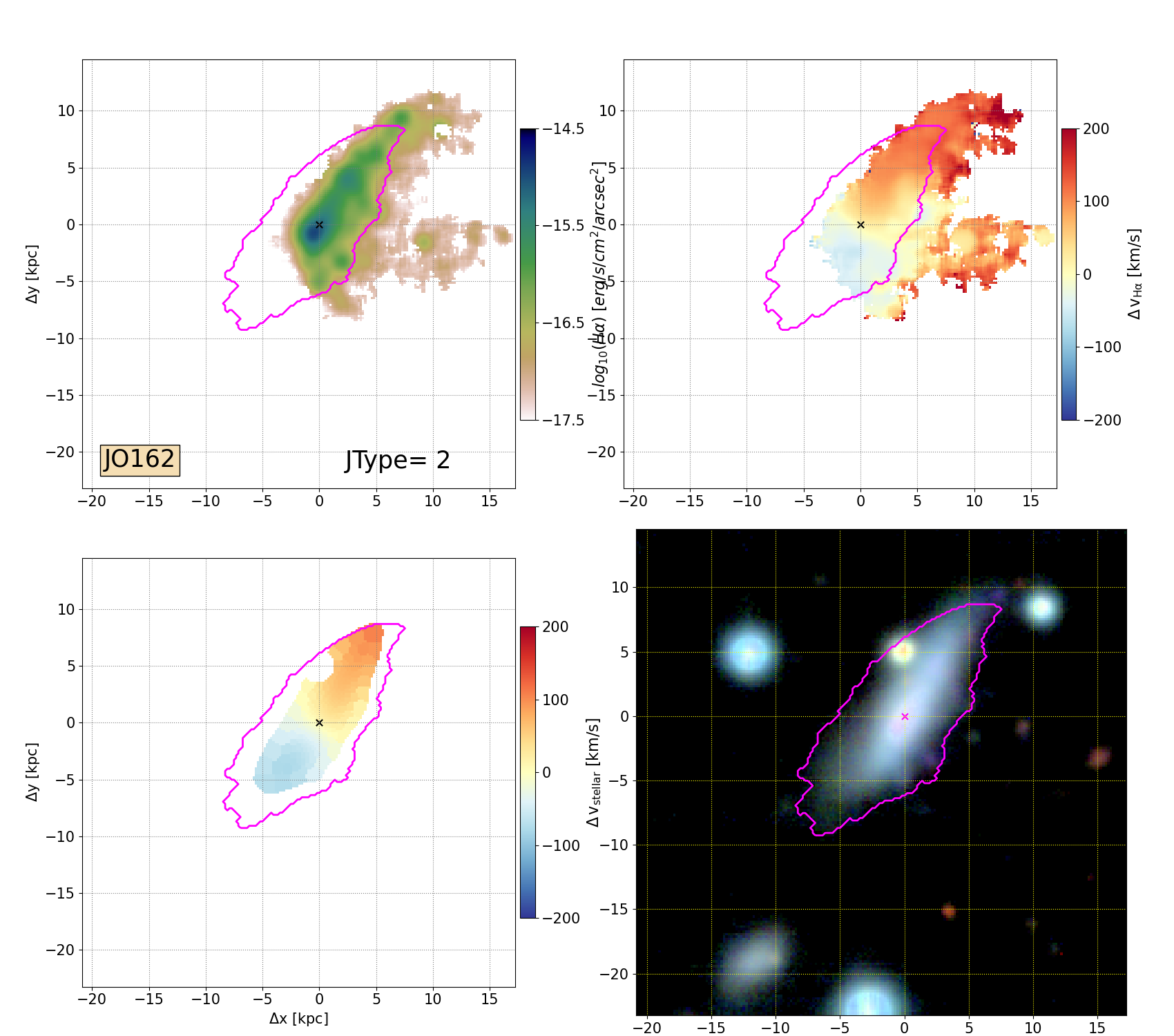}}
\caption{Continued. Stripping candidates.}
\end{figure*}

\addtocounter{figure}{-1} 

\begin{figure*}
\centerline{\includegraphics[scale=0.14]
{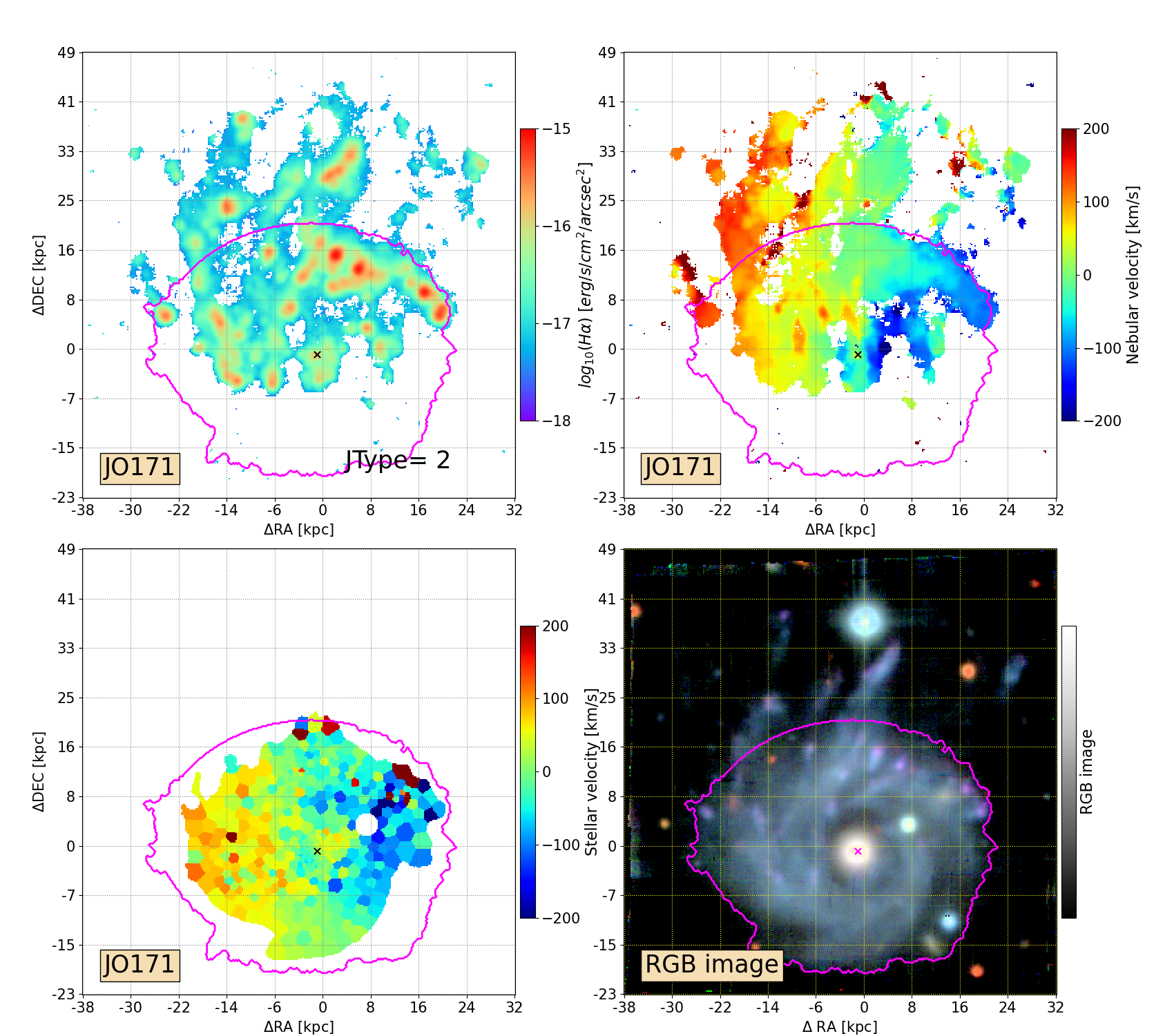}\includegraphics[scale=0.2]{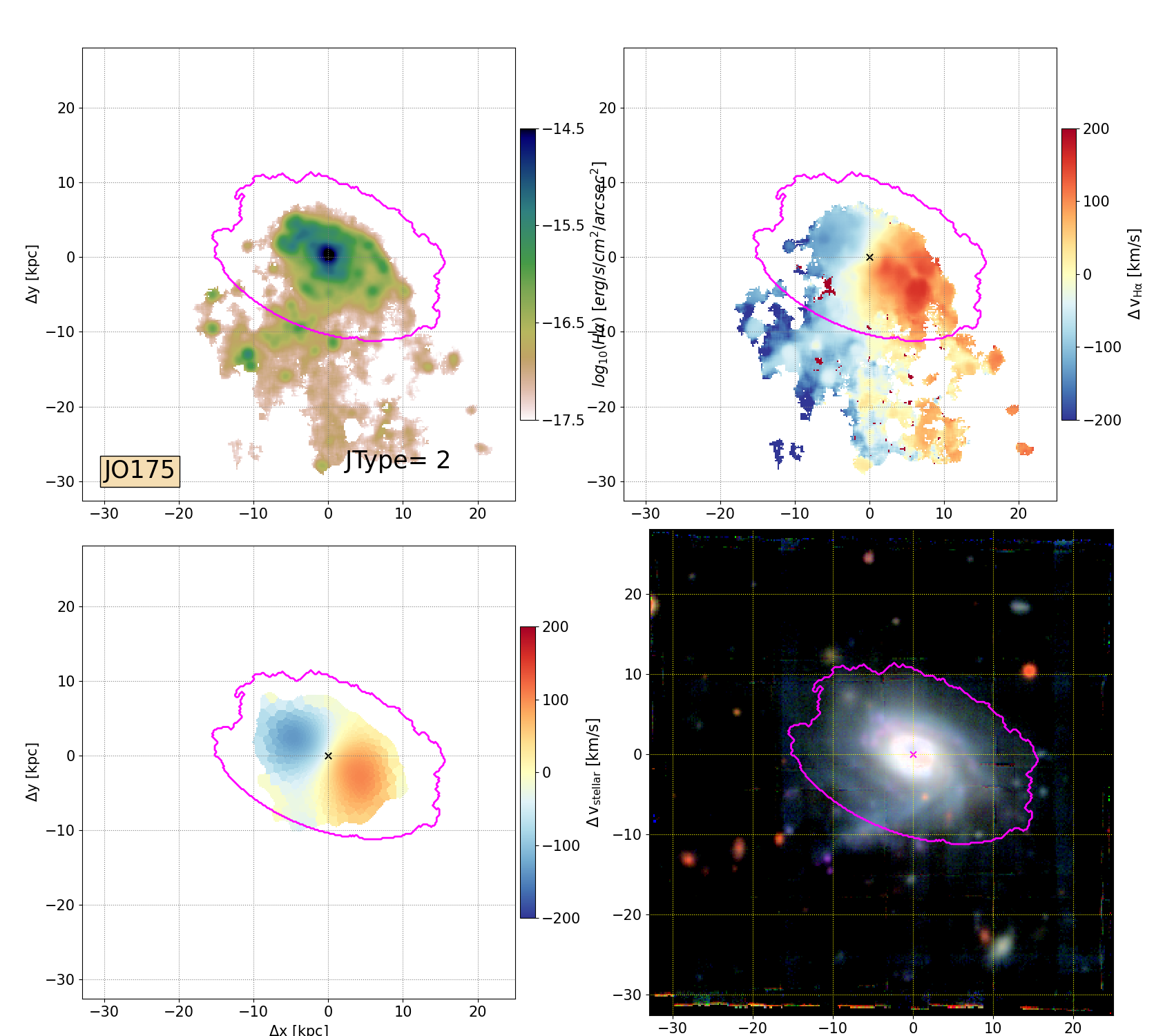}}
\centerline{\includegraphics[scale=0.2]
{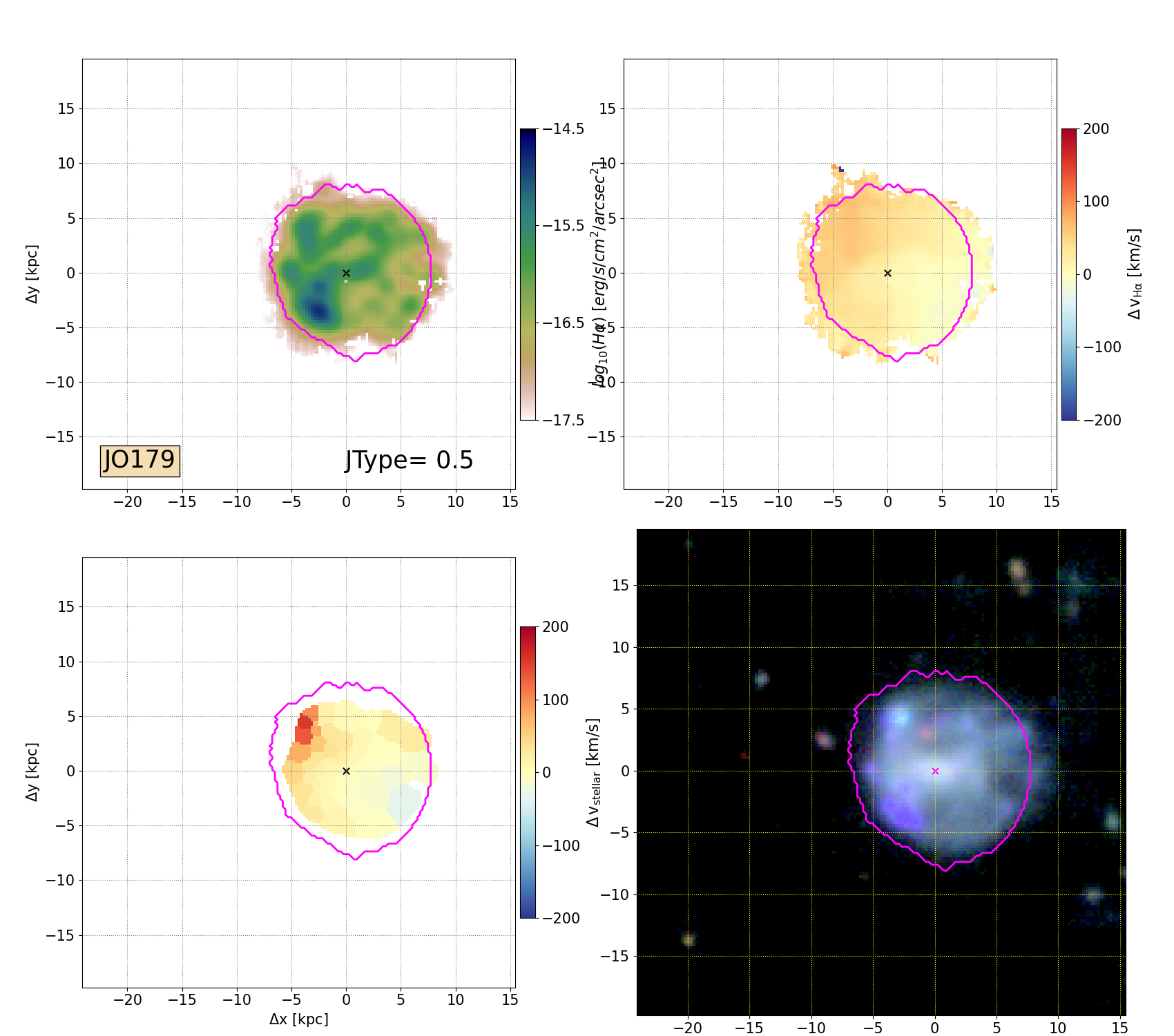}\includegraphics[scale=0.2]{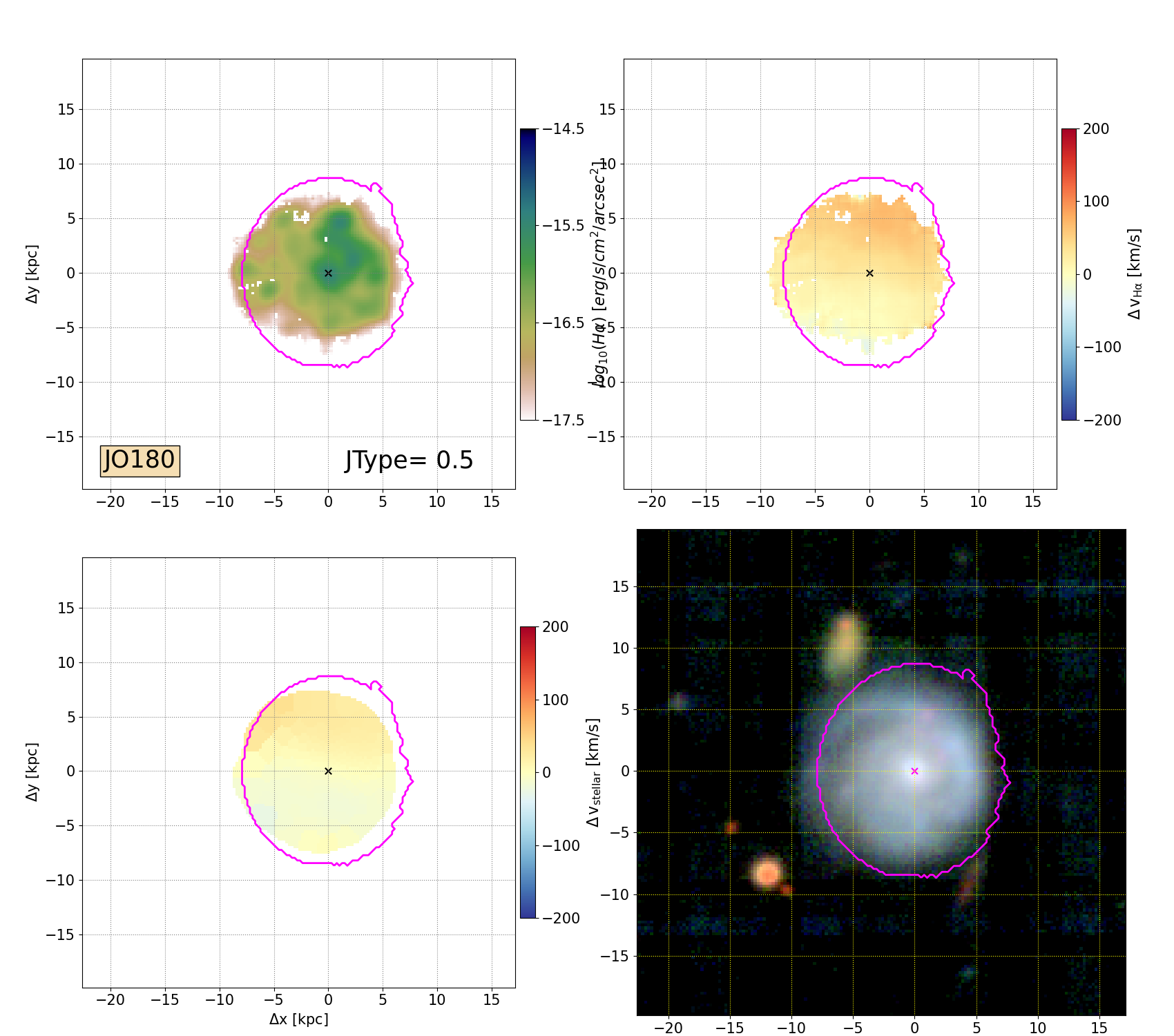}}
\centerline{\includegraphics[scale=0.2]
{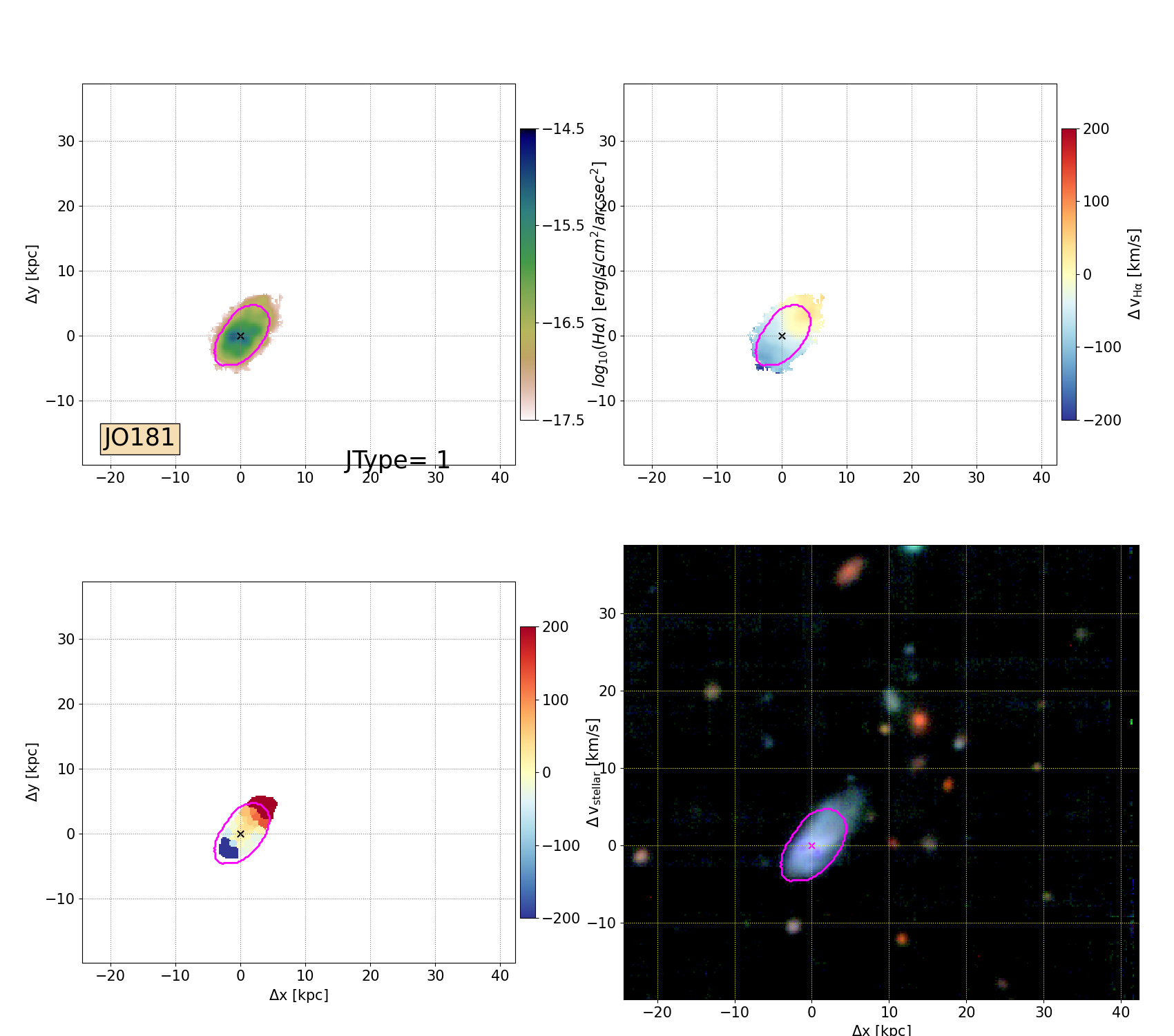}\includegraphics[scale=0.14]{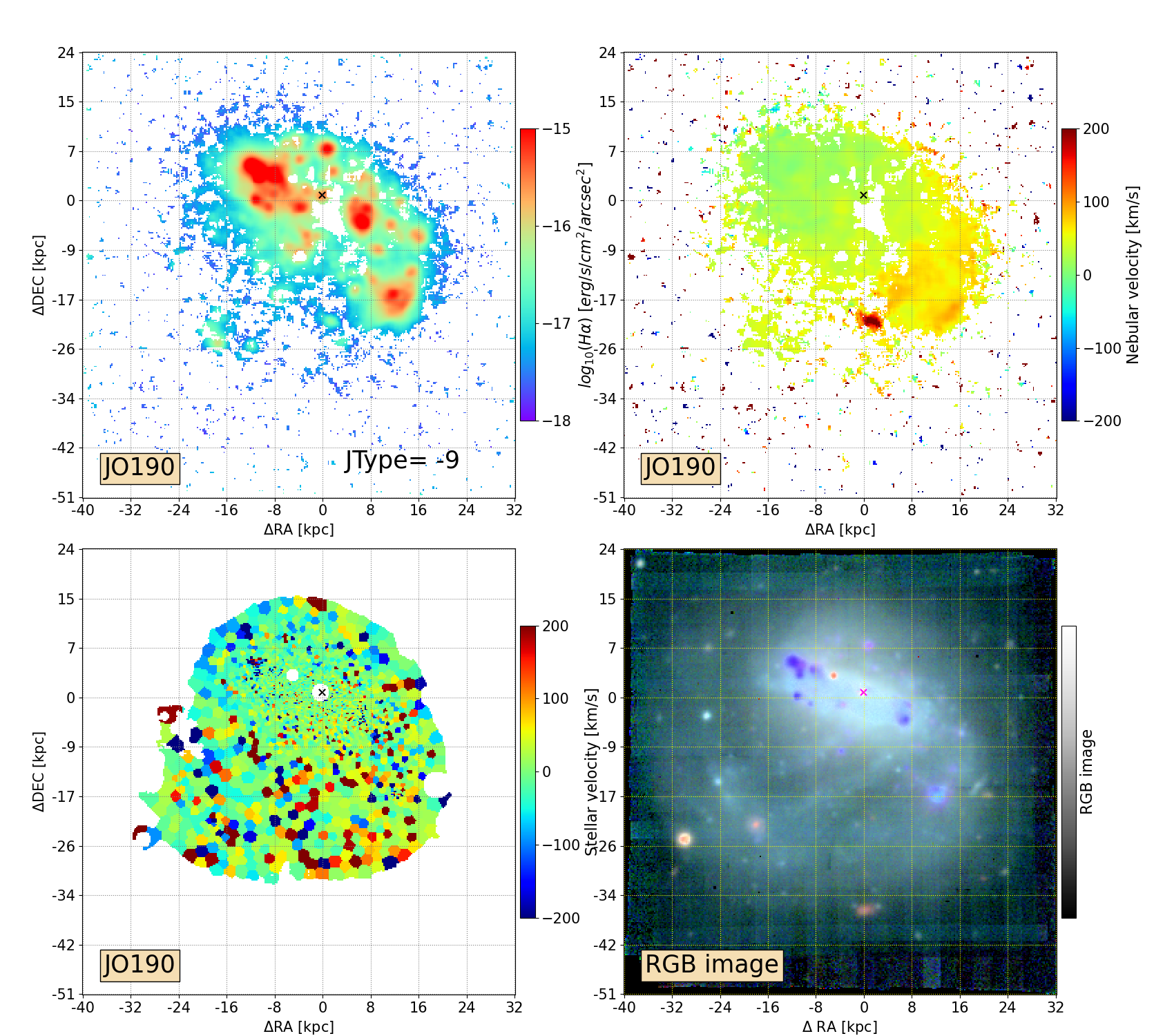}}
\caption{Continued. Stripping candidates.}
\end{figure*}

\addtocounter{figure}{-1}

\begin{figure*}
\centerline{\includegraphics[scale=0.2]
{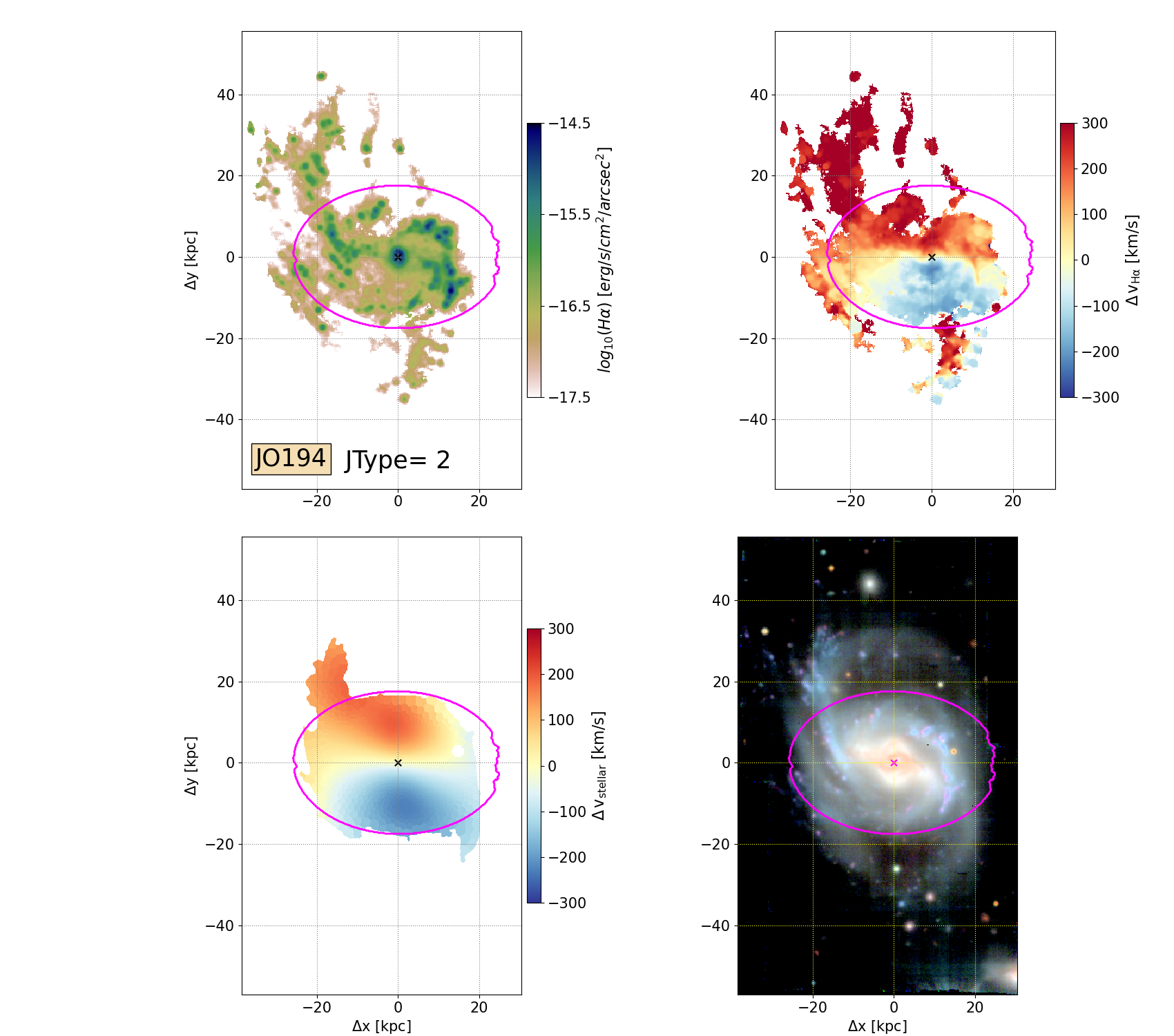}\includegraphics[scale=0.2]{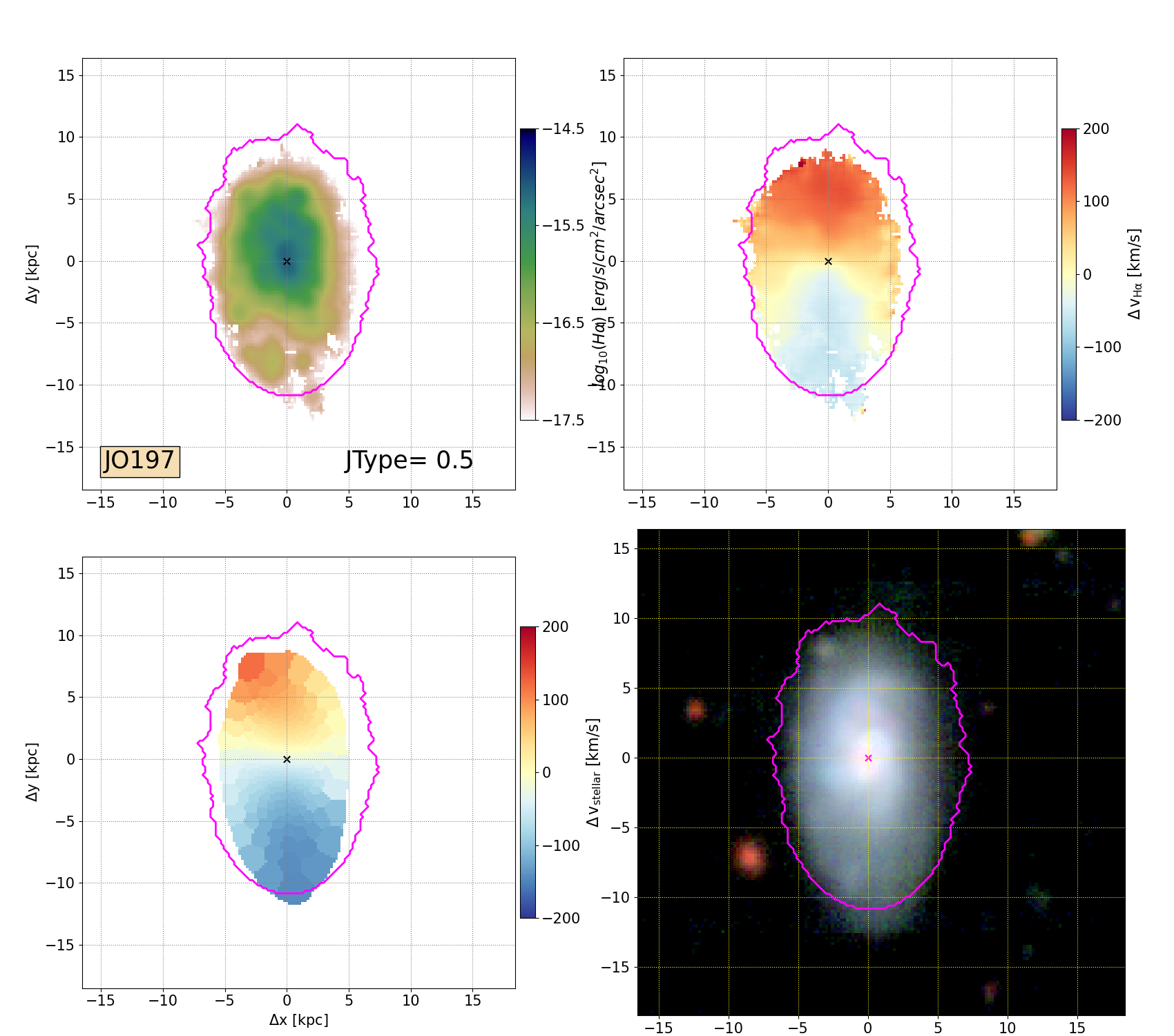}}
\centerline{\includegraphics[scale=0.2]
{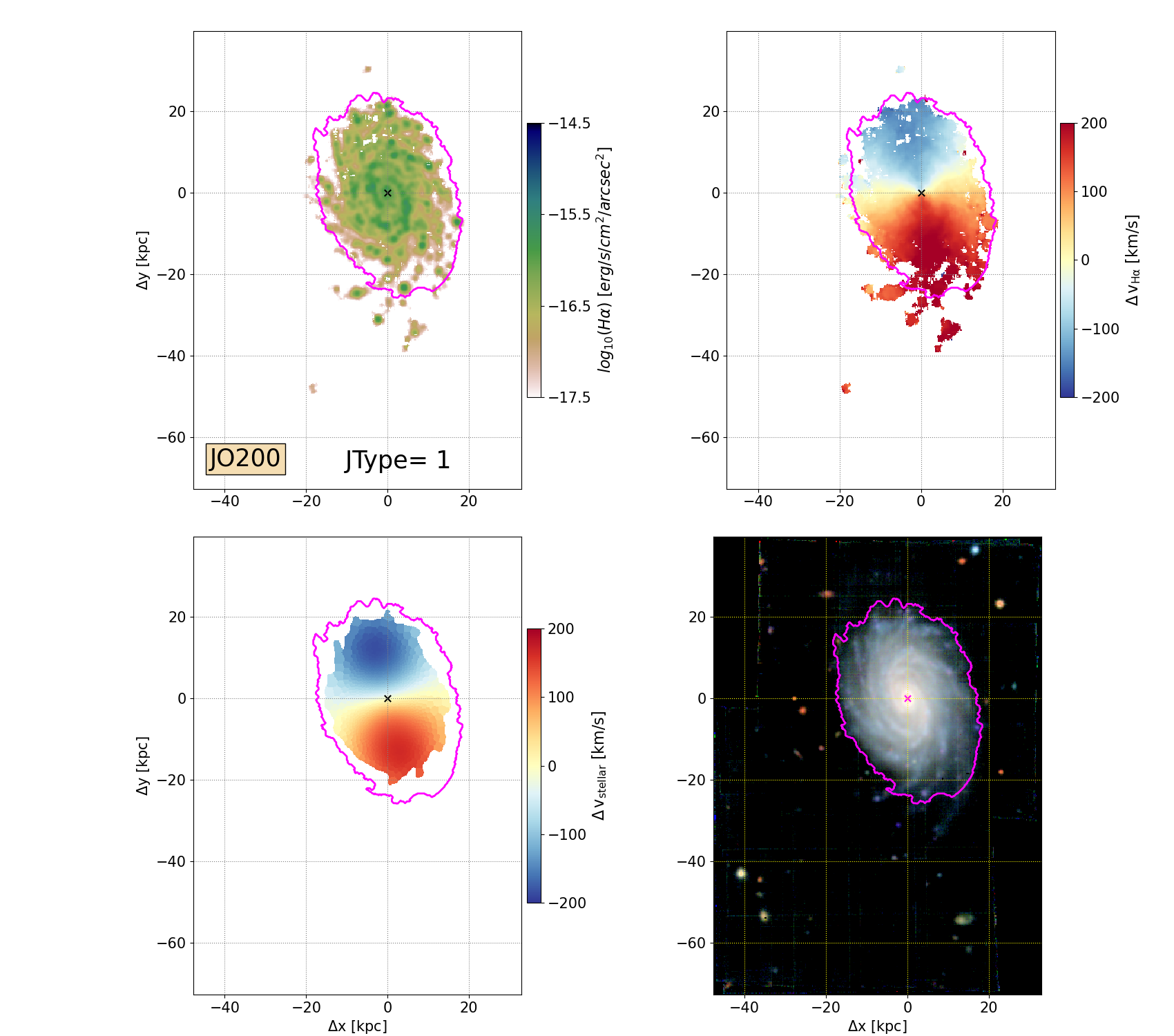}\includegraphics[scale=0.2]{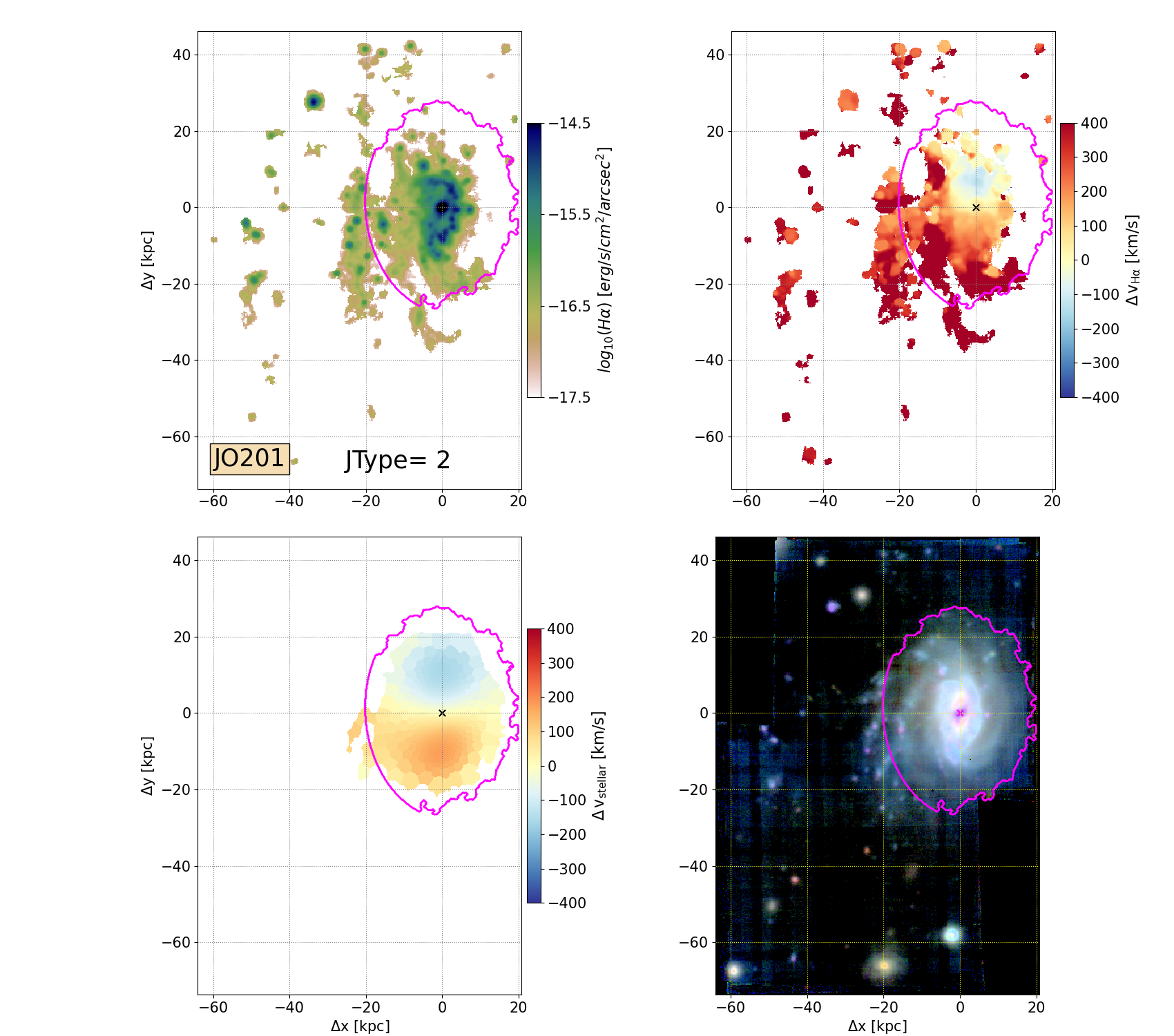}}
\centerline{\includegraphics[scale=0.2]
{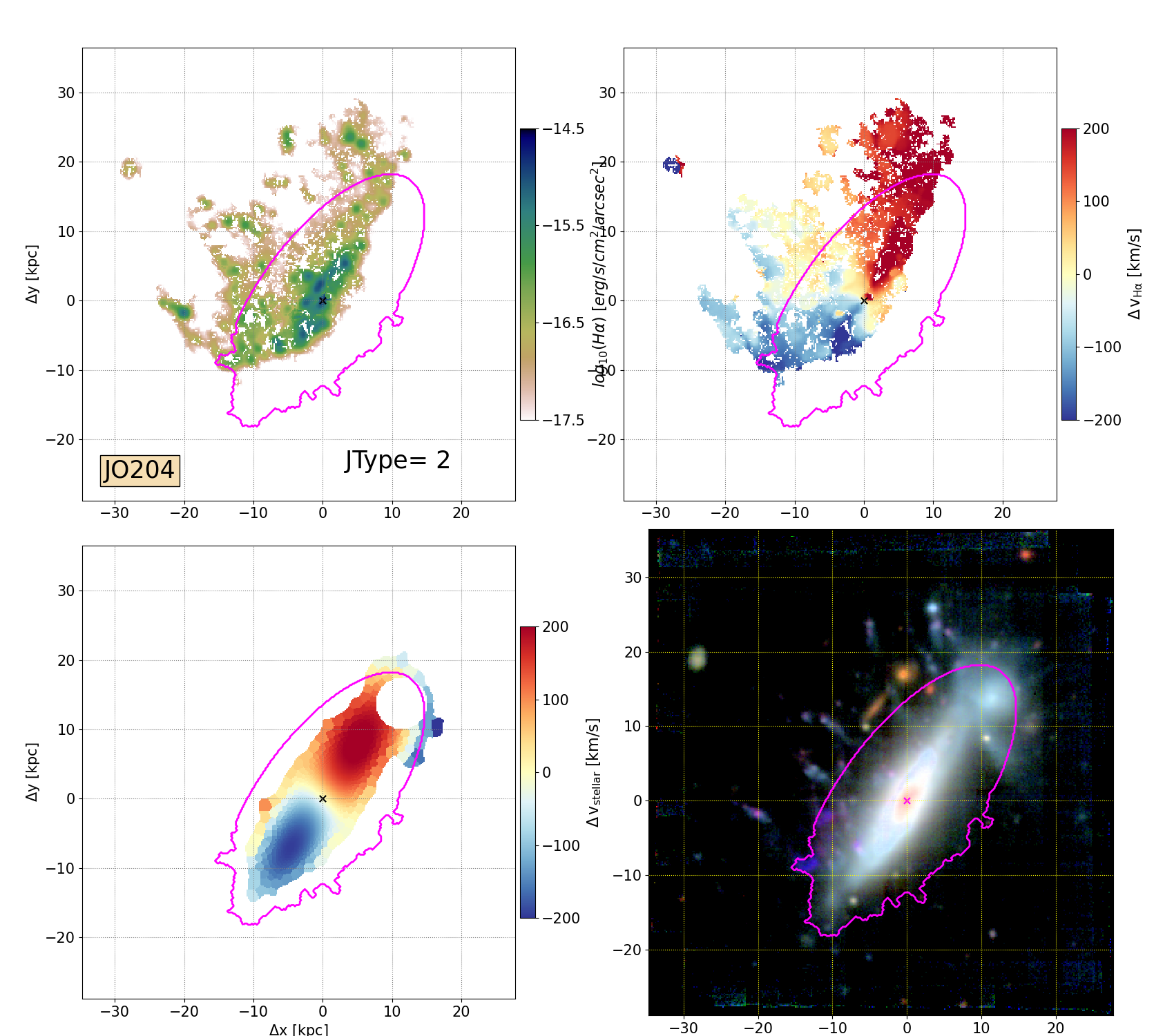}\includegraphics[scale=0.2]{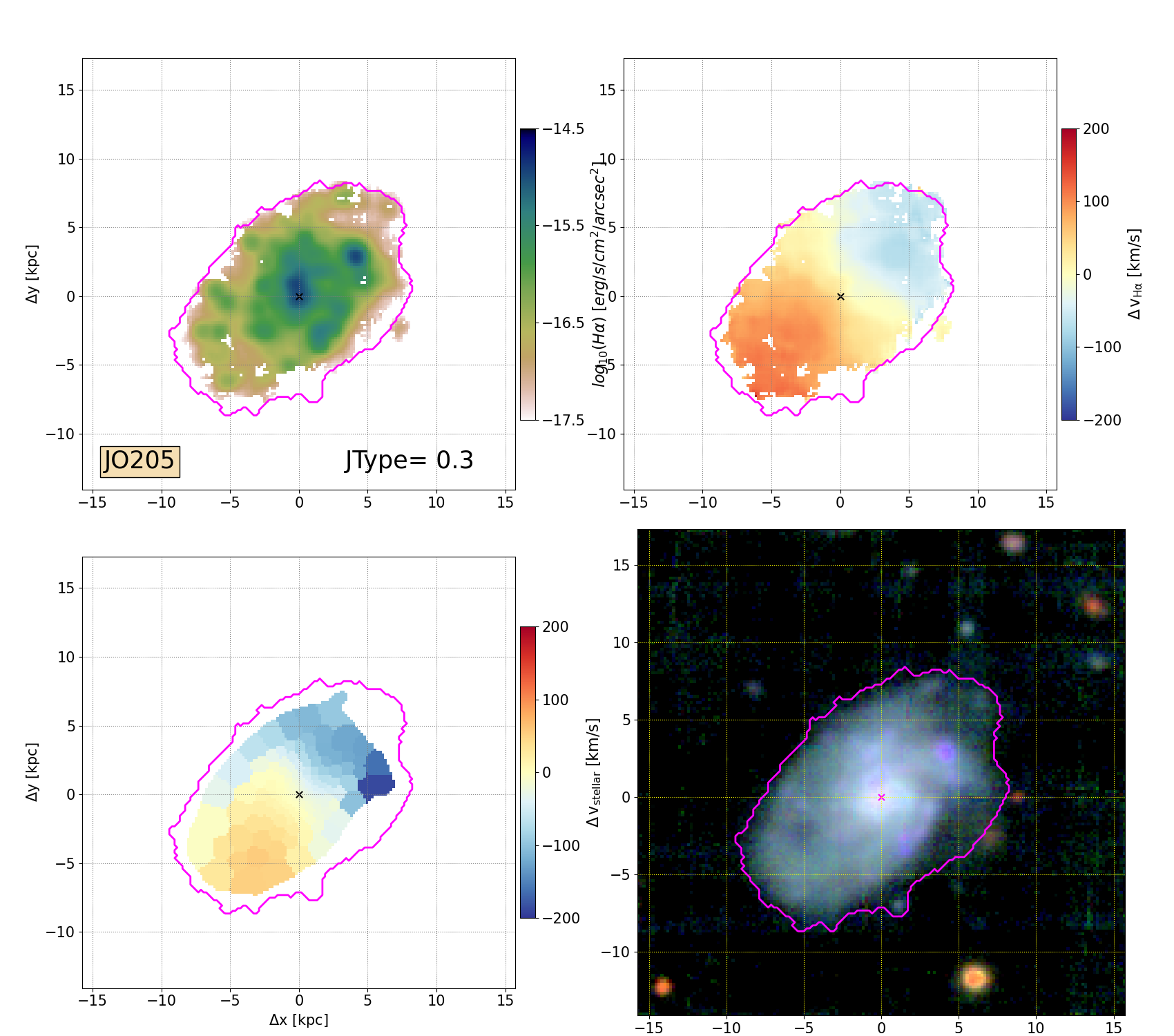}}
\caption{Continued. Stripping candidates.}
\end{figure*}

\addtocounter{figure}{-1}

\begin{figure*}
\centerline{\includegraphics[scale=0.2]
{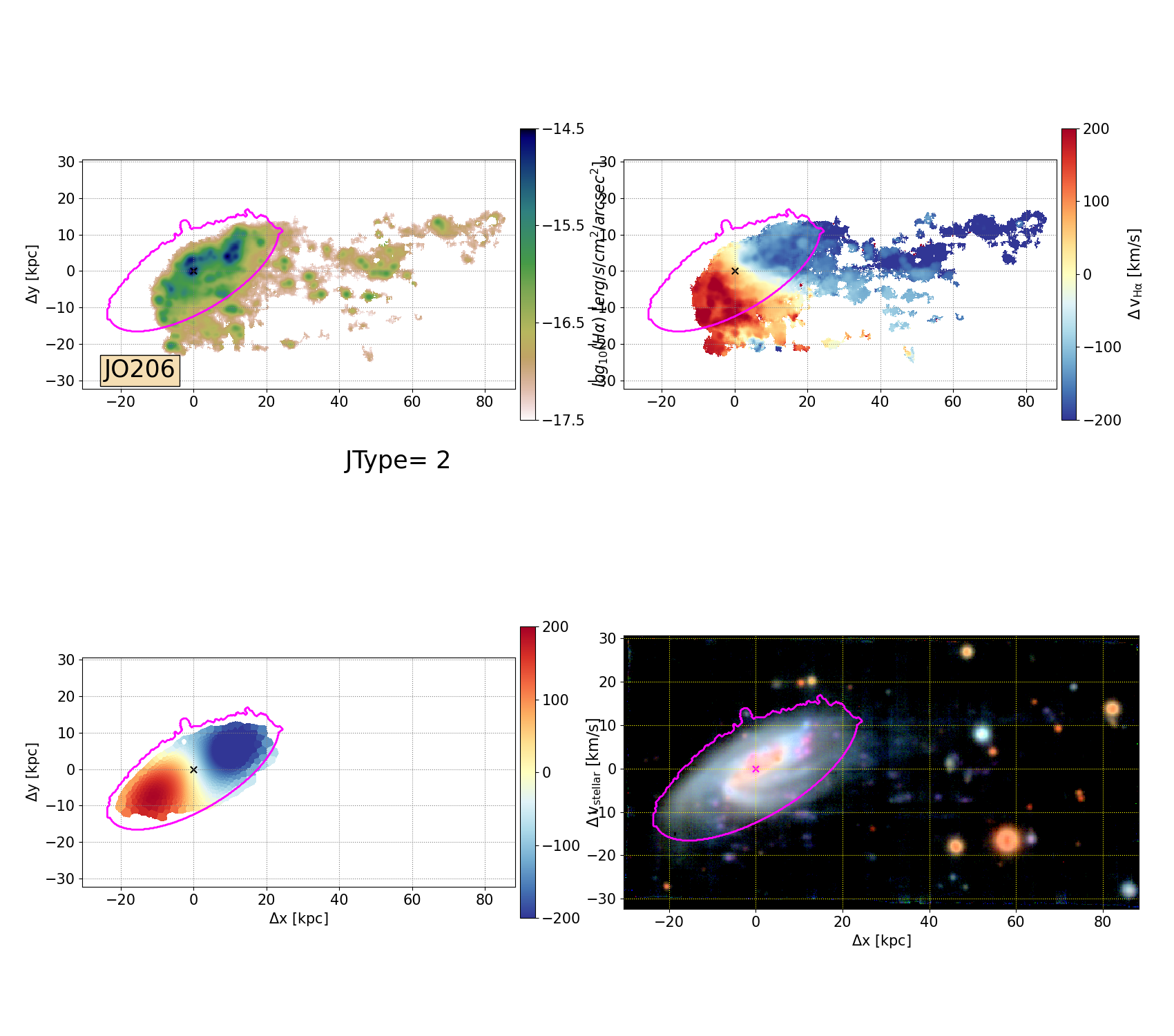}\includegraphics[scale=0.2]{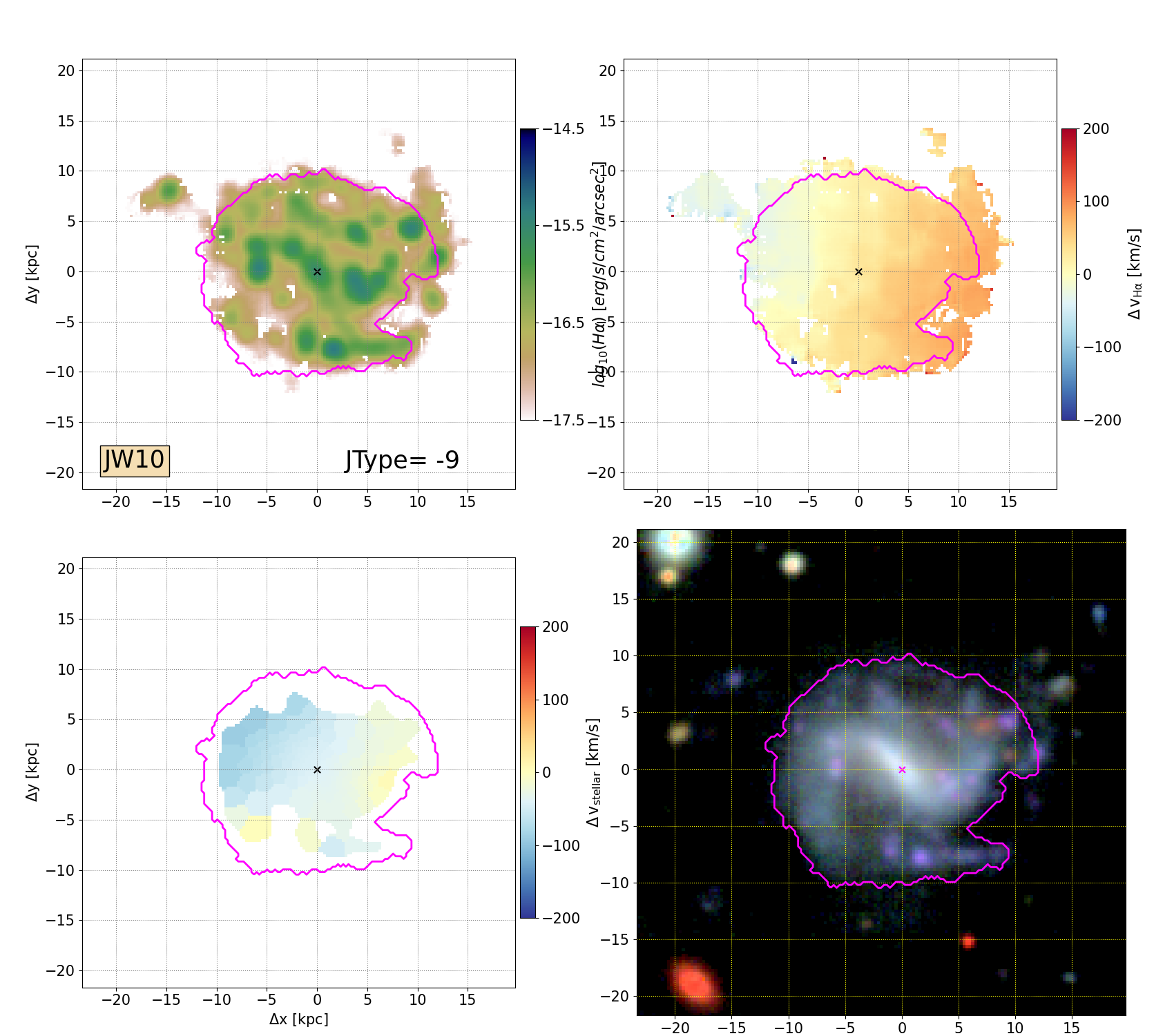}}
\centerline{\includegraphics[scale=0.2]
{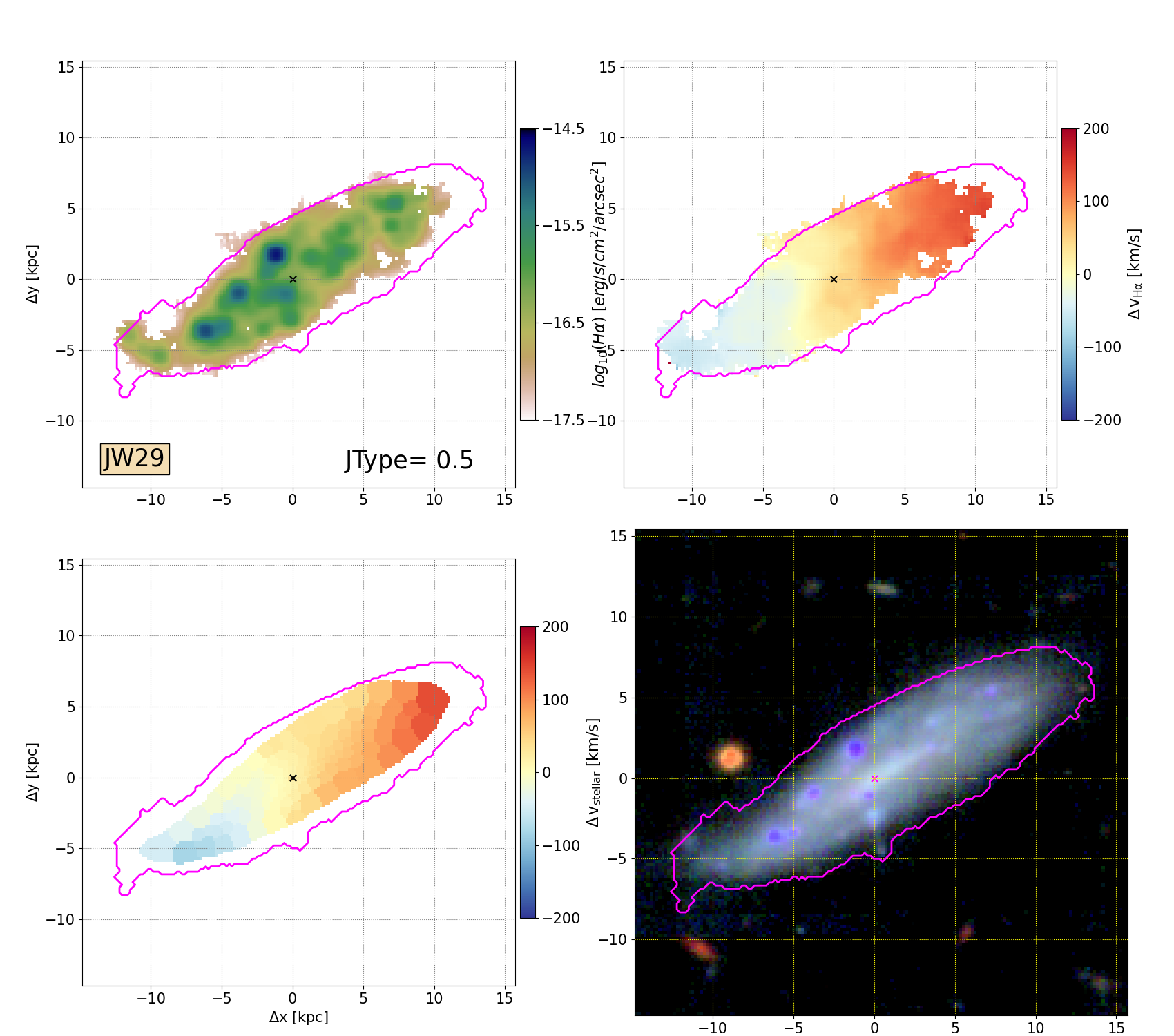}\includegraphics[scale=0.2]{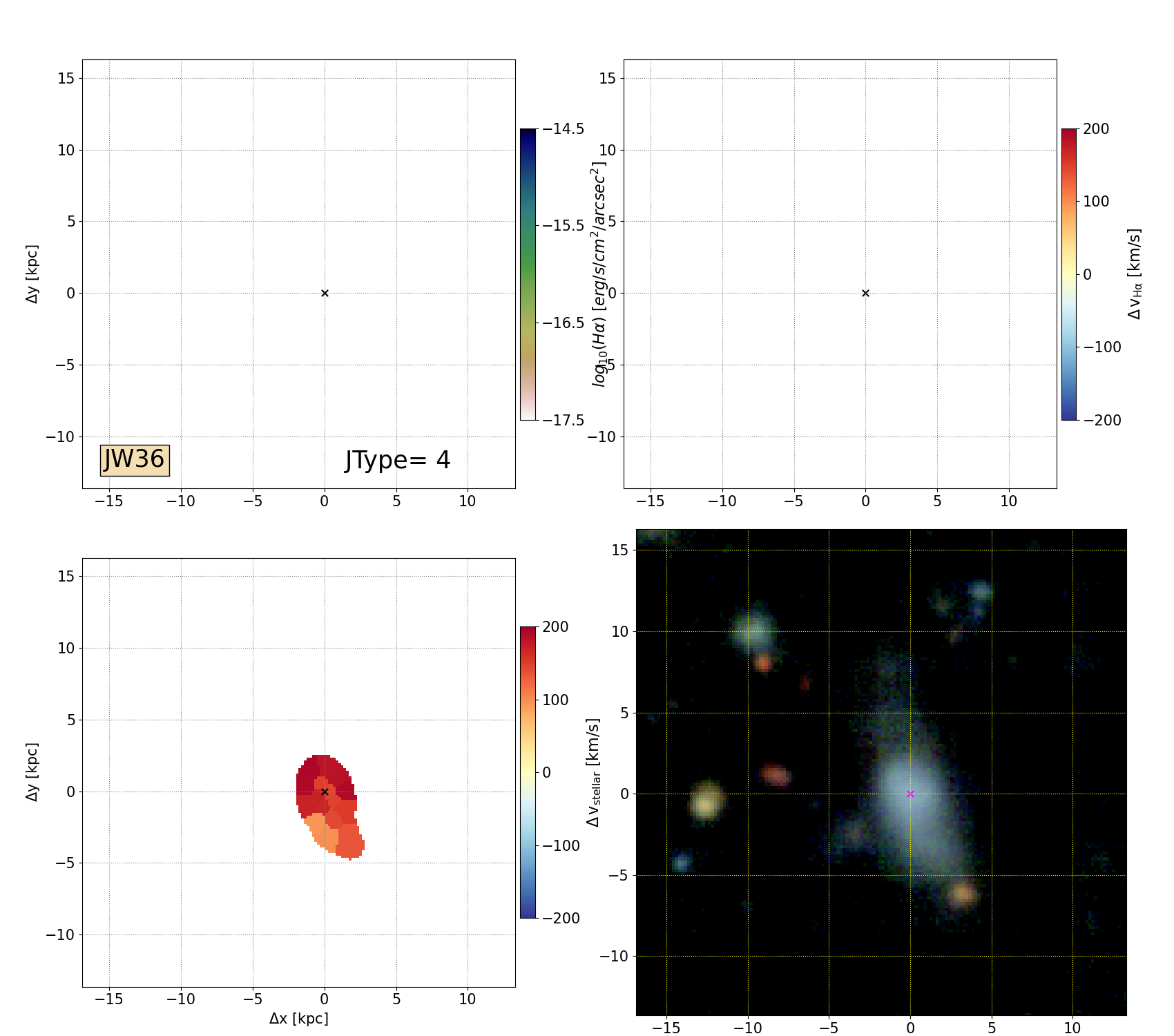}}
\centerline{\includegraphics[scale=0.2]
{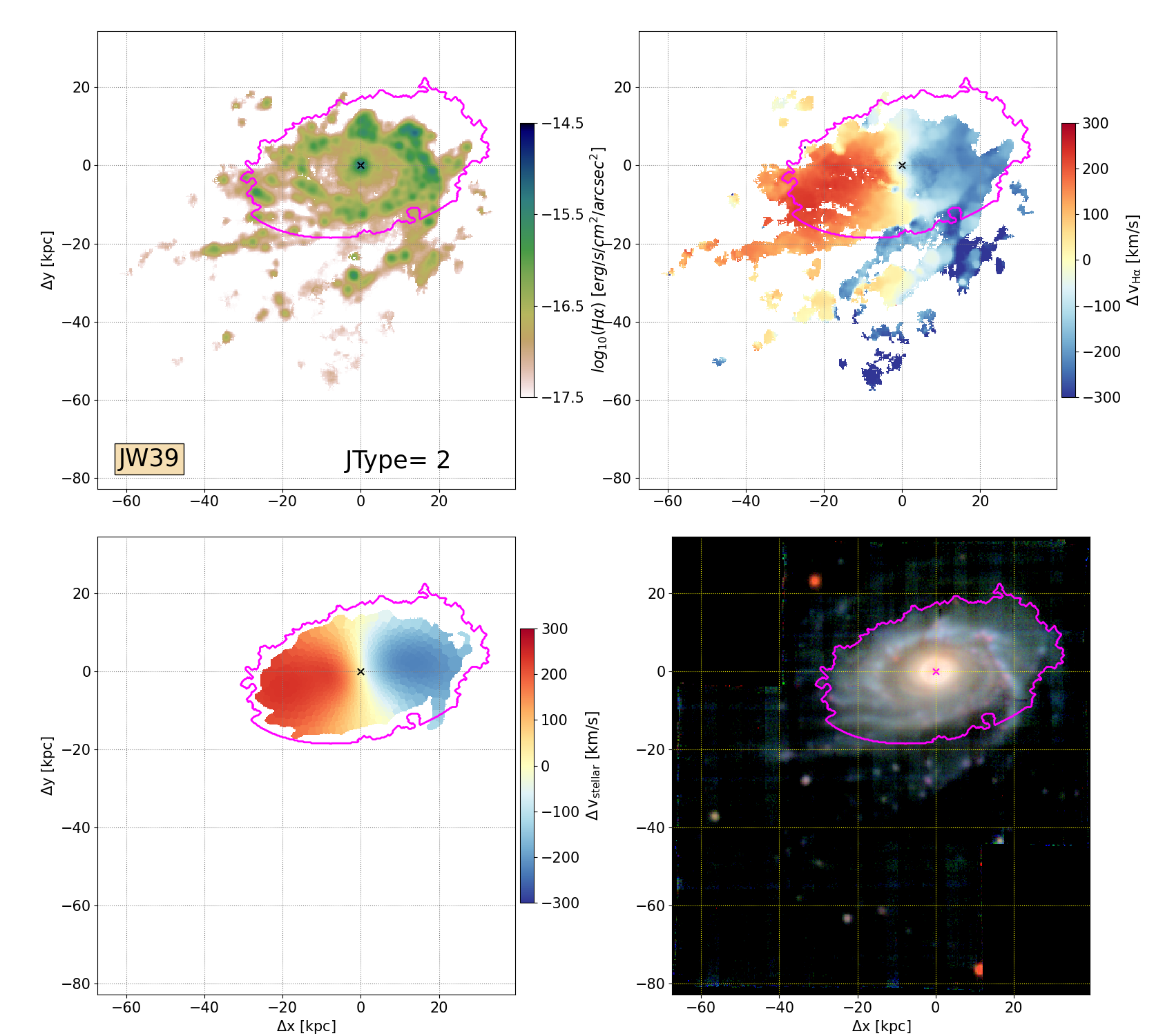}\includegraphics[scale=0.2]{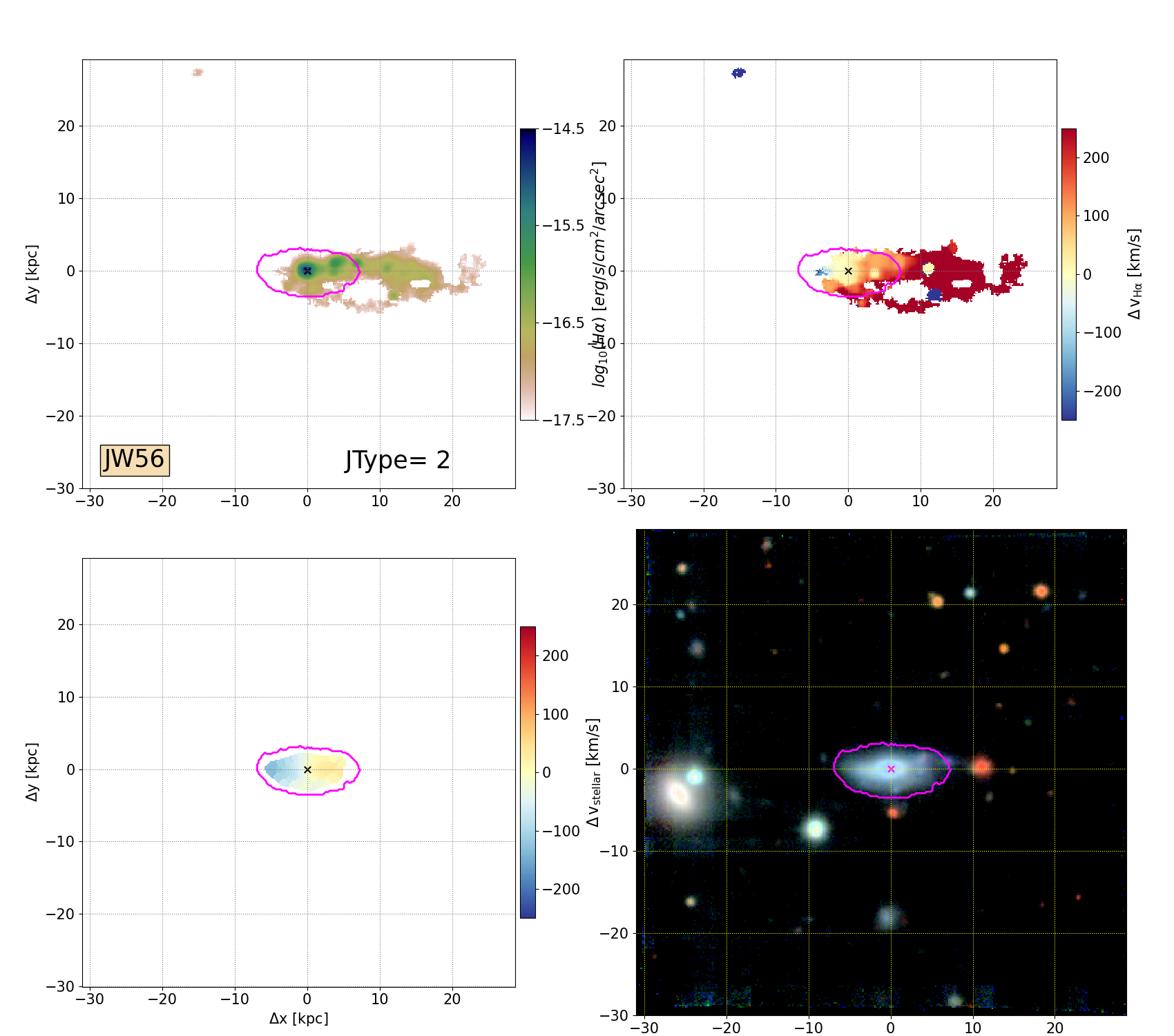}}
\caption{Continued. Stripping candidates.}
\end{figure*}

\addtocounter{figure}{-1}

\begin{figure*}
\centerline{\includegraphics[scale=0.2]
{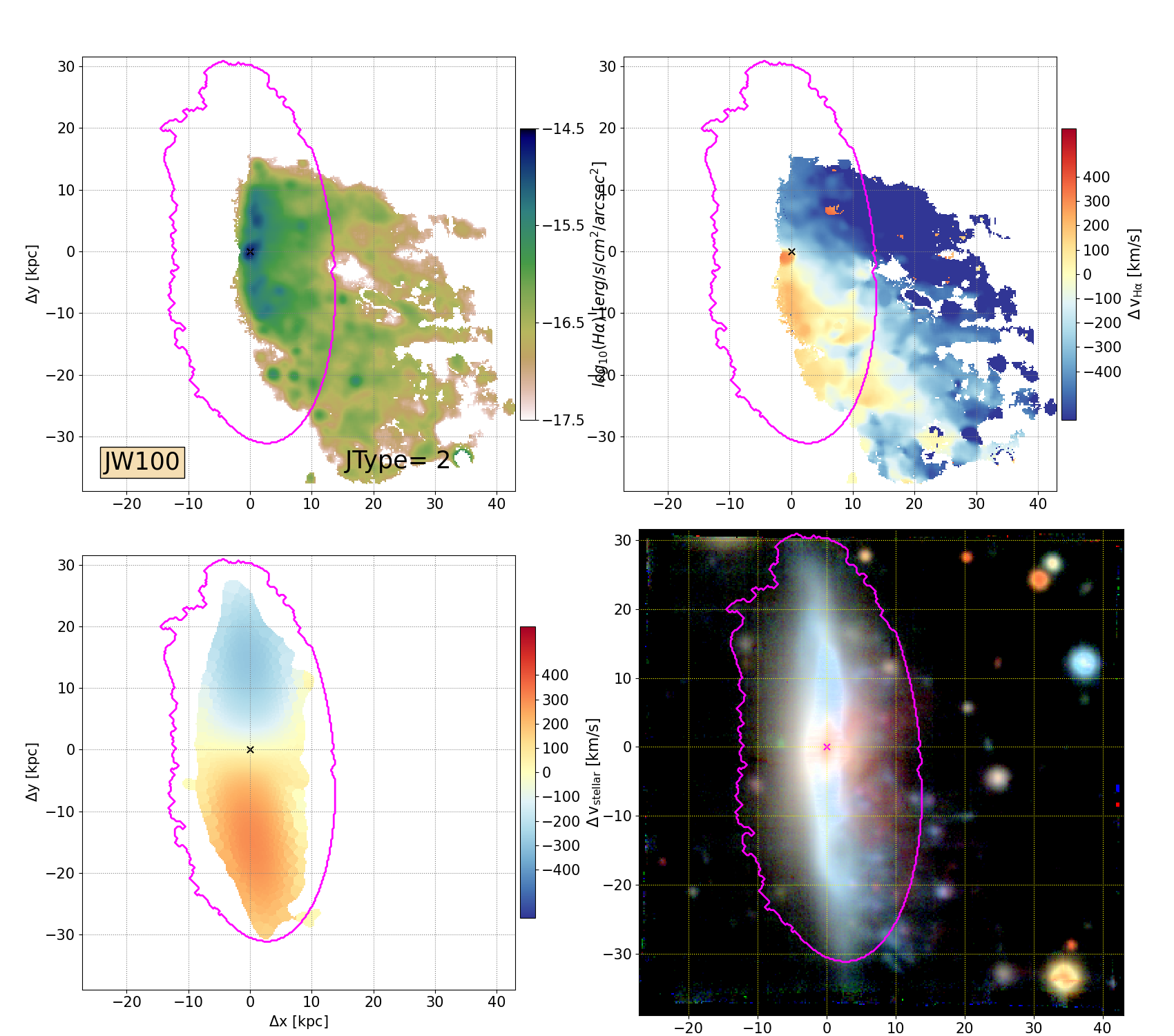}\includegraphics[scale=0.2]{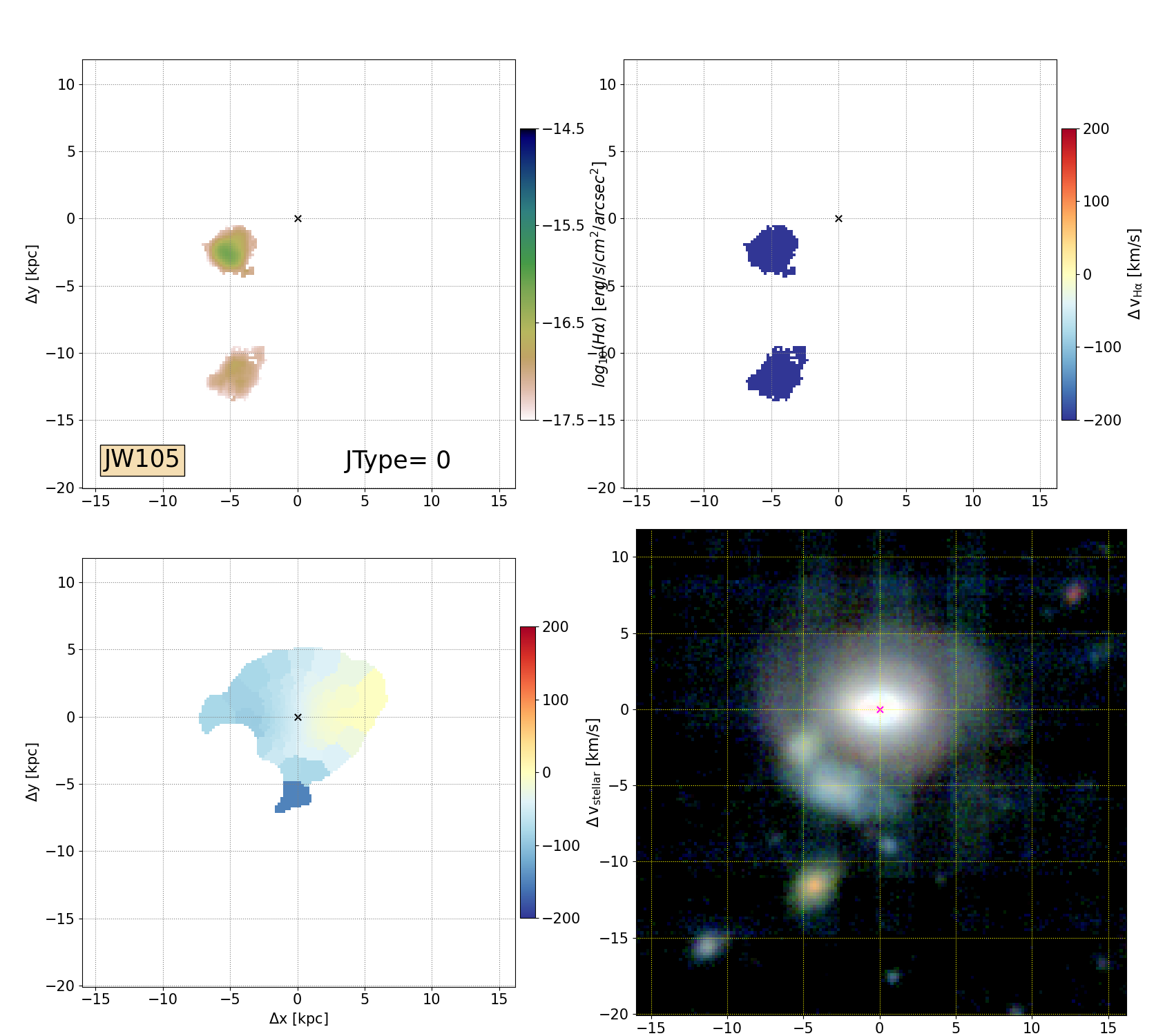}}
\centerline{\includegraphics[scale=0.2]
{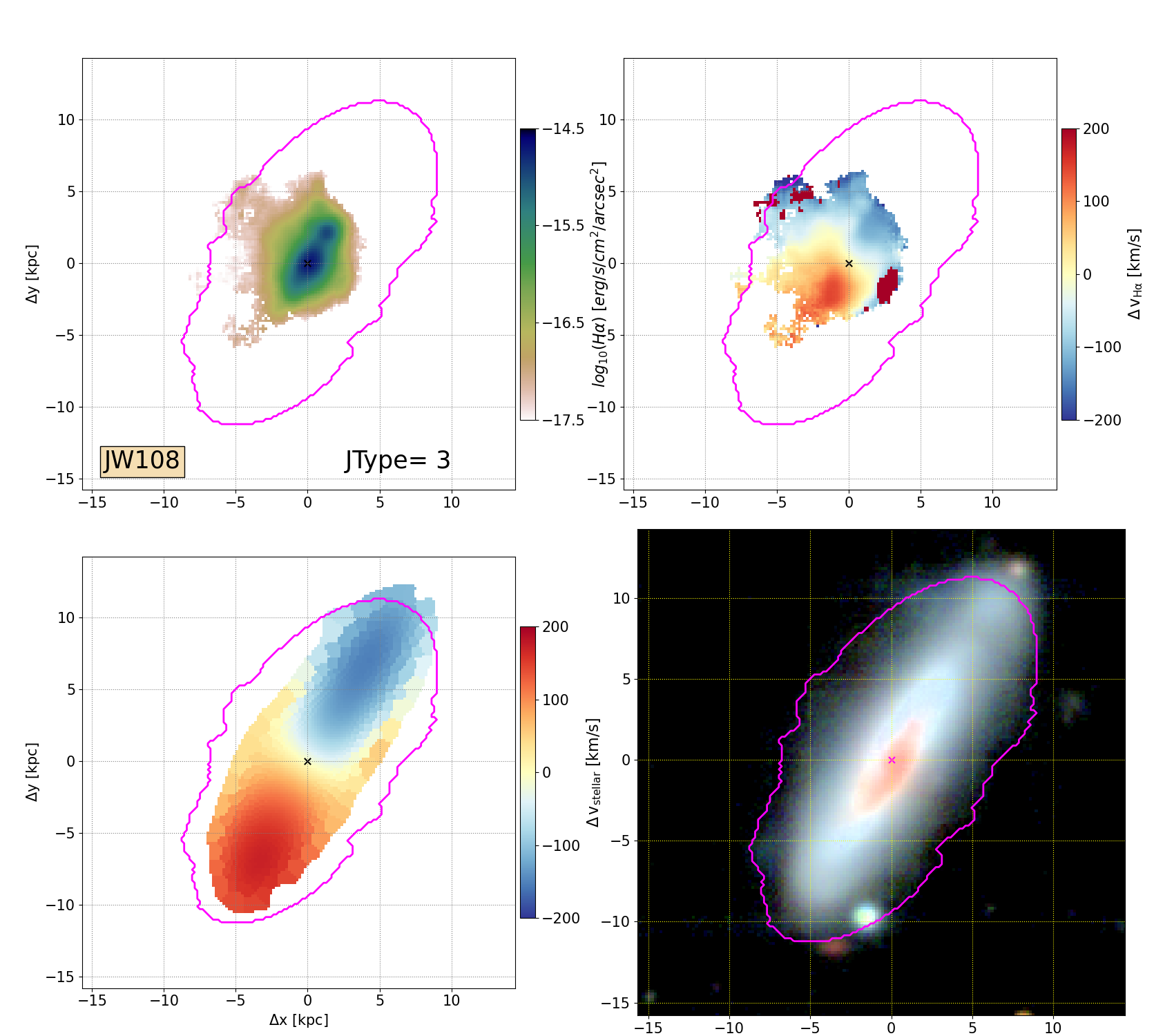}\includegraphics[scale=0.2]{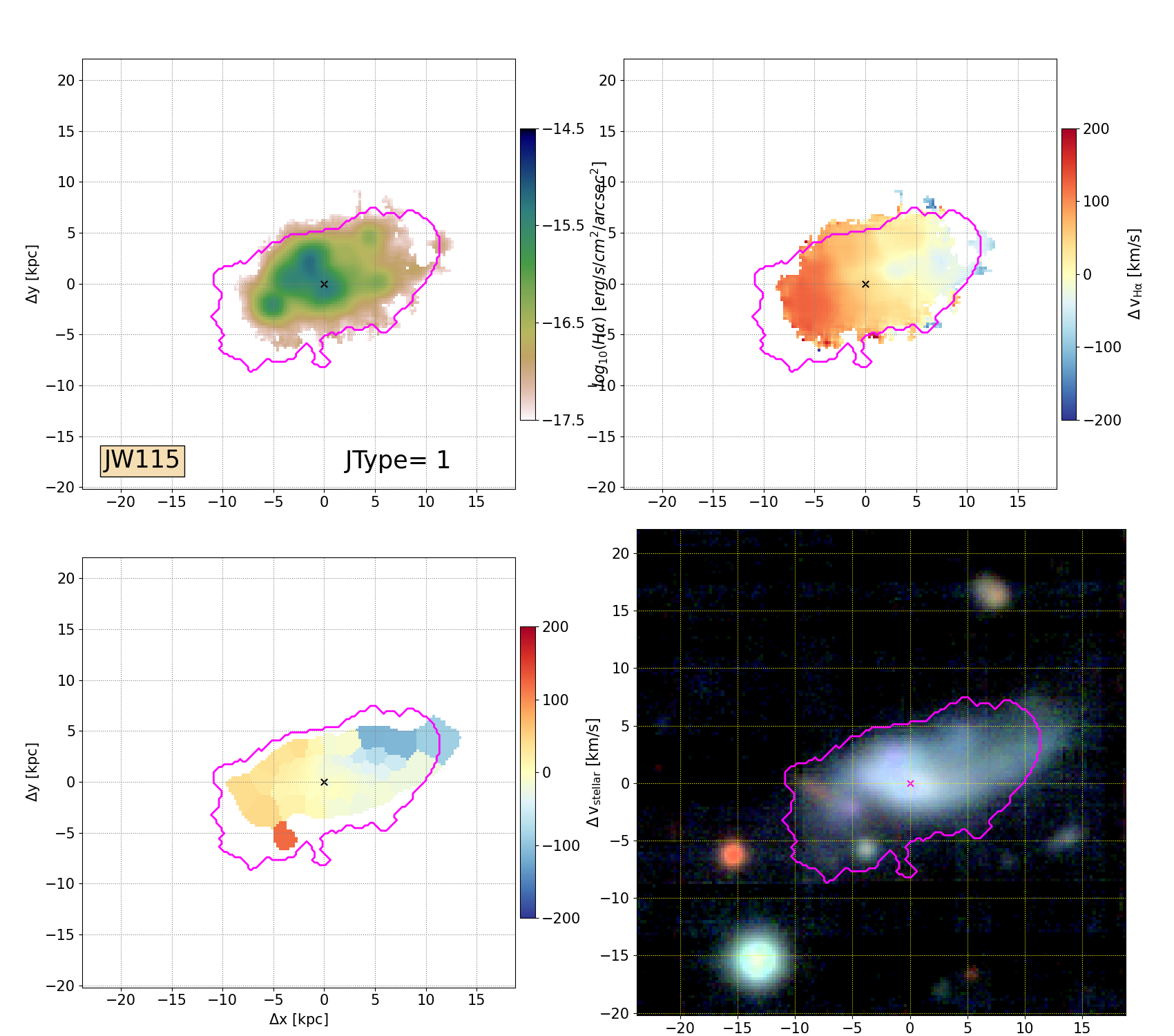}}
\caption{Continued. Stripping candidates. }
\end{figure*}

\end{appendix}

\end{document}